\documentclass[12pt]{article}
\usepackage{slashed}
\usepackage{mathrsfs}
\makeatletter \@addtoreset{equation}{section} \makeatother
\renewcommand{\theequation}{\thesection.\arabic{equation}}
\newcommand{\ba}{\begin{array}}
\newcommand{\ea}{\end{array}}
\newcommand{\beq}{\begin{equation}}
\newcommand{\eeq}{\end{equation}}
\newcommand{\bea}{\begin{eqnarray}}
\newcommand{\eea}{\end{eqnarray}}

\def\bce{\begin{center}}
\def\ece{\end{center}}

\def\nonu{\nonumber}

\def\pa{\partial}
\def\al{\alpha}
\def\be{\beta}
\def\ga{\gamma}
\def\Ga{\Gamma}
\def\de{\delta}

\def\la{\lambda}

\def\si{\sigma}

\def\eps6{{\displaystyle \mathop{\epsilon}^{6}}{}}
\def\g6{{\displaystyle \mathop{g}^{6}}{}}

\def\nab6{{\displaystyle \mathop{\nabla}^{6}}{}}

\def\0{{\sst{(0)}}}
\def\1{{\sst{(1)}}}
\def\2{{\sst{(2)}}}
\def\3{{\sst{(3)}}}
\def\4{{\sst{(4)}}}
\def\5{{\sst{(5)}}}
\def\6{{\sst{(6)}}}
\def\7{{\sst{(7)}}}
\def\8{{\sst{(8)}}}

\def\ba{\begin{array}}
\def\ea{\end{array}}
\def\beq{\begin{equation}}
\def\eeq{\end{equation}}
\def\be{\begin{equation}}
\def\ee{\end{equation}}

\def\la{\lambda}
\def\eps{\epsilon}

\def\ba{\begin{array}}
\def\ea{\end{array}}
\def\beq{\begin{equation}}
\def\eeq{\end{equation}}
\def\be{\begin{equation}}
\def\ee{\end{equation}}

\def\la{\lambda}
\def\eps{\epsilon}

\def\eps6{{\displaystyle \mathop{\epsilon}^{6}}{}}

\def\nab6{{\displaystyle \mathop{\nabla}^{6}}{}}

\newcommand{\bean}{\begin{eqnarray*}}
\newcommand{\eean}{\end{eqnarray*}}

\usepackage{amsmath}

\usepackage{amsmath}
\usepackage{graphicx}
\usepackage{amssymb}
\usepackage{amsthm}
\usepackage{grffile}
\usepackage{simplewick}
\usepackage{tikz}

\usepackage{nccmath}
\usepackage{physics}
\usepackage{wrapfig}
\usepackage{mathtools}
\usepackage{stackengine}

\usetikzlibrary{intersections,decorations.markings}

\allowdisplaybreaks

\begin{document}
\thispagestyle{empty} \addtocounter{page}{-1}
   \begin{flushright}
\end{flushright}

\vspace*{1.3cm}
  
\centerline{ \Large \bf
  A Supersymmetric $w_{1+\infty}$ Symmetry, the Extended Supergravity
}
\vspace*{0.5cm}
\centerline{ \Large \bf
and the Celestial Holography}
\vspace*{1.5cm}
\centerline{ {\bf  Changhyun Ahn$^\dagger$}
  and {\bf Man Hea Kim$^{\ddagger,\ast}$}
} 
\vspace*{1.0cm} 
\centerline{\it 
$\dagger$ Department of Physics, Kyungpook National University, Taegu
41566, Korea}
\vspace*{0.3cm}
\centerline{\it 
  $\ddagger$ Center for High Energy
  Physics, Kyungpook National University, Taegu
41566, Korea} 
\vspace*{0.3cm}
\centerline{\it 
$\ast$
Department of Physics Education,  Sunchon National University,
Sunchon 57922, Korea }
\vspace*{0.8cm} 
\centerline{\tt ahn@knu.ac.kr,
manhea.kim10000@gmail.com
} 
\vskip2cm

\centerline{\bf Abstract}
\vspace*{0.5cm}

We determine
the ${\cal N}=4$ supersymmetric $W_{1+\infty}^{2,2}[\lambda=\frac{1}{4}]$
algebra which is an extension of ${\cal N}=4$ $SO(4)$ superconformal algebra
with vanishing central charge.
We identify the soft current algebra between the graviton, the gravitinos,
the vectors, the Majorana fermions,
the scalar or the pseudoscalar, from the
${\cal N}=4$ supersymmetric
$w_{1+\infty}^{2,2}[\la=\frac{1}{4}]$ algebra,
in two dimensions
with the ${\cal N}=4$ supergravity theory with
$SO(4)$ global symmetry in four dimensions
found by
Das (at Stony Brook in 1977), via celestial holography.
Furthermore, 
the truncations of
${\cal N}=4$ supersymmetric soft  current algebra
provide the soft current algebras
for the ${\cal N}=2,3$ supergravity theories,
the ${\cal N}=2$ supergravity coupled to its
Abelian vector multiplet and the ${\cal N}=1$
supersymmetric Maxwell Einstein theory.
For the ${\cal N}=2$ supergravity theory,
the soft current algebra can be also realized from
the ${\cal N}=2$ supersymmetric
$w_{1+\infty}^{K,K}[\la=0]$ algebra.

\vspace*{2cm}

\baselineskip=18pt
\newpage
\renewcommand{\theequation}
{\arabic{section}\mbox{.}\arabic{equation}}

\tableofcontents

\section{ Introduction
  and Summary
}

In two dimensional conformal field theory,
the Virasoro algebra  \cite{Virasoro} consisting of spin $2$
generator (See also
\cite{FV1} for the central extension of Virasoro algebra)
plays an important role.
Its supersymmetric version, the ${\cal N}=1$ superconformal algebra,
by introducing the fermionic spin $\frac{3}{2}$ generator in addition to
the bosonic one of spin $2$,
can be obtained from \cite{NS,Ramond} (See also
\cite{Schwarz} for the central extension of this algebra).
Other supersymmetric version, the
${\cal N}=2$ superconformal algebra can be determined by
adding one spin $1$ and two spin $\frac{3}{2}$
as well as one spin $2$
\cite{Ademolloplb} (See \cite{Ademollonpb} for the central extension
of this algebra).
Other supersymmetric version, the 
${\cal N}=3$ superconformal algebra can be described by
considering three spin $\frac{3}{2}$, three spin $1$ and
one spin $\frac{1}{2}$ with one spin $2$ \cite{Ademolloplb}
(See \cite{RS} for the central extension of this algebra).
Other supersymmetric version,
the `small' ${\cal N}=4$ superconformal algebra
can be realized by 
introducing
four spin $\frac{3}{2}$ and three spin $1$
with one spin $2$ \cite{Ademollonpb}.
Finally,
other supersymmetric version,
the `large' ${\cal N}=4$ superconformal algebra
by
adding four spin $\frac{3}{2}$, six spin $1$, 
four spin $\frac{1}{2}$ and one spin $0$
with one spin $2$ is found in \cite{Schoutens,STV}.


The $w_{\infty}$ algebra is a particular generalization of the
Virasoro algebra with generators of spin $2, 3, \cdots, \infty$
and the $w_{1+\infty}$ algebra has
the extra spin $1$ generator in addition to the above
generators of spin \cite{Sezgin9202}.
The area preserving diffeomorphism \cite{BPS}
of a two torus (one sphere times
one sphere)
has been found in \cite{FI} where the
Virasoro generators can be obtained from the linear combination
of the infinite number of generators of area
preserving diffeomorphism of a two torus (See
\cite{ADFI} for the central extension).
In \cite{Bakas}, the Virasoro algebra can be embedded
in the area
preserving diffeomorphism of a two plane and
all the other generators of the algebra take a
representation of the Virasoro algebra and
they can be viewed as generators of spins $3,4, \cdots, \infty$.
In \cite{PRS},
there exists more convenient choice of basis to
describe the algebra of area
preserving diffeomorphism of a cylinder (one sphere times a real axis)
and the $w_{1+\infty}$ algebra appears explicitly \footnote{
Moreover, in \cite{FFZ}, the algebra having the trigonometric
structure constants is found and this leads to the one in \cite{FI}
by taking the proper limits for the two arbitrary parameters.
On the other hand, the $w_{1+\infty}$ algebra
occurs in the quantum Hall effect \cite{GMP}
in the three dimensions
\cite{CTZ,CTZplb,IKS,CM}.}.
The ${\cal N}=1$ supersymmetric extension of
$w_{1+\infty}$ algebra can be found in \cite{Sezgin1989,PS}
which is an algebra of the symplectic diffeomorphism of
a superplane with two bosonic and one fermionic dimension.
The ${\cal N}=2$ supersymmetric extension of
$w_{1+\infty}$ algebra can be obtained in \cite{PS,Sezgin9112}
which is an algebra of the symplectic diffeomorphism of
a superplane with two bosonic and two fermionic dimensions.
The similar ${\cal N}=2$ construction
has been done in \cite{FN,FL} along the line of \cite{FFZ}. 
See also the work of \cite{FN1} in the context of \cite{FFZ}.


After the
simple supergravity containing
a single gravitino (${\cal N}=1$) in four dimensions
has been found in \cite{FvF},
there were most productive results during $1977$ and $1978$. 
Among them, the ${\cal N}=4$ supergravity
having the global $SO(4)$ symmetry has been
studied by 
Das (at Stony Brook in $1977$)  \cite{Das} (See also \cite{CS1,CS2}
for the ${\cal N}=4$ supergravity
with the global $SO(4)$ symmetry and \cite{CSF}
for the  ${\cal N}=4$ supergravity
with the global $SU(4)$ symmetry).
The massless particles are given by
one spin $2$, four spin $\frac{3}{2}$,
six spin $1$, four spin $\frac{1}{2}$ and
one spin $0$ parity doublet.
The Lagrangian in the second order formalism
has thirteen terms (where the six terms are
kinetic terms) up to the linear term in the gravitational
coupling constant $\kappa$.
By taking
one spin $\frac{3}{2}$, three spin $1$,
three spin $\frac{1}{2}$
and one spin $0$ to
be zero (consistent truncations
in the sense that the variation of these fields
also vanish), the ${\cal N}=3$ supergravity \cite{FSZ,Freedman}
with the global $SO(3)$ symmetry is reproduced.
The particle contents are given by
one spin $2$, three spin $\frac{3}{2}$,
three spin $1$ and one spin $\frac{1}{2}$.
The Lagrangian is reduced to seven terms (where the four terms are
kinetic terms)
up to
the linear $\kappa$.
Similarly,
by putting
two spin $\frac{3}{2}$, four spin $1$ and 
two spin $\frac{1}{2}$
to
be zero,
the ${\cal N}=2$ supergravity  \cite{Fv}
with the global $SO(2)$ symmetry
(where the particles are given by
one spin $2$, two spin $\frac{3}{2}$ and
one spin $1$)
coupled to its Abelian vector multiplet
consisting of
one spin $1$, two spin $\frac{1}{2}$
and one spin $0$ parity doublet
is obtained. Moreover, 
the ${\cal N}=2$ supergravity 
coupled to its several Abelian vector multiplets
has been constructed in \cite{Luciani}. 
When  
one spin $\frac{3}{2}$, two spin $1$ and one spin $\frac{1}{2}$
from the above ${\cal N}=3$ supergravity are restricted to
be zero (in the context of consistent truncation), then 
the ${\cal N}=2$ supergravity with the global $SO(2)$
symmetry is reproduced and the particle contents are
given by one spin $2$, two spin $\frac{3}{2}$ and one spin $1$.
The Lagrangian has five terms (including the three kinetic terms)
up to the linear $\kappa$.
On the other hand,
by taking   
two spin $\frac{3}{2}$ and two spin $1$
from the above ${\cal N}=3$ supergravity to
be zero, then the ${\cal N}=1$ (or simple) supergravity coupled to
the vector multiplet having one spin $1$ and one spin $\frac{1}{2}$
is obtained \cite{FSv}.
There are still different five terms (including the four kinetic
terms) in the Lagrangian
\footnote{
In the ${\cal N}=8$ supergravity theory with
the $SO(8)$ global symmetry \cite{dF},
the particle contents are
one spin $2$, eight spin $\frac{3}{2}$, twenty eight spin $1$,
fifty six spin $\frac{1}{2}$ and seventy spin $0$.
The Lagrangian contains the six kinetic terms and the fourteen interaction
terms up to the linear $\kappa$.
In particular, the cubic term in the spin $\frac{1}{2}$,
the spin $\frac{1}{2}$ and the spin $1$ appears newly
compared to the ${\cal N}=4$ supergravity \cite{Das}.}.


The gravitational scattering amplitudes
in the 
asymptotically flat four dimensional spacetimes and the conformal
field theory living on the two dimensional celestial sphere
are connected by the celestial holography
\cite{Strominger,Raclariu,Pasterski,PPR,Donnay}.
A soft current algebra
between the gravitons and the gluons in the Einstein Yang-Mills theory
is obtained in \cite{Strominger2105}.
It turns out that
the wedge subalgebra of $w_{1+\infty}$ algebra \cite{Bakas}
appears.
The relevant work can be found in
\cite{GHPS} where the previous works on
\cite{FFT,PRSY} were used \footnote{
By considering the fermionic partners,
gravitinos and gluinos (as well as the above bosonic ones) in the
${\cal N}=1$
supersymmetric Einstein Yang-Mills theory, the operator product
expansions(OPEs)
of
gravitinos and gluinos with gravitons and gluons
are determined in \cite{FSTZ}, along the lines of \cite{FFT,PRSY}.
By generalizing the works of \cite{Strominger2105,GHPS},
the corresponding ${\cal N}=1$ supersymmetric soft current algebra
between the above soft particles
is obtained in \cite{Ahn2111,Ahn2202} explicitly.
In particular, the anticommutators between the fermionic
operators are vanishing.}.
On the other hand, in \cite{Prabhu},
note that the ${\cal N}=1$ superconformal algebra
\cite{Schwarz,FQS,BKT}
where there is a nontrivial anticommutator
for the modes of the fermionic current (as well as
the usual two commutators)
is reproduced from the Lie superalgebra based on the BMS symmetries
\cite{BvM,Sachs}. It is obvious to ask how we can construct
the supersymmetric soft current algebra where
the anticommutators between the fermionic
operators do not  vanish \footnote{
\label{n2walgebra}
  In \cite{AK2309}, by considering the deformation parameter $\la$,
the generalization of \cite{PRS} is studied.
The bosonic subalgebras are given by
$W^K_{1+\infty}[\la] \times W^K_{1+\infty}[\la+\frac{1}{2}]$.
The first factor is realized by $K$ values of
$(b,c)$ fermions and the
second factor is realized by $K$ values of $(\beta,\gamma)$ bosons.
Note that each number of $K$ appears in the superscript. 
Compared to the previous $w_{1+\infty}$ algebra,
the $W_{1+\infty}$ algebra contains all the possible terms
with the nonzero parameter on the
right hand side of the (anti)commutators.
By taking the zero limit of this parameter,
the latter becomes the former.
The above deformation parameter $\la$ arises in the weights
of above fields nontrivially. 
By construction, there are several anticommutators between the
modes for the fermionic currents \cite{AK2309}.
In particular, for $K=2$, the ${\cal N}=2$ supersymmetry 
is enhanced to ${\cal N}=4$ supersymmetry.}.
The asymptotic symmetry algebra of ${\cal N}=8$ supergravity
theory in four dimensions is studied in \cite{BRS1,BRS2}. 
There should appear the soft current algebra in two dimensions.

The ${\cal N}=2$ supersymmetric $W_{1+\infty}^{K,K}[\la]$ algebra,
where its bosonic subalgebra is given by
the footnote \ref{n2walgebra},
is realized by the multiple
$(b, c)$ and $(\beta, \gamma)$ system and the corresponding
${\cal N}=2$ supersymmetric $w_{1+\infty}^{K,K}[\la]$ algebra
can be determined by taking the zero limit for the nonzero parameter.
As a first step \cite{AK2407},
we can
identify the soft current algebra between the graviton, the gravitino,
the photon (the gluon), the photino (the gluino),
the scalar or the pseudoscalar, equivalent to
${\cal N}=1$ supersymmetric
$w_{1+\infty}^{K,K}[\la]$ algebra (the
${\cal N}=2$ supersymmetry is reduced to the ${\cal N}=1$
one) for generic $\la$,
in two dimensions
with the ${\cal N}=1$ supergravity theory in four dimensions
\cite{FvF,DZ}
and its matter coupled theories \cite{FSv,FFvBGS},
via celestial holography.


As a next step on this direction,
in this paper,
we obtain
the ${\cal N}=4$ supersymmetric $W_{1+\infty}^{2,2}[\lambda=\frac{1}{4}]$
algebra for fixed $\la$ which is the
extension of ${\cal N}=4$ $SO(4)$ superconformal algebra
with vanishing central charge studied by \cite{Schoutens}.
The soft current algebra between the graviton, the gravitinos,
the vectors, the Majorana fermions,
the scalar or the pseudoscalar, from the
${\cal N}=4$ supersymmetric
$w_{1+\infty}^{2,2}[\la=\frac{1}{4}]$ algebra (the previous number
$K$ is fixed by $K=2$ due to the ${\cal N}=4$ supersymmetry),
in two dimensions
is identified with the ${\cal N}=4$ supergravity theory with
$SO(4)$ global symmetry in four dimensions
found by \cite{Das}, via celestial holography.

Note that the parameter $\la$ is fixed by
$\la =\frac{1}{4}$.
The $\la$ dependence, which takes the form
$(1-4\la)$, appears at the two central terms
in the realization of ${\cal N}=4$ $SO(4)$ superconformal algebra.
Moreover,
the anticommutator between the spin $\frac{3}{2}$ generator
and itself and the commutator between the spin $\frac{3}{2}$
generator and the spin $1$ generator
contain this $\la$ dependence on the right hand side explicitly.
The former has $SO(4)$ epsilon tensor in front of
spin $1$ generator while the latter
has the Kronecker delta in front of spin $\frac{1}{2}$
generator on the right hand sides.
Of course,
the former has $\la$ independent factor in front of
spin $1$ generator while the latter
has the $\la$ independent factor with
$SO(4)$ epsilon tensor
in front of spin $\frac{1}{2}$
generator on the right hand sides.
See also (\ref{realn4sca}).

By looking at the
interaction between the gravitinos, the gravitinos and the vectors
of \cite{Das}, corresponding to the above
anticommutator, the two $SO(4)$ indices for the two gravitinos
are contracted with the ones for the vectors.
Similarly, for the interaction
between the vectors, the gravitinos and the Majoranas
of \cite{Das}, corresponding to the above commutator,
the two $SO(4)$ indices for the vectors and the gravitinos
are contracted with the ones for the Majoranas together with the presence of
$SO(4)$ epsilon tensor.
These two interactions correspond to
the $\la$ independent terms on the right hand sides of 
the above anticommutator and the commutator.
Therefore, it is necessary to put the above $\la$ dependent coefficients
$(1-4\la)$ should vanish  because there are {\it no } such interactions
of the ${\cal N}=4$ supergravity theory.
Therefore, the parameter $\la$ should be equal to $\la=\frac{1}{4}$.

It is known that the 
tree level ${\cal N}=8$ supergravity splitting amplitude
is given by the product of two tree level ${\cal N}=4$ super
Yang-Mills theory 
splitting amplitudes multiplied by
the angle bracket, the square bracket
between the two collinear particles and minus sign \cite{BDPR}.
Similarly,
it is also known that the 
tree level ${\cal N}=4$ supergravity splitting amplitude
is given by the product of one tree level ${\cal N}=4$ super
Yang-Mills theory 
splitting amplitude and
one tree level ${\cal N}=0$ super
Yang-Mills theory (or nonsupersymmetric Yang-Mills theory) 
splitting amplitude
multiplied by
the angle bracket, the square bracket
between the two collinear particles and minus sign \cite{BBJ}.
The tree level ${\cal N}=5,6$ supergravity theories
splitting amplitudes can be described by
taking the second tree level ${\cal N}=1,2$ super
Yang-Mills theory splitting amplitudes respectively
for the first common tree level ${\cal N}=4$ super
Yang-Mills theory 
splitting amplitude.

The point is that the helicities of two particles
in the ${\cal N}=8$ (or ${\cal N}=4$)
supergravity theory splitting amplitude
are given by the sum of the helicities of each particle
in the ${\cal N}=4$ super
Yang-Mills theory 
splitting amplitudes
respectively (For the ${\cal N}=4$ supergravity splitting amplitude,
we should take the second
tree level nonsupersymmetric Yang-Mills theory 
splitting amplitude as before).
Moreover, the third helicity
of the ${\cal N}=8$ (or ${\cal N}=4$)
supergravity theory splitting amplitude
is given by the sum of the third helicity of each particle
in the ${\cal N}=4$ super
Yang-Mills theory 
splitting amplitudes (For the ${\cal N}=4$ supergravity
splitting amplitude,
the second
tree level nonsupersymmetric Yang-Mills theory 
splitting amplitude is taken).
See also \cite{BRS2}.
The helicities for the gluons, the gluinos and the scalars
are given by $(\pm 1, \pm \frac{1}{2}, 0)$
in the ${\cal N}=4$ super Yang-Mills theory and
the helicities for the gluons are given by $\pm 1$
in the nonsupersymmetric Yang-Mills theory.
By linear combinations of these two sets of
helicities, we are left with the helicities $(\pm 2,
\pm \frac{3}{2}, \pm 1, \pm \frac{1}{2}, 0)$
corresponding to the helicities of gravitons,
gravitinos, vectors, Majoranas and scalars in the ${\cal N}=4$
supergravity theory whose massless particles are the sames as
the ones for ${\cal N}=8$ supergravity theory with different
multiplicities.

In \cite{Strominger2105}, the soft current algebra
between the gravitons and the gluons
having all the positive helicities in the Einstein Yang-Mills theory
is found.
In \cite{FSTZ},
the soft current algebra
between the gravitons, the gravitinos, the gluons
and the gluinos
having all the positive helicities in the
${\cal N}=1$ supersymmetric Einstein Yang-Mills theory
is described.
It is known in \cite{BE} that
the three point vertex between
two gravitinos and one graviton
in the ${\cal N}=1$ supergravity theory
\cite{FvF}
is the product of the three point vertex between
two gluinos and one gluon
in the ${\cal N}=1$ super Yang-Mills theory
and the three point vertex between the three gluons
in the nonsupersymmetric Yang-Mills theory.
Moreover, the polarization tensors of the gravitinos
are given by the product of the ones of
gluinos with helicities $\pm \frac{1}{2}$ from the first factor
and
the ones of the gluons with helicities $\pm 1$ from the second factor.
It turns out that
the three point amplitude between
two gravitinos with opposite helicities $\pm \frac{3}{2}$ and one graviton
with positive helicity $+2$ \cite{CDRR}
in the ${\cal N}=1$ supergravity theory
arises for the scaling dimension of three point vertex
$d_V=5$.
Therefore, the OPEs between the two collinear particles
having the positive, the negative and the zero helicities
will provide the soft symmetries organized from the
${\cal N}=4$ supersymmetric
$w_{1+\infty}^{2,2}[\la =\frac{1}{4}]$ algebra.

It is known that
the conformal dimension (or weight) $\Delta$
for the soft graviton contains all the conformally soft poles
$\Delta=1, 0, -1, \cdots$
appearing in the Euler beta function of the OPE between the
gravitons \cite{GHPS,Strominger}.
In particular, the $\Delta=1$ generates the supertranslation
(corresponding to the leading soft graviton),
the $\Delta=0$ generates the superrotation
(corresponding to the subleading soft graviton)
and
the $\Delta=-1$ is related to the subsubleading
soft graviton.
The $\Delta$
for the soft gravitino contains all the (conformally) soft poles
$\Delta=\frac{1}{2}, -\frac{1}{2}, \cdots$
appearing in the Euler beta function of the OPE between the
gravitinos in the ${\cal N}=8$ (or ${\cal N}=4$)
supergravity \cite{Tropper,BRS2} \footnote{Why does the ${\cal N}=4$
supergravity theory provide also the same behavior for the
Euler beta function?
The reason for this is that the ${\cal N}=8$ $SO(8)$ supergravity
theory \cite{dF} contains all the $SO({\cal N})$ extended
supergravity theories with ${\cal N} < 8$ and the
OPEs from the collinear singularities of the amplitudes
in the ${\cal N}=8$ $SO(8)$ supergravity should contain
the ones in the ${\cal N}=4$ supergravity.}.
In particular, the $\Delta=\frac{1}{2}$ generates
the soft symmetry corresponding to
the leading soft gravitinos
and the $\Delta=-\frac{1}{2}$ generates
the soft symmetry corresponding to 
the subleading soft gravitinos \cite{Tropper,BRS2} \footnote{The
supersymmetric soft theorem satisfies for any number ${\cal N}$ of
supersymmetry in \cite{Tropper} and we put the central charge
of $Z_{m}^{IJ}$ and the particle changing operators
${\cal F}_{k,m}$, ${\cal F}_{k,m}^I$,
${\cal F}_{k,m}^{I J}$, ${\cal F}_{k,m}^{I J K}$
(which can be written in terms of $SO(4)$ epsilon tensor
with the Majoranas having a single $SO(4)$ index)
and ${\cal F}_{k,m}^{I J K L}$
(which is written in terms of $SO(4)$ epsilon tensor
with the $SO(4)$  singlet scalars 
in \cite{Tropper})
acting on the $m$-th particle to zero.
Here $I,J,K,L=1, \cdots, 4$ for ${\cal N}=4$.}.
Similarly,
the $\Delta$
for the soft vectors contains all the soft poles
$\Delta=0, -1, \cdots$
appearing in (the Euler beta function of) the OPE between the
vectors in the ${\cal N}=8$ (or ${\cal N}=4$) supergravity.
The $\Delta=0$ generates
the soft symmetry corresponding to
the leading soft vectors \cite{Tropper,BRS2}.
The $\Delta$
for the soft Majoranas contains all the  soft poles
$\Delta=-\frac{1}{2}, -\frac{3}{2}, \cdots$
appearing in (the Euler beta function of) the OPE between the
Majoranas in the ${\cal N}=8$ (or ${\cal N}=4$) supergravity.
The $\Delta=-\frac{1}{2}$ generates
the soft symmetry corresponding to
the leading soft Majoranas \cite{Tropper,BRS2}.
Finally,
the $\Delta$
for the soft scalars contains all the  soft poles
$\Delta=-1,-2, \cdots$
appearing in (the Euler beta function of) the OPE between the
scalars in the ${\cal N}=8$ (or ${\cal N}=4$) supergravity.
The $\Delta=-1$ generates
the soft symmetry corresponding to
the leading soft scalars \cite{Tropper,BRS2}.

\begin{itemize}
\item[]
We present the main results of this paper as follows:

From the ${\cal N}=4$ supersymmetric
$W_{1+\infty}^{2,2}[\la =\frac{1}{4}]$ algebra
given explicitly in Appendix $A.2$,
by focusing on the lowest terms in the parameter
appearing on the right hand sides of the (anti)commutators,
we obtain
the corresponding
 ${\cal N}=4$ supersymmetric
$w_{1+\infty}^{2,2}[\la =\frac{1}{4}]$ algebra
presented in (\ref{finalanticommcomm}).
Due to the positive and negative helicities 
of the particles of ${\cal N}=4$ supergravity,
the additional terms are expected on the right hand sides
of (\ref{finalanticommcomm})
as we impose the helicities on the particles.
Moreover, from the analysis of split factors in the
${\cal N}=4$ supergravity,
some of the (anti)commutators do not appear
although the corresponding terms in the Lagrangian exist.
It turns out that
the celestial soft current algebra
in the ${\cal N}=4$ supergravity theory
is described by (\ref{SOFT})
(The superspace description for this algebra
can be found in (\ref{twosuper})).
The structure constants (or the couplings)
appearing on the right hand sides
of (\ref{SOFT})
are fixed by the Jacobi identity and they are
summarized by (\ref{kapparelation}).

\end{itemize}  

Moreover, 
the truncations of
${\cal N}=4$ supersymmetric
soft algebra
provide the soft current algebras
for the ${\cal N}=2,3$ supergravity theories \cite{Fv,FSZ,Freedman},
the ${\cal N}=2$ supergravity coupled to its
Abelian vector multiplet \cite{Das,Luciani} and the ${\cal N}=1$
supersymmetric Maxwell Einstein theory \cite{FSv}.

In the sections $2$, $3$ and $4$,
we construct the extension of ${\cal N}=4$ $SO(4)$ superconformal algebra
and as an application of it,
we determine the ${\cal N}=4$ soft current algebra (and its subalgebras)
between the graviton, the gravitinos, the vectors, the Majoranas,
the scalar or the pseudoscalar.
In the section $5$,
we determine the extension of ${\cal N}=2$ $SO(2)$
superconformal algebra by using the different free field realization.
In the section $6$,
we summarize what we have done in this paper and present some open problems.
In Appendices $A$-$B$,
we list some details from the sections $2, 3$, $4$ and $5$.
For more details, we refer to the arXiv version $1$
of this paper
\footnote{The Thielemans package \cite{Thielemans} can be used
together with a mathematica \cite{mathematica} all
the times in this paper.
We list some partial  works
\cite{AGS,33,34,35,36,37,38,39,40,41,42,43,44,45,46,47,48,49,50,BHP,
CST,Sato,ACD,Tropper2412,Sato1,Mol}
in the context of \cite{Strominger}.}.

\section{
The
${\cal N}=4$ supersymmetric $W_{1+\infty}^{2,2}[\la =\frac{1}{4}]
$
algebra
}


\subsection{ The 
${\cal N}=4$ $SO(4)$ superconformal algebra}

The ${\cal N}=4$ $SO(4)$ superconformal algebra \cite{Schoutens}
consisting of one spin $2$ generator $L_m$, four
spin $\frac{3}{2}$ generators $G^i_r$,
six spin $1$ antisymmetric generators $T^{ij}_m$,
four spin $\frac{1}{2}$ generators $\Ga^i_r$ and one spin $0$
$\Delta_m$, where $i,j, \cdots = 1,2,3,4$
are $SO(4)$ vector indices  and the Laurent mode $m$ is an integer
and the Laurent mode $r$ is an integer (or half an odd integer), is
described by
\bea
\comm{L_m}{L_n}&=& (m-n)\,L_{m+n} + \frac{1}{12} \, c_{\alpha} \,
m (m^2-1) \, \delta_{m+n}\,,
\nonu \\
\comm{L_m}{G^i_r}&= & (\tfrac{1}{2}m-r)\,G^i_{m+r}\,,
\nonu \\
\comm{L_m}{T^{ij}_n}&=& -n\,T^{ij}_{m+n}\,,
\nonu \\
\comm{L_m}{\Gamma^i_r}&=& (-\tfrac{1}{2}m-r)\,\Gamma^i_{m+r}\,,
\nonu \\
\comm{L_m}{\Delta_n}&=& -\frac{1}{6}\, c_{\alpha}'\, (m+1)\,
\delta_{m+n}+(-m-n)\,\Delta_{m+n}\,,
\nonu \\
\acomm{G^i_r}{G^j_s}&= & 2\, \delta^{ij}\,L_{r+s}-\mathrm{i}\,(r-s)\,
( T^{ij}_{r+s} + 2 \, \alpha \,  \frac{1}{2} \, \epsilon^{ijkl} \,
T^{kl}_{r+s}) +
\frac{1}{3} \, c_{\alpha} \, (r^2-\frac{1}{4})\, \delta^{ij} \,
\delta_{r+s}\,,
\nonu \\
\comm{G^i_r}{T^{jk}_m}&=&
\mathrm{i}\,\delta^{ij}\,G^k_{r+m}-\mathrm{i}\,
\delta^{ik}\,G^j_{r+m}+\epsilon^{ijkl}\,m\,\Gamma^{l}_{r+m}-
2 \, m\, \alpha ( \delta^{ij} \, \Ga^k_{r+m}-
\delta^{ik} \, \Ga^j_{r+m}) \, \,,
\nonu \\
\acomm{G^i_r}{\Gamma^j_s}&= & \frac{\mathrm{i}}{3}\,c_{\alpha}'\,
(r+\tfrac{1}{2})\,\delta^{ij} \, \delta_{r+s}
+\mathrm{i}\,(r+s)\,\delta^{ij}\,\Delta_{r+s}-
\frac{1}{2} \, \epsilon^{ijkl} \,
T^{kl}_{r+s}\,,
\nonu \\
\comm{G^i_r}{\Delta_m}&= & \mathrm{i}\,\Gamma^i_{r+m}\,,
\nonu \\
\comm{T^{ij}_m}{T^{kl}_n}&= & -
\frac{1}{3} \, c'\, \epsilon^{ijkl}\,m \, \delta_{m+n}
+\frac{1}{3} \, c \, (\delta^{ik} \, \delta^{jl}
- 
\delta^{il} \, \delta^{jk}  )\, m\, \delta_{m+n}
\nonu \\
& - &
\mathrm{i}\,\delta^{ik}\,T^{jl}_{m+n}+\mathrm{i}\,\delta^{il}\,T^{jk}_{m+n}
+ \mathrm{i}\,\delta^{jk}\,T^{il}_{m+n}-\mathrm{i}\,\delta^{jl}\,T^{ik}_{m+n}\,,
\nonu  \\
\comm{T^{ij}_m}{\Gamma^{k}_r}&=&
-\mathrm{i}\,\delta^{ik}\,\Gamma^{j}_{m+r}
+\mathrm{i}\,\delta^{jk}\,\Gamma^{i}_{m+r}\,,
\nonu \\
\acomm{\Gamma^i_r}{\Gamma^j_s}&=& \frac{1}{3}\,c \,
\delta^{ij} \, \delta_{r+s}\,,
\nonu \\
\comm{\Delta_m}{\Delta_n}&=& \frac{1}{3\, m}\, c\,
\delta_{m+n}\,,
\label{n4sca}
\eea
where the charges appearing in the
commutator of (\ref{n4sca}) between the spin $2$ generator
and itself (or the anticommutator between the
spin $\frac{3}{2}$ generators and itself) and
the
commutator between the spin $2$ generator
and the spin $0$ generator (or the anticommutator between the
spin $\frac{3}{2}$ generators and the
spin $\frac{1}{2}$ generators)
are given by
\bea
c_{\al} = c \, (1+ 4 \, \alpha^2)- 4 \, \al\, c'\, , \qquad
c_{\al}'= c' -2 \, \al\, c \, ,
\label{twocentral}
\eea
respectively.
Here the $c$ and $c'$ of (\ref{twocentral}) appearing in the
commutator between the spin $1$ generators and itself,
as two linear combinations, play the
role of the central charges (or levels) of two commuting $SU(2)$
Kac-Moody algebras. The charge $c$ also appears in the
anticommutator between the spin $\frac{1}{2}$ generators and
itself (or the commutator between the spin $0$ generator and
itself): the last two relations.
Moreover, the real parameter $\alpha$,
which is introduced in the above 
spin $2$ and spin $\frac{3}{2}$ generators \footnote{The former
consists of the original spin $2$ generator plus
the second derivative of spin $0$ generator with $\al$
coefficient and the latter consists of the original
spin $\frac{3}{2}$ generators plus
the first derivative of spin $\frac{1}{2}$ generators
with $\al$ coefficient \cite{Schoutens}.
See also (\ref{fiveh0}).}, appears on the right
hand sides of the anticommutator between the spin $\frac{3}{2}$
generators and itself (and the commutator between the spin
$\frac{3}{2}$ generators and the spin $1$ generators). 
When  the parameter $\alpha$ and the charge $c$ vanish, then 
the corresponding terms in (\ref{n4sca}) do not appear.
It is obvious from (\ref{twocentral}) that
the charges become $c$ and $c'$ respectively for the
vanishing $\al$ \footnote{As noted in \cite{STV},
  by using their equation $(6)$, the above relations
  of (\ref{n4sca}) lead to their algebra where the central
  charge is given by $(c-\frac{c' c'}{c})$ and their $\gamma$
  is $\frac{(c-c')}{2 c}$.}.

\subsection{The realization of
${\cal N}=4$ $SO(4)$ superconformal algebra approximately}

The $(\beta, \ga)$ fields are bosonic operators while
the $(b, c)$ fields are fermionic operators.
The
spins of $(\beta , \ga)$ fields are given by
$(\la,1-\la)$ and
the spins of $(b , c)$ fields are given by
$(\frac{1}{2}+\la,\frac{1}{2}-\la)$ under the stress energy tensor.
The operator product expansions
of the $(\beta , \ga)$ and $(b , c)$ systems are given by \cite{CHR,Ahn2203}
\bea
\ga^{i,\bar{a}}(z)\, \beta^{\bar{j},b}(w) =
\frac{1}{(z-w)}\, \de^{i \bar{j}}\, \de^{\bar{a} b} + \cdots\, ,
\qquad
c^{i, \bar{a}}(z) \, b^{\bar{j},b}(w) =
\frac{1}{(z-w)}\, \de^{i \bar{j}}\, \de^{\bar{a} b} + \cdots\, ,
\label{fundOPE}
\eea
where the fundamental indices $a, b $,
the antifundamental indices $\bar{a}, \bar{b}$
of $SU(2)$, the fundamental indices $i, j $
and the antifundamental indices $\bar{i}, \bar{j}$ of
$SU(N)$
appear in the $(\beta , \ga)$ and $(b , c)$ systems
\footnote{In this paper, when we describe the OPEs corresponding to
the (anti)commutators in two dimensional conformal field theory,
we assume the antiholomorphic sector with
`unusual' unbarred notations for the complex coordinates
rather than the holomorphic sector.}.
We are interested in the $SU(N)$ singlet currents
by summing over the $SU(N)$ indices
based on the work of
\cite{BVd1,BVd2} and they are given by \cite{Ahn2205,Ahn2208}
with (\ref{fundOPE})
\bea
W_{F,h}^{\lambda,\bar{a}b}(z)
& = &  (-4 q)^{h-2}\sum_{i=0}^{h-1} a^{i}
(h,\lambda+\tfrac{1}{2})\,
\partial^{h-i-1}
((\partial^{i}b^{\bar{l} b})\, \delta_{\bar{l} l}\,c^{l \bar{a}})(z),
\nonu \\
W_{B,h}^{\lambda,\bar{a}b}(z)
& = & (-4 q)^{h-2}\sum_{i=0}^{h-1} a^{i}(h,\lambda)\,
\partial^{h-i-1}
((\partial^{i}\beta^{\bar{l} b})\, \delta_{\bar{l} l}\,\gamma^{l \bar{a}})(z),
\nonu \\
Q_{h+\frac{1}{2}}^{\lambda,\bar{a}b}(z)
& = & \sqrt{2}\,(-4q)^{h-1}\sum_{i=0}^{h-1} \beta^{i}(h+1,\lambda)\,
\partial^{h-i-1}
((\partial^{i}b^{\bar{l} b})\, \delta_{\bar{l} l}\,\gamma^{l \bar{a}})(z),
\nonu \\
\bar{Q}_{h+\frac{1}{2}}^{\lambda,a\bar{b}}(z)
& = & \sqrt{2}\, (-4q)^{h-1}\sum_{i=0}^{h} \alpha^{i}(h+1,\lambda)\,
\partial^{h-i}
((\partial^{i}\beta^{\bar{l} a})\,
\delta_{\bar{l} l}\,c^{l \bar{b}})(z),
\label{fourcurrents}
\eea
where
the relative coefficients \cite{BVd1,BVd2}
are given by the binomial
coefficients denoted by parentheses and the rising Pochhammer symbols
\footnote{We use
$(a)_n \equiv a(a+1) \cdots (a+n-1)$ where $n$
is a nonnegative integer \cite{PRS}.}
\bea 
a^i(h, \la) \equiv \left(\begin{array}{c}
h-1 \\  i \\
\end{array}\right) \, \frac{(-2\la-h+2)_{h-1-i}}{(h+i)_{h-1-i}}\, ,
\qquad 0 \leq i \leq h-1\, ,
\nonu \\
\al^i(h, \la) \equiv \left(\begin{array}{c}
h-1 \\  i \\
\end{array}\right) \, \frac{(-2\la-h+2)_{h-1-i}}{(h+i-1)_{h-1-i}}\, ,
\qquad 0 \leq i \leq h-1\, ,
\nonu \\
\beta^i(h, \la) \equiv \left(\begin{array}{c}
h-2 \\  i \\
\end{array}\right) \, \frac{(-2\la-h+2)_{h-2-i}}{(h+i)_{h-2-i}}\, ,
\qquad 0 \leq i \leq h-2 \, .
\label{coeff}
\eea
There exist four bosonic currents of spin $h$,
four bosonic currents of spin $h$,
four fermionic currents of spin $(h+\frac{1}{2})$,
and
four fermionic currents of spin $(h+\frac{1}{2})$ where $h=1,2,
\cdots $. There are also four fermionic currents of spin $\frac{1}{2}$
appearing in the last of (\ref{fourcurrents})
by construction \footnote{
The parameter $q$ can be taken as any nonzero real value.}.
The $\la$ dependence in (\ref{coeff}) appears only at the numerators.

The above $16$ generators appearing in (\ref{n4sca})
can be described by  
a single ${\cal N}=4$ multiplet of superspin $h$
\cite{Schoutens,BCG,AK1509} by taking $h=0$
\bea
\mathbf{\Phi}^{(h)}(Z)
=\Phi^{(h)}_0(z)
+\theta^i\,\Phi^{(h),i}_\frac{1}{2}(z)
+\frac{1}{2} \, \theta^{4-ij}\,\Phi^{(h),ij}_1(z)
+\theta^{4-i}\,\Phi^{(h),i}_\frac{3}{2}(z)
+\theta^{4-0}\,\Phi^{(h)}_2(z),
\label{bigPhi}
\eea  
where
the Grassmann coordinate $\theta^i$
with $SO(4)$ vector index $i$
has a spin $-\frac{1}{2}$ and each term
of (\ref{bigPhi}) has the spin $h$ because
the sum of spins in the product of
Grassmann coordinates and the subscript of
component field is equal to zero.
The conventions for the ${\cal N}=4$ superspace coordinates
can be found from \cite{Schoutens,AK1509}.
The realization of ${\cal N}=4$ $SO(4)$ superconformal algebra
by using (\ref{fundOPE}) is given by \cite{Ahn2205,Ahn2208,AK2309}.
Moreover, the expression for the
${\cal N}=4$ multiplet of superspin $h$ is given explicitly
in \cite{Ahn2208,AK2309} and they are described in Appendix (\ref{Phih})
with some rescalings appearing in the overall factors.
For the ${\cal N}=4$ stress energy tensor of superspin $0$,
we identify the $16$ generators (\ref{Phih}) with $h=0$
with the corresponding $16$ generators in (\ref{n4sca})
as follows:
\bea
\Phi^{(0)}_0&= & -\frac{1}{32}\,\Delta\,,
\nonu \\
\Phi^{(0),i}_\frac{1}{2}&=&
\frac{\mathrm{i}}{32\sqrt{2}}\,\Gamma^i\,,
\nonu \\
\Phi^{(0),ij}_1&=& -\frac{\mathrm{i}}{8}\,T^{ij}\,,
\nonu \\
\Phi^{(0),i}_\frac{3}{2}&=& -\frac{1}{8\sqrt{2}}\,
\Big(G^i -\mathrm{i} \, (1-4\la) \, \pa \,
\Gamma^i \Big) \, ,
\nonu \\
\Phi^{(0)}_2&=& \Big(L -\frac{1}{2}\, (1-4\la)\, \pa^2 \, \Delta
\Big)\,.
\label{fiveh0}
\eea
In the last two of (\ref{fiveh0}),
there are $\la$ dependences.
The OPEs between the ${\cal N}=4$ stress energy tensor
and itself in the component approach is given by
Appendix $A$ of \cite{Ahn2205} and the corresponding (anti)commutators
by using the identifications of (\ref{fiveh0})
are summarized by
\bea
\comm{L_m}{L_n}&=& (m-n)\,L_{m+n} + \frac{1}{2} \,N \, (1-4\la) \,
m (m^2-1) \, \delta_{m+n}\,,
\nonu \\
\comm{L_m}{G^i_r}&= & (\tfrac{1}{2}m-r)\,G^i_{m+r}\,,
\nonu \\
\comm{L_m}{T^{ij}_n}&=& -n\,T^{ij}_{m+n}\,,
\nonu \\
\comm{L_m}{\Gamma^i_r}&=& (-\tfrac{1}{2}m-r)\,\Gamma^i_{m+r}\,,
\nonu \\
\comm{L_m}{\Delta_n}&=& \frac{N}{2}\, (m+1)\,
\delta_{m+n}+(-m-n)\,\Delta_{m+n}\,,
\nonu \\
\acomm{G^i_r}{G^j_s}&= & 2\, \delta^{ij}\,L_{r+s}-\mathrm{i}\,(r-s)\,
( T^{ij}_{r+s} + (1-4\la) \,
\frac{1}{2} \, \epsilon^{ijkl} \,
T^{kl}_{r+s}) \nonu \\
& + &
2\, N \, (1-4\la) \, (r^2-\frac{1}{4})\, \delta^{ij} \,
\delta_{r+s}\,,
\nonu \\
\comm{G^i_r}{T^{jk}_m}&=&
\mathrm{i}\,\delta^{ij}\,G^k_{r+m}-\mathrm{i}\,
\delta^{ik}\,G^j_{r+m}+\epsilon^{ijkl}\,m\,\Gamma^{l}_{r+m}-
(1-4\la)\, m\,  ( \delta^{ij} \, \Ga^k_{r+m}-
\delta^{ik} \, \Ga^j_{r+m}) \, \,,
\nonu \\
\acomm{G^i_r}{\Gamma^j_s}&= & -N\, \mathrm{i} \,
(r+\tfrac{1}{2})\,\delta^{ij} \, \delta_{r+s}
+\mathrm{i}\,(r+s)\,\delta^{ij}\,\Delta_{r+s}-
\frac{1}{2} \, \epsilon^{ijkl} \,
T^{kl}_{r+s}\,,
\nonu \\
\comm{G^i_r}{\Delta_m}&= & \mathrm{i}\,\Gamma^i_{r+m}\,,
\nonu \\
\comm{T^{ij}_m}{T^{kl}_n}&= & 
N\, \epsilon^{ijkl}\, m \, \delta_{m+n}
- 
\mathrm{i}\,\delta^{ik}\,T^{jl}_{m+n}+\mathrm{i}\,\delta^{il}\,T^{jk}_{m+n}
+ \mathrm{i}\,\delta^{jk}\,T^{il}_{m+n}-\mathrm{i}\,\delta^{jl}\,T^{ik}_{m+n}\,,
\nonu  \\
\comm{T^{ij}_m}{\Gamma^{k}_r}&=&
-\mathrm{i}\,\delta^{ik}\,\Gamma^{j}_{m+r}
+\mathrm{i}\,\delta^{jk}\,\Gamma^{i}_{m+r}\,.
\label{realn4sca}
\eea
By comparing with the ones in (\ref{n4sca}),
we observe that the central terms having $c$
in the commutator between spin $1$ and itself,
the anticommutator between the spin $\frac{1}{2}$
and itself, and the commutator between the
spin $0$ and itself do not appear in this free field realization.
Also the $\la$ plays the role of
deformation parameter $\al$
\bea
\al =\frac{1}{2}\, (1-4\la)\, .
\label{alphala}
\eea
It is easy to see  that
\bea
c' = -3 \, N = c_{\al}', \qquad c_{\al} = 6 \, N \, (1-4\la)\, . 
\label{calphaalpha'}
\eea
This implies that 
the above (anti)commutators
become the ones in (\ref{n4sca})
for $
c=0$ together with (\ref{calphaalpha'}) (in this sense
we put the subtitle of this subsection).
Note that in the description of \cite{AK1509},
we consider the case of $c_{\al}'=0$ \footnote{
  This is the reason why the corresponding two terms
  (appearing in the commutator
between the spin $2$ and spin $0$ and in the anticommutator
between the spin $\frac{3}{2}$ and the spin $\frac{1}{2}$)
in (\ref{n4sca})
do not appear in Appendix $A$ of \cite{AK1509}.}.
On the other hand, the free field realization with (\ref{fundOPE})
provides the nonzero $c_{\al}'$ because of nonzero $N$ from
(\ref{calphaalpha'}).
In the language of OPE, the additional terms of
above commutator and anticommutator
(corresponding to the fifth and the eighth of (\ref{realn4sca}))
appear in the third and second order poles of Appendix $A$
of \cite{Ahn2205} respectively. 
Also for $\la =\frac{1}{4}$ (or equivalently $\al=0$)
from (\ref{alphala}),
the $c_{\al}$ vanishes according to (\ref{calphaalpha'})
\footnote{For $\alpha = \pm \frac{1}{2}$
(or equivalently $\la=0$ or $\la=\frac{1}{2}$), the finite
Lie superalgebra $OSp(4|2)$ occurs \cite{Schoutens}.
On the other hand, in this paper (sections $2$-$4$)
we do not have this finite
Lie superalgebra.}.


\subsection{The 
${\cal N}=4$ $SO(4)$ superconformal algebra with vanishing
central charge}

In this subsection, we will consider the particular case
where there is no deformation by the previous $\al$ (that is,
$\al=0$ or $\la =\frac{1}{4}$).
This will be clearer when we consider the soft current algebra
associated with the ${\cal N}=4$ supergravity theory later.
After substituting the value $\la =\frac{1}{4}$ into the
ones of (\ref{realn4sca}), we obtain the following
(anti)commutators
\bea
\comm{L_m}{L_n}&=& (m-n)\,L_{m+n}\,,
\nonu \\
\comm{L_m}{G^i_r}&= & (\tfrac{1}{2}m-r)\,G^i_{m+r}\,,
\nonu \\
\comm{L_m}{T^{ij}_n}&=& -n\,T^{ij}_{m+n}\,,
\nonu \\
\comm{L_m}{\Gamma^i_r}&=& (-\tfrac{1}{2}m-r)\,\Gamma^i_{m+r}\,,
\nonu \\
\comm{L_m}{\Delta_n}&=& \frac{N}{2}\, (m+1)\,
\delta_{m+n}+(-m-n)\,\Delta_{m+n}\,,
\nonu \\
\acomm{G^i_r}{G^j_s}&= & 2\, \delta^{ij}\,L_{r+s}-\mathrm{i}\,(r-s)\,
T^{ij}_{r+s}\,,
\nonu \\
\comm{G^i_r}{T^{jk}_m}&=&
\mathrm{i}\,\delta^{ij}\,G^k_{r+m}-\mathrm{i}\,
\delta^{ik}\,G^j_{r+m}+\epsilon^{ijkl}\,m\,\Gamma^{l}_{r+m} \,,
\nonu \\
\acomm{G^i_r}{\Gamma^j_s}&= & -N\, \mathrm{i}  \,
(r+\tfrac{1}{2})\,\delta^{ij} \, \delta_{r+s}
+\mathrm{i}\,(r+s)\,\delta^{ij}\,\Delta_{r+s}-
\frac{1}{2} \, \epsilon^{ijkl} \,
T^{kl}_{r+s}\,,
\nonu \\
\comm{G^i_r}{\Delta_m}&= & \mathrm{i}\,\Gamma^i_{r+m}\,,
\nonu \\
\comm{T^{ij}_m}{T^{kl}_n}&= & 
N\, \epsilon^{ijkl}\, m \, \delta_{m+n}
- 
\mathrm{i}\,\delta^{ik}\,T^{jl}_{m+n}+\mathrm{i}\,\delta^{il}\,T^{jk}_{m+n}
+ \mathrm{i}\,\delta^{jk}\,T^{il}_{m+n}-\mathrm{i}\,\delta^{jl}\,T^{ik}_{m+n}\,,
\nonu  \\
\comm{T^{ij}_m}{\Gamma^{k}_r}&=&
-\mathrm{i}\,\delta^{ik}\,\Gamma^{j}_{m+r}
+\mathrm{i}\,\delta^{jk}\,\Gamma^{i}_{m+r}\,.
\label{n4commanticomm}
\eea
Then it is obvious to see that
the (anti)commutators given by (\ref{n4commanticomm})
become the ones in (\ref{n4sca}) with
$c' = -3 \, N = c_{\al}'$ from (\ref{calphaalpha'}) and
$c = c_{\al}=0$ according to (\ref{twocentral}) \footnote{
Note that for $\la \neq \frac{1}{4}$,
there are two nontrivial terms coming from
the anticommutator between the spin $\frac{3}{2}$ generator
and itself and the commutator between the spin $\frac{3}{2}$
generator and the spin $1$ generator: epsilon term and delta terms
having the spin $\frac{1}{2}$ generators in (\ref{realn4sca}).
In other words, at the nondeformation $\al =0$,
the epsilon term vanishes in the former while
the epsilon term survives (or delta terms appearing in the
spin $\frac{1}{2}$ generators vanish) in the latter.}.
Furthermore, the two central terms in
(\ref{n4commanticomm})
corresponding to
$c_{\alpha}$ vanish.
Therefore, the free field realization characterized by
Appendix (\ref{Phih}) and (\ref{fiveh0}) together with (\ref{fundOPE}) 
provides the above algebra (\ref{n4commanticomm}) where
there are three nontrivial central terms. 

\subsection{The realization of
${\cal N}=4$ $SO(4)$ superconformal algebra with vanishing central charge}


In order to
obtain the generalization of the above
${\cal N}=4$ $SO(4)$ superconfromal algebra with vanishing
central charge, we need to use the ${\cal N}=4$ multiplet
of superspin $h$ (the reason for the change of
notation) and according to (\ref{fiveh0}),
the above (anti)commutators in (\ref{n4commanticomm}) in different orders
can be written as
\bea
\comm{(\Phi^{(0)}_{0})_m}{(\Phi^{(0),i}_{\frac{3}{2}})_r}
&=&-\frac{1}{8}\,(\Phi^{(0),i}_{\frac{1}{2}})_{m+r}\,,
\nonu \\
\comm{(\Phi^{(0)}_{0})_m}{(\Phi^{(0)}_{2})_n}
&=&-\frac{1}{4^3}\,N\,(m-1)\,\delta_{m+n}+(m+n)\,(\Phi^{(0)}_{0})_{m+n}\,,
\nonu \\
\comm{(\Phi^{(0),i}_{\frac{1}{2}})_r}{(\Phi^{(0),jk}_{1})_m}
&=&\frac{1}{8}\,\delta^{ij}\,(\Phi^{(0),k}_{\frac{1}{2}})_{r+m}-
\frac{1}{8}\,\delta^{ik}\,(\Phi^{(0),j}_{\frac{1}{2}})_{r+m}\,,
\nonu \\
\acomm{(\Phi^{(0),i}_{\frac{1}{2}})_r}{(\Phi^{(0),j}_{\frac{3}{2}})_s}
&=&\delta^{ij}\,\bigg[\frac{N}{2\cdot 4^4}\,(r-\tfrac{1}{2})\,\delta_{r+s}
-\frac{1}{4^2}\,(r+s)\,(\Phi^{(0)}_{0})_{r+s} \bigg]
+\frac{1}{4^3}\, \frac{1}{2} \, \epsilon^{ijkl}
\, (\Phi^{(0),kl}_{1})_{r+s}\,,
\nonu \\
\comm{(\Phi^{(0),i}_{\frac{1}{2}})_r}{(\Phi^{(0)}_{2})_m}
&=&(r+\tfrac{1}{2}\,m)\,(\Phi^{(0),i}_{\frac{1}{2}})_{r+m}\,,
\nonu \\
\comm{(\Phi^{(0),ij}_{1})_m}{(\Phi^{(0),kl}_{1})_n}
&=&-\frac{N}{4^3}\,\epsilon^{ijkl}\,m\,\delta_{m+n}
+\frac{1}{8} \, \bigg[\, -\delta^{ik}\,(\Phi^{(0),jl}_{1})_{m+n}
+\delta^{il}\,(\Phi^{(0),jk}_{1})_{m+n}
\nonu \\
& + & \delta^{jk}\,(\Phi^{(0),il}_{1})_{m+n}
-\delta^{jl}\,(\Phi^{(0),ik}_{1})_{m+n}\,\bigg]\,,
\nonu \\
\comm{(\Phi^{(0),ij}_{1})_m}{(\Phi^{(0),k}_{\frac{3}{2}})_r}
&=&-\frac{1}{8}\,\delta^{ik}\,(\Phi^{(0),j}_{\frac{3}{2}})_{m+r}
+\frac{1}{8}\,\delta^{jk}\,(\Phi^{(0),i}_{\frac{3}{2}})_{m+r}
-\frac{1}{2}\,\epsilon^{ijkl}\,m\,(\Phi^{(0),l}_{\frac{1}{2}})_{r+m}\,,
\nonu \\
\comm{(\Phi^{(0),ij}_{1})_m}{(\Phi^{(0)}_{2})_n}
&=&
m\,(\Phi^{(0),ij}_{1})_{m+n}\,,
\nonu \\
\acomm{(\Phi^{(0),i}_{\frac{3}{2}})_r}{(\Phi^{(0),j}_{\frac{3}{2}})_s}
&=&\frac{1}{4^3}\,\delta^{ij}\,(\Phi^{(0)}_{2})_{r+s}+\frac{1}{4^2}(r-s)\,(\Phi^{(0),ij}_{1})_{r+s}\,,
\nonu \\
\comm{(\Phi^{(0),i}_{\frac{3}{2}})_r}{(\Phi^{(0)}_{2})_m}
&=&
(r-\tfrac{1}{2}\,m)\,(\Phi^{(0),i}_{\frac{3}{2}})_{r+m}\,,
\nonu \\
\comm{(\Phi^{(0)}_{2})_m}{(\Phi^{(0)}_{2})_n}
&=&
(m-n)\,(\Phi^{(0)}_{2})_{m+n}\,.
\label{n4phi}
\eea
Note that there is an antisymmetric property
between the commutator between the spin $1$ and itself
by interchanging each $SO(4)$ index and the modes
while there is a symmetric property
between the anticommutator between the spin $\frac{3}{2}$ and itself
by interchanging each $SO(4)$ index and the modes.
We should determine the generalization of these
(anti)commutators for generic superspin $h$.
We expect that for nonzero $h_1$ and $h_2$ appearing on the
left hand sides of any (anti)commutator,
the above results in (\ref{n4phi}) should reappear
in (\ref{finalanticommcomm})
by taking $h_1$ and $h_2$ to be zero \footnote{
By considering the following ${\cal N}=4$ multiplet
\bea
{\bf J}(Z) & = &
32\,\Phi^{(0)}_0(z)
+32\sqrt{2}\,\theta^{j}\, \Phi^{(0),j}_\frac{1}{2}(z)
+4\,\theta^{4-jk}\, \Phi^{(0),jk}_1(z)
+8\sqrt{2}\,\theta^{4-j} \, \Phi^{(0),j}_\frac{3}{2}(z)
+2\,\theta^{4-0}  \, \Phi^{(0)}_2(z),
\nonu
\eea
the above (\ref{n4phi})
can be summarized by
\bea
{\bf J}(Z_{1}) \, {\bf J}(Z_{2}) & = &
-\frac{\theta_{12}^{4-0}}{z_{12}^{2}}\, N
+
\frac{\theta_{12}^{4-i}}{z_{12}} \, D^{i} {\bf J}(Z_{2})+
\frac{\theta_{12}^{4-0}}{z_{12}} \, 2 \, \partial {\bf J}(Z_{2})
+ \cdots
\nonu
\eea
with the ${\cal N}=4$ superspace notation \cite{AK2309}.
There are different coefficients
in the above ${\cal N}=4$ multiplet compared to (\ref{bigPhi}).
Note that the above five components
with (\ref{fiveh0}) are the same as the ones in \cite{AK2309}.}.

\subsection{The extension of ${\cal N}=4$
$SO(4)$  superconformal algebra with vanishing central charge}

Although the explicit results
for the (anti)commutators with generic $h_1$
and $h_2$ are given in Appendix $B$ of \cite{AK2309},
the structure constants
appearing in the
(anti)commutators are implicit in the sense that
they do depend on two other dummy variables in the double
summations.
Of course, once $h_1$ and $h_2$ are fixed,
then they appear as the numerical values together with
mode dependent function (see also Appendix $B$)
which has definite symmetric or
antisymmetric properties under the exchange of $h_1$ and $h_2$
and the exchange of two modes appearing on the left hand sides.
In order to obtain the leading behaviors of the (anti)commutators
on the right hand sides, we introduce the parameter $q$
which will be small but nonzero 
and consider the following rescalings in the $16$ generators of
one spin $h$, four spin $(h+\frac{1}{2})$,
six spin $(h+1)$, four spin $(h+\frac{3}{2})$
and one spin $(h+2)$ as follows:
\bea
\Phi^{(h)}_{0} & \longrightarrow  & q^h \, \Phi^{(h)}_{0} \, ,
\nonu \\
\Phi^{(h),i}_{\frac{1}{2}} & \longrightarrow &
q^h\, \Phi^{(h),i}_{\frac{1}{2}} \, ,
\nonu \\
\Phi^{(h),ij}_{1} & \longrightarrow & q^h \, \Phi^{(h),ij}_{1} \, ,
\nonu \\
\Phi^{(h),i}_\frac{3}{2} & \longrightarrow &
q^h \, \Phi^{(h),i}_\frac{3}{2} \, ,
\nonu  \\
\Phi^{(h)}_2 & \longrightarrow  & q^h\, \Phi^{(h)}_2 \, .
\label{qrescalings}
\eea
In general, we can introduce the additional unknown parameters
in the exponents of $q$ in (\ref{qrescalings}) but they do vanish
by imposing that for the vanishing $h_1$ and $h_2$
the above (anti)commutators become the ones in (\ref{n4phi}).
Note that in the expression of Appendix (\ref{Phih}),
there is no $q$ dependence because the various $q$ factors
are cancelled by those appearing in the four different kinds of
currents in (\ref{fourcurrents}).
From Appendix $B$ of \cite{AK2309},
the whole (anti)commutators can be read off and we try to
determine the various structure constants for generic
$h_1$ and $h_2$ by starting with $h_1, h_2= 1,2,3, \cdots, 10$.
It turns out that
\bea
\comm{(\Phi^{(h_1)}_{0})_m}{(\Phi^{(h_2)}_{0})_n}
& = &
q^{4}\,\Big((h_2-1)m-(h_1-1)n \Big)\, (\Phi^{(h_1+h_2-4)}_2)_{m+n}
\, ,
\nonu \\
\comm{(\Phi^{(h_1)}_{0})_m}{(\Phi^{(h_2),i}_{\frac{1}{2}})_r}
& = &
-\frac{1}{8}\,q^{2}\,
(\Phi^{(h_1+h_2-2),i}_\frac{3}{2})_{m+r} \, ,
\nonu \\
\comm{(\Phi^{(h_1)}_{0})_m}{(\Phi^{(h_2),ij}_{1})_n} 
& = &
-q^{2}\,\Big(h_2\,m-(h_1-1)n \Big)
\,
\frac{1}{2} \, \epsilon^{ijkl} (\Phi^{(h_1+h_2-2),kl}_1)_{m+n} \, ,
\nonu \\
\comm{(\Phi^{(h_1)}_{0})_m}{(\Phi^{(h_2),i}_{\frac{3}{2}})_r}
& = &
-\frac{1}{8}\,(\Phi^{(h_1+h_2),i}_\frac{1}{2})_{m+r} \, ,
\nonu \\
\comm{(\Phi^{(h_1)}_{0})_m}{(\Phi^{(h_2)}_{2})_n}
& = & \Big((h_2+1)m-(h_1-1)r \Big)\,(\Phi^{(h_1+h_2)}_0)_{m+n}\, ,
\nonu \\
\acomm{(\Phi^{(h_1),i}_{\frac{1}{2}})_r}{(\Phi^{(h_2),j}_{\frac{1}{2}})_s}
& = & -\frac{1}{64}\,q^2 \, \delta^{ij}\,
(\Phi^{(h_1+h_2-2)}_{2})_{r+s} \, 
\nonu \\
&-& \frac{1}{8}
\,
q^{2}\,\Big((h_2-\tfrac{1}{2})r-(h_1-\tfrac{1}{2})s \Big)\,(\Phi^{(h_1+h_2-2),ij}_{1})_{r+s} \, ,
\nonu \\
\comm{(\Phi^{(h_1),i}_{\frac{1}{2}})_r}{(\Phi^{(h_2),jk}_{1})_m}
& = & \frac{1}{8}\,\delta^{ij}\,
(\Phi^{(h_1+h_2),k}_{\frac{1}{2}})_{r+m}
-\frac{1}{8}\,\delta^{ik}\,
(\Phi^{(h_1+h_2),j}_{\frac{1}{2}})_{r+m} \, 
\nonu \\
&- & q^2 \, \epsilon^{ijkl}\,
\Big(h_2\,r-(h_1-\tfrac{1}{2})m \Big)\, (\Phi^{(h_1+h_2-2),l}_{\frac{3}{2}})_{r+m}
\, ,
\nonu \\
\acomm{(\Phi^{(h_1),i}_{\frac{1}{2}})_r}{(\Phi^{(h_2),j}_{\frac{3}{2}})_s}
& = & \frac{1}{64}\,
\frac{1}{2} \, \epsilon^{ijkl} \,
(\Phi_{1}^{(h_1+h_2),kl})_{r+s} \, 
\nonu \\
&- &\frac{1}{8}\, \delta^{ij}\,
\Big((h_2+\tfrac{1}{2})r-(h_1-\tfrac{1}{2})s\Big)\, (\Phi^{(h_1+h_2)}_{0})_{r+s} \, ,
\nonu \\
\comm{(\Phi^{(h_1)}_{2})_m}{(\Phi^{(h_2),i}_{\frac{1}{2}})_r}
& = & 
\Big((h_2-\tfrac{1}{2})m-(h_1+1)r \Big)\,(\Phi^{(h_1+h_2),i}_\frac{1}{2})_{m+r} \, ,
\nonu \\
\comm{(\Phi^{(h_1),ij}_{1})_m}{(\Phi^{(h_2),kl}_{1})_n}
& = & -q^{2}\,(\delta^{ik}\delta^{jl}-\delta^{il}\delta^{jk})
\,\Big(h_2\,m-h_1\,n\Big)\,(\Phi^{(h_1+h_2-2)}_2)_{m+n} 
\nonu \\
&+ & \epsilon^{ijkl}\,\Big(h_2\,m-h_1\,n\Big)\,
(\Phi^{(h_1+h_2)}_0)_{m+n}
\nonu \\
& + & \frac{1}{8} \, \bigg[\, -\delta^{ik}\,(\Phi^{(h_1+h_2),jl}_{1})_{m+n}
+\delta^{il}\,(\Phi^{(h_1+h_2),jk}_{1})_{m+n}
\nonu \\
& + & \delta^{jk}\,(\Phi^{(h_1+h_2),il}_{1})_{m+n}
-\delta^{jl}\,(\Phi^{(h_1+h_2),ik}_{1})_{m+n}\,\bigg]\,,
\nonu \\
\comm{(\Phi^{(h_1),ij}_{1})_m}{(\Phi^{(h_2),k}_{\frac{3}{2}})_r}
& = &
-\frac{1}{8}\,\delta^{ik}\,(\Phi^{(h_1+h_2),j}_{\frac{3}{2}})_{m+r}
+\frac{1}{8}\,\delta^{jk}\,(\Phi^{(h_1+h_2),i}_{\frac{3}{2}})_{m+r}
\nonu \\
&- & \epsilon^{ijkl}\,
\Big((h_2+\tfrac{1}{2})m-h_1\,r \Big)\,(\Phi^{(h_1+h_2),l}_{\frac{1}{2}})_{m+r} \, ,
\nonu \\
\comm{(\Phi^{(h_1),ij}_{1})_m}{(\Phi^{(h_2)}_{2})_n}
& = & \Big((h_2+1)m-h_1\,n\Big)\,(\Phi^{(h_1+h_2),ij}_1)_{m+n}\, ,
\nonu \\
\acomm{(\Phi^{(h_1),i}_{\frac{3}{2}})_r}{(\Phi^{(h_2),j}_{\frac{3}{2}})_s}
&= & \frac{1}{64}\,\delta^{ij}\,
(\Phi^{(h_1+h_2)}_{2})_{r+s}
\nonu \\
&+ & \frac{1}{8}\,\Big((h_2+\tfrac{1}{2})r-(h_1+\tfrac{1}{2})s\Big)\,
(\Phi^{(h_1+h_2),ij}_{1})_{r+s} \, ,
\nonu \\
\comm{(\Phi^{(h_1)}_{2})_m}{(\Phi^{(h_2),i}_{\frac{3}{2}})_r}
& = & \Big((h_2+\tfrac{1}{2})m-(h_1+1)r \Big)\,
(\Phi^{(h_1+h_2),i}_\frac{3}{2})_{m+r}
\, ,
\nonu \\
\comm{(\Phi^{(h_1)}_{2})_m}{(\Phi^{(h_2)}_{2})_n}
& = &
\Big((h_2+1)m-(h_1+1)n\Big)\,(\Phi^{(h_1+h_2)}_2)_{m+n}\, ,
\label{finalanticommcomm}
\eea
where the higher order terms in the parameter $q$ are ignored.
See also (\ref{finalanticommcomm1}) for subleading terms.
All the $q$ independent terms on the right hand sides
of (\ref{finalanticommcomm}) can be seen from (\ref{n4phi})
except the epsilon term in the commutator between the spin $1$ and
itself where each  mode dependent term contains $h_1$ and $h_2$.
The power of $q$ on the right hand sides is related to
the upper indices for the spins such that the sum of
the power of $q$ and the spin in the superscript
at each term is equal to $(h_1+h_2)$
\footnote{
\label{diffordering}
We can rewrite some of the commutators by changing
the two modes appearing on the left hand sides as follows:
\bea
\comm{(\Phi^{(h_1),i}_{\frac{1}{2}})_r}{(\Phi^{(h_2)}_{2})_m}
& = & \Big((h_2+1)r-(h_1-\tfrac{1}{2})m \Big)\,
(\Phi^{(h_1+h_2),i}_\frac{1}{2})_{m+r} \,,
\nonu \\
\comm{(\Phi^{(h_1),i}_{\frac{3}{2}})_r}{(\Phi^{(h_2)}_{2})_m}
& = &
\Big((h_2+1)r-(h_1+\tfrac{1}{2})m\Big)\,(\Phi^{(h_1+h_2),i}_\frac{3}{2})_{m+r}.
\nonu
\eea}.
The 
various central terms which are not present in
(\ref{finalanticommcomm}) can be obtained
from Appendix $B$ of \cite{AK2309} or (\ref{finalanticommcomm1})
\footnote{We denote the above algebra (\ref{finalanticommcomm})
by the ${\cal N}=4$ supersymmetric
$w_{1+\infty}^{2,2}[\la=\frac{1}{4}]$
algebra (in the abstract) which is new. The last relation of
(\ref{finalanticommcomm}) contains
$w_{\infty}$ algebra \cite{Bakas}. Recall that $\Phi_2^{(-1)}$
of spin $1$
is proportional to the derivative of $\Delta$
of spin $0$ appearing in
(\ref{n4sca}). See also the $w_{1+\infty}^{2,2}[\la]$ algebra
in Appendix $A.6$.}.

\subsection{The
${\cal N}=4$ supersymmetric $W_{1+\infty}^{2,2}[\la =\frac{1}{4}]
$
algebra
}
%

By considering all the other terms in (\ref{finalanticommcomm}),
we present
the complete
${\cal N}=4$ supersymmetric $W_{1+\infty}^{2,2}[\la =\frac{1}{4}]
$
algebra in (\ref{N4algebrafor}) with (\ref{bigC}) and (\ref{STRUCT})
which can be obtained
from the results in \cite{AK2309} after inserting the
particular value for $\la=\frac{1}{4}$ \footnote{
Although the algebra (\ref{finalanticommcomm})
cannot be combined as the single ${\cal N}=4$
supersymmetric OPE between the two ${\cal N}=4$ multiplets
(For example, the sixth relation of (\ref{finalanticommcomm})
having the Kronecker delta produces the unwanted
higher oder terms in this  ${\cal N}=4$
supersymmetric OPE. Here the order is greater than or equal to
$2$),
the above algebra (\ref{N4algebrafor}), similar to
$(2.30)$ of \cite{AK2309} where $q=1$,
can be written in terms of
the single ${\cal N}=4$
supersymmetric OPE after the simple numerical
rescalings in the component
fields in the ${\cal N}=4$ multiplet.}.
By construction of free field realization in (\ref{fundOPE}),
the above algebra satisfies the Jacobi identity
(See also the section $1$ of \cite{BSphysreport}).
In principle we can check the Jacobi identity from the various
(anti)commutators presented in  (\ref{N4algebrafor})
although we did not do it.
We present a particular example of Jacobi identity
in Appendix $A.4$.
In (\ref{finalanticommcomm1}), we present the higher order terms in $q$ up to
the $q^4$ beyond the lowest orders in $q$ given by
(\ref{finalanticommcomm}).

\section{The connection with
  celestial holography: the soft current algebra and ${\cal N}=4$
  supergravity theory
\label{sugra}}

In this section, 
we construct the soft current algebra
corresponding to the ${\cal N}=4$ supergravity by Das in
\cite{Das}.

We expect that the above $15$ (anti)commutators
between the five operators should provide the soft
current algebra in the ${\cal N}=4$ supergravity theory.
We would like to construct the abstract (closed) algebra
in which the right hand sides should contain the right hand
sides of (\ref{finalanticommcomm}) and
each term on the right hand sides of this algebra (the `additional'
thirteen  terms from mode independent terms)
by keeping the $SO(4)$ symmetry should
have the mode dependent terms linearly (in terms of OPEs,
this implies that there are the second and the first order poles).
The reason for this is that when we consider the helicities
for the particles, then
the independent operators is increased by ten.
From the conformal field theory analysis,
the anticommutator between the same fermions
should not contain the mode dependent terms from the second order pole.
However, the anticommutator between the different fermions
can contain the mode dependent terms from the second order pole.
For the time being, the quadratic and higher order
terms in the modes
(the third and the
higher order poles in the OPE) are ignored from simplicity.
Moreover, we introduce the various couplings
appearing on the right hand sides of this algebra.
They can be determined, in principle, by using the Jacobi identity.
The free field realization in this abstract algebra
is no longer valid any more.
Some couplings can vanish from other requirements in the
splitting amplitudes below.

\subsection{The soft current algebra}

The ${\cal N}=4$ supergravity is found in \cite{Das}.
The Lagrangian consists of
a vierbein $ e_{\mu}^a$, four spin $\frac{3}{2}$ Majoranas
$ \psi_{\mu}^i$, six antisymmetric vectors $ A^{i j}_{\mu}$, four
spin $\frac{1}{2}$ Majoranas $\chi^i$, a scalar $A$ and
a pseudoscalar $B$.
We claim that the soft current algebra between the graviton
(helicity $\pm 2$),
the gravitinos with the helicity $\pm \frac{3}{2}$,
the vectors with the helicity $\pm 1$, the Majoranas
with the helicity $\pm \frac{1}{2}$, scalar  and
pseudoscalar (with a complex linear combination of
the helicity $+0$ and the helicity $-0$ and see (\ref{CORR}))
can be summarized by \footnote{In general, there are holomorphic
and antiholomorphic mode expansions for the soft current.
We consider the particular holomorphic mode leading to
the simple pole on the holomorphic coordinate, along the line of
\cite{HPS,Ahn2111}. This implies that there is no loop counting parameter
appeared in \cite{AMS}, in our soft current algebra,
because the holomorphic mode is not arbitrary but fixed by
one minus the left conformal weight.

For example, in \cite{Strominger2105, GHPS,HPS}, the OPE
of two conformal primary gravitons of arbitrary weight contains
the Euler beta function.
For the supersymmetric case, see also \cite{Tropper2412}.
Then the conformally soft gravitons are
introduced by removing the divergence appearing in
the Euler beta function.
The corresponding OPE between the
soft gravitons can be used to obtain the soft current algebra
for gravity in terms of
commutator by computing the various contour integrals \cite{GHPS}.
The helicities for the particles in the three point
scattering amplitude, in general,
are given by $(s_1,s_2, -s_1-s_2+2)$ \cite{HPS,Tropper2412}
for $d_V=5$
where $s_1, s_2 = \pm 2, \pm \frac{3}{2}, \pm 1, \pm \frac{1}{2}, 0$.
On the other hand,
the explicit OPEs having
the above
Euler beta function from the collinear singularities of the amplitudes
between the particles in the ${\cal N}=4$
$SO(4)$ supergravity are not known so far, although they are known in the
${\cal N}=8$ $SO(8)$ supergravity \cite{BRS2}.
Instead of following the above procedures,
we use the above condition for the helicities directly
to determine
the soft current algebra (\ref{SOFT}) from (\ref{finalanticommcomm})
obtained in the two dimensional conformal field theory.
Note that the above $s_1$ and $s_2$ can be positive, negative or zero
helicity. It would be interesting to obtain
the OPEs
between the conformal primaries having
the Euler beta function in the ${\cal N}=4$ $SO(4)$ supergravity theory,
along the line of \cite{BRS2}.}
\bea
\comm{(\Phi^{(h_1)}_{0, + 0})_m}{(\Phi^{(h_2)}_{0, - 0})_n}
& = &
\kappa_{+0,-0,+2}\,\Big((h_2-1)m-(h_1-1)n \Big)\, (\Phi^{(h_1+h_2-4)}_{2,-2})_{m+n}
\, : \mbox {eq}. 1,
\nonu \\
\comm{(\Phi^{(h_1)}_{0, - 0})_m}{(\Phi^{(h_2),i}_{\frac{1}{2}, +\frac{1}{2}})_r}
& = &
\kappa_{-0,+\frac{1}{2},+\frac{3}{2}}\,\Big((h_2-\tfrac{1}{2})m-(h_1-1)r \Big)\,
(\Phi^{(h_1+h_2-3),i}_{\frac{3}{2}, -\frac{3}{2}})_{m+r} \, :
\mbox {eq}. 2,
\nonu \\
\comm{(\Phi^{(h_1)}_{0, - 0})_m}{(\Phi^{(h_2),ij}_{1, +1})_n} 
& = &
\kappa_{-0,+1,+1}\,\Big(h_2\,m-(h_1-1)n \Big)
\,
\frac{1}{2} \, \epsilon^{ijkl} (\Phi^{(h_1+h_2-2),kl}_{1,-1})_{m+n} \,
: \mbox {eq}. 3,
\nonu \\
\comm{(\Phi^{(h_1)}_{0, - 0})_m}{(\Phi^{(h_2),i}_{\frac{3}{2},+\frac{3}{2}})_r}
& = &
\kappa_{-0,+\frac{3}{2},+\frac{1}{2}}\,
\Big((h_2+\tfrac{1}{2})m-(h_1-1)r \Big)\,
(\Phi^{(h_1+h_2-1),i}_{\frac{1}{2},-\frac{1}{2}})_{m+r} \,
: \mbox {eq}. 4,
\nonu \\
\comm{(\Phi^{(h_1)}_{0, + 0})_m}{(\Phi^{(h_2)}_{2, +2})_n}
& = & \kappa_{+0,+2,-0}\,\Big((h_2+1)m-(h_1-1)n \Big)\,(\Phi^{(h_1+h_2)}_{0,+ 0})_{m+n}\,
: \mbox {eq}. 5-1,
\nonu \\
\comm{(\Phi^{(h_1)}_{0, - 0})_m}{(\Phi^{(h_2)}_{2, +2})_n}
& = & \kappa_{-0,+2,+ 0}\,\Big((h_2+1)m-(h_1-1)n \Big)\,(\Phi^{(h_1+h_2)}_{0,- 0})_{m+n}\,
: \mbox {eq}. 5-2,
\nonu \\
\acomm{(\Phi^{(h_1),i}_{\frac{1}{2},+\frac{1}{2}})_r}{
(\Phi^{(h_2),j}_{\frac{1}{2},-\frac{1}{2}})_s}
& = & \kappa_{+\frac{1}{2},-\frac{1}{2},+2} \, \delta^{ij}\,
\Big((h_2-\tfrac{1}{2})r-(h_1-\tfrac{1}{2})s \Big)\,
(\Phi^{(h_1+h_2-3)}_{2,-2})_{r+s} \, : \mbox {eq}. 6,
\nonu \\
\comm{(\Phi^{(h_1),i}_{\frac{1}{2},-\frac{1}{2}})_r}{(\Phi^{(h_2),jk}_{1,+1})_m}
& = & \kappa_{-\frac{1}{2},+1,+\frac{3}{2}} \, \epsilon^{ijkl}\,
\Big(h_2\,r-(h_1-\tfrac{1}{2})m \Big)\, (\Phi^{(h_1+h_2-2),l}_{\frac{3}{2},
- \frac{3}{2}})_{r+m}
\, : \mbox {eq}. 9,
\nonu \\
\acomm{(\Phi^{(h_1),i}_{\frac{1}{2}, -\frac{1}{2}})_r}{
(\Phi^{(h_2),j}_{\frac{3}{2},+\frac{3}{2}})_s}
& = & \kappa_{-\frac{1}{2},+\frac{3}{2},+1}\,
\Big((h_2+\tfrac{1}{2})\,r-(h_1-\tfrac{1}{2})s \Big)\,\frac{1}{2} \, \epsilon^{ijkl} \,
(\Phi_{1, -1}^{(h_1+h_2-1),kl})_{r+s} \, : \mbox {eq}. 10,
\nonu \\
\acomm{(\Phi^{(h_1),i}_{\frac{1}{2}, +\frac{1}{2}})_r}{
(\Phi^{(h_2),j}_{\frac{3}{2},+\frac{3}{2}})_s}
&= & \kappa_{+\frac{1}{2},+\frac{3}{2},-0}\,\, \delta^{ij}\,
\Big((h_2+\tfrac{1}{2})r-(h_1-\tfrac{1}{2})s\Big)\, (\Phi^{(h_1+h_2)}_{0,+0})_{r+s} \, : \mbox {eq}. 11,
\nonu \\
\comm{(\Phi^{(h_1)}_{2, +2})_m}{(\Phi^{(h_2),i}_{\frac{1}{2}, + \frac{1}{2}})_r}
& = & 
\kappa_{+2,+\frac{1}{2},-\frac{1}{2}}\,
\Big((h_2-\tfrac{1}{2})m-(h_1+1)r \Big)\,(\Phi^{(h_1+h_2),i}_{\frac{1}{2},
+ \frac{1}{2}})_{m+r} \, : \mbox {eq}. 12-1,
\nonu \\
\comm{(\Phi^{(h_1)}_{2, +2})_m}{(\Phi^{(h_2),i}_{\frac{1}{2}, - \frac{1}{2}})_r}
& = & 
\kappa_{+2,-\frac{1}{2},+\frac{1}{2}}\,
\Big((h_2-\tfrac{1}{2})m-(h_1+1)r \Big)\,(\Phi^{(h_1+h_2),i}_{\frac{1}{2},
- \frac{1}{2}})_{m+r} \, : \mbox {eq}. 12-2,
\nonu \\
\comm{(\Phi^{(h_1),ij}_{1,+1})_m}{(\Phi^{(h_2),kl}_{1,-1})_n}
& = &\kappa_{+1,-1,+2}\,\,(\delta^{ik}\delta^{jl}-\delta^{il}\delta^{jk})
\,\Big(h_2\,m-h_1\,n\Big)\,(\Phi^{(h_1+h_2-2)}_{2,-2})_{m+n} \,
: \mbox {eq}. 13,
\nonu \\
\comm{(\Phi^{(h_1),ij}_{1,+1})_m}{(\Phi^{(h_2),kl}_{1,+1})_n}
&= & \kappa_{+1,+1,-0}\,\epsilon^{ijkl}\,\Big(h_2\,m-h_1\,n\Big)\,
(\Phi^{(h_1+h_2)}_{0, +0})_{m+n} \, : \mbox {eq}. 14,
\nonu \\
\comm{(\Phi^{(h_1),ij}_{1,-1})_m}{(\Phi^{(h_2),k}_{\frac{3}{2},+\frac{3}{2}})_r}
& = &
\kappa_{-1,+\frac{3}{2},+\frac{3}{2}}\,
\Big((h_2+\tfrac{1}{2})\,m-h_1\,r\Big)
\nonu
\\
&&\times \bigg[\delta^{ik}\,(\Phi^{(h_1+h_2-1),j}_{\frac{3}{2},-\frac{3}{2}})_{m+r}
-\delta^{jk}\,(\Phi^{(h_1+h_2-1),i}_{\frac{3}{2},-\frac{3}{2}})_{m+r}\bigg]
\, : \mbox {eq}. 15,
\nonu \\
\comm{(\Phi^{(h_1),ij}_{1,+1})_m}{(\Phi^{(h_2),k}_{\frac{3}{2},+\frac{3}{2}})_r}
&= & \kappa_{+1,+\frac{3}{2},-\frac{1}{2}}\,  \epsilon^{ijkl}\,
\Big((h_2+\tfrac{1}{2})m-h_1\,r \Big)\,(\Phi^{(h_1+h_2),l}_{\frac{1}{2},
+\frac{1}{2}})_{m+r} \, : \mbox {eq}. 16,
\nonu \\
\comm{(\Phi^{(h_1),ij}_{1, + 1})_m}{(\Phi^{(h_2)}_{2, +2})_n}
& = & \kappa_{+1,+2,- 1}\Big((h_2+1)m-h_1\,n\Big)\,(\Phi^{(h_1+h_2),ij}_{1, + 1})_{m+n}\,
: \mbox {eq}. 17-1,
\nonu \\
\comm{(\Phi^{(h_1),ij}_{1, - 1})_m}{(\Phi^{(h_2)}_{2, +2})_n}
& = & \kappa_{-1,+2,+ 1}\Big((h_2+1)m-h_1\,n\Big)\,(\Phi^{(h_1+h_2),ij}_{1, - 1})_{m+n}\,
: \mbox {eq}. 17-2,
\nonu \\
\acomm{(\Phi^{(h_1),i}_{\frac{3}{2},+\frac{3}{2}})_r}{(\Phi^{(h_2),j}_{\frac{3}{2},
-\frac{3}{2}})_s}
&= & \kappa_{+\frac{3}{2},-\frac{3}{2},+2}\,\delta^{ij}\,
\Big((h_2+\tfrac{1}{2})\,r-(h_1+\tfrac{1}{2})\,s\Big)\,
(\Phi^{(h_1+h_2-1)}_{2, -2})_{r+s} \, : \mbox {eq}. 18,
\nonu \\
\acomm{(\Phi^{(h_1),i}_{\frac{3}{2}, +\frac{3}{2}})_r}{(\Phi^{(h_2),j}_{
\frac{3}{2}, +\frac{3}{2}})_s}
&= & \kappa_{+\frac{3}{2},+\frac{3}{2},-1}\,\Big((h_2+\tfrac{1}{2})r-(h_1+\tfrac{1}{2})s\Big)\,
(\Phi^{(h_1+h_2),ij}_{1, +1})_{r+s} \, : \mbox {eq}. 19,
\nonu \\
\comm{(\Phi^{(h_1)}_{2, +2})_m}{(\Phi^{(h_2),i}_{\frac{3}{2}, + \frac{3}{2}})_r}
& = & 
\kappa_{+2,+\frac{3}{2},-\frac{3}{2}}\,\Big((h_2+\tfrac{1}{2})m-(h_1+1)r \Big)\,
(\Phi^{(h_1+h_2),i}_{\frac{3}{2}, + \frac{3}{2}})_{m+r}
\, : \mbox {eq}. 20-1,
\nonu \\
\comm{(\Phi^{(h_1)}_{2, +2})_m}{(\Phi^{(h_2),i}_{\frac{3}{2}, - \frac{3}{2}})_r}
& = & 
\kappa_{+2,-\frac{3}{2},+\frac{3}{2}}\,\Big((h_2+\tfrac{1}{2})m-(h_1+1)r \Big)\,
(\Phi^{(h_1+h_2),i}_{\frac{3}{2}, - \frac{3}{2}})_{m+r}
\, : \mbox {eq}. 20-2,
\nonu \\
\comm{(\Phi^{(h_1)}_{2, + 2})_m}{(\Phi^{(h_2)}_{2, + 2})_n}
& = &
\kappa_{+2,+2 ,-2}\Big((h_2+1)m-(h_1+1)n\Big)\,(\Phi^{(h_1+h_2)}_{2, + 2})_{m+n}\,
: \mbox {eq}. 21-1,
\nonu
\\
\comm{(\Phi^{(h_1)}_{2, + 2})_m}{(\Phi^{(h_2)}_{2, - 2})_n}
& = &
\kappa_{+2,-2 ,+2}\Big((h_2+1)m-(h_1+1)n\Big)\,(\Phi^{(h_1+h_2)}_{2, -2})_{m+n}\,
: \mbox {eq}. 21-2,
\label{SOFT}
\eea
where the field contents in the ${\cal N}=4$ supergravity
have the following correspondences 
\bea
A + i B & \longleftrightarrow  &  \Phi^{(h)}_{0,+0} \, ,
\nonu \\
A- i B & \longleftrightarrow  &  \Phi^{(h)}_{0,-0} \, ,
\nonu \\
\chi^{i} & \longleftrightarrow &
\Phi^{(h),i}_{\frac{1}{2},\pm \frac{1}{2}} \, ,
\nonu \\
A^{ij}_{\mu} & \longleftrightarrow &  \Phi^{(h),ij}_{1, \pm 1} \, ,
\nonu \\
\psi^{i}_{\mu} & \longleftrightarrow &
 \Phi^{(h),i}_{\frac{3}{2},\pm \frac{3}{2}} \, ,
\nonu  \\
e_{\mu}^a & \longleftrightarrow  &  \Phi^{(h)}_{2,\pm 2} \, .
\label{CORR}
\eea
Here, the various helicities given by
$(\pm 0, \pm \frac{1}{2}, \pm 1, \pm \frac{3}{2}, \pm 2)$
are denoted at the second elements of
subscripts
\footnote{
As in the footnote \ref{diffordering}, the following
celestial
commutators, by interchanging the two modes on the left hand sides, satisfy  
\bea
\comm{(\Phi^{(h_1),i}_{\frac{1}{2},\pm \frac{1}{2}})_r}{(\Phi^{(h_2)}_{2,+2})_m}
& = &
\Big((h_2+1)r-(h_1-\tfrac{1}{2})m \Big)\,
(\Phi^{(h_1+h_2),i}_{\frac{1}{2}, \pm \frac{1}{2}})_{m+r} \,,
\nonu \\
\comm{(\Phi^{(h_1),i}_{\frac{3}{2},\pm \frac{3}{2}})_r}{(\Phi^{(h_2)}_{2,+2})_m}
& = &
\Big((h_2+1)r-(h_1+\tfrac{1}{2})m\Big)\,(\Phi^{(h_1+h_2),i}_{
\frac{3}{2}, \pm \frac{3}{2}})_{m+r}.
\nonu
\eea
These correspond to twelfth and twentieth of (\ref{SOFT}).

 As noted by \cite{Tropper2412},
  there is no ${\cal N}=1$ super Virasoro subalgebra
  in the celestial soft algebra because
  there is no global superconformal algebra
  generated by five generators (three global conformal generators
  and two supercharges). In the present case,
  the finite dimensional global
  ${\cal N}=4$ superconformal subalgebra
  \cite{Schoutens}
  is generated by
$
  L_{0, \pm 1}$,  $ G^i_{\pm \frac{1}{2}}$ , and $T^{ij}_{0}$
  and their algebra (denoted by the exceptional superalgebra
  $D(2,1| \alpha)$) can be found in (\ref{n4commanticomm}).
We have checked that such subalgebra does not exist  
in (\ref{SOFT}) for the corresponding modes having
positive helicities, $(\Phi^{(0)}_{2, +2})_{0, \pm 1}$,
$(\Phi^{(0),i}_{\frac{3}{2}, +\frac{3}{2}})_{\pm \frac{1}{2}}$,
and $(\Phi^{(0),i j}_{1, +1})_{0}$.
}.

The split factor $\mbox{Split}^{SG}_{-(h+\tilde{h})}(1^{h_1+\tilde{h}_1},
2^{h_2+\tilde{h}_2})$ in the ${\cal N}=4$ supergravity theory,
from the collinear limit between the particle $1$ and the
particle $2$ in the relation of amplitudes,
contains the split factor $\mbox{Split}^{SYM}_{-h}(1^{h_1},
2^{h_2})$
in the ${\cal N}=4$ super Yang-Mills (SYM)
theory multiplied by
the split factor
$\mbox{Split}^{YM}_{-\tilde{h}}(2^{\tilde{h}_2},
1^{\tilde{h}_1})$
in the ${\cal N}=0$  Yang-Mills (YM)
theory \cite{BBJ}. Then we can calculate the following
eight
vanishing  split factors
$\mbox{Split}^{SG}_{-(h+\tilde{h})}(1^{h_1+\tilde{h}_1},
2^{h_2+\tilde{h}_2})$ in the ${\cal N}=4$ supergravity theory
by realizing that either the first split factor or the second split factor
becomes zero and
it turns out that 
there are eight vanishing relations between the split factors
\bea
&& \mbox{Split}_{0}^{SYM}(1^{-\frac{1}{2}},2^{-\frac{1}{2}}) \,
{\it Split_{+ 1}^{YM}(2^{+1},1^{+1})}  \rightarrow 
\mbox{Split}_{+1}^{SG}(1^{-\frac{1}{2}+1},2^{-\frac{1}{2}+1}):
\mbox{helicity} (+\tfrac{1}{2}, +\tfrac{1}{2},+1), 
\nonu \\
&& \mbox{Split}_{-\frac{1}{2}}^{SYM}(1^{-\frac{1}{2}},2^{0}) \,
{\it Split_{+ 1}^{YM}(2^{+1},1^{+1})}  \rightarrow 
\mbox{Split}_{+\frac{1}{2}}^{SG}(1^{-\frac{1}{2}+1},2^{0+1}):
\mbox{helicity} (+\tfrac{1}{2}, +1, +\tfrac{1}{2}), 
\nonu \\
&&
\mbox{Split}_{\pm 1}^{SYM}(1^{0},2^{0}) \,
\mbox{Split}_{- 1}^{YM}(2^{+1},1^{+1})  \rightarrow 
\mbox{trivial} \,\,\, \mbox{Split}_{\pm 1}^{SG}(1^{0+1},2^{0+1}):
\mbox{helicity} (+1, +1, \pm 1), 
\nonu \\
&&
\mbox{Split}_{+ 1}^{SYM}(1^{-1},2^{-1}) \,
{\it Split_{+ 1}^{YM}(2^{+1},1^{+1})}  \rightarrow 
\mbox{Split}_{+2}^{SG}(1^{-1+1},2^{-1+1}):
\mbox{helicity} (+0, +0,+2), 
\nonu \\
&&
{\it Split_{+ 1}^{SYM}(1^{+1},2^{+1})} \,
\mbox{Split}_{+ 1}^{YM}(2^{-1},1^{-1})  \rightarrow 
\mbox{Split}_{+2}^{SG}(1^{1-1},2^{1-1}):
\mbox{helicity} (-0, -0,+2),
\label{split}
\\
&&
\mbox{Split}_{+ \frac{1}{2}}^{SYM}(1^{-\frac{1}{2}},2^{-1}) \,
{\it Split_{+ 1}^{YM}(2^{+1},1^{+1})}  \rightarrow 
\mbox{Split}_{+\frac{3}{2}}^{SG}(1^{-\frac{1}{2}+1},2^{-1+1}):
\mbox{helicity} (+\tfrac{1}{2}, +0, +\tfrac{3}{2}), 
\nonu \\
&&
\mbox{Split}_{0}^{SYM}(1^{0},2^{-1}) \,
{\it Split_{+ 1}^{YM}(2^{+1},1^{+1})}  \rightarrow 
\mbox{Split}_{+1}^{SG}(1^{0+1},2^{-1+1}):
\mbox{helicity} (+1, +0,+1), 
\nonu \\
&&
\mbox{Split}_{-\frac{1}{2}}^{SYM}(1^{+\frac{1}{2}},2^{-1}) \,
{\it Split_{+ 1}^{YM}(2^{+1},1^{+1})}  \rightarrow 
\mbox{Split}_{+\frac{1}{2}}^{SG}(1^{\frac{1}{2}+1},2^{-1+1}):
\mbox{helicity} (+\tfrac{3}{2}, +0, +\tfrac{1}{2}),
\nonu 
\eea
where we follow the notations of \cite{BRS1,BRS2}.
On the left hand side of (\ref{split}),
the split factors for collinear gluons with italic fonts vanish.
On the third of (\ref{split}),
among all possible combinations between the
nonzero split factors from the ${\cal N}=4$ super Yang-Mills theory and the
${\cal N}=0$
Yang-Mills theory,
the split factors corresponding to the helicity $(+1,+1,\pm 1)$
become trivial because the only nonzero $-(h+\tilde{h})$
helicity
is given by $-2$ or $0$ corresponding to the eq. $13$ and the eq. $14$
of (\ref{SOFT}).
We have checked that the remaining split factors for the
$24$ (anti)commutators appearing in (\ref{SOFT}) are nonvanishing
and they can be extracted also from \cite{BRS1,BRS2}.
In particular, when the $SO(4)$ indices
satisfy $(k,l) =(i,j)$ in eq. $14$ of (\ref{SOFT}),
the right hand side of the commutator vanishes
(that is, there is no first order pole coming from a new primary
operator in the corresponding OPE).
Similarly, in eq. $21-1$, there is no new primary operator
in the first order pole.
Moreover,
when the $SO(4)$ indices
satisfy $i=j$
in eq. $19$ of (\ref{SOFT}),
there is no second order pole in the corresponding OPE
\footnote{ We can calculate
  the following (anti)commutator (ignoring $SO(4)$ indices)
  between the soft currents
  from the celestial OPE
  \cite{HPS,MRSV} with $s_1+s_2+s_3=2$ by absorbing the infinities
appearing in the Euler beta function  
\bea
&&
{\cal O}_{\Delta_1,s_1}(z_1,\bar{z}_1) {\cal O}_{\Delta_2,s_2}(z_2,\bar{z}_2)
=
 \frac{1}{z_{12}}
\sum_{\al=0}^{\infty}
\binom{-\Delta_1-\Delta_2+s_1+s_2-\al-2}{-\Delta_2+s_2-1}\,
\frac{\bar{z}_{12}^{1+\al}}{\al!}
\, \partial^{\al}_{\bar{z}_2}\, {\cal O}_{\Delta_3,-s_3}(z_2,\bar{z}_2) + \cdots,
\nonumber
\eea
by using the procedure in \cite{Strominger2105,GHPS}
and it turns out (See also \cite{Tropper2412})
that the corresponding (anti)commutator leads to
\bea
&& \bigg[ ({\cal O}_{\Delta_1,s_1})_m, ({\cal O}_{\Delta_2,s_2})_n \bigg\}  
=
2\,\bigg(\tfrac{\Delta_2-s_2}{2}\,m-\tfrac{\Delta_1-s_1}{2}\,n\bigg)
\nonu \\
&& \times \Bigg[
  \frac{(-\tfrac{\Delta_1-s_1}{2}-\tfrac{\Delta_2-s_2}{2}-m-n-1)!(
    -\tfrac{\Delta_1-s_1}{2}-\tfrac{\Delta_2-s_2}{2}+m+n-1)!}{
    (-\tfrac{\Delta_1-s_1}{2}-m)!(-\tfrac{\Delta_2-s_2}{2}-n)!(-
    \tfrac{\Delta_1-s_1}{2}+m)!(-\tfrac{\Delta_2-s_2}{2}+n)!}
\Bigg]
({\cal O}_{\Delta_1+\Delta_2,s_1+s_2-2})_{m+n}.
\label{anticommfootnote}
\eea
Finally, by multiplying the denominator of the right hand in this equation
and redefining each two factorials multiplied each celestial operator
as the new celestial operator,
we obtain the final (anti)commutators in (\ref{SOFT})
by adding the couplings. Note that
the coefficients (the right conformal weights) in the modes $m$ and $n$
for the first factor on the right hand side of (\ref{anticommfootnote})
are replaced by $(\tfrac{\Delta_2-s_2}{2}-1)$ and $(\tfrac{\Delta_1-s_1}{2}-1)$
respectively in (\ref{SOFT}) with proper $SO(4)$ indices.
}.

The Jacobi identities can be calculated from (\ref{SOFT})
and there exist the following relations between the couplings
\bea
\kappa_{+0, -0, +2} & = &
-\frac{
  \kappa_{-\frac{1}{2},+ 1, +\frac{3}{2}}\, \kappa_{+\frac{3}{2},
    -\frac{3}{2}, +2}\, \kappa_{-0, +\frac{3}{2}, +\frac{1}{2}}}{\kappa_{+1,+ 1, -0} \,\kappa_{+\frac{3}{2}, +\frac{3}{2}, -1}}\,, 
\qquad
\kappa_{+0, +2, -0}=\kappa_{-0, +2,+0}\,, 
\nonu \\
\kappa_{+\frac{1}{2}, -\frac{1}{2},+ 2}
& = & -\frac{  \kappa_{+\frac{3}{2}, -\frac{3}{2}, +2}\, \kappa_{-0, +\frac{1}{2},
    +\frac{3}{2}}} { \kappa_{-0, +\frac{3}{2}, +\frac{1}{2}}}\,, 
\qquad
\kappa_{-\frac{1}{2}, +\frac{3}{2}, +1}
=\frac{\kappa_{+\frac{3}{2}, +\frac{3}{2}, -1}\, \kappa_{-0,+ 1, +1}}{
 \kappa_{-0, +\frac{3}{2}, +\frac{1}{2}}}\,,
\nonu \\
\kappa_{+\frac{1}{2}, +\frac{3}{2}, -0} & = & \frac{
  \kappa_{+1,+ 1, -0} \,\kappa_{+\frac{3}{2}, +\frac{3}{2}, -1}\, \kappa_{-0,+
    \frac{1}{2}, +\frac{3}{2}}}{
  \kappa_{-\frac{1}{2},+ 1, +\frac{3}{2}} \,\kappa_{-0, +\frac{3}{2},+
    \frac{1}{2}}}\,,
\qquad
\kappa_{+2, +\frac{1}{2}, -\frac{1}{2}}=\kappa_{-0, +2,+0}\,, 
\nonu \\
\kappa_{+2, -\frac{1}{2}, +\frac{1}{2}} & = & \kappa_{-0, +2,+0}\,, 
\qquad
\kappa_{+1, -1, +2}=-\frac{
  \kappa_{-\frac{1}{2},+ 1, +\frac{3}{2}}\,
  \kappa_{+\frac{3}{2}, -\frac{3}{2}, +2}\,
  \kappa_{-0, +\frac{3}{2}, +\frac{1}{2}}}{\kappa_{+\frac{3}{2}, +\frac{3}{2},
    -1}
  \, \kappa_{-0, +1, +1}}\,, 
\nonu \\
\kappa_{-1, +\frac{3}{2}, +\frac{3}{2}}
&  = & \frac{\kappa_{-\frac{1}{2},+1, +\frac{3}{2}}\, \kappa_{-0, +\frac{3}{2},
      +\frac{1}{2}}}{
 \kappa_{-0, +1, +1}}\,,
\qquad
\kappa_{+1, +\frac{3}{2}, -\frac{1}{2}}=\frac{
  \kappa_{-\frac{1}{2}, +1, +\frac{3}{2}}\, \kappa_{-0, +\frac{3}{2}, +
    \frac{1}{2}}}{
 \kappa_{-0, +\frac{1}{2}, +\frac{3}{2}}}\,,
\nonu \\
 \kappa_{+1, +2, -1}& = & \kappa_{-0, +2,+0}, 
\qquad
 \kappa_{-1, +2, +1}=\kappa_{-0,+2,+0}\,, 
\qquad
\kappa_{+2, +\frac{3}{2}, -\frac{3}{2}}=\kappa_{-0, +2,+0}\,, 
\nonu \\
\kappa_{+2, -\frac{3}{2}, +\frac{3}{2}} & = & \kappa_{-0, +2,+0}\,, 
\qquad
\kappa_{+2, +2, -2}=\kappa_{-0, +2,+0}\,,
\qquad
\kappa_{+2, -2, +2}=\kappa_{-0, +2,+0}.
\label{kapparelation}
\eea
We have checked the Jacobi identities for $ 1 \leq h_1, h_2 \leq 6$ but
we can do this for general $h_1$ and $h_2$.
The sixteen couplings in the $24$ (anti)commutators
can be written in terms of
the eight arbitrary couplings.
The three minus signs appearing on the right hand sides of (\ref{kapparelation})
imply that the simplest solution satisfying (\ref{kapparelation})
is given by $ \kappa_{+\frac{3}{2},
  -\frac{3}{2}, +2}=-1$
which appears in the above three places
with the remaining $\kappa_{s_1,s_2, -s_3}$ having $+1$
\footnote{
  Note that there are relations between the couplings
  appearing in the eqs. $6,11,18$ of (\ref{SOFT})
  as follows:
  $\kappa_{-\frac{1}{2}, +\frac{1}{2},+2}= -
  \kappa_{+\frac{1}{2}, -\frac{1}{2},+2}$, $\kappa_{+\frac{3}{2},+\frac{1}{2},-0}=
  - \kappa_{+\frac{1}{2},+\frac{3}{2},-0}$ and $
  \kappa_{-\frac{3}{2},+\frac{3}{2},+2}= -  \kappa_{+\frac{3}{2},-\frac{3}{2},+2}$.
  Moreover, there exist the relations
  between the couplings
$\kappa_{s_2,s_1,-s_3}=\kappa_{s_1,s_2,-s_3}$
  in the remaining eqs. of (\ref{SOFT}).
  Note that there are no sign changes
in  the anticommutators appearing in the eqs. $10$ and $19$
  of (\ref{SOFT}) compared to the previous anticommutators.
  
  The $U(1)$ charges \cite{BPR} for
  the graviton with helicity $\pm 2$,
 the gravitinos with helicity $\pm \frac{3}{2}$,
  the vectors with helicity $\pm 1$, the Majoranas with helicity $\pm
  \frac{1}{2}$,
  the complex scalar with helicity $+0$,
  and the conjugate scalar with helicity $-0$ are given by
  $0, \pm \frac{1}{2}, \pm 1, \pm \frac{3}{2}, +2, -2$.
  We observe that this $U(1)$ charge is conserved in (\ref{SOFT}).
  In other words, the complete algebra
  (consisting of $24$ (anti)commutators among all possible $55$ ones from
$10$ operators in (\ref{CORR}))  satisfying both the condition $s_1+s_2+s_3=2$
  and the $U(1)$ charge is given by (\ref{SOFT}).}.

\subsection{The ${\cal N}=4$ supergravity analysis}

Let us describe how we obtain the above soft current algebra
(\ref{SOFT})
by analyzing the Lagrangian of \cite{Das} and using the relations
(\ref{CORR}).

$\bullet$ The first $\frac{1}{\kappa^2} \, e \, R$ term

For this term of Lagrangian in \cite{Das}, $e$ is the determinant
of vierbein $e \equiv \mbox{det} \, e_{\mu}^a$,
$R$ is a scalar curvature and $\kappa$ is a gravitational coupling.
The metric (inverse metric, its determinant, affine connection
and a scalar curvature) and 
the vierbein (and its determinant) 
are expanded around the flat Minkowski spacetime
\cite{DeWitt,BG,Woodard,CSS}.
The cubic gravitons with two derivatives
appear in the linear $\kappa$ term of the expansion of
this term (due to the overall factor
$\frac{1}{\kappa^2}$ in this term).
Then the scaling dimension of
three point vertex $d_V$ is given by
the sum of three (coming from the three gravitons) and
two (coming from two derivatives) \cite{PRSY}.
The graviton has the scaling dimension one.
Moreover, the sum of helicities for
these gravitons is given by $(d_V-3)$ \cite{HPS}
from the identifications of the left and right conformal weights
of both sides in the celestial OPE.
This implies that by substituting the above $d_V=5$ into this
relation, the sum of helicities of those gravitons becomes two
\footnote{In this paper, we focus on the $d_V=5$ case (i.e., the sum of
helicities is given by two) mainly.}.  
Therefore, we can easily see that
the helicities $(+2,\pm 2, \mp 2)$ for three gravitons
with two derivatives 
should appear in the coupling of this three point graviton
amplitude and 
the corresponding two celestial commutators are given by the twenty first
relation of (\ref{SOFT}) \footnote{
\label{21relation}
The helicities on the
right hand sides of (\ref{SOFT}) appear negatively \cite{PRSY}.
When we write down the twenty first
relation of (\ref{SOFT}) as the OPE, then
the right hand side of OPE in the
antiholomorphic sector contains the second and the first order
poles \cite{Ahn2111}. The right conformal
weight of the left hand side
$(h_1+2+h_2+2)$ is consistent with the one of the right hand side.
See also the last of (\ref{opeversion}).}.

$\bullet$ The second 
$\epsilon^{\mu \nu \sigma \rho} \, \bar{\psi}^i_{\mu} \, \ga_5\, \ga_{\nu}\,
D_{\rho}\, \psi_{\si}^i$  term

The Majorana conjugate of the spin $\frac{3}{2}$ spinor (the gravitino)
is given by its transpose and the charge conjugation matrix
\footnote{Sometimes the factor
$\epsilon^{\mu \nu \sigma \rho}  \, \ga_5\, \ga_{\nu}$
can be written in terms of
the linear combination of triple product of gamma matrices \cite{FV}.}.
The $\ga_5$ matrix is a constant \cite{VanNieuwenhuizen}.
Because the $\ga_{\nu}$ matrix is given by
the contraction between the $\ga^a$ and above vierbein,
its expansion around the flat Minkowski spacetime \cite{Woodard}
provides the linear $\kappa$ term having a graviton.
Furthermore, the covariant derivative
acting on the gravitinos 
consists of the partial derivative term,
the Christoffel symbol term and the vierbein (or spin) connection term.
However, due to the presence of epsilon tensor in this term,
the contribution from the  Christoffel symbol term does not contribute.
We can compute the scaling dimension of the three point vertex between
the graviton and two gravitinos together with a
single derivative and it is given by
$d_V = 1+ 2 \times \frac{3}{2}+1=5$ where the number $\frac{3}{2}$
comes from the scaling dimension of  a gravitino.
Again, this 
leads to the fact that the sum of helicities should be equal to
two according to the previous analysis.
The helicities $(+2,\pm \frac{3}{2}, \mp \frac{3}{2})$
for the graviton and two gravitinos
with a single derivative 
should appear in the coupling of this three point amplitude and 
the corresponding celestial commutator
is given by the twentieth relations of (\ref{SOFT})
\footnote{
\label{commexample}
When we write down the twentieth
relation of (\ref{SOFT}) in terms of  the OPE, 
the right hand side of OPE in the
antiholomorphic sector contains the second and the first order
poles \cite{Ahn2111}. The right conformal
weight of the left hand side
$(h_1+2+h_2+\frac{3}{2})$ is
consistent with the one of the right hand side of this OPE.
See also the second from the below in (\ref{opeversion}).}
while
the helicities $(+ \frac{3}{2}, - \frac{3}{2},+2)$
for these appear in the anticommutator given by
the eighteenth relation of (\ref{SOFT}) \footnote{
For the OPE, 
the right hand side of OPE in the antiholomorphic sector
contains the first order
pole.
See also the third from below in (\ref{opeversion}).
The weight of the left hand side
$(h_1+\frac{3}{2}+h_2+\frac{3}{2})$
is consistent with the one of the right hand side.
We can easily observe that the OPE between the
operator corresponding to the
mode for the helicity $-\frac{3}{2}$ and
the operator corresponding to 
the mode
for the helicity $+\frac{3}{2}$ is the same as the right hand side
of the eighteenth relation of (\ref{SOFT}) because there exists an extra
minus sign for the fermionic property and there is other extra
minus sign for the interchange of two complex arguments appearing on the
right hand side.
\label{anticommexample}}.
Because the $SO(4)$ indices are summed over the gravitinos,
if we ignore them with $i=2,3,4$, then
the previous results in \cite{AK2407} can be reproduced
as long as this particular interaction is concerned.

$\bullet$ The third
$e \, g^{\mu \rho} \, g^{\nu \si}\, F_{\mu \nu}^{ij} \, F_{\rho \si}^{ i j}$
term

Here $F_{\mu \nu}^{ij}$ is given by
$F_{\mu \nu}^{ij} \equiv \pa_{\mu} \, A_{\nu}^{ij} - \pa_{\nu}\, A_{\mu}^{ij}$.
As for three gravitons case, the determinant of vierbein $e$
contains the linear term in the $\kappa$.
Of course, the $\kappa$ independent term (the lowest order term in $e$) is
the kinetic term for the vectors.
We can calculate the scaling dimension of three point vertex
between the graviton and two vectors together with two
derivatives and it is given by
$d_V=5$ (three from the contributions of graviton and two vectors and
two from the two derivatives). Then the sum of helicities
is given by two again. The helicities $(+1,-1,+2)$ for two vectors
and graviton arise in the celestial commutator 
of thirteenth relation of (\ref{SOFT})
\footnote{When the mode independent product of two deltas
appearing on the right hand side is contracted with the second
mode having $h_2$ on the left hand side, then the $SO(4)$ indices
become $i j$.} and
similarly the helicities $(\pm 1, +2, \mp 1)$ appear in the
seventeenth relation of (\ref{SOFT}) \footnote{
Because the $SO(4)$ indices are summed over the vectors
in this third term,
if we keep only the first contribution
having $(i,j)=(1,2)$ indices
by ignoring the remaining terms, then
the previous results in \cite{AK2407} can be reproduced
if we are interested in this particular interaction.}.

$\bullet$ The fourth
$e \, \bar{\chi}^i \, \ga^{\mu} \, D_{\mu}\, \chi^i$
term

After extracting the
usual kinetic term, there exists a linear term in the
$\kappa$ from the expansion of the determinant
of vierbein around the flat Minkowski spacetime as before.
We can calculate the scaling dimension of the three point vertex between
the graviton and two Majoranas together with a
single derivative and it is given, as before, by
$d_V = 1+ 2 \times \frac{3}{2}+1=5$ where the number $\frac{3}{2}$
comes from the scaling dimension from a Majorana.
Again, the sum of helicities should be equal to
two.
The helicities $(+ \frac{1}{2}, - \frac{1}{2},+2)$
for  the graviton and two Majoranas
with a single derivative  should  appear in the celestial
anticommutator given by
the sixth relation of (\ref{SOFT}) \footnote{
The corresponding OPE looks very similar to
the one described in the footnote \ref{anticommexample}
associated with the eighteenth relation of (\ref{SOFT}).
See also the sixth of (\ref{opeversion}).}.
Moreover, 
the helicities $(+2,\pm \frac{1}{2}, \mp \frac{1}{2})$
for these
should appear in the coupling of this three point amplitude and 
the corresponding celestial commutator
is given by the twelfth relations of (\ref{SOFT}) 
\footnote{
As before, the corresponding OPE looks similar to
the one described in the footnote \ref{commexample}
associated with the twentieth relation of (\ref{SOFT}).
Because the $SO(4)$ indices are summed over the Majoranas
in this fourth term,
if we keep the contribution from the first term, then
the previous analysis in \cite{AK2407} can be reproduced
for this particular interaction. See also the ninth of
(\ref{opeversion}).}.

$\bullet$ The fifth
$e \, g^{\mu \nu}\, (\pa_{\mu}\, A) \, (\pa_{\nu}\, A)$
and sixth 
$e \, g^{\mu \nu}\, (\pa_{\mu}\, B) \, (\pa_{\nu}\, B)$
terms

After identifying
the lowest order term
in the expansion of the determinant of vierbein
around the flat Minkowski spacetime
with the kinetic term,
the next linear $\kappa$ term provides the interaction
between the graviton and two scalars (or two pseudoscalars)
with two derivatives.
The sum of helicities should be
equal to two as before and 
we observe that the first celestial commutator
in (\ref{SOFT}) shows the helicities $(+ 0, - 0, +2)$
for complex scalar, complex conjugated scalar and graviton.
Similarly,  
the fifth celestial commutator of (\ref{SOFT}) shows 
the helicities $(\pm 0, +2, \mp 0)$
for  complex scalar (or conjugated one), graviton and
conjugated scalar (or complex one)  \footnote{
The corresponding OPEs can be determined by
using the description given in the footnote \ref{21relation}.  
Because there is no $SO(4)$ index in this fifth term of Lagrangian,
the analysis of this interaction can be seen in the previous work in
\cite{AK2407}.
See also the first and the fifth of (\ref{opeversion}).}.

$\bullet$ The seventh
$\kappa \,  e\, g^{\mu \rho} \, g^{\nu \si}\,
\bar{\psi}^i_{\mu} \,  F_{\rho \si}^{ i j}\, \psi_{\nu}^j$
and eighth
$\kappa \,   \epsilon^{\mu \nu \rho \si} \,
\bar{\psi}^i_{\mu} \,  \ga_5\, F_{\rho \si}^{ i j}\, \psi_{\nu}^j$
terms

In this case, because of the overall factor $\kappa$,
we consider the interactions between two gravitinos and the vectors.
Due to the several indices of $SO(4)$, we do not see
these interactions in the ${\cal N}=1$ supergravity (and its matters)
described by \cite{AK2407}.
Note that the $SO(4)$ indices for the two gravitinos
are contracted with the ones for the vectors.
There is no epsilon tensor for the $SO(4)$.
This feature allows us to consider the particular case of
the vanishing parameter $\al=0$ in (\ref{n4sca}).
The sum of helicities should be
equal to two as before and 
the fifteenth celestial commutator
in (\ref{SOFT}) provides the helicities $(-1, +\frac{3}{2}, +
\frac{3}{2})$
for vector and two gravitinos \footnote{On the right hand side
of the commutator,
the Kronecker deltas provide this particular interactions of these terms
because the relative signs between two terms
are opposite.}.
Similarly,  
the nineteenth celestial anticommutator of (\ref{SOFT}) gives 
the helicities $(+\frac{3}{2}, +\frac{3}{2}, -1)$
for  two gravitinos and vector
\footnote{We observe that the second order pole of
  the corresponding OPE (see also the third from the below
of (\ref{opeversion})) between the
operator corresponding to
the mode for the helicity $+\frac{3}{2}$ with an index $j$ having
the weight $h_2$ and the operator corresponding to the mode
for the helicity $+\frac{3}{2}$ with an index $i$
having the weight
$h_1$ is the same as the minus of
the second order pole of the right hand side
of the OPE corresponding to the
nineteenth relation of (\ref{SOFT}). This is
because there exists an extra
minus sign for the fermionic property and the interchange of two
arguments in the complex coordinates appearing on the left hand side
remains the same for the quadratic behavior of their difference.
The first order pole has the similar behavior but the
coefficient contains the $h_2$ dependence rather than the $h_1$
\cite{Ahn2111}. Or
for given OPE on the nineteenth relation, we can exchange
$h_1 \leftrightarrow h_2$ and $i \leftrightarrow j$ and obtain the
above OPE. This implies that compared to the previous case
(an odd number of pole)
in the footnote \ref{anticommexample}, the second order pole
(an even number of pole)
occurs for the antisymmetric combination of $SO(4)$ indices
on the right hand side of the anticommutator.}. 

$\bullet$ The ninth
$\kappa \, e \, \epsilon^{i j k l} \, \bar{\psi}^i_{\la} \,
\sigma^{\mu \nu} \, F^{k l}_{\mu \nu}\, \gamma^{\la}\, \chi^j$
term

There exists a single derivative from the vector.
The $\sigma^{\mu \nu}$ is the antisymmetric
gamma matrices and is contracted with the vector.
The single gamma matrix is contracted with the gravitino \footnote{
If we restrict to consider the only one single term
in the $SO(4)$ indices, then we observe that this kind of
interaction appeared in previous description in \cite{AK2407}.}.
Here we consider other remaining terms by extending it to
the $SO(4)$ case with the help of epsilon tensor.
The ninth celestial commutator of (\ref{SOFT})
contains the helicities $(-\frac{1}{2},+1,+\frac{3}{2})$
for a Majorana, vector and gravitino.
The tenth celestial anticommutator of (\ref{SOFT})
provides the helicities $(-\frac{1}{2}, +\frac{3}{2},+1)$
between the Majorana, gravitino and vector \footnote{
\label{antisymmandsymm} Note that
compared to the previous case described in the footnote
\ref{anticommexample} (the same kinds of
fermions with different helicities), the right hand side of this
anticommutator, where
the left hand side has different kinds of
fermions, is antisymmetric in the $SO(4)$  indices.
Furthermore, the right hand side of
eleventh celestial anticommutator in (\ref{SOFT}) is symmetric under the
interchange of two $SO(4)$ indices.}.
Similarly, the sixteenth celestial
commutator of (\ref{SOFT}) gives
the helicities $(+1, +\frac{3}{2},-\frac{1}{2})$
for vector, gravitino and a Majorana.
The epsilon tensor appears on the right hand sides of these
celestial (anti)commutators.

$\bullet$ The tenth
$\kappa \, e\, \bar{\psi}^i_{\mu} \, (\pa_{\nu} \, A) \, \ga^{\nu}\,
\ga^{\mu} \, \chi^i$ and eleventh
$\kappa \, e\, \bar{\psi}^i_{\mu} \, (\pa_{\nu} \, \ga_5\, B) \, \ga^{\nu}\,
\ga^{\mu} \, \chi^i$
terms

There is a single derivative \footnote{
If we consider the only one single term
in the $SO(4)$ indices, then this kind of
interaction can be seen from the
previous description in \cite{AK2407}.}.
We consider other remaining terms by extending it to
the $SO(4)$ case with the help of Kronecker delta tensor
which is invariant under the $SO(4)$.
The helicities of second celestial commutator of (\ref{SOFT})
are given by $(- 0, +\frac{1}{2},+\frac{3}{2})$
for the conjugated scalar, a Majorana and the gravitino.
The helicities of fourth celestial commutator
of (\ref{SOFT}) are
$(- 0, +\frac{3}{2},+\frac{1}{2})$
between the conjugated scalar, the gravitino
and the Majorana.
The helicities of eleventh celestial anticommutator
of (\ref{SOFT}) are denoted by
$(+\frac{1}{2},+\frac{3}{2}, - 0)$
for the Majorana, the gravitino and the conjugated scalar
\footnote{As noticed in the footnote \ref{antisymmandsymm},
the right hand side is symmetric under the
interchange of two $SO(4)$ indices.}.

$\bullet$ The twelfth
$\kappa \, e \, g^{\mu \rho} \, g^{\nu \si}\,
\epsilon^{i j k l} \, A\,
F_{\mu \nu}^{ij} \, F_{\rho \si}^{ k l}$
and thirteenth
$\kappa \, e \, g^{\mu \rho} \, g^{\nu \si}\,
\epsilon^{i j k l} \, \epsilon^{\mu \nu \rho \si} \, B\,
F_{\mu \nu}^{ij} \, F_{\rho \si}^{ k l}$
terms

In this case also, due to the presence of
the overall factor $\kappa$,
we consider the interactions between two vectors and scalar
(or pseudoscalar).
Due to the several indices of $SO(4)$, 
these interactions in the ${\cal N}=1$ supergravity (and its matters)
described by \cite{AK2407} are not described before.
Note that the $SO(4)$ indices in the vectors are contracted with
epsilon tensors.
The helicities of third celestial commutator
of (\ref{SOFT}) are
$(- 0, +1,+1)$
between the conjugated scalar and two vectors.
Similarly,
the helicities of fourteenth celestial commutator
of (\ref{SOFT}) are
$(+1,+1, - 0)$
between two vectors and  the conjugated scalar.
The epsilon tensor appears on the right hand sides of these commutators.

In summary, the thirteen terms of
the Lagrangian in \cite{Das}
have their soft current algebra
characterized by (\ref{SOFT})
\footnote{In \cite{dF}, there exists
a term $\kappa \, e \, \epsilon^{i j k l m n p q}\,
\bar{\chi}^{i j k}\, \sigma^{\mu \nu}\,
\chi^{l m n}\, F_{\mu \nu}^{p q}$ which is so called
``Pauli moment'' coupling in the Lagrangian of
${\cal N}=8$ $SO(8)$ supergravity theory.
By considering $ \chi^{i j k}= \epsilon^{i j k l} \, \chi^l$
from the truncation,
the term
$\kappa \, e \, \bar{\chi}^i \,
\sigma^{\mu \nu} \,  \chi^j \, F^{i j}_{\mu \nu}$ cannot appear
in the ${\cal N}=4$ $SO(4)$ supergravity 
because the factor $\epsilon^{i j k l m n p q}$ is identically zero
after imposing that the indices are restricted to the values
$1,2,3$ and $4$.
On the other hand, the helicities
in these (anti)commutators can take
$(\pm \frac{1}{2},\mp \frac{1}{2},+1)$
or $(\pm \frac{1}{2},+1, \mp \frac{1}{2})$
by reducing the total number of helicities.
This leads to 
$d_V=4$
corresponding to the interaction with
two Majoranas and the vectors without a derivative.
See also the equation $(6.48)$ of \cite{FV} for this kind of
interaction.
According to the observation of \cite{PRSY}, for example,
the OPE
between the Majorana
operator at $(z_1, \bar{z}_1)$ and the Majorana
at $(z_2, \bar{z}_2)$
on the celestial sphere
contains the singular term
$\frac{\bar{z}_{12}^{d_V-4}}{z_{12}}\Big|_{d_V=4}=
\frac{1}{z_{12}}$ 
on the right hand side where $z_{12} \equiv z_1-z_2$
and $\bar{z}_{12} \equiv \bar{z}_1-\bar{z}_2$.
Furthermore, there is no such term on the
right hand side of the corresponding OPE
for the ${\cal N}=8$ $SO(8)$ supergravity theory
\cite{BRS2}
because  there exist only
the singular terms
$\frac{\bar{z}_{12}^{d_V-4}}{z_{12}}\Big|_{d_V=5}=
\frac{\bar{z}_{12}}{z_{12}}$ in the limit of $z_{12} \rightarrow 0
$ with fixed $\bar{z}_1$ and $\bar{z}_2$.
Therefore, 
the above Pauli moment coupling
with the helicities $(+\frac{1}{2},+\frac{1}{2},+1)$
and $(+\frac{1}{2},+1,+\frac{1}{2})$
should appear in the ${\cal N}=8$ $SO(8)$ supergravity.
Recall that in (\ref{split}), the first two split factors
become zero in the ${\cal N}=4$ $SO(4)$ supergravity.
}.
Moreover, from the tenth commutator of
(\ref{finalanticommcomm}), there are also the $SO(4)$ vectors
appearing at the last four terms on the right hand side.
If we consider three $+1$ helicities, then
the $d_V$ becomes $d_V=6$
\footnote{In this case, there is the singular  term
$\frac{\bar{z}_{12}^{d_V-4}}{z_{12}}\Big|_{d_V=6}=
\frac{\bar{z}_{12}^2}{z_{12}}$ on the
right hand side in the
corresponding OPE. However,
note that  in (\ref{split}), the third split factor
becomes zero in the ${\cal N}=4$ $SO(4)$ supergravity theory.

Let us introduce the modes \cite{Tropper2412} for the two
${\cal N}=4$ multiplets \cite{BPR}
\bea
{\bf (\Phi^{+}_{h,s})}_m(  \eta) & = &
\Phi^{(h-2)}_{2,+2,m}
+\eta^i\,\Phi^{(h-2),i}_{\frac{3}{2},+\frac{3}{2},m}
+\frac{1}{2!} \, \eta^j\, \eta^i \,\Phi^{(h-2),ij}_{1,+1,m}
+\frac{1}{3!}\, \eta^k\, \eta^j \, \eta^i \,
\epsilon^{i j k l} \, \Phi^{(h-2),l}_{\frac{1}{2},+\frac{1}{2},m}
\nonu \\
& + & \frac{1}{4!} \, \eta^{l}\, \eta^{k}\, \eta^{j}\,
\eta^i \, \epsilon^{i j k l} \, \Phi^{(h-2)}_{0, +0,m},
\nonu \\
{\bf (\Phi^{-}_{h,s})}_m(\eta) & = &
\Phi^{(h)}_{0,-0,m}
+\eta^i\,\Phi^{(h-1),i}_{\frac{1}{2},-\frac{1}{2},m}
+\frac{1}{2!} \, \eta^j\, \eta^i \,\epsilon^{i j k l} \,
\Phi^{(h-2),k l}_{1,-1,m}
+\frac{1}{3!}\, \eta^k\, \eta^j \, \eta^i \,
\epsilon^{i j k l} \, \Phi^{(h-3),l}_{\frac{3}{2},-\frac{3}{2},m}
\nonu \\
& + & \frac{1}{4!} \, \eta^{l}\, \eta^{k}\, \eta^{j}\,
\eta^i \, \epsilon^{i j k l} \, \Phi^{(h-4)}_{2, -2,m}.
\label{phi+phi-}
\eea
The $\eta^i$ with $i=1,2,3,4$ are the Grassmann coordinates.
The spins and helicities for the first multiplet are given by
${\bf h}=(h,h-\frac{1}{2},h-1,h-\frac{3}{2},h-2)$ and
${\bf s}=(+2,+\frac{3}{2},+1,+\frac{1}{2},+0)$ respectively
while
those for the second one are
${\bf h} =(h,h-\frac{1}{2},h-1,h-\frac{3}{2},h-2)$ and
${\bf s}=(-0,-\frac{1}{2},-1,-\frac{3}{2},-2)$.
Note the $SO(4)$ indices in the third term of the second multiplet
consistent with the eq. of $13$ and $15$ of (\ref{SOFT})
compared to the one of \cite{BPR}.

Then the above (anti)commutators (\ref{SOFT}) can be written as  
\bea
\comm{{\bf (\Phi^{+}_{h_1,s_1})}_m(\eta_1)}{{\bf
    (\Phi^{+}_{h_2,s_2})}_n(\eta_2)}
& = &
\kappa_{{\bf s_1,s_2 ,-s_1-s_2+2}}\Big(({\bf {h}_2}-1) m-
({\bf {h}_1}-1) n\Big)\,
{\bf (\Phi^{+}_{h_1+h_2-2,s_1+s_2-2})}_{m+n}(\eta_1+\eta_2),
\nonu \\
\comm{{\bf (\Phi^{+}_{h_1,s_1})}_m(\eta_1)}{{\bf
    (\Phi^{-}_{h_2,s_2})}_n(\eta_2)}
& = &
\kappa_{{\bf s_1,s_2 ,-s_1-s_2+2}}\Big(({\bf {h}_2}-1) m-
({\bf {h}_1}-1) n\Big)\,
      {\bf (\Phi^{-}_{h_1+h_2-2,s_1+s_2-2})}_{m+n}(\eta_1+\eta_2).
      \nonu \\
\label{twosuper}
\eea
We emphasize that
the quantities ${\bf s_1}$, ${\bf s_2}$, ${\bf h_1}$
and ${\bf h_2}$ on the right hand sides of (\ref{twosuper})
are multivalued.
They can be fixed by choosing the specific locations of the components
appearing on the left hand sides of (\ref{twosuper}).
For example, when we choose the third and second components
on the left hand side of the first relation in (\ref{twosuper}), then
from the previous paragraph, ${\bf h_1}=h_1-1,
{\bf s_1}=+1$ and ${\bf h_2}=h_2-\frac{1}{2}, {\bf s_2}=+\frac{3}{2}$
corresponding to the eq. $16$ of (\ref{SOFT}).
Note that each mode of the components in (\ref{phi+phi-})
can be determined by
\bea
{\bf (\Phi^{+}_{h,s})}_m(  \eta)\Big|_{\eta^i=0} & = &
\Phi_{2, +2,m}^{(h-2)}, \qquad
\frac{\pa}{\pa \eta^i} {\bf (\Phi^{+}_{h,s})}_m(  \eta)\Big|_{\eta^j=0}=
\Phi_{\frac{3}{2}, +\frac{3}{2},m}^{(h-2),i},
\nonu \\
\frac{\pa}{\pa \eta^i} \frac{\pa}{\pa \eta^j}
{\bf (\Phi^{+}_{h,s})}_m(  \eta)\Big|_{\eta^k=0} & = &
\Phi_{1, +1,m}^{(h-2),i j},    
\nonu \\
\frac{\pa}{\pa \eta^i} \frac{\pa}{\pa \eta^j}
\frac{\pa}{\pa \eta^k}
{\bf (\Phi^{+}_{h,s})}_m(  \eta)\Big|_{\eta^p=0}& = & \epsilon^{i j k l}
\Phi_{\frac{1}{2}, +\frac{1}{2},m}^{(h-2),l}, \qquad
\frac{\pa}{\pa \eta^l} \frac{\pa}{\pa \eta^k}
\frac{\pa}{\pa \eta^j} \frac{\pa}{\pa \eta^i} 
{\bf (\Phi^{+}_{h,s})}_m(  \eta)\Big|_{\eta^p=0} =  \epsilon^{i j k l}\Phi_{0, +0,m}^{(h-2)}.
    \nonu
    \eea
    Similarly, those for the second multiplet
can be obtained by taking the Grassmann derivatives.    
Therefore, by taking the Grassmann derivatives into
both sides of (\ref{twosuper}), the previous relations
(\ref{SOFT}) can be determined.
Also the relations (\ref{split}) are consistent with the above
construction in the sense that the simple Grassmann derivatives
provide the corresponding trivial (anti)commutators (Either the higher
Grassmann derivatives morn than five or equal to four
where the right hand side of (\ref{twosuper}) vanishes.).}.
\section{
The truncated
${\cal N}=4$ supersymmetric
$W_{1+\infty}^{2,2}[\la =\frac{1}{4}]$ algebra
}

In this section, by realizing that
there exists a consistent truncation described in
\cite{Das}, we construct the soft current subalgebra.

\subsection{The soft current algebra and ${\cal N}=3$
  supergravity theory
\label{N3}}

The ${\cal N}=3$ supergravity is studied in \cite{FSZ,Freedman}.
The Lagrangian consists of
a vierbein $ e_{\mu}^a$, three spin $\frac{3}{2}$ Majoranas
$ \psi_{\mu}^i$, three vectors $ A^{i }_{\mu}$ and a single
spin $\frac{1}{2}$ Majoranas $\chi$.
We observe that the soft current algebra between the graviton
(helicity $\pm 2$),
the gravitinos with the helicity $\pm \frac{3}{2}$,
the vectors with the helicity $\pm 1$ and the Majorana
with the helicity $\pm \frac{1}{2}$
can be summarized by
\bea
\acomm{(\Phi^{(h_1),4}_{\frac{1}{2}, +\frac{1}{2}})_r}{
(\Phi^{(h_2),4}_{\frac{1}{2},-\frac{1}{2}})_s}&=
& \kappa_{+\frac{1}{2},-\frac{1}{2},+2}\,\Big((h_2-\tfrac{1}{2})r-(h_1-\tfrac{1}{2})s\Big)\,(\Phi^{(h_1+h_2-3)}_{2,-2})_{r+s}\, : \mbox {eq}. 1,
\nonu \\
\comm{(\Phi^{(h_1),4}_{\frac{1}{2}, -\frac{1}{2}})_r}{(\Phi^{(h_2),ab}_{1,+1})_m}&=&
-\kappa_{-\frac{1}{2},+1,+\frac{3}{2}}\,\Big(h_2\,r-(h_1-\tfrac{1}{2})m\Big)\,\epsilon^{abc}\,
(\Phi^{(h_1+h_2-2),c}_{\frac{3}{2},-\frac{3}{2}})_{r+m}\, : \mbox {eq}. 2,
\nonu \\
\acomm{(\Phi^{(h_1),4}_{\frac{1}{2},-\frac{1}{2}})_r}{(\Phi^{(h_2),a}_{\frac{3}{2},
+\frac{3}{2}})_s}
&=& 
-\kappa_{-\frac{1}{2},+\frac{3}{2},+1}\,\Big((h_2+\tfrac{1}{2})r-(h_1-\tfrac{1}{2})s\Big)\,\epsilon^{abc}\,\frac{1}{2}\,
(\Phi^{(h_1+h_2-1),bc}_{1,-1})_{r+s}\, : \mbox {eq}. 3,
\nonu \\
\comm{(\Phi^{(h_1),4}_{\frac{1}{2},+ \frac{1}{2}})_r}{(\Phi^{(h_2)}_{2,+2})_m}&=
& 
\kappa_{+\frac{1}{2},+2,-\frac{1}{2}}\,\Big((h_2+1)r-(h_1-\tfrac{1}{2})m\Big)\,
(\Phi^{(h_1+h_2),4}_{\frac{1}{2},+ \frac{1}{2}})_{r+m}\, : \mbox {eq}. 4-1,
\nonu \\
\comm{(\Phi^{(h_1),4}_{\frac{1}{2},- \frac{1}{2}})_r}{(\Phi^{(h_2)}_{2,+2})_m}&=
& 
\kappa_{-\frac{1}{2},+2,+\frac{1}{2}}\,\Big((h_2+1)r-(h_1-\tfrac{1}{2})m\Big)\,
(\Phi^{(h_1+h_2),4}_{\frac{1}{2},-\frac{1}{2}})_{r+m}\, : \mbox {eq}. 4-2,
\nonu  \\
\comm{(\Phi^{(h_1),ab}_{1,+1})_m}{(\Phi^{(h_2),cd}_{1,-1})_n}&=&
\kappa_{+1,-1,+2}\,(\delta^{ac}\delta^{bd}-\delta^{ad}\delta^{bc})\,\Big(h_2\,m-h_1\,n\Big)\,(\Phi^{(h_1+h_2-2)}_{2,-2})_{m+n} \, : \mbox {eq}. 5,
\nonu \\
\comm{(\Phi^{(h_1),ab}_{1,+1})_m}{(\Phi^{(h_2),c}_{\frac{3}{2},+\frac{3}{2}})_r}&=&
\kappa_{+1,+\frac{3}{2},-\frac{1}{2}}\,\epsilon^{abc}\Big((h_2+\tfrac{1}{2})\,m-h_1\,r\Big)\,
(\Phi^{(h_1+h_2),4}_{\frac{1}{2},+\frac{1}{2}})_{m+r} \, : \mbox {eq}. 6,
\nonu \\
\comm{(\Phi^{(h_1),ab}_{1,-1})_m}{(\Phi^{(h_2),c}_{\frac{3}{2},+\frac{3}{2}})_r}&=&
\kappa_{-1,+\frac{3}{2},+\frac{3}{2}}\,\Big((h_2+\tfrac{1}{2})m-h_1\,r\Big)
\label{n3soft} \\
&&
\times \bigg[\delta^{ac}\,(\Phi^{(h_1+h_2-1),b}_{\frac{3}{2},-\frac{3}{2}})_{m+r}
-\delta^{bc}\,(\Phi^{(h_1+h_2-1),a}_{\frac{3}{2},-\frac{3}{2}})_{m+r}\bigg]\,
: \mbox {eq}. 7,
\nonu    \\
\comm{(\Phi^{(h_1),ab}_{1,+ 1})_m}{(\Phi^{(h_2)}_{2,+2})_n}&=&
\kappa_{+1,+2,-1}\,\Big((h_2+1)\,m-h_1\,n\Big)\,(\Phi^{(h_1+h_2),ab}_{1, + 1})_{m+n}\,
: \mbox {eq}. 8-1,
\nonu \\
\comm{(\Phi^{(h_1),ab}_{1,-1})_m}{(\Phi^{(h_2)}_{2,+2})_n}&=&
\kappa_{-1,+2,+1}\,\Big((h_2+1)\,m-h_1\,n\Big)\,(\Phi^{(h_1+h_2),ab}_{1, - 1})_{m+n}\,
: \mbox {eq}. 8-2,
\nonu \\
\acomm{(\Phi^{(h_1),a}_{\frac{3}{2},-\frac{3}{2}})_r}{
(\Phi^{(h_2),b}_{\frac{3}{2},+\frac{3}{2}})_s}&=&
\kappa_{-\frac{3}{2},+\frac{3}{2},+2}\,\Big((h_2+\tfrac{1}{2})r-(h_1+\tfrac{1}{2})s\Big)\, \delta^{ab}\,(\Phi^{(h_1+h_2-1)}_{2,-2})_{r+s} \,
: \mbox {eq}. 9,
\nonu  \\
\acomm{(\Phi^{(h_1),a}_{\frac{3}{2},+\frac{3}{2}})_r}{
(\Phi^{(h_2),b}_{\frac{3}{2},+\frac{3}{2}})_s}&=&
\kappa_{+\frac{3}{2},+\frac{3}{2},-1}\,\Big((h_2+\tfrac{1}{2})\,r-(h_1+\tfrac{1}{2})\,s\Big)\,
(\Phi^{(h_1+h_2),ab}_{1,+1})_{r+s}\, : \mbox {eq}. 10,
\nonu  \\
\comm{(\Phi^{(h_1),a}_{\frac{3}{2},+ \frac{3}{2}})_r}{(\Phi^{(h_2)}_{2,+2})_m}&=&
\kappa_{+\frac{3}{2},+2,-\frac{3}{2}}\,\Big((h_2+1)\,r-(h_1+\tfrac{1}{2})\,m\Big)\,
(\Phi^{(h_1+h_2),a}_{\frac{3}{2},+ \frac{3}{2}})_{r+m}\, : \mbox {eq}. 11-1,
\nonu \\
\comm{(\Phi^{(h_1),a}_{\frac{3}{2},-\frac{3}{2}})_r}{(\Phi^{(h_2)}_{2,+2})_m}&=&
\kappa_{-\frac{3}{2},+2,+\frac{3}{2}}\,\Big((h_2+1)\,r-(h_1+\tfrac{1}{2})\,m\Big)\,
(\Phi^{(h_1+h_2),a}_{\frac{3}{2},- \frac{3}{2}})_{r+m}\, : \mbox {eq}. 11-2,
\nonu \\
\comm{(\Phi^{(h_1)}_{2,+2})_m}{(\Phi^{(h_2)}_{2,+ 2})_n}&=&
\kappa_{+2,+2,-2}\,
\Big((h_2+1)\,m-(h_1+1)\,n\Big)\,(\Phi^{(h_1+h_2)}_{2,+ 2})_{nm+n}\,
: \mbox {eq}. 12-1,
\nonu \\
\comm{(\Phi^{(h_1)}_{2,+2})_m}{(\Phi^{(h_2)}_{2,- 2})_n}&=&
\kappa_{+2,-2,+2}\,\Big((h_2+1)\,m-(h_1+1)\,n\Big)\,(\Phi^{(h_1+h_2)}_{2,- 2})_{nm+n}\,
: \mbox {eq}. 12-2.
\nonu 
\eea
From the consistent truncation in \cite{Das}, we put
the scalar and the pseudoscalar,
three spin $\frac{1}{2}$ Majoranas,
three spin $1$ vectors and
one spin $\frac{3}{2}$ to be zero
\bea
\qquad A  & = & 0= B,
\qquad
\chi^{i =1,2,3} =0,
\qquad
F_{\mu \nu}^{i j=14,24,34} =0,
\qquad
\psi_{\mu}^{i=4}  =  0,
\nonu \\
\chi^{i=4} & \equiv & \chi,
\qquad
F_{\mu \nu}^{i j}  \equiv  \epsilon^{i j k}\, F_{\mu \nu}^{k}=
\epsilon^{i j k}\, (\pa_{\mu} \, A_{\nu}^k- \pa_{\nu} \, A_{\mu}^k) \, ,
\label{trunc}
\eea
leading to a single spin $\frac{1}{2}$ Majorana,
three spin $1$ vectors and three gravitinos and a graviton
\footnote{
The Jacobi identities satisfy the following relations between the
couplings
\bea
\kappa_{-\frac{1}{2}, +\frac{3}{2}, +1} & = & \frac{
  \kappa_{+\frac{1}{2}, -\frac{1}{2}, +2} \,\kappa_{+1, +\frac{3}{2}, -\frac{1}{2}}}{\kappa_{+1, -1, +2}}\,, 
\qquad
\kappa_{+\frac{1}{2}, +2, -\frac{1}{2}}= \kappa_{+2, +2, -2}\,,
\qquad
\kappa_{-\frac{1}{2}, +2, +\frac{1}{2}}= \kappa_{+2, +2, -2}\,, 
\nonu\\
\kappa_{-1, +\frac{3}{2}, +\frac{3}{2}} & = & \frac{
  \kappa_{-\frac{1}{2}, +1, +\frac{3}{2}}\, \kappa_{+1, -1, +2} \,
  \kappa_{+\frac{3}{2}, +\frac{3}{2}, -1}}{  \kappa_{+\frac{1}{2}, -\frac{1}{2}, +2} \kappa_{+1, +\frac{3}{2}, -\frac{1}{2}}}\,, 
\qquad
\kappa_{+1, +2, -1}= \kappa_{+2, +2, -2}\,, 
\qquad
\kappa_{-1, +2, +1}= \kappa_{+2, +2, -2}\,, 
\nonu\\
\kappa_{-\frac{3}{2}, +\frac{3}{2}, +2} & = & \frac{
  \kappa_{+\frac{1}{2}, -\frac{1}{2}, +2} \,\kappa_{+1, +\frac{3}{2}, -\frac{1}{2}}}{\kappa_{-\frac{1}{2}, +1, +\frac{3}{2}}}\,,
\qquad
\kappa_{+\frac{3}{2}, +2, -\frac{3}{2}}= \kappa_{+2, +2, -2}\,, 
\qquad
 \kappa_{-\frac{3}{2}, +2, +\frac{3}{2}}= \kappa_{+2, +2, -2}\,, 
 \nonu\\
\kappa_{+2, -2, +2} & = & \kappa_{+2, +2, -2}\,.
\nonu
\eea
}.

$\bullet$ $\frac{1}{\kappa^2} \, e \, R$
 term

As done in
the section \ref{sugra},
we can analyze the corresponding
soft current algebra.
Because
the truncation (\ref{trunc}) does not change the previous
description in the ${\cal N}=4$ supergravity,
the twelfth relation of (\ref{n3soft})
provides the celestial commutator between the gravitons
having the helicities $(+2, \pm 2, \mp 2)$.
 
$\bullet$
$\epsilon^{\mu \nu \sigma \rho} \, \bar{\psi}^i_{\mu} \, \ga_5\, \ga_{\nu}\,
D_{\rho}\, \psi_{\si}^i$
term

Here the summation over the
$SO(4)$ indices $i$ is given by the vectors of
$SO(3)$ with $i =1,2,3$.
The ninth relation of (\ref{n3soft})
with helicities $(-\frac{3}{2}, +\frac{3}{2},+2)$
gives the celestial anticommutator
between the gravitinos and the graviton while
the eleventh of (\ref{n3soft}) with helicities
$(\pm \frac{3}{2}, +2, \mp \frac{3}{2})$ leads to the celestial
commutator
between the gravitino, the graviton and the gravitino.

$\bullet$
$e \, g^{\mu \rho} \, g^{\nu \si}\, F_{\mu \nu}^{i} \, F_{\rho \si}^{ i }$
term

Compared to the description of the section
\ref{sugra},
half of them can survive.
The fifth of (\ref{n3soft}) with helicities $(+1,-1,+2)$
describes the celestial commutator between the vectors and the graviton
while
the eighth of (\ref{n3soft}) with helicities $(\pm 1, +2, \mp 1)$
implies the celestial
commutator between the vector, the graviton and the vector
\footnote{It is obvious to see that
the product of two Kronecker deltas on the right hand side of this
fifth celestial commutator provides the same
$SO(4)$ indices for the two vectors.}.

$\bullet$
$e \, \bar{\chi} \, \ga^{\mu} \, D_{\mu}\, \chi$
term

In this case, the previous three terms in the
section \ref{sugra} are disappeared.
The first relation of (\ref{n3soft}) with the
helicities $(+\frac{1}{2},-\frac{1}{2},+2)$ provides the
celestial anticommutator
between the two Majoranas and the graviton and
the fourth relation of (\ref{n3soft}) with helicities
$(\pm \frac{1}{2}, +2, \mp \frac{1}{2})$
leads to the celestial
commutator between the Majorana, the graviton and
the Majorana.

$\bullet$
$\kappa \,  e\, g^{\mu \rho} \, g^{\nu \si}\,
\epsilon^{i j k}
\bar{\psi}^i_{\mu} \,  F_{\rho \si}^{ j}\, \psi_{\nu}^k$
and
$\kappa \,   \epsilon^{\mu \nu \rho \si} \,
\epsilon^{i j k} \,
\bar{\psi}^i_{\mu} \,  \ga_5\, F_{\rho \si}^{  j}\, \psi_{\nu}^k$
terms

Compared to the description of the section
\ref{sugra},
half of them can survive.
The seventh relation of (\ref{n3soft}) with the helicities $(-1, +
\frac{3}{2}, +\frac{3}{2})$ gives the celestial commutator
between the vector, the two gravitinos while 
the tenth relation of (\ref{n3soft}) with the helicities
$(+\frac{3}{2}, +\frac{3}{2},-1)$ gives the
celestial anticommutator between the
two gravitinos and the vector \footnote{According to
(\ref{trunc}), the product of epsilon tensor and the field strength
having a single $SO(3)$ index can be written in terms of
the field strength with two $SO(3)$ indices. They are contracted with
the $SO(3)$ indices for the two gravitinos. In
the seventh relation of (\ref{n3soft}), we also observe that
the $SO(3)$ indices for the two gravitinos
are contracted with those for the vector from the Kronecker deltas.}.

$\bullet$
$\kappa \, e   \, \bar{\psi}^i_{\mu} \,
\ga^{\nu}\, \ga^{\rho} \, \ga^{\mu} \,
\chi\, F^{i}_{\nu \rho}$
term

In this case,
the half of previous analysis in the section \ref{sugra}
survives.
The second relation of (\ref{n3soft})
with helicities $(-\frac{1}{2},+1, +\frac{3}{2})$
provides the celestial commutator between the Majorana, the vector
and the gravitino,
the third  relation of (\ref{n3soft})
with helicities $(-\frac{1}{2},+\frac{3}{2},+1)$
provides the celestial anticommutator between the Majorana, 
the gravitino and the vector and
the sixth  relation of (\ref{n3soft})
with helicities $(+1, +\frac{3}{2}, -\frac{1}{2})$
provides the celestial commutator between the vector, the gravitino
and Majorana \footnote{According to (\ref{trunc}),
the right hand side of the third  relation of (\ref{n3soft})
can be interpreted as the field strength having
a single $SO(3)$ index. For the other celestial (anti)commutators,
we observe this particular interaction by multiplying
other epsilon tensor both sides of these celestial
commutators.}.

\subsection{The soft current algebra and ${\cal N}=2$
supergravity theory}

The ${\cal N}=2$ supergravity is studied in \cite{Fv}.
The Lagrangian consists of
a vierbein $ e_{\mu}^a$, two spin $\frac{3}{2}$ Majoranas
$ \psi_{\mu}^i$ and  a vector $ A_{\mu}$.
The soft current algebra between the graviton
(helicity $\pm 2$),
the gravitinos with the helicity $\pm \frac{3}{2}$ and 
the vector with the helicity $\pm 1$ 
can be described by
\bea
\comm{(\Phi^{(h_1),12}_{1,+1})_m}{(\Phi^{(h_2),12}_{1,-1})_n}&=
& \kappa_{+1,-1,+2}\,\Big(h_2\,m-h_1\,n\Big)\,(\Phi^{(h_1+h_2-2)}_{2,-2})_{m+n}\,
: \mbox {eq}. 1,
\nonu \\
\comm{(\Phi^{(h_1),12}_{1,-1})_m}{(\Phi^{(h_2),1}_{\frac{3}{2},+\frac{3}{2}})_r}&=&
\kappa_{-1,+\frac{3}{2},+\frac{3}{2}}\,\Big((h_2+\tfrac{1}{2})m-h_1\,r\Big)\,
(\Phi^{(h_1+h_2-1),2}_{\frac{3}{2},-\frac{3}{2}})_{m+r}\,
: \mbox {eq}. 2,
\nonu \\
\comm{(\Phi^{(h_1),12}_{1,-1})_m}{(\Phi^{(h_2),2}_{\frac{3}{2},+\frac{3}{2}})_r}&=&
-\kappa_{-1,+\frac{3}{2},+\frac{3}{2}}\,\Big((h_2+\tfrac{1}{2})m-h_1\,r\Big)\,(\Phi^{(h_1+h_2-1),1}_{\frac{3}{2},-\frac{3}{2}})_{m+r}\,
: \mbox {eq}. 3,
\nonu \\
\comm{(\Phi^{(h_1),12}_{1,+ 1})_m}{(\Phi^{(h_2)}_{2,+2})_n}&=&
\kappa_{+1,+2,-1}\,\Big((h_2+1)m-h_1\,n\Big)\,(\Phi^{(h_1+h_2),12}_{1,+ 1})_{m+n}\,
: \mbox {eq}. 4-1,
\nonu  \\
\comm{(\Phi^{(h_1),12}_{1,- 1})_m}{(\Phi^{(h_2)}_{2,+2})_n}&=&
\kappa_{-1,+2,+1}\,\Big((h_2+1)m-h_1\,n\Big)\,(\Phi^{(h_1+h_2),12}_{1,- 1})_{m+n}\,
: \mbox {eq}. 4-2,
\nonu  \\
\acomm{(\Phi^{(h_1),a}_{\frac{3}{2},-\frac{3}{2}})_r}{
(\Phi^{(h_2),b}_{\frac{3}{2},+\frac{3}{2}})_s}&=&
\kappa_{-\frac{3}{2},+\frac{3}{2},+2}\,\delta^{ab}\,\Big((h_2+\tfrac{1}{2})r-(h_1+\tfrac{1}{2})s\Big)\,(\Phi^{(h_1+h_2-1)}_{2,-2})_{r+s} \,
: \mbox {eq}. 5,
\label{n2soft}
   \\
\acomm{(\Phi^{(h_1),a}_{\frac{3}{2},+\frac{3}{2}})_r}{
(\Phi^{(h_2),b}_{\frac{3}{2},+\frac{3}{2}})_s}&=&
\kappa_{+\frac{3}{2},+\frac{3}{2},-1}\,\Big((h_2+\tfrac{1}{2})\,r-(h_1+\tfrac{1}{2})\,s\Big)\,
(\Phi^{(h_1+h_2),ab}_{1,+1})_{r+s}\, : \mbox {eq}. 6,
\nonu   \\
\comm{(\Phi^{(h_1),a}_{\frac{3}{2},+ \frac{3}{2}})_r}{(\Phi^{(h_2)}_{2,+2})_m}&=&
\kappa_{+\frac{3}{2},+2,-\frac{3}{2}}\,\Big((h_2+1)\,r-(h_1+\tfrac{1}{2})\,m\Big)\,
(\Phi^{(h_1+h_2),a}_{\frac{3}{2},+ \frac{3}{2}})_{r+m}\,
: \mbox {eq}. 7-1,
\nonu \\
\comm{(\Phi^{(h_1),a}_{\frac{3}{2},- \frac{3}{2}})_r}{(\Phi^{(h_2)}_{2,+2})_m}&=&
\kappa_{-\frac{3}{2},+2,+\frac{3}{2}}\,\Big((h_2+1)\,r-(h_1+\tfrac{1}{2})\,m\Big)\,
(\Phi^{(h_1+h_2),a}_{\frac{3}{2},-\frac{3}{2}})_{r+m}\,
: \mbox {eq}. 7-2,
\nonu \\
\comm{(\Phi^{(h_1)}_{2,+2})_m}{(\Phi^{(h_2)}_{2,+ 2})_n}&=&
\kappa_{+2,+2,-2}\,\Big((h_2+1)\,m-(h_1+1)\,n\Big)\,(\Phi^{(h_1+h_2)}_{2,+ 2})_{m+n}\,
: \mbox {eq}. 8-1.
\nonu\\
\comm{(\Phi^{(h_1)}_{2,+2})_m}{(\Phi^{(h_2)}_{2,- 2})_n}&=&
\kappa_{+2,-2,+2}\,\Big((h_2+1)\,m-(h_1+1)\,n\Big)\,(\Phi^{(h_1+h_2)}_{2,- 2})_{m+n}\,
: \mbox {eq}. 8-2.
\nonu
\eea
From the consistent truncation in \cite{FSZ}, we put
a single spin $\frac{1}{2}$ Majorana,
two spin $1$ vectors and
one spin $\frac{3}{2}$ to be zero among the generators in the
subsection \ref{N3}
\bea
\psi^{i=3}_{\mu}=0, \qquad \chi =0,
\qquad A_{\mu}^{i=1,2}=0,
\qquad
A_{\mu}^{i=3} \equiv A_{\mu} \, ,
\label{COND}
\eea
which leads to a single spin $1$ vector, two gravitinos and a graviton
\footnote{ The following relations come from the
Jacobi identities  
\bea
\kappa_{+1, +2, -1} & = & \kappa_{+2, +2, -2}\,, 
\qquad
\kappa_{-1, +2, +1}= \kappa_{+2, +2, -2}\,, 
\qquad
\kappa_{-\frac{3}{2}, +\frac{3}{2}, +2}= -\frac{
  \kappa_{+1, -1, +2}\, \kappa_{+\frac{3}{2}, +\frac{3}{2}, -1}}{\kappa_{-1,  +\frac{3}{2},+\frac{3}{2}}}\,,
\nonu  \\
\kappa_{+\frac{3}{2}, +2, -\frac{3}{2}} & = & \kappa_{+2, +2, -2}\,, 
\qquad
 \kappa_{-\frac{3}{2}, +2, +\frac{3}{2}}= \kappa_{+2, +2, -2}\,, 
\qquad
\kappa_{+2, -2, +2}= \kappa_{+2, +2, -2}\,.
\nonu
\eea}.

$\bullet$ $\frac{1}{\kappa^2} \, e \, R$

As found in
the section \ref{sugra},
the corresponding
soft current algebra can be analyzed.
The eighth relation of (\ref{n2soft})
gives the celestial commutator between the gravitons
having the helicities $(+2, \pm 2, \mp 2)$.

$\bullet$
$\epsilon^{\mu \nu \sigma \rho} \, \bar{\psi}^i_{\mu} \, \ga_5\, \ga_{\nu}\,
D_{\rho}\, \psi_{\si}^i$
term

The summation over the
$SO(4)$ indices $i$ is reduced to the vectors of
$SO(2)$ with $i =1,2$ according to (\ref{COND}).
The fifth relation of (\ref{n2soft})
with helicities $(-\frac{3}{2}, +\frac{3}{2},+2)$
gives the celestial anticommutator
between the gravitinos and the graviton while
the seventh of (\ref{n2soft}) with helicities
$(\pm \frac{3}{2}, +2, \mp \frac{3}{2})$ leads to the
celestial commutator
between the gravitino, the graviton and the gravitino.

$\bullet$
$e \, g^{\mu \rho} \, g^{\nu \si}\, F_{\mu \nu} \, F_{\rho \si}$
term

From (\ref{COND}), we can identify
$F_{\mu \nu}  \equiv  
(\pa_{\mu} \, A_{\nu}- \pa_{\nu} \, A_{\mu})$.
The first relation of (\ref{n2soft}) with helicities $(+1,-1,+2)$
describes the celestial commutator between the vectors and the graviton
while
the fourth relation
of (\ref{n2soft}) with helicities $(\pm 1, +2, \mp 1)$
implies the celestial
commutator between the vector, the graviton and the vector.

$\bullet$
$\kappa \,  e\, g^{\mu \rho} \, g^{\nu \si}\,
\epsilon^{i j }
\bar{\psi}^i_{\mu} \,  F_{\rho \si}\, \psi_{\nu}^j$
and
$\kappa \,   \epsilon^{\mu \nu \rho \si} \,
\epsilon^{i j } \,
\bar{\psi}^i_{\mu} \,  \ga_5\, F_{\rho \si}\, \psi_{\nu}^j$
terms

The second and third relations of (\ref{n2soft}) with
the helicities $(-1, +
\frac{3}{2}, +\frac{3}{2})$ give the celestial commutator
between the vector, the two gravitinos while 
the sixth relation of (\ref{n2soft}) with the helicities
$(+\frac{3}{2}, +\frac{3}{2},-1)$ gives the
celestial anticommutator between the
two gravitinos and the vector \footnote{For these celestial
(anti)commutators, the corresponding $SO(2)$ indices
for the two gravitinos (and for the vector)
are different from each other.}.

\subsection{The soft current algebra and ${\cal N}=1$
supersymmetric Maxwell Einstein theory}

The ${\cal N}=1$ supergravity coupled to
a matter multiplet is studied in \cite{FSv}.
The Lagrangian consists of
a vierbein $ e_{\mu}^a$, a spin $\frac{3}{2}$ Majorana
$ \psi_{\mu}$, a vector $ A_{\mu}$ and a spin $\frac{1}{2}$ Majorana.
The soft current algebra between the graviton
(helicity $\pm 2$),
the gravitino with the helicity $\pm \frac{3}{2}$, 
the vector with the helicity $\pm 1$
and a spin $\frac{1}{2}$ with the helicity $\pm \frac{1}{2}$
can be obtained by
\bea
\acomm{(\Phi^{(h_1),4}_{\frac{1}{2}, +\frac{1}{2}})_r}{
(\Phi^{(h_2),4}_{\frac{1}{2},-\frac{1}{2}})_s}&=
& \kappa_{+\frac{1}{2},-\frac{1}{2},+2} \,\Big((h_2-\tfrac{1}{2})r-(h_1-\tfrac{1}{2})s\Big)\, (\Phi^{(h_1+h_2-3)}_{2,-2})_{r+s}\,
: \mbox {eq}. 1,
\nonu  \\
\comm{(\Phi^{(h_1),4}_{\frac{1}{2},-\frac{1}{2}})_r}{(\Phi^{(h_2),23}_{1,+1})_m}&=&
-\kappa_{-\frac{1}{2},+1,+\frac{3}{2}}\,\Big(h_2\,r-(h_1-\tfrac{1}{2})m\Big)\,
(\Phi^{(h_1+h_2-2),1}_{\frac{3}{2},-\frac{3}{2}})_{r+m}\,
: \mbox {eq}. 2,
\nonu \\
\acomm{(\Phi^{(h_1),4}_{\frac{1}{2},-\frac{1}{2}})_r}{
(\Phi^{(h_2),1}_{\frac{3}{2},+\frac{3}{2}})_s}&
=&- \kappa_{-\frac{1}{2},+\frac{3}{2},+1}\,
\Big((h_2+\tfrac{1}{2})r-(h_1-\tfrac{1}{2})m\Big)\,(\Phi^{(h_1+h_2-1),23}_{1,-1})_{r+s}\,
: \mbox {eq}. 3,
\nonu \\
\comm{(\Phi^{(h_1),4}_{\frac{1}{2},+ \frac{1}{2}})_r}{(\Phi^{(h_2)}_{2,+2})_m}&=
&\kappa_{+\frac{1}{2},+2,-\frac{1}{2}}\, \Big((h_2+1)r-(h_1-\tfrac{1}{2})m\Big)\,
(\Phi^{(h_1+h_2),4}_{\frac{1}{2},+\frac{1}{2}})_{r+m}\,
: \mbox {eq}. 4-1,
\nonu \\
\comm{(\Phi^{(h_1),4}_{\frac{1}{2},-\frac{1}{2}})_r}{(\Phi^{(h_2)}_{2,+2})_m}&=
&\kappa_{-\frac{1}{2},+2,+\frac{1}{2}}\, \Big((h_2+1)r-(h_1-\tfrac{1}{2})m\Big)\,
(\Phi^{(h_1+h_2),4}_{\frac{1}{2},- \frac{1}{2}})_{r+m}\,
: \mbox {eq}. 4-2,
\nonu \\
\comm{(\Phi^{(h_1),23}_{1,+1})_m}{(\Phi^{(h_2),23}_{1,-1})_n}&=
& \kappa_{+1,-1,+2}\,\Big(h_2\,m-h_1\,n\Big)\,(\Phi^{(h_1+h_2-2)}_{2,-2})_{m+n}\,
: \mbox {eq}. 5,
\label{n1soft}
\\
\comm{(\Phi^{(h_1),23}_{1,+1})_m}{(\Phi^{(h_2),1}_{\frac{3}{2},+\frac{3}{2}})_r}&=&
\kappa_{+1,+\frac{3}{2},-\frac{1}{2}}\,\Big((h_2+\tfrac{1}{2})\,m-h_1\,r\Big)\,
(\Phi^{(h_1+h_2),4}_{\frac{1}{2},+\frac{1}{2}})_{m+r} \, : \mbox {eq}. 6,
\nonu \\
\comm{(\Phi^{(h_1),23}_{1, + 1})_m}{(\Phi^{(h_2)}_{2,+2})_n}&=
& 
\kappa_{+1,+2,-1}\,\Big((h_2+1)m-h_1\,n\Big)\,(\Phi^{(h_1+h_2),23}_{1,+ 1})_{m+n}\,
: \mbox {eq}. 7-1,
\nonu \\
\comm{(\Phi^{(h_1),23}_{1, - 1})_m}{(\Phi^{(h_2)}_{2,+2})_n}&=
& 
\kappa_{-1,+2,+1}\,\Big((h_2+1)m-h_1\,n\Big)\,(\Phi^{(h_1+h_2),23}_{1,- 1})_{m+n}\,
: \mbox {eq}. 7-2,
\nonu \\
\acomm{(\Phi^{(h_1),1}_{\frac{3}{2},-\frac{3}{2}})_r}{
(\Phi^{(h_2),1}_{\frac{3}{2},+\frac{3}{2}})_s}&= &
\kappa_{-\frac{3}{2},+\frac{3}{2},+2}\,
\Big((h_2+\tfrac{1}{2})r-(h_1+\tfrac{1}{2})s\Big)\,
(\Phi^{(h_1+h_2-1)}_{2,-2})_{r+s}
\, : \mbox {eq}. 8,
\nonu \\
\comm{(\Phi^{(h_1),1}_{\frac{3}{2},+ \frac{3}{2}})_r}{
(\Phi^{(h_2)}_{2,+2})_m}&=&
\kappa_{+\frac{3}{2},+2,-\frac{3}{2}}\,
\Big((h_2+1)\,r-(h_1+\tfrac{1}{2})\,m\Big)\,
(\Phi^{(h_1+h_2),1}_{\frac{3}{2},+\frac{3}{2}})_{r+m}\,
: \mbox {eq}. 9-1,
\nonu  \\
\comm{(\Phi^{(h_1),1}_{\frac{3}{2},- \frac{3}{2}})_r}{
(\Phi^{(h_2)}_{2,+2})_m}&=&
\kappa_{-\frac{3}{2},+2,+\frac{3}{2}}\,
\Big((h_2+1)\,r-(h_1+\tfrac{1}{2})\,m\Big)\,
(\Phi^{(h_1+h_2),1}_{\frac{3}{2},- \frac{3}{2}})_{r+m}\,
: \mbox {eq}. 9-2,
\nonu  \\
\comm{(\Phi^{(h_1)}_{2,+2})_m}{(\Phi^{(h_2)}_{2, + 2})_n}&=&
\kappa_{+2,+2,-2}\,\Big((h_2+1)\,m-(h_1+1)\,n\Big)\,(\Phi^{(h_1+h_2)}_{2,+ 2})_{m+n}\,
: \mbox {eq}. 10-1.
\nonu  \\
\comm{(\Phi^{(h_1)}_{2,+2})_m}{(\Phi^{(h_2)}_{2, - 2})_n}&=&
\kappa_{+2,-2,+2}\,\Big((h_2+1)\,m-(h_1+1)\,n\Big)\,(\Phi^{(h_1+h_2)}_{2,- 2})_{m+n}\,
: \mbox {eq}. 10-2.
\nonu
\eea
From the consistent truncation in \cite{FSZ}, we put
two spin $1$ vectors and
two spin $\frac{3}{2}$ gravitinos
to be zero among the generators in the
subsection \ref{N3}
\bea
\psi^{i=2,3}_{\mu}=0, \qquad 
\qquad A_{\mu}^{i=2,3}=0 \, ,
\label{cond}
\eea
implying that
there are a graviton, a gravitino, spin $1$ vector and a spin
$\frac{1}{2}$ Majorana
\footnote{
  From the Jacobi identities, the couplings satisfy
\bea
  \kappa_{-\frac{1}{2}, +\frac{3}{2}, +1} & = & \frac{
  \kappa_{+\frac{1}{2}, -\frac{1}{2}, +2} \,\kappa_{+1, +\frac{3}{2}, -\frac{1}{2}}}{\kappa_{+1, -1, +2}}\,, 
\qquad
\kappa_{+\frac{1}{2}, +2, -\frac{1}{2}}= \kappa_{+2, +2, -2}\,,
\qquad
\kappa_{-\frac{1}{2}, +2, +\frac{1}{2}}= \kappa_{+2, +2, -2}\,, 
\nonu\\
\kappa_{+1, +2, -1} & = & \kappa_{+2, +2, -2}\,, 
\qquad
\kappa_{-1, +2, +1}= \kappa_{+2, +2, -2}\,, 
\qquad
\kappa_{-\frac{3}{2}, +\frac{3}{2}, +2}= \frac{
  \kappa_{+\frac{1}{2}, -\frac{1}{2}, +2} \,\kappa_{+1, +\frac{3}{2}, -\frac{1}{2}}}{\kappa_{-\frac{1}{2}, +1, +\frac{3}{2}}}\,,
\nonu  \\
\kappa_{+\frac{3}{2}, +2, -\frac{3}{2}} & = & \kappa_{+2, +2, -2}\,, 
\qquad
 \kappa_{-\frac{3}{2}, +2, +\frac{3}{2}}= \kappa_{+2, +2, -2}\,, 
\qquad
\kappa_{+2, -2, +2}= \kappa_{+2, +2, -2}\,.
\nonu
\eea}. 


$\bullet$ $\frac{1}{\kappa^2} \, e \, R$
term

The tenth relation of (\ref{n1soft})
provides the celestial commutator between the gravitons
having the helicities $(+2, \pm 2, \mp 2)$.

$\bullet$
$\epsilon^{\mu \nu \sigma \rho} \, \bar{\psi}^i_{\mu} \, \ga_5\, \ga_{\nu}\,
D_{\rho}\, \psi_{\si}^i$
term

The summation over the
$SO(4)$ indices $i$ is reduced to the vector of
$SO(1)$ with $i =1$ with (\ref{cond}).
The eighth relation of (\ref{n1soft})
with helicities $(-\frac{3}{2}, +\frac{3}{2},+2)$
gives the celestial anticommutator
between the gravitinos and the graviton while
similarly the ninth of (\ref{n1soft}) with helicities
$(\pm \frac{3}{2}, +2, \mp \frac{3}{2})$ implies the
celestial commutator
between the gravitino, the graviton and the gravitino.

$\bullet$
$e \, g^{\mu \rho} \, g^{\nu \si}\, F_{\mu \nu} \, F_{\rho \si}$
term

Here $F_{\mu \nu}$ is the same as before.
The fifth relation of (\ref{n1soft}) with helicities $(+1,-1,+2)$
gives the celestial
commutator between the vectors and the graviton
while
the seventh relation
of (\ref{n1soft}) with helicities $(\pm 1, +2, \mp 1)$
gives the celestial
commutator between the vector, the graviton and the vector.

$\bullet$
$e \, \bar{\chi} \, \ga^{\mu} \, D_{\mu}\, \chi$
term

As in the section \ref{sugra},
the similar analysis can be done.
The first relation of (\ref{n1soft}) with helicities
$(+\frac{1}{2},-\frac{1}{2},+2)$
gives the celestial
anticommutator between the two Majorana and the graviton
and
the fourth relation
of (\ref{n1soft}) with helicities $(\pm \frac{1}{2}, +2,
\mp \frac{1}{2})$
gives the celestial
commutator between the Majorana, the graviton and the Majorana.

$\bullet$
$\kappa \, e   \, \bar{\psi}^{i=1}_{\mu} \,
\ga^{\nu}\, \ga^{\rho} \, \ga^{\mu} \,
\chi\, F_{\nu \rho}$
term 

The second relation of (\ref{n1soft}) with helicities
$(-\frac{1}{2},+1, +\frac{3}{2})$
gives the celestial
commutator between the Majorana, the vector and the gravitino,
the third 
of (\ref{n1soft}) with helicities $(- \frac{1}{2}, +\frac{3}{2},
+1)$
gives the celestial
anticommutator between the Majorana, the gravitino and the
vector
and the sixth relation
of (\ref{n1soft}) with helicities $(+1, + \frac{3}{2},
- \frac{1}{2})$
gives the celestial commutator between the vector, the gravitino and
the Majorana.

\subsection{The soft current algebra and ${\cal N}=2$
supergravity theory coupled to its Abelian vector multiplet}

The ${\cal N}=2$ supergravity theory
coupled to its Abelian vector multiplet \cite{Das,Luciani}
contains
the ${\cal N}=2$ supergravity multiplet
of spins $(1, \frac{3}{2}, \frac{3}{2}, 2)$
and its
Abelian vector multiplet of spins
$(0^{\pm}, \frac{1}{2}, \frac{1}{2}, 1)$.
The soft current algebra between these
generators are given by
\bea
\comm{(\Phi^{(h_1)}_{0, + 0})_m}{(\Phi^{(h_2)}_{0,- 0})_n}
& = &
\kappa_{+0,-0,+2}\,\Big((h_2-1)m-(h_1-1)n \Big)\, (\Phi^{(h_1+h_2-4)}_{2,-2})_{m+n}\,
: \mbox {eq}. 1,
\nonu    \\
\comm{(\Phi^{(h_1)}_{{0,- 0}})_m}{(\Phi^{(h_2),a}_{\frac{1}{2},+
\frac{1}{2}})_r}
& = &
\kappa_{-0,+\frac{1}{2},+\frac{3}{2}} \, \Big((h_2-\tfrac{1}{2})m-(h_1-1)r\Big)\,
(\Phi^{(h_1+h_2-3),a}_{\frac{3}{2},-\frac{3}{2}})_{m+r}
\, : \mbox {eq}. 2,
\nonu \\
\comm{(\Phi^{(h_1)}_{0, -0})_m}{(\Phi^{(h_2),12}_{1,+1})_n} 
& = &
\kappa_{-0,+1,+1}\,\Big(h_2\,m-(h_1-1)n \Big)
\,
 (\Phi^{(h_1+h_2-2),34}_{1,-1})_{m+n}\, : \mbox {eq}. 3,
\nonu \\
\comm{(\Phi^{(h_1)}_{0, - 0})_m}{(\Phi^{(h_2),34}_{1,+1})_n} 
& = &
\kappa_{-0,+1,+1}\,\Big(h_2\,m-(h_1-1)n \Big)
\,
 (\Phi^{(h_1+h_2-2),12}_{1,-1})_{m+n} \, : \mbox {eq}. 4,
\nonu \\
\comm{(\Phi^{(h_1)}_{0,- 0})_m}{(\Phi^{(h_2),a}_{\frac{3}{2},+\frac{3}{2}})_r}
& = & \kappa_{-0,+\frac{3}{2},+\frac{1}{2}}\,\Big((h_2+\tfrac{1}{2})m-(h_1-1)r\Big)\,(\Phi^{(h_1+h_2-1),a}_{\frac{1}{2},-\frac{1}{2}})_{m+r} \,
: \mbox {eq}. 5,
\nonu \\
\comm{(\Phi^{(h_1)}_{0, +0})_m}{(\Phi^{(h_2)}_{2,+2})_n}
& = & \kappa_{+0,+2,-0}\,\Big((h_2+1)m-(h_1-1)n \Big)\,(\Phi^{(h_1+h_2)}_{0,+0})_{m+n} \,
: \mbox {eq}. 6-1,
\nonu \\
\comm{(\Phi^{(h_1)}_{0, - 0})_m}{(\Phi^{(h_2)}_{2,+2})_n}
& = & \kappa_{-0,+2,+0}\,\Big((h_2+1)m-(h_1-1)n \Big)\,(\Phi^{(h_1+h_2)}_{0,- 0})_{m+n} \,
: \mbox {eq}. 6-2,
\nonu \\
\acomm{(\Phi^{(h_1),a}_{\frac{1}{2},+\frac{1}{2}})_r}{(\Phi^{(h_2),b}_{\frac{1}{2},
-\frac{1}{2}})_s}
& = & \kappa_{+\frac{1}{2},-\frac{1}{2},+2}\,
\Big((h_2-\tfrac{1}{2})r-(h_1-\tfrac{1}{2})s \Big)
\, \delta^{ab}\,
(\Phi^{(h_1+h_2-3)}_{2,-2})_{r+s} \, : \mbox {eq}. 7,
\nonu \\
\comm{(\Phi^{(h_1),a}_{\frac{1}{2},-\frac{1}{2}})_r}{(\Phi^{(h_2),34}_{1,+1})_m}
& = & \kappa_{-\frac{1}{2},+1,+\frac{3}{2}}\,\Big(h_2\,r-(h_1-\tfrac{1}{2})m
\Big)\,
\epsilon^{a b}\,
(\Phi^{(h_1+h_2-2),b}_{\frac{3}{2},-\frac{3}{2}})_{r+m}
\, : \mbox {eq}. 10,
\nonu \\
\acomm{(\Phi^{(h_1),a}_{\frac{1}{2},-\frac{1}{2}})_r}{(\Phi^{(h_2),b}_{
\frac{3}{2},+\frac{3}{2}})_s}
& = &
\kappa_{-\frac{1}{2},+\frac{3}{2},+1}\,\Big((h_2+\tfrac{1}{2})\,r-(h_1-\tfrac{1}{2})s \Big)\,
\epsilon^{ab}\, (\Phi_{1,-1}^{(h_1+h_2-1),34})_{r+s} \,
: \mbox {eq}. 11,
\nonu \\
\acomm{(\Phi^{(h_1),a}_{\frac{1}{2},+\frac{1}{2}})_r}{(\Phi^{(h_2),b}_{
\frac{3}{2}, +\frac{3}{2}})_s}
& = &
\kappa_{+\frac{1}{2},+\frac{3}{2},-0}\, \delta^{ab}\,
\Big((h_2+\tfrac{1}{2})r-(h_1-\tfrac{1}{2})s\Big)\,
(\Phi^{(h_1+h_2)}_{0, + 0})_{r+s} \, : \mbox {eq}. 12,
\nonu \\
\comm{(\Phi^{(h_1)}_{2,+2})_m}{(\Phi^{(h_2),a}_{\frac{1}{2},+ \frac{1}{2}})_r}
& = & 
\kappa_{+2,+\frac{1}{2},-\frac{1}{2}}\,\Big((h_2-\tfrac{1}{2})m-(h_1+1)r \Big)\,(\Phi^{(h_1+h_2),a}_{\frac{1}{2},+\frac{1}{2}})_{m+r} \, : \mbox {eq}. 13-1,
\nonu \\
\comm{(\Phi^{(h_1)}_{2,+2})_m}{(\Phi^{(h_2),a}_{\frac{1}{2},- \frac{1}{2}})_r}
& = & 
\kappa_{+2,-\frac{1}{2},+\frac{1}{2}}\,\Big((h_2-\tfrac{1}{2})m-(h_1+1)r \Big)\,(\Phi^{(h_1+h_2),a}_{\frac{1}{2},-\frac{1}{2}})_{m+r} \, : \mbox {eq}. 13-2,
\nonu \\
\comm{(\Phi^{(h_1),12}_{1,+1})_m}{(\Phi^{(h_2),12}_{1,-1})_n}
& = & 
\kappa_{+1,-1,+2}\,\Big(h_2\,m-h_1\,n\Big)\,(\Phi^{(h_1+h_2-2)}_{2,-2})_{m+n}
\, : \mbox {eq}. 14,
\nonu \\
\comm{(\Phi^{(h_1),34}_{1,+1})_m}{(\Phi^{(h_2),34}_{1,-1})_n}
& = &
\kappa_{+1,-1,+2}\,\Big(h_2\,m-h_1\,n\Big)\,(\Phi^{(h_1+h_2-2)}_{2,-2})_{m+n}
\, : \mbox {eq}. 15,
\nonu \\
\comm{(\Phi^{(h_1),12}_{1,+1})_m}{(\Phi^{(h_2),34}_{1,+1})_n}
& = & 
\kappa_{+1,+1,-0}\,\Big(h_2\,m-h_1\,n\Big)\,
(\Phi^{(h_1+h_2)}_{0,+ 0})_{m+n}
\, : \mbox {eq}. 16,
\label{n2softmatter}
\\
\comm{(\Phi^{(h_1),12}_{1,-1})_m}{(\Phi^{(h_2),a}_{\frac{3}{2},+\frac{3}{2}})_r}
& = &
\kappa_{-1,+\frac{3}{2},+\frac{3}{2}}\,\Big((h_2+\tfrac{1}{2})m-h_1\,r \Big)\,
\epsilon^{a b}\,
(\Phi^{(h_1+h_2-1),b}_{\frac{3}{2},-\frac{3}{2}})_{m+r}
\, : \mbox {eq}. 17,
\nonu \\
\comm{(\Phi^{(h_1),34}_{1,+1})_m}{(\Phi^{(h_2),a}_{\frac{3}{2},+\frac{3}{2}})_r}
& = &
\kappa_{+1,+\frac{3}{2},-\frac{1}{2}}\,\Big((h_2+\tfrac{1}{2})m-h_1\,r \Big)\,
\epsilon^{a b}\,
(\Phi^{(h_1+h_2),b}_{\frac{1}{2},+\frac{1}{2}})_{m+r}
\, 
:  \mbox {eq}. 18,
\nonu \\
\comm{(\Phi^{(h_1),12}_{1, + 1})_m}{(\Phi^{(h_2)}_{2,+2})_n}
& = & 
\kappa_{+1,+2,-1}\,\Big((h_2+1)m-h_1\,n\Big)\,(\Phi^{(h_1+h_2),12}_{1,+ 1})_{m+n} \,
: \mbox {eq}. 19-1,
\nonu \\
\comm{(\Phi^{(h_1),12}_{1, - 1})_m}{(\Phi^{(h_2)}_{2,+2})_n}
& = & 
\kappa_{-1,+2,+1}\,\Big((h_2+1)m-h_1\,n\Big)\,(\Phi^{(h_1+h_2),12}_{1,- 1})_{m+n} \,
: \mbox {eq}. 19-2,
\nonu \\
\comm{(\Phi^{(h_1),34}_{1, + 1})_m}{(\Phi^{(h_2)}_{2, +2})_n}
& = &
\kappa_{+1,+2,-1}\,\Big((h_2+1)m-h_1\,n\Big)\,(\Phi^{(h_1+h_2),34}_{1,+ 1})_{m+n}\,
: \mbox {eq}. 20-1,
\nonu 
\\
\comm{(\Phi^{(h_1),34}_{1, - 1})_m}{(\Phi^{(h_2)}_{2, +2})_n}
& = &
\kappa_{-1,+2,+1}\,\Big((h_2+1)m-h_1\,n\Big)\,(\Phi^{(h_1+h_2),34}_{1,- 1})_{m+n}\,
: \mbox {eq}. 20-2,
\nonu 
\\
\acomm{(\Phi^{(h_1),a}_{\frac{3}{2},+\frac{3}{2}})_r}{(\Phi^{(h_2),b}_{\frac{3}{2},
-\frac{3}{2}})_s}
& = & 
\kappa_{+\frac{3}{2},-\frac{3}{2},+2}\,\delta^{ab}\,
\Big((h_2+\tfrac{1}{2})r-(h_1+\tfrac{1}{2})s\Big)\,
(\Phi^{(h_1+h_2-1)}_{2,-2})_{r+s} \, : \mbox {eq}. 21,
\nonu \\
\acomm{(\Phi^{(h_1),a}_{\frac{3}{2},+\frac{3}{2}})_r}{(\Phi^{(h_2),b}_{\frac{3}{2},+
\frac{3}{2}})_s}
& = &
\kappa_{+\frac{3}{2},+\frac{3}{2},-1}\,\Big((h_2+\tfrac{1}{2})r-(h_1+\tfrac{1}{2})s\Big)\,
(\Phi^{(h_1+h_2),ab}_{1,+1})_{r+s} \, : \mbox {eq}. 22,
\nonu \\
\comm{(\Phi^{(h_1)}_{2,+2})_m}{(\Phi^{(h_2),a}_{\frac{3}{2}, + \frac{3}{2}})_r}
& = & 
\kappa_{+2,+\frac{3}{2},-\frac{3}{2}}\,\Big((h_2+\tfrac{1}{2})m-(h_1+1)r \Big)\,
(\Phi^{(h_1+h_2),a}_{\frac{3}{2},+ \frac{3}{2}})_{m+r} \, : \mbox {eq}. 23-1,
\nonu \\
\comm{(\Phi^{(h_1)}_{2,+2})_m}{(\Phi^{(h_2),a}_{\frac{3}{2}, - \frac{3}{2}})_r}
& = & 
\kappa_{+2,-\frac{3}{2},+\frac{3}{2}}\,\Big((h_2+\tfrac{1}{2})m-(h_1+1)r \Big)\,
(\Phi^{(h_1+h_2),a}_{\frac{3}{2},-\frac{3}{2}})_{m+r} \, : \mbox {eq}. 23-2,
\nonu \\
\comm{(\Phi^{(h_1)}_{2,+2})_m}{(\Phi^{(h_2)}_{2, + 2})_n}
& = & 
\kappa_{+2,+2,-2}\,\Big((h_2+1)m-(h_1+1)n\Big)\,(\Phi^{(h_1+h_2)}_{2,+ 2})_{m+n}
\, : \mbox {eq}. 24-1.
\nonu \\
\comm{(\Phi^{(h_1)}_{2,+2})_m}{(\Phi^{(h_2)}_{2, - 2})_n}
& = & 
\kappa_{+2,-2,+2}\,\Big((h_2+1)m-(h_1+1)n\Big)\,(\Phi^{(h_1+h_2)}_{2,- 2})_{m+n}
\, : \mbox {eq}. 24-2.
\nonu 
\eea
From the consistent truncation in \cite{Das}, we put
two Majoranas,
four spin $1$ vectors and
two spin $\frac{3}{2}$ gravitinos
to be zero among the generators in the
section \ref{sugra}
\bea
\psi_{\mu}^{i=3,4} =0, 
\qquad
\chi^{i =3,4} =0,
\qquad 
F_{\mu \nu}^{i j=13,14,23,24}=0,
\qquad
F_{\mu \nu}^{i j=12}  \equiv G_{\mu \nu},
\quad
F_{\mu \nu}^{ij =3 4} \equiv F_{\mu \nu},
\label{n2softcond}
\eea
which leads to the graviton, two gravitinos,
two vectors, two Majoranas, the scalar and the pseudoscalar.
The relations between the couplings are given previously by
(\ref{kapparelation})
\footnote{
When we interchange the ordering of the
modes on the left hand sides,
the following celestial commutators hold
  \bea
\comm{(\Phi^{(h_1),a}_{\frac{1}{2},\pm \frac{1}{2}})_r}{(\Phi^{(h_2)}_{2,+2})_m}
& = &
\Big((h_2+1)r-(h_1-\tfrac{1}{2})m \Big)\,
(\Phi^{(h_1+h_2),a}_{\frac{1}{2},\pm\frac{1}{2}})_{m+r} \, ,
\nonu \\
\comm{(\Phi^{(h_1),a}_{\frac{3}{2},\pm \frac{3}{2}})_r}{(\Phi^{(h_2}_{2,+2})_m}
& = &
\Big((h_2+1)r-(h_1+\tfrac{1}{2})m\Big)\,
(\Phi^{(h_1+h_2),a}_{\frac{3}{2},\pm\frac{3}{2}})_{m+r} \, .
\nonu
\eea
}. 

$\bullet$ $\frac{1}{\kappa^2} \, e \, R$
term

As before, 
the twenty fourth relation of (\ref{n2softmatter})
gives the celestial commutator between the three gravitons
with helicities $(+2, \pm 2, \mp 2)$.

$\bullet$
$\epsilon^{\mu \nu \sigma \rho} \, \bar{\psi}^i_{\mu} \, \ga_5\, \ga_{\nu}\,
D_{\rho}\, \psi_{\si}^i$
term

The summation over the index $i$
is given by $i=1,2$ which is $SO(2)$ vector index with (\ref{n2softcond}).
The twenty first relation of (\ref{n2softmatter})
provides the celestial anticommutator between the two gravitinos and the
graviton with the helicities $(-\frac{3}{2}, +\frac{3}{2},+2)$
while the twenty third relation of (\ref{n2softmatter})
gives the celestial commutator between the graviton and the two gravitinos
with the helicities $(+2, \pm \frac{3}{2},\mp \frac{3}{2})$.

$\bullet$
$e \, g^{\mu \rho} \, g^{\nu \si}\, G_{\mu \nu} \, G_{\rho \si}$
term

The fourteenth relation of (\ref{n2softmatter})
describes the celestial commutator between
two vectors with (\ref{n2softcond})
and the graviton
with the helicities $(+1, -1, +2)$ and 
the nineteenth relation of (\ref{n2softmatter})
explains the celestial commutator between the vector,
the graviton and the
vector where the helicities are given by $(\pm 1, +2, \mp 1)$.

$\bullet$
$\kappa \,  e\, g^{\mu \rho} \, g^{\nu \si}\,
\epsilon^{i j }
\bar{\psi}^i_{\mu} \,  G_{\rho \si}\, \psi_{\nu}^j$
and
$\kappa \,   \epsilon^{\mu \nu \rho \si} \,
\epsilon^{i j } \,
\bar{\psi}^i_{\mu} \,  \ga_5\, G_{\rho \si}\, \psi_{\nu}^j$
terms

Here, the indices $i,j$ are $SO(2)$ vector indices with
(\ref{n2softmatter}).
The seventeenth relation of (\ref{n2softmatter})
can give the celestial commutator between the vector and two gravitinos
with the helicities $(-1, +\frac{3}{2},+\frac{3}{2})$ 
while the twenty second relation of (\ref{n2softmatter})
describes the celestial
anticommutator between the two gravitinos and the vector
with the helicities $(+\frac{3}{2}, +\frac{3}{2},-1)$
\footnote{In these celestial (anti)commutators, the
two $SO(2)$ indices for the two gravitinos are different
from each other.}.


$\bullet$
$e \, g^{\mu \rho} \, g^{\nu \si}\, F_{\mu \nu} \, F_{\rho \si}$
term

As before, 
the fifteenth of (\ref{n2softmatter}) shows
the celestial commutator between the two vectors and the graviton
with the helicities $(+1,-1,+2)$
and
the twentieth of (\ref{n2softmatter}) describes
the celestial commutator between the vector, the graviton and the vector
with the helicities $(\pm 1, +2, \mp 1)$.

$\bullet$
$e \, \bar{\chi}^i \, \ga^{\mu} \, D_{\mu}\, \chi^i$
term

Again, there exists a summation over $SO(2)$ indices.
The seventh of (\ref{n2softmatter})
shows the celestial
anticommutator between the two Majoranas and the graviton
with the helicities $(+\frac{1}{2},-\frac{1}{2},+2)$
and 
the thirteenth of (\ref{n2softmatter})
provides the celestial
commutator between the graviton and the two
Majoranas with the helicities $(+2, \pm \frac{1}{2}, \mp \frac{1}{2})$.

$\bullet$
$e \, g^{\mu \nu}\, (\pa_{\mu}\, A) \, (\pa_{\nu}\, A)$
and 
$e \, g^{\mu \nu}\, (\pa_{\mu}\, B) \, (\pa_{\nu}\, B)$
terms

The first of (\ref{n2softmatter})
deals with the celestial commutator between the
complex scalar, conjugated scalar and the graviton with the helicities
$(+ 0, - 0, +2)$ and
the sixth of (\ref{n2softmatter})
deals with the celestial commutator between the
complex scalar (or conjugated scalar), the graviton and the
conjugated scalar (or complex scalar) with the helicities
$(\pm 0, +2, \mp 0)$.

$\bullet$
$\kappa \, e \, \epsilon^{i j } \, \bar{\psi}^i_{\la} \,
\sigma^{\mu \nu} \, F_{\mu \nu}\, \gamma^{\la}\, \chi^j$
term

There exists a summation over $SO(2)$ indices with (\ref{n2softcond}).
The tenth of (\ref{n2softmatter}) explains
the celestial commutator between the Majorana, the vector and the gravitino
with the helicities $(-\frac{1}{2},+1, +\frac{3}{2})$,
the eleventh of (\ref{n2softmatter}) describes the
celestial anticommutator between the Majorana, the gravitino and the
vector with the helicities $(-\frac{1}{2},+\frac{3}{2}, +1)$ and
the eighteenth of (\ref{n2softmatter})
provides the celestial
commutator between the vector, the gravitino and the
Majorana with the helicities $(+1, +\frac{3}{2}, -\frac{1}{2})$
\footnote{In these celestial commutators,
the $SO(2)$ indices for the gravitino and the Majorana are
different from each other.}.

$\bullet$
$\kappa \, e\, \bar{\psi}^i_{\mu} \, (\pa_{\nu} \, A) \, \ga^{\nu}\,
\ga^{\mu} \, \chi^i$ and
$\kappa \, e\, \bar{\psi}^i_{\mu} \, (\pa_{\nu} \, \ga_5\, B) \, \ga^{\nu}\,
\ga^{\mu} \, \chi^i$
terms

The summation over the index $i$
is given by $i=1,2$ for $SO(2)$ vector.
The second of (\ref{n2softmatter}) shows the celestial commutator
between the conjugated scalar,  the
Majorana and the gravitino with the helicities $(- 0,
+\frac{1}{2}, +\frac{3}{2})$,
the fifth of (\ref{n2softmatter}) describes
the celestial commutator
between the conjugated scalar, the gravitino
and the Majorana with the helicities $(- 0, +\frac{3}{2},
+\frac{1}{2})$ and 
the twelfth of (\ref{n2softmatter}) deals with
celestial anticommutator between the Majorana, the gravitino and
the conjugated scalar with the helicities
$(+\frac{1}{2}, +\frac{3}{2},- 0)$.

$\bullet$
$\kappa \, e \, g^{\mu \rho} \, g^{\nu \si}\,
A\,
F_{\mu \nu} \, G_{\rho \si}$
and 
$\kappa \, e \, g^{\mu \rho} \, g^{\nu \si} \, B\,
F_{\mu \nu} \, G_{\rho \si}$
terms

The third of (\ref{n2softmatter}) describes the
celestial commutator between the conjugated scalar, the two vectors
with the helicities $(- 0, +1, +1)$,
the fourth of (\ref{n2softmatter}) describes the
celestial commutator between the conjugated scalar,
the two vectors with the helicities $(- 0, +1, +1)$,
the sixteenth of (\ref{n2softmatter}) gives the
celestial commutator between the two vectors and the
conjugated scalar with the helicities $(+1,+1, - 0)$.

\subsection{The soft current algebra and ${\cal N}=2$
supergravity theory coupled to its several Abelian vector multiplets}

In \cite{Luciani}, the Lagrangian is written for
the ${\cal N}=2$
supergravity theory coupled to its several Abelian vector multiplets.
The previous single vector multiplet of spins 
$(0^{\pm}, \frac{1}{2}, \frac{1}{2}, 1)$
denoted by the matter
$(A,B, \chi^i, F_{\mu \nu})$ is generalized to the multiple
vector multiplets. It would be interesting to observe how the
corresponding soft current algebra arises. For the
${\cal N}=8$ supergravity theory \cite{dF}, there exist
the twenty eight vectors. It is an open problem to check whether
there exists the ${\cal N}=2$ supergravity theory with
the several vector multiplets from the ${\cal N}=8$ supergravity theory,
by truncation or not
\footnote{It is an open problem to check whether
the relations (\ref{n3soft}),
(\ref{n2soft}), (\ref{n1soft}) and (\ref{n2softmatter})
can be written as the superspace description in
(\ref{twosuper}) with the corresponding truncations or not.}.

\section{
The 
${\cal N}=2$ supersymmetric
$W_{1+\infty}^{K,K}[\la =0]$ algebra
}

In this section,
by using the different free field realization,
we describe the corresponding soft current algebra
for ${\cal N}=2$ supergravity theory.

\subsection{The realization of
${\cal N}=2$ $SO(2)$ superconformal algebra}


By introducing the simplified notations \footnote{
Here the previous
$SU(2)$ fundamental $a$ (and antifundamental $\bar{a}$) indices
are generalized to the ones in $SU(K)$. The number of free fields
in the $(\beta,\ga)$ and $(b,c)$ system around (\ref{fundOPE})
is given by $K$.} 
on the currents in \cite{AK2309}
\bea
W^{\la=0,\bar{a} a}_{\mathrm{F},h}\,\delta_{ a \bar{a}} & \equiv &
W^{\la=0}_{\mathrm{F},h}\,,
\qquad
W^{\la=0,\bar{a} a}_{\mathrm{B},h}\,\delta_{ a \bar{a}} \equiv
W^{\la=0}_{\mathrm{B},h}\,,
\nonu \\
Q^{\la=0,\bar{a} a}_{h+\frac{1}{2}}\,\delta_{ a \bar{a}} & \equiv &
Q^{\la=0}_{h+\frac{1}{2}}\,,
\qquad
\bar{Q}^{\la=0,\bar{a} a}_{h+\frac{1}{2}}\,\delta_{ a \bar{a}} \equiv
\bar{Q}^{\la=0}_{h+\frac{1}{2}}\,,
\label{simple}
\eea
we can identify the currents appearing in \cite{BPRSS} with
the ones in \cite{AK2309} as follows:
\bea
V^i(z)&= & (-1)^i\,W^{\la=0}_{\mathrm{B},i+2}\,,\quad i= 0, 1,
\cdots, 
\nonu \\
\tilde{V}^i(z)&=&
(-1)^i\,W^{\la=0}_{\mathrm{F},i+2}\,,\quad i=-1, 0, 1, \cdots,
\nonu \\
G^\alpha(z)&=&
(-1)^{\alpha}\,\frac{1}{\sqrt{2}}\,Q^{\la=0}_{\alpha+\frac{3}{2}}\,,\quad\alpha= 0, 1, 2, \cdots,
\nonu \\
\bar{G}^\alpha(z)&=&
(-1)^{\alpha}\,\frac{1}{\sqrt{2}}\,\bar{Q}^{\la=0}_{\alpha+\frac{3}{2}}\,,\quad \alpha =  0, 1, \cdots.
\label{REL}
\eea
Moreover, the currents appearing in \cite{BPRSS} can be associated with
the ones in \cite{Sezgin9112} as follows:
\bea
V^{l}_{\,\,\,m}& \longrightarrow &
\frac{1}{2}\,\Big( (1+\de^{l,-1})\,
v^{l}_{\,\,\,m}-\frac{1}{2} \,q^{-1}\,J^{l}_{\,\,\,m}\Big)\,,
\qquad
\tilde{V}^{l}_{\,\,\,m} \longrightarrow 
\frac{1}{2}\,\Big( v^{l}_{\,\,\,m}+\frac{1}{2} \,q^{-1}\,J^{l}_{\,\,\,m}\Big)\,,
\nonu \\
\tilde{V}^{-1}_{\,\,\,m}  & \longrightarrow & 
\frac{1}{4} \,q^{-1}\,J^{-1}_{\,\,\,m}\,,
\qquad G^{\alpha}_{\,\,\,r} \longrightarrow 
\frac{1}{\sqrt{2}}\,G^{\alpha,+}_{\,\,\,r} \,,
\qquad
\bar{G}^{\alpha}_{\,\,\,r} \longrightarrow 
\frac{1}{\sqrt{2}}\,G^{\alpha,-}_{\, \,\,r}\,.
\label{REL1}
\eea
Then it turns out, from (\ref{simple}), (\ref{REL}),
(\ref{REL1}) and (\ref{Phih}), that 
we have the explicit relations between
the currents in \cite{Sezgin9112}
and the currents in \cite{AK2309} at $\la=0$ as follows:
\bea
 v^i\quad &\longleftrightarrow & \quad
 \Phi^{(i)}_2 \equiv 4^{-i}\,\tilde{\Phi}^{(i)}_2
\,, \nonu \\
 J^{i-1}\quad &\longleftrightarrow  & \quad
 \Phi^{(i)}_1 \equiv
-2\, (4)^{-i+1} \Big( q^{-1}\,\Phi^{(i+1)}_0
+\frac{q}{2i+1}\,\tilde{\Phi}^{(i-1)}_2 \Big)\,, \nonu \\
 J^{-1}\quad &\longleftrightarrow  & \quad
 \Phi^{(0)}_1 \equiv
-2\, (4) \, q^{-1}\,\Phi^{(1)}_0 \,, \nonu \\
 G^{\alpha,\pm}\quad &\longleftrightarrow &
\quad  \Phi^{(\alpha),\pm}_\frac{3}{2} \equiv
-
2\sqrt{2}(4)^{-\alpha} \nonu \\
&\times & \bigg[
\Big(\tilde{\Phi}^{(\alpha),1}_\frac{3}{2}
-\mathrm{i}\,\tilde{\Phi}^{(\alpha),2}_\frac{3}{2}
-\mathrm{i}\,\tilde{\Phi}^{(\alpha),3}_\frac{3}{2}
+3\,\tilde{\Phi}^{(\alpha),4}_\frac{3}{2}
\Big)
\nonu \\
& \pm & q^{-1}
\Big(
\Phi^{(\alpha+1),1}_\frac{1}{2}
-\mathrm{i}\,\Phi^{(\alpha+1),2}_\frac{1}{2}
-\mathrm{i}\,\Phi^{(\alpha+1),3}_\frac{1}{2}
+3\,\Phi^{(\alpha+1),4}_\frac{1}{2}
\Big)
\bigg]\,,
\label{relationinsection3}
\eea
where the final expressions appearing on the right hand sides
of (\ref{relationinsection3}) hold for $K=2$.

Then the standard ${\cal N}=2$ superconformal algebra, by using
(\ref{relationinsection3}),
is realized by 
\bea
\comm{(\Phi^{(0)}_2)_{m}}{(\Phi^{(0)}_2)_{n}}&=&
\frac{c}{24}\,K\,(m^3-m)\,\delta_{m+n}+(m-n)\,(\Phi^{(0)}_2)_{m+n}\,,
\nonu \\
\comm{(\Phi^{(0)}_2)_{m}}{(\Phi^{(0)}_1)_{n}}&= &
-n\,(\Phi^{(0)}_1)_{m+n}
\,,
\nonu \\
\acomm{(\Phi^{(0),-}_\frac{3}{2})_{r}}{(\Phi^{(0),+}_\frac{3}{2})_{s}}&= &
\frac{c}{6}\,K\,(r^2-\tfrac{1}{4})\,\delta_{r+s}+2\,(\Phi^{(0)}_2)_{r+s}-
(r-s)\,(\Phi^{(0)}_1)_{r+s} \, ,
\nonu \\
\comm{(\Phi^{(0)}_2)_{m}}{(\Phi^{(0),\pm}_\frac{3}{2})_{r}}&=&
(\tfrac{1}{2}m-r )\,(\Phi^{(0),\pm}_\frac{3}{2})_{m+r}
\,,
\nonu \\
\comm{(\Phi^{(0)}_1)_{m}}{(\Phi^{(0),\pm}_\frac{3}{2})_{r}}&=&
\pm (\Phi^{(0),\pm}_\frac{3}{2})_{m+r} \, ,
\nonu \\
\comm{(\Phi^{(h_1)}_1)_{m}}{(\Phi^{(h_2)}_1)_{n}}&= &
\frac{c}{6}\,K\,m \,\delta_{m+n} \, ,
\label{n2scadiff}
\eea
where the central term $c$ is given by $c = 6 N$.
In other words, the standard central term is given by
$3 N K$.
Note that the Jacobi identities for
(\ref{n2scadiff}) are satisfied
\footnote{
\label{commsection7}
  As before, 
we can rewrite some of the commutators,
by changing the ordering
of the modes on the left hand sides, as
\bea
\comm{(\Phi^{(0)}_1)_{m}}{(\Phi^{(0)}_2)_{n}} & = & m \, (\Phi^{(0)}_1)_{m+n},
\qquad
\comm{(\Phi^{(0),\pm}_\frac{3}{2})_{r}}{(\Phi^{(0)}_2)_{m}}  = 
\pm (\Phi^{(0),\pm}_\frac{3}{2})_{m+r}.
\nonu
\eea
These correspond to the second and the fourth relations of
(\ref{n2scadiff}).}.



\subsection{The extension of ${\cal N}=2$
superconformal algebra}

Then we can try to calculate the above algebra
for nonzero $h_1$ and $h_2$
and it turns out that 
\bea
\comm{(\Phi^{(h_1)}_2)_{m}}{(\Phi^{(h_2)}_2)_{n}} & = &
\frac{c}{24}\,K\,(m^3-m)\,\de^{h_1,0} \, \de^{h_2,0}\, \delta_{m+n}
\nonu \\
& + &
\Big((h_2+1)m-(h_1+1)n\Big)\,(\Phi^{(h_1+h_2)}_2)_{m+n} \,,
\nonu \\
\comm{(\Phi^{(h_1)}_2)_{m}}{(\Phi^{(h_2)}_1)_{n}} & = &
\Big(h_2\,m-(h_1+1)n\Big)\,(\Phi^{(h_1+h_2)}_1)_{m+n} \,,
\nonu 
\\
\acomm{(\Phi^{(h_1),-}_\frac{3}{2})_{r}}{(\Phi^{(h_2),+}_\frac{3}{2})_{s}}
& = &
\frac{c}{6}\,K\,(r^2-\frac{1}{4})\,\de^{h_1,0} \, \de^{h_2,0}\,
\de_{r+s}
\nonu \\
& + &
2\,(\Phi^{(h_1+h_2)}_2)_{r+s}
-  2\,\Big(  (h_2+\tfrac{1}{2})r-(h_1+\tfrac{1}{2})s\Big)\,
(\Phi^{(h_1+h_2)}_1)_{r+s} \, ,
\nonu \\
\comm{(\Phi^{(h_1)}_2)_{m}}{(\Phi^{(h_2),\pm}_\frac{3}{2})_{r}} & = &
\Big((h_2+\tfrac{1}{2})\,m-(h_1+1)r\Big)\,
(\Phi^{(h_1+h_2),\pm}_\frac{3}{2})_{\,\,\,m+r}
\,,
\nonu \\
\comm{(\Phi^{(h_1)}_1)_{m}}{(\Phi^{(h_2),\pm}_\frac{3}{2})_{r}} & = &
\pm (\Phi^{(h_1+h_2),\pm}_\frac{3}{2})_{m+r} \, ,
\nonu \\
\comm{(\Phi^{(h_1)}_1)_{m}}{(\Phi^{(h_2)}_1)_{n}}
& = &
\frac{c}{6}\,K\,m \,\de^{h_1,0}\, \de^{h_2,0}\, \delta_{m+n} \nonu \\
& + &
q^2\,4\Big(h_2\,m-h_1\,n\Big)\,(\Phi^{(h_1+h_2-2)}_2)_{m+n} \, .
\label{genn2case} 
\eea
Note that the last relation of (\ref{genn2case})
implies the nonzero contribution for nonzero $h_1$ and $h_2$
\footnote{
\label{prevfootnote}
For the commutators presented in previous footnote
\ref{commsection7},
we have the following commutators
for nonzero $h_1$ and $h_2$ in different ordering
\bea
\comm{(\Phi^{(h_1)}_1)_{m}}{(\Phi^{(h_2)}_2)_{n}} & = &
\Big((h_2+1)\,m-h_1\, n\Big)\,(\Phi^{(h_1+h_2)}_1)_{m+n},
\nonu \\
\comm{(\Phi^{(h_1),\pm}_\frac{3}{2})_{r}}{(\Phi^{(h_2)}_2)_{m}} & = &
-\Big((h_1+\tfrac{1}{2})\,m-(h_2+1)r\Big)\,
(\Phi^{(h_1+h_2),\pm}_\frac{3}{2})_{\,\,\,m+r}.
\nonu
\eea
}.

\subsection{The 
${\cal N}=2$ supersymmetric
$W_{1+\infty}^{K,K}[\la =0]$ algebra}

By considering all the terms ignored in
(\ref{genn2case}), the extension of  standard (anti)commutators
for the ${\cal N}=2$ superconformal algebra,
the 
${\cal N}=2$ supersymmetric
$W_{1+\infty}^{K,K}[\la =0]$ algebra,
can be obtained and is given by 
(\ref{n2appendix}).
We list the higher order terms in $q$ in (\ref{subleading}).
A particular Jacobi identity in order to
observe how the $q^2$ terms in the Jacobi identity vanish,
by adding the higher order terms in $q$ to (\ref{genn2case}),
is described in (\ref{Jacobiexp}),
(\ref{firstterm}) and (\ref{thirdterm}).
Note that under the conditions of (\ref{COND1}),
the structure constants can be written in terms of
the known ones in the literatures (For example, \cite{PRSnpb}).


\subsection{The soft current algebra and ${\cal N}=2$
supergravity theory}

We can write down the soft current algebra
based on (\ref{genn2case}) by putting the corresponding
helicities properly
\bea
\comm{(\Phi^{(h_1)}_{2,+2})_m}{(\Phi^{(h_2)}_{2,+ 2})_n}&=&
\kappa_{+2,+2,-2}\,\Big((h_2+1)\,m-(h_1+1)\,n\Big)\,(\Phi^{(h_1+h_2)}_{2,+ 2})_{m+n}\,
,
\nonu\\
\comm{(\Phi^{(h_1)}_{2,+2})_m}{(\Phi^{(h_2)}_{2,- 2})_n}&=&
\kappa_{+2,-2,+2}\,\Big((h_2+1)\,m-(h_1+1)\,n\Big)\,(\Phi^{(h_1+h_2)}_{2,- 2})_{m+n}\,
,
\nonu\\
\comm{(\Phi^{(h_1)}_{2,+2})_m}{(\Phi^{(h_2)}_{1,+ 1})_n}&=&
\kappa_{+1,+2,-1}\,\Big(h_2\,m-(h_1+1)\,n\Big)\,(\Phi^{(h_1+h_2)}_{1,+ 1})_{m+n}\,
,
\nonu  \\
\comm{(\Phi^{(h_1)}_{2,+2})_m}{(\Phi^{(h_2)}_{1,- 1})_n}&=&
\kappa_{-1,+2,+1}\,\Big(h_2\,m-(h_1+1)\,n\Big)\,(\Phi^{(h_1+h_2)}_{1,- 1})_{m+n}\,
,
\nonu  \\
\acomm{(\Phi^{(h_1),-}_{\frac{3}{2},-\frac{3}{2}})_r}{
(\Phi^{(h_2),+}_{\frac{3}{2},+\frac{3}{2}})_s}&=&
2\,\kappa_{-\frac{3}{2},+\frac{3}{2},+2}\,\Big((h_2+\tfrac{1}{2})r-(h_1+\tfrac{1}{2})s\Big)\,(\Phi^{(h_1+h_2-1)}_{2,-2})_{r+s} \,
,
\nonu   \\
\acomm{(\Phi^{(h_1),-}_{\frac{3}{2},+\frac{3}{2}})_r}{
(\Phi^{(h_2),+}_{\frac{3}{2},-\frac{3}{2}})_s}&=&
-2\,\kappa_{-\frac{3}{2},+\frac{3}{2},+2}\,\Big((h_2+\tfrac{1}{2})r-(h_1+\tfrac{1}{2})s\Big)\,(\Phi^{(h_1+h_2-1)}_{2,-2})_{r+s} \,
,
\nonu   \\
\acomm{(\Phi^{(h_1),-}_{\frac{3}{2},+\frac{3}{2}})_r}{
(\Phi^{(h_2),+}_{\frac{3}{2},+\frac{3}{2}})_s}&=&
2\mathrm{i}\,\kappa_{+\frac{3}{2},+\frac{3}{2},-1}\,\Big((h_2+\tfrac{1}{2})\,r-(h_1+\tfrac{1}{2})\,s\Big)\,
(\Phi^{(h_1+h_2)}_{1,+1})_{r+s}\, ,
\nonu   \\
\comm{(\Phi^{(h_1)}_{2,+2})_m}{(\Phi^{(h_2),+}_{\frac{3}{2},+ \frac{3}{2}})_r}&=&
\kappa_{+\frac{3}{2},+2,-\frac{3}{2}}\,\Big((h_2+\tfrac{1}{2})\,r-(h_1+1)\,m\Big)\,
(\Phi^{(h_1+h_2),+}_{\frac{3}{2},+ \frac{3}{2}})_{r+m}\,
,
\nonu \\
\comm{(\Phi^{(h_1)}_{2,+2})_m}{(\Phi^{(h_2),-}_{\frac{3}{2},- \frac{3}{2}})_r}&=&
\kappa_{-\frac{3}{2},+2,+\frac{3}{2}}\,\Big((h_2+\tfrac{1}{2})\,r-(h_1+1)\,m\Big)\,
(\Phi^{(h_1+h_2),-}_{\frac{3}{2},- \frac{3}{2}})_{r+m}\,
,
\nonu \\
\comm{(\Phi^{(h_1)}_{1,-1})_m}{(\Phi^{(h_2),+}_{\frac{3}{2},+\frac{3}{2}})_r}&=&
-\mathrm{i}\,\kappa_{-1,+\frac{3}{2},+\frac{3}{2}}\,\Big((h_2+\tfrac{1}{2})m-h_1\,r\Big)\,
(\Phi^{(h_1+h_2-1),+}_{\frac{3}{2},-\frac{3}{2}})_{m+r}\,
,
\nonu \\
\comm{(\Phi^{(h_1)}_{1,-1})_m}{(\Phi^{(h_2),-}_{\frac{3}{2},+\frac{3}{2}})_r}&=&
\mathrm{i}\,\kappa_{-1,+\frac{3}{2},+\frac{3}{2}}\,\Big((h_2+\tfrac{1}{2})m-h_1\,r\Big)\,
(\Phi^{(h_1+h_2-1),-}_{\frac{3}{2},-\frac{3}{2}})_{m+r}\,
,
\nonu \\
\comm{(\Phi^{(h_1)}_{1,+1})_m}{(\Phi^{(h_2)}_{1,-1})_n}&=
& \kappa_{+1,-1,+2}\,\Big(h_2\,m-h_1\,n\Big)\,(\Phi^{(h_1+h_2-2)}_{2,-2})_{m+n}\,
,
\label{softsoft}
\eea
This is equivalent to the previous results in (\ref{n2soft}).
Therefore, we can do the similar analysis for the
soft current algebra corresponding to the ${\cal N}=2$ supergravity
as before \footnote{
Moreover, 
we can rewrite the above celestial
commutators (the second and the fifth
relations in (\ref{softsoft})) as 
\bea
\comm{(\Phi^{(h_1)}_{1,\pm 1})_{m}}{(\Phi^{(h_2)}_{2,+2})_{n}} & = &
\Big((h_2+1)\,m-h_1\, n\Big)\,(\Phi^{(h_1+h_2)}_{1, \pm 1})_{m+n},
\nonu \\
\comm{(\Phi^{(h_1),\pm}_{\frac{3}{2},\pm \frac{3}{2}})_{r}}{(\Phi^{(h_2)}_{2,+2})_{m}}& = & -\Big((h_1+\tfrac{1}{2})\,m-(h_2+1)r\Big)\,
(\Phi^{(h_1+h_2),\pm}_{\frac{3}{2},\pm \frac{3}{2}})_{\,\,\,m+r},
\nonu
\eea
which originate from the ones in the footnote \ref{prevfootnote}.
}.

%
%
%
%

\section{ Conclusions and outlook}

The main results of this paper are described in the abstract.
The most important equations are given by (\ref{finalanticommcomm})
(originated from (\ref{N4algebrafor})) and (\ref{SOFT}).
In the latter, the helicities of the particles occur.
By using the celestial holography, the soft current algebra given by
(\ref{SOFT}) can be interpreted as the three point amplitudes
of the cubic terms in the
Lagrangian
of the ${\cal N}=4$ supergravity theory by Das.
In doing this, the condition of $\la =\frac{1}{4}$ is crucial.

It is an open problem to obtain the
soft current algebra corresponding to the ${\cal N}=8$ supergravity
as a next step.
So far, we have considered the particular $\la=\frac{1}{4}$
or $\la=0$. On the other hand, for general $\la$,
it is also interesting to examine how we can associate the
given two dimensional conformal field theory with the bulk theory
in four dimensions.
It is also open problem to discuss about the
results of this paper in the context of the supersymmetric soft theorem.
Although we expect that the Jacobi identity
is satisfied, it is also nontrivial to see this
from the explicit (anti)commutators in Appendix $A$.
We have considered only the case of $d_V=5$
and it is an open problem to
observe
how the higher order derivative terms, where
$d_V=6,7,8,9$ (or the sum of
helicities is given by $3,4,5,6$)
in the context of the supersymmetric celestial algebra, 
occur.
It is not clear how the current results of this paper
can be  associated with the supersymmetric asymptotic symmetries.
Moreover, it is an open problem to
check whether the extension of
small ${\cal N}=4$ superconformal algebra
is realized or not.
It is not clear how the results of this paper can associate
with the algebra having the trigonometric structure constants
described in the abstract.


\vspace{.7cm}

\centerline{\bf Acknowledgments}

CA thanks M. Pate for  the discussion on
the general aspects of the celestial holography, 
A. Tropper for the discussion on
the module of the Lie algebra,
the global superconformal subalgebra and 
the three point coefficients in the context of 
\cite{Tropper2412} and
on the supersymmetric soft theorems \cite{Tropper},
M. Rocek for the discussion on the
${\cal N}=4$ $SO(4)$ supergravity and
A. Das for the discussion on his paper \cite{Das}.
We thank the referees for the discussion on the improving this draft.
This work was supported by the National
Research Foundation of Korea(NRF) grant funded by the
Korea government(MSIT) 
(No. 2023R1A2C1003750).
This research was supported by Basic Science Research
Program through the National Research Foundation of Korea(NRF)
funded by the Ministry of Education(RS-2024-00446084).

\newpage

\appendix

\renewcommand{\theequation}{\Alph{section}\mbox{.}\arabic{equation}}

\section{The extension of ${\cal N}=4$ $SO(4)$
superconformal algebra: the
${\cal N}=4$ supersymmetric
$W_{1+\infty}^{2,2}[\la =\frac{1}{4}]$ algebra
}

Some details appearing in sections $2$ and $3$ are
presented in this Appendix.

+
\subsection{The $16$ generators in terms of free fields}

The component fields in (\ref{bigPhi})
in terms of (\ref{fourcurrents}) are given by
\bea
\Phi_0^{(h)} & = &2 \, \frac{q^{2-h}(-4)^{h-2}}{(2h-1)}\,
\Bigg[ -(h-2\la)\,
  W^{\la,\bar{a} a }_{\mathrm{F},h}+
  (h-1+2\la) \, W^{\la,\bar{a} a}_{\mathrm{B},h} \Bigg],
\nonu \\
\Phi^{(h),1}_{\frac{1}{2}}
&
=& - \sqrt{2} \, q^{1-h}\, (-4)^{h-3}\,
\Bigg[-\frac{1}{2}\,(
Q^{\la,11}_{h+\frac{1}{2}}
+i\sqrt{2}\,Q^{\la,12}_{h+\frac{1}{2}}
+2i \sqrt{2}\,Q^{\la,21}_{h+\frac{1}{2}}
-2\,Q^{\la,22}_{h+\frac{1}{2}}
\nonu \\
& + & 2\,\bar{Q}^{\la,11}_{h+\frac{1}{2}}
+2i \sqrt{2}\, \bar{Q}^{\la,12}_{h+\frac{1}{2}}
+i\sqrt{2}\,\bar{Q}^{\la,21}_{h+\frac{1}{2}}
-\bar{Q}^{\la,22}_{h+\frac{1}{2}}
)\,\Bigg],
\nonu\\
\Phi^{(h),2}_{\frac{1}{2}}
&
=& -  \sqrt{2}  q^{1-h} (-4)^{h-3}\Bigg[
\frac{i}{2}(
Q^{\la,11}_{h+\frac{1}{2}}
+2i\sqrt{2} Q^{\la,21}_{h+\frac{1}{2}}
-2 Q^{\la,22}_{\frac{3}{2}}
+2 \bar{Q}^{\la,11}_{h+\frac{1}{2}}
+2i\sqrt{2}  \bar{Q}^{\la,12}_{h+\frac{1}{2}}
-\bar{Q}^{\la,22}_{h+\frac{1}{2}}
) \Bigg],
\nonu\\
\Phi^{(h),3}_{\frac{1}{2}}
&
=& -  \sqrt{2}  q^{1-h} (-4)^{h-3}\Bigg[
\frac{i}{2}(
Q^{\la,11}_{h+\frac{1}{2}}
+i\sqrt{2} Q^{\la,12}_{h+\frac{1}{2}}
-2Q^{\la,22}_{h+\frac{1}{2}}
+2 \bar{Q}^{\la,11}_{h+\frac{1}{2}}
+i \sqrt{2}  \bar{Q}^{\la,21}_{h+\frac{1}{2}}
-\bar{Q}^{\la,22}_{h+\frac{1}{2}}
) \Bigg],
\nonu\\
\Phi^{(h),4}_{\frac{1}{2}}
&
=& -  \sqrt{2} \, q^{1-h}\, (-4)^{h-3}\,\Bigg[
\frac{1}{2}\,Q^{\la,11}_{h+\frac{1}{2}}
+Q^{\la,22}_{h+\frac{1}{2}}
-\bar{Q}^{\la,11}_{h+\frac{1}{2}}
-\frac{1}{2}\,\bar{Q}^{\la,22}_{h+\frac{1}{2}} \Bigg],
\nonu \\
\Phi^{(h),12}_{1}
&
=&  8\, q^{1-h}\, (-4)^{h-3}\,\Bigg[
2i\,W^{\la,11}_{\mathrm{B},h+1}
-\sqrt{2}\,W^{\la,12}_{\mathrm{B},h+1}
-2i\,\,W^{\la,22}_{\mathrm{B},h+1}
\nonu \\
& + & 2i\,W^{\la,11}_{\mathrm{F},h+1}
-2\sqrt{2}\,W^{\la,12}_{\mathrm{F},h+1}
-2i\,W^{\la,22}_{\mathrm{F},h+1}\, \Bigg],
\nonu\\
\Phi^{(h),13}_{1}
&
=& 8 \,
  q^{1-h}\, (-4)^{h-3}\,\Bigg[
-2i\,W^{\la,11}_{\mathrm{B},h+1}
+4\sqrt{2}\,W^{\la,21}_{\mathrm{B},h+1}
+2i\,\,W^{\la,22}_{\mathrm{B},h+1}
\nonu \\
& - & 2i\,W^{\la,11}_{\mathrm{F},h+1}
+2\sqrt{2}\,W^{\la,21}_{\mathrm{F},h+1}
+2i\,W^{\la,22}_{\mathrm{F}+h+1}\, \Bigg],
\nonu\\
\Phi^{(h),14}_{1}
&
=&  8\,  q^{1-h}\, (-4)^{h-3}\,\Bigg[
2\,W^{\la,11}_{\mathrm{B},h+1}
+i\sqrt{2}\,W^{\la,12}_{\mathrm{B},h+1}
+4i\sqrt{2}\,\,W^{\la,21}_{\mathrm{B},h+1}
-2\,W^{\la,22}_{\mathrm{B},h+1}
\nonu \\
& - & 2\,W^{\la,11}_{\mathrm{F},h+1}
-2i\sqrt{2}\,W^{\la,12}_{\mathrm{F},h+1}
 -  2i\sqrt{2}\,W^{\la,21}_{\mathrm{F},h+1}
+2\,W^{\la,22}_{\mathrm{F},h+1}\, \Bigg],
\nonu\\
\Phi^{(h),23}_{1}
&
=&  8\,  q^{1-h}\, (-4)^{h-3}\,\Bigg[
-2\,W^{\la,11}_{\mathrm{B},h+1}
-i\sqrt{2}\,W^{\la,12}_{\mathrm{B},h+1}
-4i\sqrt{2}\,\,W^{\la,21}_{\mathrm{B},h+1}
+2\,W^{\la,22}_{\mathrm{B},h+1}
\nonu \\
& - & 2\,W^{\la,11}_{\mathrm{F},h+1}
-2i\sqrt{2}\,W^{\la,12}_{\mathrm{F},h+1}
-  2i\sqrt{2}\,W^{\la,21}_{\mathrm{F},h+1}
+2\,W^{\la,22}_{\mathrm{F},h+1}\, \Bigg],
\nonu\\
\Phi^{(h),24}_{1}
&
=&  8\, q^{1-h}\, (-4)^{h-3}\,\Bigg[
-2i\,W^{\la,11}_{\mathrm{B},h+1}
+4\sqrt{2}\,W^{\la,21}_{\mathrm{B},h+1}
+2i\,\,W^{\la,22}_{\mathrm{B},h+1}
\nonu \\
& + & 2i\,W^{\la,11}_{\mathrm{F},h+1}
-2\sqrt{2}\,W^{\la,21}_{\mathrm{F},h+1}
-2i\,W^{\la,22}_{\mathrm{F},h+1}\, \Bigg],
\nonu\\
\Phi^{(h),34}_{1}
&
=&  8\,  q^{1-h}\, (-4)^{h-3}\,\Bigg[
-2i\,W^{\la,11}_{\mathrm{B},h+1}
+\sqrt{2}\,W^{\la,12}_{\mathrm{B},h+1}
+2i\,\,W^{\la,22}_{\mathrm{B},h+1}
\nonu \\
& + & 2i\,W^{\la,11}_{\mathrm{F},h+1}
-2\sqrt{2}\,W^{\la,12}_{\mathrm{F},h+1}
-2i\,W^{\la,22}_{\mathrm{F},h+1}\, \Bigg],
\nonu \\
\tilde{\Phi}^{(h),1}_{\frac{3}{2}}
&
\equiv &
\Big( \Phi^{(h),1}_{\frac{3}{2}} -\frac{1}{(2h+1)}\, (1-4\la)\,
\pa \,\Phi^{(h),1}_{\frac{1}{2}} \Big)
\nonu \\
&=&  \sqrt{32} \, q^{-h}\, (-4)^{h-3}\,\Bigg[
  -\frac{1}{2}\,(
Q^{\la,11}_{h+\frac{3}{2}}
+i\sqrt{2}\,Q^{\la,12}_{h+\frac{3}{2}}
+2i\sqrt{2}\,Q^{\la,21}_{h+\frac{3}{2}}
-2\,Q^{\la,22}_{h+\frac{3}{2}}
\nonu \\
& - & 2\,\bar{Q}^{\la,11}_{h+\frac{3}{2}}
-2i\sqrt{2}\,\bar{Q}^{\la,12}_{h+\frac{3}{2}}
-i\sqrt{2}\,\bar{Q}^{\la,21}_{h+\frac{3}{2}}
+\bar{Q}^{\la,22}_{h+\frac{3}{2}}
)\, \Bigg],
\nonu\\
\tilde{\Phi}^{(h),2}_{\frac{3}{2}}
&
\equiv &
\Big (\Phi^{(h),2}_{\frac{3}{2}} -\frac{1}{(2h+1)}\, (1-4\la)\,
\pa \,\Phi^{(h),2}_{\frac{1}{2}} \Big)
=  \sqrt{32} \,  q^{-h}\, (-4)^{h-3}\,
\nonu \\
& \times & \Bigg[
\frac{i}{2}\,(\,
Q^{\la,11}_{h+\frac{3}{2}}
+2i\sqrt{2}\,Q^{\la,21}_{h+\frac{3}{2}}
-2\,Q^{\la,22}_{h+\frac{3}{2}}
-2\,\bar{Q}^{\la,11}_{h+\frac{3}{2}}
-2i\sqrt{2}\,\bar{Q}^{\la,12}_{h+\frac{3}{2}}
+\bar{Q}^{\la,22}_{h+\frac{3}{2}}
)\, \Bigg],
\nonu\\
\tilde{\Phi}^{(h),3}_{\frac{3}{2}}
&
\equiv &
\Big(
\Phi^{(h),3}_{\frac{3}{2}} -\frac{1}{(2h+1)}\, (1-4\la)\,
\pa \,\Phi^{(h),3}_{\frac{1}{2}} \Big)
=  \sqrt{32}\, q^{-h}\, (-4)^{h-3}\,
\nonu \\
& \times & \Bigg[
\frac{i}{2}\,(
Q^{\la,11}_{h+\frac{3}{2}}
+i\sqrt{2}\,Q^{\la,12}_{h+\frac{3}{2}}
-2\,Q^{\la,22}_{h+\frac{3}{2}}
-2\,\bar{Q}^{\la,11}_{h+\frac{3}{2}}
-i\sqrt{2}\,\bar{Q}^{\la,21}_{h+\frac{3}{2}}
+\bar{Q}^{\la,22}_{h+\frac{3}{2}}
)\, \Bigg],
\nonu\\
\tilde{\Phi}^{(h),4}_{\frac{3}{2}}
&
\equiv &
\Big(
\Phi^{(h),4}_{\frac{3}{2}} -\frac{1}{(2h+1)}\, (1-4\la)\,
\pa \,\Phi^{(h),4}_{\frac{1}{2}} \Big)
\nonu \\
&
=&  \sqrt{32}\,  q^{-h}\, (-4)^{h-3}\,\Bigg[
\frac{1}{2}\,
(
Q^{\la,11}_{h+\frac{3}{2}}
+2\,Q^{\la,22}_{h+\frac{3}{2}}
+2\,\bar{Q}^{\la,11}_{h+\frac{3}{2}}
+\bar{Q}^{\la,22}_{h+\frac{3}{2}}
)\, \Bigg],
\nonu \\
\tilde{\Phi}^{(h)}_{2}
&
\equiv &
\Big(
\Phi^{(h)}_{2} -\frac{1}{(2h+1)} (1-4\la)
\pa^2 \Phi^{(h)}_{0} \Big)
=
 -  32 q^{-h} (-4)^{h-3} \Bigg[
2(
W^{\la,\bar{a} a}_{\mathrm{B},h+2}
+W^{\la, \bar{a} a}_{\mathrm{F},h+2}
) \Bigg].
\label{Phih}
\eea
We have also its inverse relations as follows:
\bea
W^{\la,11 }_{\mathrm{F},h}&=&
q^{h-2}\frac{(-1)^{h+1}}{2^{2(h-1)}}\,\bigg[\Phi_0^{(h)}
-\frac{(4\la+2h-2)}{2h-1}\,\tilde{\Phi}_2^{(h-2)}
+\Big(\mathrm{i}\,\Phi_1^{(h-1),12} -\mathrm{i}\,\Phi_1^{(h-1),13}
-\Phi_1^{(h-1),14}
\nonu \\
&&-\Phi_1^{(h-1),23}+\mathrm{i}\,\Phi_1^{(h-1),24}+\mathrm{i}\,\Phi_1^{(h-1),34}     \Big)
\bigg]\,,
\nonu \\
W^{\la,22 }_{\mathrm{F},h}&=&
q^{h-2}\frac{(-1)^{h+1}}{2^{2(h-1)}}\,\bigg[\Phi_0^{(h)}
-\frac{(4\la+2h-2)}{2h-1}\,\tilde{\Phi}_2^{(h-2)}
-\Big(\mathrm{i}\,\Phi_1^{(h-1),12} -\mathrm{i}\,\Phi_1^{(h-1),13}
-\Phi_1^{(h-1),14}
\nonu \\
&&-\Phi_1^{(h-1),23}+\mathrm{i}\,\Phi_1^{(h-1),24}+\mathrm{i}\,\Phi_1^{(h-1),34}     \Big)
\bigg]\,,
\nonu \\
W^{\la,11 }_{\mathrm{B},h}&=&
q^{h-2}\frac{(-1)^{h}}{2^{2(h-1)}}\,\bigg[\Phi_0^{(h)}
-\frac{(4\la+2h)}{2h-1}\,\tilde{\Phi}_2^{(h-2)}
-\Big(\mathrm{i}\,\Phi_1^{(h-1),12} -\mathrm{i}\,\Phi_1^{(h-1),13}
+\Phi_1^{(h-1),14}
\nonu \\
&&-\Phi_1^{(h-1),23}-\mathrm{i}\,\Phi_1^{(h-1),24}-\mathrm{i}\,\Phi_1^{(h-1),34}     \Big)
\bigg]\,,
\nonu \\
W^{\la,22 }_{\mathrm{B},h}&=&
q^{h-2}\frac{(-1)^{h}}{2^{2(h-1)}}\,\bigg[\Phi_0^{(h)}
-\frac{(4\la+2h)}{2h-1}\,\tilde{\Phi}_2^{(h-2)}
+\Big(\mathrm{i}\,\Phi_1^{(h-1),12} -\mathrm{i}\,\Phi_1^{(h-1),13}
+\Phi_1^{(h-1),14}
\nonu \\
&&-\Phi_1^{(h-1),23}-\mathrm{i}\,\Phi_1^{(h-1),24}-\mathrm{i}\,\Phi_1^{(h-1),34}     \Big)
\bigg]\,,
\nonu \\
W^{\la,12 }_{\mathrm{F},h}&=&
q^{h-2}\frac{(-1)^{h+1}}{2^{2(h-1)-\frac{1}{2}}}\,\Big(
\Phi_1^{(h-1),13}
-\mathrm{i}\,\Phi_1^{(h-1),14}
-\mathrm{i}\,\Phi_1^{(h-1),23}-\Phi_1^{(h-1),24}    \Big)
\,,
\nonu \\
W^{\la,21 }_{\mathrm{F},h}&=&
q^{h-2}\frac{(-1)^{h}}{2^{2(h-1)-\frac{1}{2}}}\,\Big(
\Phi_1^{(h-1),12}
+\mathrm{i}\,\Phi_1^{(h-1),14}
+\mathrm{i}\,\Phi_1^{(h-1),23}+\Phi_1^{(h-1),34}    \Big)
\,,
\nonu \\
W^{\la,12 }_{\mathrm{B},h}&=&
q^{h-2}\frac{(-1)^{h+1}}{2^{2(h-2)+\frac{1}{2}}}\,\Big(
\Phi_1^{(h-1),13}
+\mathrm{i}\,\Phi_1^{(h-1),14}
-\mathrm{i}\,\Phi_1^{(h-1),23}+\Phi_1^{(h-1),24}    \Big)
\,,
\nonu \\
W^{\la,21 }_{\mathrm{B},h}&=&
q^{h-2}\frac{(-1)^{h}}{2^{2(h-1)+\frac{1}{2}}}\,\Big(
\Phi_1^{(h-1),12}
-\mathrm{i}\,\Phi_1^{(h-1),14}
+\mathrm{i}\,\Phi_1^{(h-1),23}-\Phi_1^{(h-1),34}    \Big)
\,,
\nonu \\
Q^{\la,11}_{h+\frac{1}{2}}
&=&
q^{h-1}\frac{(-1)^{h}}{2^{2(h-2)-\frac{1}{2}}}\,
\bigg[
\Big(
\Phi^{(h),1}_\frac{1}{2}-\mathrm{i}\,\Phi^{(h),2}_\frac{1}{2}
-\mathrm{i}\,\Phi^{(h),3}_\frac{1}{2}
+\Phi^{(h),4}_\frac{1}{2}
\Big)
\nonu \\
& + & \Big(
\tilde{\Phi}^{(h-1),1}_\frac{3}{2}-\mathrm{i}\,\tilde{\Phi}^{(h-1),2}_\frac{3}{2}
-\mathrm{i}\,\tilde{\Phi}^{(h-1),3}_\frac{3}{2}
+\tilde{\Phi}^{(h-1),4}_\frac{3}{2}
\Big)
\bigg]\,,
\nonu \\
Q^{\la,22}_{h+\frac{1}{2}}
&=&
q^{h-1}\frac{(-1)^{h+1}}{2^{2(h-2)+\frac{1}{2}}}\,
\bigg[
\Big(
\Phi^{(h),1}_\frac{1}{2}-\mathrm{i}\,\Phi^{(h),2}_\frac{1}{2}
-\mathrm{i}\,\Phi^{(h),3}_\frac{1}{2}
-\Phi^{(h),4}_\frac{1}{2}
\Big)
\nonu \\
& + & \Big(
\tilde{\Phi}^{(h-1),1}_\frac{3}{2}-\mathrm{i}\,\tilde{\Phi}^{(h-1),2}_\frac{3}{2}
-\mathrm{i}\,\tilde{\Phi}^{(h-1),3}_\frac{3}{2}
-\tilde{\Phi}^{(h-1),4}_\frac{3}{2}
\Big)
\bigg]\,,
\nonu \\
Q^{\la,12}_{h+\frac{1}{2}}
&=&
q^{h-1}\frac{(-1)^{h}}{2^{2(h-2)-1}}\,
\bigg[
\Big(
\mathrm{i}\,\Phi^{(h),1}_\frac{1}{2}
+\Phi^{(h),2}_\frac{1}{2}
\Big)
+\Big(
\mathrm{i}\,\tilde{\Phi}^{(h-1),1}_\frac{3}{2}
+\tilde{\Phi}^{(h-1),2}_\frac{3}{2}
\Big)
\bigg]\,,
\nonu \\
Q^{\la,21}_{h+\frac{1}{2}}
&=&
q^{h-1}\frac{(-1)^{h}}{2^{2(h-2)}}\,
\bigg[
\Big(
\mathrm{i}\,\Phi^{(h),1}_\frac{1}{2}
+\Phi^{(h),3}_\frac{1}{2}
\Big)
+\Big(
\mathrm{i}\,\tilde{\Phi}^{(h-1),1}_\frac{3}{2}
+\tilde{\Phi}^{(h-1),3}_\frac{3}{2}
\Big)
\bigg]\,,
\nonu \\
\bar{Q}^{\la,11}_{h+\frac{1}{2}}
&=&
q^{h-1}\frac{(-1)^{h}(1-\frac{1}{3}\delta^{h,0})}{2^{2(h-2)+\frac{1}{2}}}\,
\bigg[
\Big(
\Phi^{(h),1}_\frac{1}{2}-\mathrm{i}\,\Phi^{(h),2}_\frac{1}{2}
-\mathrm{i}\,\Phi^{(h),3}_\frac{1}{2}
-\Phi^{(h),4}_\frac{1}{2}
\Big)
\nonu \\
&&
-\Big(
\tilde{\Phi}^{(h-1),1}_\frac{3}{2}-\mathrm{i}\,\tilde{\Phi}^{(h-1),2}_\frac{3}{2}
-\mathrm{i}\,\tilde{\Phi}^{(h-1),3}_\frac{3}{2}
-\tilde{\Phi}^{(h-1),4}_\frac{3}{2}
\Big)
\bigg]\,,
\nonu \\
\bar{Q}^{\la,22}_{h+\frac{1}{2}}
&=&
q^{h-1}\frac{(-1)^{h+1}(1-\frac{1}{3}\delta^{h,0})}{2^{2(h-2)-\frac{1}{2}}}\,
\bigg[
\Big(
\Phi^{(h),1}_\frac{1}{2}-\mathrm{i}\,\Phi^{(h),2}_\frac{1}{2}
-\mathrm{i}\,\Phi^{(h),3}_\frac{1}{2}
+\Phi^{(h),4}_\frac{1}{2}
\Big)
\nonu \\
&&
-\Big(
\tilde{\Phi}^{(h-1),1}_\frac{3}{2}-\mathrm{i}\,\tilde{\Phi}^{(h-1),2}_\frac{3}{2}
-\mathrm{i}\,\tilde{\Phi}^{(h-1),3}_\frac{3}{2}
+\tilde{\Phi}^{(h-1),4}_\frac{3}{2}
\Big)
\bigg]\,,
\nonu \\
\bar{Q}^{\la,12}_{h+\frac{1}{2}}
&=&
q^{h-1}\frac{(-1)^{h}(1-\frac{1}{3}\delta^{h,0})}{2^{2(h-2)}}\,
\bigg[
\Big(
\mathrm{i}\,\Phi^{(h),1}_\frac{1}{2}
+\Phi^{(h),3}_\frac{1}{2}
\Big)
-\Big(
\mathrm{i}\,\tilde{\Phi}^{(h-1),1}_\frac{3}{2}
+\tilde{\Phi}^{(h-1),3}_\frac{3}{2}
\Big)
\bigg]\,,
\nonu \\
\bar{Q}^{\la,21}_{h+\frac{1}{2}}
&=&
q^{h-1}\frac{(-1)^{h}(1-\frac{1}{3}\delta^{h,0})}{2^{2(h-2)-1}}\,
\bigg[
\Big(
\mathrm{i}\,\Phi^{(h),1}_\frac{1}{2}
+\Phi^{(h),2}_\frac{1}{2}
\Big)
+\Big(
\mathrm{i}\,\tilde{\Phi}^{(h-1),1}_\frac{3}{2}
+\tilde{\Phi}^{(h-1),2}_\frac{3}{2}
\Big)
\bigg]\,.
\label{inverse}
\eea
Note that for the last four currents in (\ref{inverse}),
there exist $\de^{h,0}$ terms where the weights are given by
$\frac{1}{2}$.

\subsection{ The
${\cal N}=4$ supersymmetric
$W_{1+\infty}^{2,2}[\la =\frac{1}{4}]$ algebra}

For convenience, we present
the complete ${\cal N}=4$ supersymmetric
$W_{1+\infty}^{2,2}[\la =\frac{1}{4}]$ algebra
which can be obtained from \cite{AK2309} with
$\la =\frac{1}{4}$
\bea
&& \comm{(\Phi^{(h_1)}_{0})_m}{(\Phi^{(h_2)}_{0})_n}
=q^{h_1+h_2}\, 4\, C_{0,0}^{h_1,h_2}(\tfrac{1}{4})\,
[m+h_1-1]_{h_1+h_2-1}\,\delta_{m+n}
\nonu \\
&&+ \sum_{h_3=1}^{h_1+h_2-1\,\,}\sum_{k=0}^{h_1+h_2-h_3-1}
q^{h_1+h_2-h_3}\, \frac{2(-1)^{h_1+h_2+1}\,4^{2(h_1+h_2-h_3)-4}}{(2h_1-1)(2h_2-1)}
(2h_3-1)! (h_1-\tfrac{1}{2})(h_2-\tfrac{1}{2})\nonu \\
& & 
\times  \Bigg[ \Big(
S^{\,\,h_1,h_2,h_3,k}_{F,\,R}(\tfrac{1}{4})
-S^{\,\,h_1,h_2,h_3,k}_{B,\,R}(\tfrac{1}{4})
\Big)  [m+h_1-1]_{h_1+h_2-h_3-k-1}[n+h_2-1]_k
\nonu \\
&& - 
\Big(
S^{\,\,h_1,h_2,h_3,k}_{F,\,L}(\tfrac{1}{4})
-S^{\,\,h_1,h_2,h_3,k}_{B,\,L}(\tfrac{1}{4})
\Big)
 [m+h_1-1]_k [n+h_2-1]_{h_1+h_2-h_3-k-1}\,\,
\Bigg](\Phi^{(h_3)}_0)_{m+n}
\nonu \\
& & + \sum_{h_3=-1}^{h_1+h_2-3\,}\sum_{k=0}^{\,h_1+h_2-h_3-3}
q^{h_1+h_2-h_3}\, \frac{(-1)^{h_1+h_2}\,4^{2(h_1+h_2-h_3)-7}}{(2h_1-1)(2h_2-1)}
\,
(2h_3+2)! (h_1-\tfrac{1}{2})(h_2-\tfrac{1}{2})(h_3+\tfrac{3}{2})
\nonu \\
& & 
\times \Bigg[\Big(
   S^{\,\,h_1,h_2,h_3+2,k}_{F,\,R}(\tfrac{1}{4})
+  
S^{\,\,h_1,h_2,h_3+2,k}_{B,\,R}(\tfrac{1}{4})\Big)
[m+h_1-1]_{h_1+h_2-h_3-k-3}[n+h_2-1]_k
\nonu \\
& &   - 
\Big(
   S^{\,\,h_1,h_2,h_3+2,k}_{F,\,L}(\tfrac{1}{4})+ 
   S^{\,\,h_1,h_2,h_3+2,k}_{B,\,L}(\tfrac{1}{4})
\Big)  [m+h_1-1]_k [n+h_2-1]_{h_1+h_2-h_3-k-3}
\,\,\Bigg] (\Phi^{(h_3)}_2)_{m+n},
\nonu \\
&& \comm{(\Phi^{(h_1)}_{0})_m}{(\Phi^{(h_2),i}_{\frac{1}{2}})_r}
=
\nonu \\
&& \sum_{h_3=0}^{h_1+h_2-1\,\,}\sum_{k=0}^{h_1+h_2-h_3-1}
q^{h_1+h_2-h_3}\,
\frac{(-1)^{h_1+h_2+1}\,4^{2(h_1+h_2-h_3)-4}}{(2h_1-1)}
(2h_3)! (h_1-\tfrac{1}{2})\nonu \\
&&
\times \Bigg[\Big(
T^{\,\,h_1,h_2,h_3,k}_{F}(\tfrac{1}{4})
+\bar{T}^{\,\,h_1,h_2,h_3,k}_{B}(\tfrac{1}{4})
\Big)[m+h_1-1]_{h_1+h_2-h_3-k-1}[r+h_2-\tfrac{1}{2}]_k
\nonu \\
&&
+
\Big(
\bar{T}^{\,\,h_1,h_2,h_3,k}_{F}(\tfrac{1}{4})
+T^{\,\,h_1,h_2,h_3,k}_{B}(\tfrac{1}{4})
\Big)
[m+h_1-1]_k [r+h_2-\tfrac{1}{2}]_{h_1+h_2-h_3-k-1}\,\,
\Bigg](\Phi^{(h_3),i}_\frac{1}{2})_{m+r}
\nonu \\
&& +\sum_{h_3=0}^{h_1+h_2-2\,\,}\sum_{k=0}^{h_1+h_2-h_3-2}
q^{h_1+h_2-h_3}\,
\frac{(-1)^{h_1+h_2+1}\,4^{2(h_1+h_2-h_3)-6}}{(2h_1-1)}
\,(2h_3+2)!(h_1-\tfrac{1}{2}) \nonu \\
&& 
\times\Bigg[
\Big(
T^{\,\,h_1,h_2,h_3+1,k}_{F}
-
\bar{T}^{\,\,h_1,h_2,h_3+1,k}_{B}\Big) 
[m+h_1-1]_{h_1+h_2-h_3-k-2}[r+h_2-\tfrac{1}{2}]_k
\nonu \\
&&
-
\Big(
\bar{T}^{\,\,h_1,h_2,h_3+1,k}_{F}
-T^{\,\,h_1,h_2,h_3+1,k}_{B}
\Big)\times[m+h_1-1]_k [r+h_2-\tfrac{1}{2}]_{h_1+h_2-h_3-k-2}
\,\,\Bigg](\Phi^{(h_3),i}_\frac{3}{2})_{m+r},
\nonu \\
&& \comm{(\Phi^{(h_1)}_{0})_m}{(\Phi^{(h_2),ij}_{1})_n} 
=
\nonu \\
&&\sum_{h_3=0}^{h_1+h_2-1\,\,}\sum_{k=0}^{h_1+h_2-h_3-1}
q^{h_1+h_2-h_3}\,\frac{(-1)^{h_1+h_2}\,4^{2(h_1+h_2-h_3)-4}}{(2h_1-1)}
(2h_3+1)! (h_1-\tfrac{1}{2}) \nonu \\
&&
\times \Bigg[\Big(
S^{\,\,h_1,h_2+1,h_3+1,k}_{F,\,R}(\tfrac{1}{4})
-S^{\,\,h_1,h_2+1,h_3+1,k}_{B,\,R}(\tfrac{1}{4})
\Big) [m+h_1-1]_{h_1+h_2-h_3-k-1}[n+h_2]_k
\nonu \\
&& 
-
\Big(
S^{\,\,h_1,h_2+1,h_3+1,k}_{F,\,L}(\tfrac{1}{4})
-S^{\,\,h_1,h_2+1,h_3+1,k}_{B,\,L}(\tfrac{1}{4})
\Big)
[m+h_1-1]_k [n+h_2]_{h_1+h_2-h_3-k-1}\,\,
\Bigg](\Phi^{(h_3),ij}_1)_{m+n}
\nonu \\
&& +\sum_{h_3=0}^{h_1+h_2-1\,\,}\sum_{k=0}^{h_1+h_2-h_3-1}
q^{h_1+h_2-h_3}\,\frac{(-1)^{h_1+h_2}\,4^{2(h_1+h_2-h_3)-4}}{(2h_1-1)}
\,(2h_3+1)!(h_1-\tfrac{1}{2}) \nonu \\
&& 
\times \Bigg[\Big(
S^{\,\,h_1,h_2+1,h_3+1,k}_{F,\,R}(\tfrac{1}{4})
+
S^{\,\,h_1,h_2+1,h_3+1,k}_{B,\,R}(\tfrac{1}{4}) \Big)
[m+h_1-1]_{h_1+h_2-h_3-k-1}[n+h_2]_k
\nonu \\
&& 
-
\Big(
S^{\,\,h_1,h_2+1,h_3+1,k}_{F,\,L}(\tfrac{1}{4})
+S^{\,\,h_1,h_2+1,h_3+1,k}_{B,\,L}(\tfrac{1}{4})
\Big) [m+h_1-1]_k [n+h_2]_{h_1+h_2-h_3-k-1}
\,\,\Bigg](\tilde{\Phi}^{(h_3),ij}_1)_{m+n},
\nonu 
\\
&& \comm{(\Phi^{(h_1)}_{0})_m}{(\Phi^{(h_2),i}_{\frac{3}{2}})_r}
=
\nonu \\
&&
\sum_{h_3=0}^{h_1+h_2\,\,}\sum_{k=0}^{h_1+h_2-h_3}
q^{h_1+h_2-h_3}\,
\frac{(-1)^{h_1+h_2}\,4^{2(h_1+h_2-h_3)-2}}{(2h_1-1)}
(2h_3)! (h_1-\tfrac{1}{2})\nonu \\
&&
\times \Bigg[\Big(
T^{\,\,h_1,h_2+1,h_3,k}_{F}(\tfrac{1}{4})
-\bar{T}^{\,\,h_1,h_2+1,h_3,k}_{B}(\tfrac{1}{4})
\Big) [m+h_1-1]_{h_1+h_2-h_3-k}[r+h_2+\tfrac{1}{2}]_k
\nonu \\
&&  
-
\Big(
\bar{T}^{\,\,h_1,h_2+1,h_3,k}_{F}(\tfrac{1}{4})
-T^{\,\,h_1,h_2+1,h_3,k}_{B}(\tfrac{1}{4})
\Big)
[m+h_1-1]_k [r+h_2+\tfrac{1}{2}]_{h_1+h_2-h_3-k}\,\,
\Bigg](\Phi^{(h_3),i}_\frac{1}{2})_{m+r}
\nonu \\
&& +\sum_{h_3=0}^{h_1+h_2-1\,\,}\sum_{k=0}^{h_1+h_2-h_3-1}q^{h_1+h_2-h_3}\,
\frac{(-1)^{h_1+h_2}\,4^{2(h_1+h_2-h_3)-4}}{(2h_1-1)}
\,(2h_3+2)! (h_1-\tfrac{1}{2}) \nonu \\
&& 
\times \Bigg[\Big(
T^{\,\,h_1,h_2+1,h_3+1,k}_{F}(\tfrac{1}{4})
+
\bar{T}^{\,\,h_1,h_2+1,h_3+1,k}_{B}(\tfrac{1}{4})\Big) 
[m+h_1-1]_{h_1+h_2-h_3-k-1}[r+h_2+\tfrac{1}{2}]_k
\nonu \\
&& 
+
\Big(
\bar{T}^{\,\,h_1,h_2+1,h_3+1,k}_{F}(\tfrac{1}{4})
+T^{\,\,h_1,h_2+1,h_3+1,k}_{B}(\tfrac{1}{4})
\Big)[m+h_1-1]_k [r+h_2+\tfrac{1}{2}]_{h_1+h_2-h_3-k-1}
\,\,\Bigg]\nonu \\
&& \times (\Phi^{(h_3),i}_\frac{3}{2})_{m+r},
\nonu \\
&& \comm{(\Phi^{(h_1)}_{0})_m}{(\Phi^{(h_2)}_{2})_n}
=q^{h_1+h_2}\,4^3\,C_{0,2}^{h_1,h_2}(\tfrac{1}{4})\,[m+h_1-1]_{h_1+h_2+1}\,\delta_{m+n}
\nonu
\\
&& +\sum_{h_3=1}^{h_1+h_2+1\,\,}\sum_{k=0}^{h_1+h_2-h_3+1}
q^{h_1+h_2-h_3}\,\frac{(-1)^{h_1+h_2}\,4^{2(h_1+h_2-h_3)}}{(2h_1-1)}
(2h_3-1)!(h_1-\tfrac{1}{2}) \nonu \\
&& \times \Bigg[ \Big(
S^{\,\,h_1,h_2+2,h_3,k}_{F,\,R}(\tfrac{1}{4})
+S^{\,\,h_1,h_2+2,h_3,k}_{B,\,R}(\tfrac{1}{4})
\Big) [m+h_1-1]_{h_1+h_2-h_3-k+1}[n+h_2+1]_k
\nonu \\
&& -
\Big(
S^{\,\,h_1,h_2+2,h_3,k}_{F,\,L}(\tfrac{1}{4})
+S^{\,\,h_1,h_2+2,h_3,k}_{B,\,L}(\tfrac{1}{4})
\Big)
[m+h_1-1]_k [n+h_2+1]_{h_1+h_2-h_3-k+1}\,\,
\Bigg]\nonu \\
&& \times (\Phi^{(h_3)}_0)_{m+n}
\nonu \\
&& 
+\sum_{h_3=-1}^{h_1+h_2-1\,\,}\sum_{k=0}^{h_1+h_2-h_3-1}
q^{h_1+h_2-h_3}\,\frac{2(-1)^{h_1+h_2+1}\,4^{2(h_1+h_2-h_3)-4}}{(2h_1-1)}
\,(2h_3+2)! (h_1-\tfrac{1}{2})(h_3+\tfrac{3}{2})\nonu \\
&& \times \Bigg[ \Big(
S^{\,\,h_1,h_2+2,h_3+2,k}_{F,\,R}(\tfrac{1}{4})
-S^{\,\,h_1,h_2+2,h_3+2,k}_{B,\,R}(\tfrac{1}{4})\Big)
[m+h_1-1]_{h_1+h_2-h_3-k-1}[n+h_2+1]_k
\nonu \\
&&  
-
\Big(
  S^{\,\,h_1,h_2+2,h_3+2,k}_{F,\,L}(\tfrac{1}{4})
 -S^{\,\,h_1,h_2+2,h_3+2,k}_{B,\,L}(\tfrac{1}{4})
\Big) [m+h_1-1]_k [n+h_2+1]_{h_1+h_2-h_3-k-1}
\,\,\Bigg]\nonu \\
&& \times (\Phi^{(h_3)}_2)_{m+n},
\nonu \\ 
&& \acomm{(\Phi^{(h_1),i}_{\frac{1}{2}})_r}{(\Phi^{(h_2),j}_{\frac{1}{2}})_s}
=
q^{h_1+h_2}\,\delta^{ij}\,2\,C_{\frac{1}{2},\frac{1}{2}}^{h_1,h_2}(\tfrac{1}{4})\,
      [r+h_1-\tfrac{1}{2}]_{h_1+h_2}\,\delta_{r+s}
      \nonu \\
&& +
\delta^{ij}
\sum_{h_3=1}^{h_1+h_2\,\,}\sum_{k=0}^{h_1+h_2-h_3}
q^{h_1+h_2-h_3}\,(-1)^{h_1+h_2}\,4^{2(h_1+h_2-h_3)-4}\,
(2h_3-1)!
\nonu \\
&& 
\times\Bigg[
\Big(
U^{\,\,h_2,h_1,h_3,k}_{B}(\tfrac{1}{4})-U^{\,\,h_1,h_2,h_3,k}_{F}(\tfrac{1}{4})\Big)
[r+h_1-\tfrac{1}{2}]_{h_1+h_2-h_3-k}
[s+h_2-\tfrac{1}{2}]_{k}
\nonu \\
&&
+\Big(
U^{\,\,h_1,h_2,h_3,k}_{B}(\tfrac{1}{4})-U^{\,\,h_2,h_1,h_3,k}_{F}(\tfrac{1}{4})\Big)
[r+h_1-\tfrac{1}{2}]_{k}
[s+h_2-\tfrac{1}{2}]_{h_1+h_2-h_3-k}
\,\Bigg](\Phi^{(h_3)}_{0})_{r+s}
\nonu \\
&& 
+\delta^{ij}
\sum_{h_3=-1}^{h_1+h_2-2\,\,}\sum_{k=0}^{h_1+h_2-h_3-2}
q^{h_1+h_2-h_3}\,2\,(-1)^{h_1+h_2}\,4^{2(h_1+h_2-h_3)-8}\,
(2h_3+2)!(h_3+\tfrac{3}{2})
\nonu \\
&& \times\Bigg[
\Big(
U^{\,\,h_1,h_2,h_3+2,k}_{F}(\tfrac{1}{4})
+U^{\,\,h_2,h_1,h_3+2,k}_{B}(\tfrac{1}{4})
\Big)
[r+h_1-\tfrac{1}{2}]_{h_1+h_2-h_3-k-2}
[s+h_2-\tfrac{1}{2}]_{k} \nonu
\\
&& 
+
\Big(
U^{\,\,h_2,h_1,h_3+2,k}_{F}(\tfrac{1}{4})
+U^{\,\,h_1,h_2,h_3+2,k}_{B}(\tfrac{1}{4})
\Big)
[r+h_1-\tfrac{1}{2}]_{k}
[s+h_2-\tfrac{1}{2}]_{h_1+h_2-h_3-k-2} 
\,\Bigg](\Phi^{(h_3)}_{2})_{r+s}
\nonu \\
&& +
\sum_{h_3=0}^{h_1+h_2-1\,\,}\sum_{k=0}^{h_1+h_2-h_3-1}
q^{h_1+h_2-h_3}\,(-1)^{h_1+h_2}\,4^{2(h_1+h_2-h_3)-6}\,(2h_3+1)!\,
\nonu\\
&&\times\Bigg(
\Bigg[
\Big(
U^{\,\,h_1,h_2,h_3+1,k}_{F}(\tfrac{1}{4})+U^{\,\,h_2,h_1,h_3+1,k}_{B}(\tfrac{1}{4})
\Big)
[r+h_1-\tfrac{1}{2}]_{h_1+h_2-h_3-k-1}
[s+h_2-\tfrac{1}{2}]_{k}
\nonu \\
&&
-
\Big(
U^{\,\,h_2,h_1,h_3+1,k}_{F}(\tfrac{1}{4})+U^{\,\,h_1,h_2,h_3+1,k}_{B}(\tfrac{1}{4})
\Big)
[r+h_1-\tfrac{1}{2}]_{k}
[s+h_2-\tfrac{1}{2}]_{h_1+h_2-h_3-k-1}
\Bigg](\Phi^{(h_3),ij}_{1})_{r+s}
\nonu \\
&&
\,
+\Bigg[
\Big(U^{\,\,h_1,h_2,h_3+1,k}_{F}(\tfrac{1}{4})-U^{\,\,h_2,h_1,h_3+1,k}_{B}(\tfrac{1}{4})
\Big)
[r+h_1-\tfrac{1}{2}]_{h_1+h_2-h_3-k-1}
[s+h_2-\tfrac{1}{2}]_{k}
\nonu \\
&&
-\Big(
U^{\,\,h_2,h_1,h_3+1,k}_{F}(\tfrac{1}{4})-U^{\,\,h_1,h_2,h_3+1,k}_{B}(\tfrac{1}{4})\Big)
[r+h_1-\tfrac{1}{2}]_{k}
[s+h_2-\tfrac{1}{2}]_{h_1+h_2-h_3-k-1}
 \Bigg] (\tilde{\Phi}^{(h_3),ij}_{1})_{r+s}\Bigg),
\nonu \\
&& \comm{(\Phi^{(h_1),i}_{\frac{1}{2}})_r}{(\Phi^{(h_2),jk}_{1})_m}
=\delta^{ij}\sum_{h_3=0}^{h_1+h_2\,\,}\sum_{t=0}^{h_1+h_2-h_3}
q^{h_1+h_2-h_3}\,2(-1)^{h_1+h_2+1}\,4^{2(h_1+h_2-h_3)-3}\,(2h_3)!
\nonu \\
&& \times\Bigg[
\Big(
T^{\,\,h_2+1,h_1,h_3,t}_{F}(\tfrac{1}{4})-\bar{T}^{\,\,h_2+1,h_1,h_3,t}_{B}(\tfrac{1}{4})
\Big)
[r+h_1-\tfrac{1}{2}]_t [m+h_2]_{h_1+h_2-h_3-t}
\nonu \\
&&
 +
\Big(
T^{\,\,h_2+1,h_1,h_3,t}_{B}(\tfrac{1}{4})-\bar{T}^{\,\,h_2+1,h_1,h_3,t}_{F}(\tfrac{1}{4})
\Big)
[r+h_1-\tfrac{1}{2}]_{h_1+h_2-h_3-t} [m+h_2]_{t}
\Bigg](\Phi^{(h_3),k}_{\frac{1}{2}})_{r+m}
\nonu \\
&& + \Bigg(
\delta^{ij}\sum_{h_3=0}^{h_1+h_2-1\,\,}\sum_{t=0}^{h_1+h_2-h_3-1}
q^{h_1+h_2-h_3}\,2(-1)^{h_1+h_2+1}\,4^{2(h_1+h_2-h_3)-5}\,(2h_3+2)!
\nonu \\
&& \times\Bigg[
\Big(
T^{\,\,h_2+1,h_1,h_3+1,t}_{F}(\tfrac{1}{4})+\bar{T}^{\,\,h_2+1,h_1,h_3+1,t}_{B}(\tfrac{1}{4})
\Big)
[r+h_1-\tfrac{1}{2}]_t [m+h_2]_{h_1+h_2-h_3-t-1}
\nonu \\
&&
+
\Big(
T^{\,\,h_2+1,h_1,h_3+1,t}_{B}(\tfrac{1}{4})+\bar{T}^{\,\,h_2+1,h_1,h_3+1,t}_{F}(\tfrac{1}{4})
\Big)
[r+h_1-\tfrac{1}{2}]_{h_1+h_2-h_3-t-1} [m+h_2]_{t}
\Bigg](\Phi^{(h_3),k}_{\frac{3}{2}})_{r+m}
\nonu \\
&& -\delta^{ik}\Bigg(\,\, j\leftrightarrow k\,\,\Bigg) \Bigg)
\nonu \\
&&
+\epsilon^{ijkl}\sum_{h_3=0}^{h_1+h_2\,\,}\sum_{t=0}^{h_1+h_2-h_3}
q^{h_1+h_2-h_3}\,2(-1)^{h_1+h_2+1}\,4^{2(h_1+h_2-h_3)-3}\,(2h_3)!
\nonu \\
&& 
\times 
\Bigg[
\Big(
T^{\,\,h_2+1,h_1,h_3,t}_{F}(\tfrac{1}{4})+\bar{T}^{\,\,h_2+1,h_1,h_3,t}_{B}(\tfrac{1}{4})
\Big) [r+h_1-\tfrac{1}{2}]_t [m+h_2]_{h_1+h_2-h_3-t}
\nonu \\
&&
-
\Big(
T^{\,\,h_2+1,h_1,h_3,t}_{B}(\tfrac{1}{4})+\bar{T}^{\,\,h_2+1,h_1,h_3,t}_{F}(\tfrac{1}{4})
\Big)
[r+h_1-\tfrac{1}{2}]_{h_1+h_2-h_3-t} [m+h_2]_{t}
\Bigg](\Phi^{(h_3),l}_{\frac{1}{2}})_{r+m}
\nonu \\
&&
+\epsilon^{ijkl}
\sum_{h_3=0}^{h_1+h_2-1\,\,}\sum_{t=0}^{h_1+h_2-h_3-1}
q^{h_1+h_2-h_3}\,2(-1)^{h_1+h_2+1}\,4^{2(h_1+h_2-h_3)-5}\,(2h_3+2)!
\nonu \\
&&
\times 
\Bigg[
\Big(
T^{\,\,h_2+1,h_1,h_3+1,t}_{F}(\tfrac{1}{4})-\bar{T}^{\,\,h_2+1,h_1,h_3+1,t}_{B}(\tfrac{1}{4})
\Big)
[r+h_1-\tfrac{1}{2}]_t [m+h_2]_{h_1+h_2-h_3-t-1}
\nonu \\
&&
-
\Big(
T^{\,\,h_2+1,h_1,h_3+1,t}_{B}(\tfrac{1}{4})-\bar{T}^{\,\,h_2+1,h_1,h_3+1,t}_{F}(\tfrac{1}{4})
\Big)
[r+h_1-\tfrac{1}{2}]_{h_1+h_2-h_3-t-1} [m+h_2]_{t}
\Bigg](\Phi^{(h_3),l}_{\frac{3}{2}})_{r+m},
\nonu \\
&&
\acomm{(\Phi^{(h_1),i}_{\frac{1}{2}})_r}{(\Phi^{(h_2),j}_{\frac{3}{2}})_s}
=
\delta^{ij}\,q^{h_1+h_2}\,8\,C_{\frac{1}{2},\frac{3}{2}}^{h_1,h_2}(\tfrac{1}{4})\,
      [r+h_1-\tfrac{1}{2}]_{h_1+h_2+1}\,\delta_{r+s}
\nonu \\     
&& +
\delta^{ij}
\sum_{h_3=1}^{h_1+h_2+1\,\,}\sum_{k=0}^{h_1+h_2-h_3+1}
q^{h_1+h_2-h_3}\,(-1)^{h_1+h_2+1}\,4^{2(h_1+h_2-h_3)-2}\,
(2h_3-1)!
\nonu \\
&&
\times\Bigg[
\Big(U^{\,\,h_1,h_2+1,h_3,k}_{F}(\tfrac{1}{4})+U^{\,\,h_2+1,h_1,h_3,k}_{B}(\tfrac{1}{4})\Big)
[r+h_1-\tfrac{1}{2}]_{h_1+h_2-h_3-k+1}
[s+h_2+\tfrac{1}{2}]_{k}
\nonu \\
&&
-\Big(U^{\,\,h_1,h_2+1,h_3,k}_{B}(\tfrac{1}{4})+U^{\,\,h_2+1,h_1,h_3,k}_{F}(\tfrac{1}{4})\Big)
[r+h_1-\tfrac{1}{2}]_{k}
[s+h_2+\tfrac{1}{2}]_{h_1+h_2-h_3-k+1}
\,\Bigg](\Phi^{(h_3)}_{0})_{r+s}
\nonu \\
&&
+\delta^{ij}
\sum_{h_3=-1}^{h_1+h_2-1\,\,}\sum_{k=0}^{h_1+h_2-h_3-1}
q^{h_1+h_2-h_3}\,2\,(-1)^{h_1+h_2}\,4^{2(h_1+h_2-h_3)-6}\,
(2h_3+2)!(h_3+\tfrac{3}{2})
\nonu \\
&&
\times\Bigg[
\Big(U^{\,\,h_1,h_2+1,h_3+2,k}_{F}(\tfrac{1}{4})
-U^{\,\,h_2+1,h_1,h_3+2,k}_{B}(\tfrac{1}{4})\Big)
[r+h_1-\tfrac{1}{2}]_{h_1+h_2-h_3-k-1}
[s+h_2+\tfrac{1}{2}]_{k} 
\nonu \\
&&
+
\Big( U^{\,\,h_1,h_2+1,h_3+2,k}_{B}(\tfrac{1}{4})
-U^{\,\,h_2+1,h_1,h_3+2,k}_{F}(\tfrac{1}{4})
\Big)
[r+h_1-\tfrac{1}{2}]_{k}
[s+h_2+\tfrac{1}{2}]_{h_1+h_2-h_3-k-1} 
\,\Bigg](\Phi^{(h_3)}_{2})_{r+s}
\nonu \\
&& +
\sum_{h_3=0}^{h_1+h_2\,\,}\sum_{k=0}^{h_1+h_2-h_3}
q^{h_1+h_2-h_3}\,(-1)^{h_1+h_2}\,4^{2(h_1+h_2-h_3)-4}\,(2h_3+1)!\,
\nonu \\
&&
\times
\Bigg(
\Bigg[
\Big(U^{\,\,h_1,h_2+1,h_3+1,k}_{F}(\tfrac{1}{4})-U^{\,\,h_2+1,h_1,h_3+1,k}_{B}(\tfrac{1}{4})\Big)
[r+h_1-\tfrac{1}{2}]_{h_1+h_2-h_3-k}
[s+h_2+\tfrac{1}{2}]_{k}
\nonu \\
&& 
+\Big(U^{\,\,h_2+1,h_1,h_3+1,k}_{F}(\tfrac{1}{4})-U^{\,\,h_1,h_2+1,h_3+1,k}_{B}(\tfrac{1}{4})\Big)
[r+h_1-\tfrac{1}{2}]_{k}
[s+h_2+\tfrac{1}{2}]_{h_1+h_2-h_3-k}
 \Bigg] (\Phi^{(h_3),ij}_{1})_{r+s}
\nonu \\
&&
+\Bigg[
\Big(U^{\,\,h_1,h_2+1,h_3+1,k}_{F}(\tfrac{1}{4})+U^{\,\,h_2+1,h_1,h_3+1,k}_{B}(\tfrac{1}{4})\Big)
[r+h_1-\tfrac{1}{2}]_{h_1+h_2-h_3-k}
[s+h_2+\tfrac{1}{2}]_{k}
\nonu \\
&&
+\Big(U^{\,\,h_2+1,h_1,h_3+1,k}_{F}(\tfrac{1}{4})+U^{\,\,h_1,h_2+1,h_3+1,k}_{B}(\tfrac{1}{4})\Big)
[r+h_1-\tfrac{1}{2}]_{k}
[s+h_2+\tfrac{1}{2}]_{h_1+h_2-h_3-k}
\Bigg]\nonu \\
&& \times (\tilde{\Phi}^{(h_3),ij}_{1})_{r+s}
\,\Bigg),
\nonu \\    
&& \comm{(\Phi^{(h_1),i}_{\frac{1}{2}})_r}{(\Phi^{(h_2)}_{2})_m}
=
\sum_{h_3=0}^{h_1+h_2+1\,\,}\sum_{k=0}^{h_1+h_2-h_3+1}
q^{h_1+h_2-h_3}\,2(-1)^{h_1+h_2+1}\,4^{2(h_1+h_2-h_3)-1}\,(2h_3)!
\nonu \\
&&
\times\Bigg[
\Big(T^{\,\,h_2+2,h_1,h_3,k}_{F}(\tfrac{1}{4})-\bar{T}^{\,\,h_2+2,h_1,h_3,k}_{B}(\tfrac{1}{4})\Big)
[r+h_1-\tfrac{1}{2}]_{k}
[m+h_2+1]_{h_1+h_2-h_3-k+1}
\nonu \\
&&
+\Big(\bar{T}^{\,\,h_2+2,h_1,h_3,k}_{F}(\tfrac{1}{4})-T^{\,\,h_2+2,h_1,h_3,k}_{B}(\tfrac{1}{4})\Big)
[r+h_1-\tfrac{1}{2}]_{h_1+h_2-h_3-k+1}
[m+h_2+1]_{k}
 \Bigg] (\Phi^{(h_3),i}_{\frac{1}{2}})_{r+m}
\nonu \\
&& +
\sum_{h_3=0}^{h_1+h_2\,\,}\sum_{k=0}^{h_1+h_2-h_3}
q^{h_1+h_2-h_3}\,2(-1)^{h_1+h_2+1}\,4^{2(h_1+h_2-h_3)-3}\,(2h_3+2)!
\nonu \\
&&
\times\Bigg[
\Big(T^{\,\,h_2+2,h_1,h_3+1,k}_{F}(\tfrac{1}{4})+\bar{T}^{\,\,h_2+2,h_1,h_3+1,k}_{B}(\tfrac{1}{4})\Big)
[r+h_1-\tfrac{1}{2}]_{k}
[m+h_2+1]_{h_1+h_2-h_3-k}
\nonu \\
&&
-\Big(\bar{T}^{\,\,h_2+2,h_1,h_3+1,k}_{F}(\tfrac{1}{4})+T^{\,\,h_2+2,h_1,h_3+1,k}_{B}(\tfrac{1}{4})\Big)
[r+h_1-\tfrac{1}{2}]_{h_1+h_2-h_3-k}
[m+h_2+1]_{k}
 \Bigg] (\Phi^{(h_3),i}_{\frac{3}{2}})_{r+m},
\nonu \\
&& \comm{(\Phi^{(h_1),ij}_{1})_m}{(\Phi^{(h_2),kl}_{1})_n}
=\Big(\delta^{ik}\delta^{jl}-\delta^{il}\delta^{jk}\Big)
\,q^{h_1+h_2}\,4^3\,C_{1,1,\,-}^{h_1,h_2}(\tfrac{1}{4})\,[m+h_1]_{h_1+h_2+1}\,\delta_{m+n}
\nonu \\
&&+
\epsilon^{ijkl}\,q^{h_1+h_2}\,4^3\,C_{1,1,\,+}^{h_1,h_2}(\tfrac{1}{4})\,[m+h_1]_{h_1+h_2+1}\,\delta_{m+n}
\nonu
\\
&&
+\Big(\delta^{ik}\delta^{jl}-\delta^{il}\delta^{jk}\Big)
\sum_{h_3=1}^{h_1+h_2+1\,\,}
\sum_{t=0}^{h_1+h_2-h_3+1}
q^{h_1+h_2-h_3}\,2(-1)^{h_1+h_2}\,4^{2(h_1+h_2-h_3)-1}\,(2h_3-1)!
\nonu \\
&& 
\times\Bigg[ 
\Big(
S^{\,\,h_1+1,h_2+1,h_3,t}_{F,\,R}(\tfrac{1}{4})
-S^{\,\,h_1+1,h_2+1,h_3,t}_{B,\,R}(\tfrac{1}{4})
\Big)
[m+h_1]_{h_1+h_2-h_3-t+1}[n+h_2,t]
\nonu \\
&&
-
\Big(
S^{\,\,h_1+1,h_2+1,h_3,t}_{F,\,L}(\tfrac{1}{4})
-S^{\,\,h_1+1,h_2+1,h_3,t}_{B,\,L}(\tfrac{1}{4})
\Big)
[m+h_1]_{t}[n+h_2]_{h_1+h_2-h_3-t+1}
\Bigg](\Phi^{(h_3)}_0)_{m+n}
\nonu \\
&&
+\Big(\delta^{ik}\delta^{jl}-\delta^{il}\delta^{jk}\Big)
\sum_{h_3=-1}^{h_1+h_2-1\,\,}
\sum_{t=0}^{h_1+h_2-h_3-1}
q^{h_1+h_2-h_3}\,(-1)^{h_1+h_2+1}\,4^{2(h_1+h_2-h_3)-4}\,(2h_3+2)!(h_3 +\tfrac{3}{2}) \nonu \\
&&
\times \Bigg[
\Big(
S^{\,\,h_1+1,h_2+1,h_3+2,t}_{F,\,R}(\tfrac{1}{4})
+S^{\,\,h_1+1,h_2+1,h_3+2,t}_{B,\,R}(\tfrac{1}{4})
\Big)
[m+h_1]_{h_1+h_2-h_3-t-1}[n+h_2]_t \nonu \\
&&
-
\Big(
S^{\,\,h_1+1,h_2+1,h_3+2,t}_{F,\,L}(\tfrac{1}{4})
+S^{\,\,h_1+1,h_2+1,h_3+2,t}_{B,\,L}(\tfrac{1}{4})
\Big)
[m+h_1]_{t}[n+h_2]_{h_1+h_2-h_3-t-1}\,
\Bigg](\Phi^{(h_3)}_2)_{m+n}
\nonu \\
&& +
\epsilon^{ijkl}
\sum_{h_3=0}^{h_1+h_2+1\,\,}
\sum_{t=0}^{h_1+h_2-h_3+1}
q^{h_1+h_2-h_3}\,2(-1)^{h_1+h_2}\,4^{2(h_1+h_2-h_3)-1}\,(2h_3-1)! \nonu \\
&& 
\times \Bigg[
\Big(
S^{\,\,h_1+1,h_2+1,h_3,t}_{F,\,R}
+S^{\,\,h_1+1,h_2+1,h_3,t}_{B,\,R}
\Big)
[m+h_1]_{h_1+h_2-h_3-t+1}[n+h_2]_t \nonu \\
&&
-
\Big(
S^{\,\,h_1+1,h_2+1,h_3,t}_{F,\,L}
+S^{\,\,h_1+1,h_2+1,h_3,t}_{B,\,L}
\Big)
[m+h_1]_{t}[n+h_2]_{h_1+h_2-h_3-t+1}\,
\Bigg](\Phi^{(h_3)}_0)_{m+n}
\nonu \\
&& +\epsilon^{ijkl}
\sum_{h_3=-1}^{h_1+h_2-1\,\,}
\sum_{t=0}^{h_1+h_2-h_3-1}
q^{h_1+h_2-h_3}\,(-1)^{h_1+h_2+1}\,4^{2(h_1+h_2-h_3)-4}\,(2h_3+2)!
(h_3 +\tfrac{3}{2})
\nonu \\
&& 
\times \Bigg[
\Big(
S^{\,\,h_1+1,h_2+1,h_3+2,t}_{F,\,R}
-S^{\,\,h_1+1,h_2+1,h_3+2,t}_{B,\,R}\Big)
[m+h_1]_{h_1+h_2-h_3-t-1}[n+h_2]_t
\nonu \\
&&
-
\Big(
S^{\,\,h_1+1,h_2+1,h_3+2,t}_{F,\,L}
-S^{\,\,h_1+1,h_2+1,h_3+2,t}_{B,\,L}
\Big)
[m+h_1]_{t}[n+h_2]_{h_1+h_2-h_3-t-1}\,
\Bigg](\Phi^{(h_3)}_2)_{m+n}
\nonu \\
&& + \Bigg(
\delta^{ik}\sum_{h_3=0}^{h_1+h_2\,\,}\sum_{t=0}^{h_1+h_2-h_3}
q^{h_1+h_2-h_3}\,2\,(-1)^{h_1+h_2+1}\,4^{2(h_1+h_2-h_3)-3}\,(2h_3+1)!
\nonu\\
&&\times\Bigg(
 \Bigg[
\Big(
S^{\,\,h_1+1,h_2+1,h_3+1,t}_{F,\,R}
+S^{\,\,h_1+1,h_2+1,h_3+1,t}_{B,\,R}
\Big)
[m+h_1]_{h_1+h_2-h_3-t}[n+h_2]_t \nonu \\
&&
+
\Big(
S^{\,\,h_1+1,h_2+1,h_3+1,t}_{F,\,L}
+S^{\,\,h_1+1,h_2+1,h_3+1,t}_{B,\,L}
\Big)
[m+h_1]_{t}[n+h_2]_{h_1+h_2-h_3-t}\,
\Bigg](\Phi^{(h_3),jl}_1)_{m+n}
\nonu \\
&& +
\Bigg[
\Big(
S^{\,\,h_1+1,h_2+1,h_3+1,t}_{F,\,R}
-S^{\,\,h_1+1,h_2+1,h_3+1,t}_{B,\,R}
\Big)
[m+h_1]_{h_1+h_2-h_3-t}[n+h_2]_t\nonu \\
&&
+
\Big(
S^{\,\,h_1+1,h_2+1,h_3+1,t}_{F,\,L}
-S^{\,\,h_1+1,h_2+1,h_3+1,t}_{B,\,L}
\Big)
[m+h_1]_{t}[n+h_2]_{h_1+h_2-h_3-t}\,
\Bigg](\tilde{\Phi}^{(h_3),jl}_1)_{m+n}\,\Bigg)
\nonu
\\
&&
-\delta^{il}\Bigg(\,\, j\leftrightarrow k\,\,\Bigg)
-\delta^{jk}\Bigg(\,\, i\leftrightarrow j\,\,\Bigg)
+\delta^{jl}\Bigg(\,\, i\leftrightarrow j\,,\,\,k\leftrightarrow l\,\,
  \Bigg) \Bigg),
\nonu \\
&& \comm{(\Phi^{(h_1),ij}_{1})_m}{(\Phi^{(h_2),k}_{\frac{3}{2}})_r}
=\Bigg( \delta^{ik}
\sum_{h_3=0}^{h_1+h_2+1\,\,}
\sum_{t=0}^{h_1+h_2-h_3+1}
q^{h_1+h_2-h_3}\,2\,(-1)^{h_1+h_2+1}\,4^{2(h_1+h_2-h_3)-1}\,(2h_3)!
\nonu \\
&& \times
\Bigg[
\Big(
T^{\,\,h_1+1,h_2+1,h_3,t}_{F}(\tfrac{1}{4})+\bar{T}^{\,\,h_1+1,h_2+1,h_3,t}_{B}(\tfrac{1}{4})
\Big)[m+h_1]_{h_1+h_2-h_3-t+1}[r+h_2+\tfrac{1}{2}]_{t}
\nonu \\
&&
+
\Big(
T^{\,\,h_1+1,h_2+1,h_3,t}_{B}(\tfrac{1}{4})+\bar{T}^{\,\,h_1+1,h_2+1,h_3,t}_{F}(\tfrac{1}{4})
\Big)[m+h_1]_{t}[r+h_2+\tfrac{1}{2}]_{h_1+h_2-h_3-t+1}
\Bigg](\Phi^{(h_3),j}_{\frac{1}{2}})_{m+r}
\nonu \\
&& + 
\delta^{ik}
\sum_{h_3=0}^{h_1+h_2\,\,}
\sum_{t=0}^{h_1+h_2-h_3}
q^{h_1+h_2-h_3}\,2\,(-1)^{h_1+h_2+1}\,4^{2(h_1+h_2-h_3)-3}\,(2h_3+2)!
\nonu \\
&&
\times
\Bigg[
\Big(
T^{\,\,h_1+1,h_2+1,h_3+1,t}_{F}(\tfrac{1}{4})-\bar{T}^{\,\,h_1+1,h_2+1,h_3+1,t}_{B}(\tfrac{1}{4})
\Big)[m+h_1]_{h_1+h_2-h_3-t}[r+h_2+\tfrac{1}{2}]_{t}
\nonu \\
&&
+
\Big(
T^{\,\,h_1+1,h_2+1,h_3+1,t}_{B}(\tfrac{1}{4})-\bar{T}^{\,\,h_1+1,h_2+1,h_3+1,t}_{F}(\tfrac{1}{4})
\Big)[m+h_1]_{t}[r+h_2+\tfrac{1}{2}]_{h_1+h_2-h_3-t}
\Bigg](\Phi^{(h_3),j}_{\frac{3}{2}})_{m+r}
\nonu
\\
&& -\delta^{jk}\Bigg(\,\, i\leftrightarrow j\,\,\Bigg) \Bigg)
\nonu \\
&&
+\epsilon^{ijkl}
\sum_{h_3=0}^{h_1+h_2+1\,\,}
\sum_{t=0}^{h_1+h_2-h_3+1}
q^{h_1+h_2-h_3}\,2\,(-1)^{h_1+h_2+1}\,4^{2(h_1+h_2-h_3)-1}\,(2h_3)!
\nonu \\
&&
\times
\Bigg[
\Big(
T^{\,\,h_1+1,h_2+1,h_3,t}_{F}(\tfrac{1}{4})-\bar{T}^{\,\,h_1+1,h_2+1,h_3,t}_{B}(\tfrac{1}{4})
\Big)[m+h_1]_{h_1+h_2-h_3-t+1}[r+h_2+\tfrac{1}{2}]_{t}
\nonu \\
&&
-
\Big(
T^{\,\,h_1+1,h_2+1,h_3,t}_{B}(\tfrac{1}{4})-\bar{T}^{\,\,h_1+1,h_2+1,h_3,t}_{F}(\tfrac{1}{4})
\Big)[m+h_1]_{t}[r+h_2+\tfrac{1}{2}]_{h_1+h_2-h_3-t+1}
\Bigg](\Phi^{(h_3),l}_{\frac{1}{2}})_{m+r}
\nonu \\
&&
+
\epsilon^{ijkl}
\sum_{h_3=0}^{h_1+h_2\,\,}
\sum_{t=0}^{h_1+h_2-h_3}
q^{h_1+h_2-h_3}\,2\,(-1)^{h_1+h_2+1}\,4^{2(h_1+h_2-h_3)-3}\,(2h_3+2)!
\nonu \\
&&
\times
\Bigg[
\Big(
T^{\,\,h_1+1,h_2+1,h_3+1,t}_{F}(\tfrac{1}{4})+\bar{T}^{\,\,h_1+1,h_2+1,h_3+1,t}_{B}(\tfrac{1}{4})
\Big)[m+h_1]_{h_1+h_2-h_3-t}[r+h_2+\tfrac{1}{2}]_{t}
\nonu \\
&&
-
\Big(
T^{\,\,h_1+1,h_2+1,h_3+1,t}_{B}(\tfrac{1}{4})+\bar{T}^{\,\,h_1+1,h_2+1,h_3+1,t}_{F}(\tfrac{1}{4})
\Big)[m+h_1]_{t}[r+h_2+\tfrac{1}{2}]_{h_1+h_2-h_3-t}
\Bigg](\Phi^{(h_3),l}_{\frac{3}{2}})_{m+r},
\nonu \\ 
&& \comm{(\Phi^{(h_1),ij}_{1})_m}{(\Phi^{(h_2)}_{2})_n}
=
\sum_{h_3=0}^{h_1+h_2+1\,\,}
\sum_{k=0}^{h_1+h_2-h_3+1}
q^{h_1+h_2-h_3}\,2\,(-1)^{h_1+h_2+1}\,4^{2(h_1+h_2-h_3)-1}\,(2h_3+1)!\,\nonu \\
&& \times \Bigg(
 \Bigg[
\Big(
S^{\,\,h_1+1,h_2+2,h_3+1,k}_{F,\,R}(\tfrac{1}{4})
+S^{\,\,h_1+1,h_2+2,h_3+1,k}_{B,\,R}(\tfrac{1}{4})
\Big)
[m+h_1]_{h_1+h_2-h_3-k+1}[n+h_2+1]_k \nonu \\
&& 
-
\Big(
S^{\,\,h_1+1,h_2+2,h_3+1,k}_{F,\,L}(\tfrac{1}{4})
+S^{\,\,h_1+1,h_2+2,h_3+1,k}_{B,\,L}(\tfrac{1}{4})
\Big)
[m+h_1]_{k}[n+h_2+1]_{h_1+h_2-h_3-k+1}\,
\Bigg]\nonu \\
 && \times (\Phi^{(h_3),ij}_1)_{m+n}
\nonu \\
&& 
+
 \Bigg[
\Big(
S^{\,\,h_1+1,h_2+2,h_3+1,k}_{F,\,R}(\tfrac{1}{4})
-S^{\,\,h_1+1,h_2+2,h_3+1,k}_{B,\,R}(\tfrac{1}{4})
\Big)
[m+h_1]_{h_1+h_2-h_3-k+1}[n+h_2+1]_k \nonu \\
&&
-
\Big(
S^{h_1+1,h_2+2,h_3+1,k}_{F,\,L}(\tfrac{1}{4})
-S^{h_1+1,h_2+2,h_3+1,k}_{B,\,L}(\tfrac{1}{4})
\Big)
[m+h_1]_{k}[n+h_2+1]_{h_1+h_2-h_3-k+1}
\Bigg]\nonu \\
 && \times (\tilde{\Phi}^{(h_3),ij}_1)_{m+n}
\Bigg),
\nonu \\
&&\acomm{(\Phi^{(h_2),i}_{\frac{3}{2}})_r}{(\Phi^{(h_2),j}_{\frac{3}{2}})_s}
=
\delta^{ij}\,q^{h_1+h_2}\,2\times 4^2\,C_{\frac{3}{2},\frac{3}{2}}^{h_1,h_2}(\tfrac{1}{4})\,
      [r+h_1+\tfrac{1}{2}]_{h_1+h_2+2}\,\delta_{r+s}
      \nonu
\\
&& +
\delta^{ij}
\sum_{h_3=1}^{h_1+h_2+2\,\,}\sum_{k=0}^{h_1+h_2-h_3+2}
q^{h_1+h_2-h_3}\,(-1)^{h_1+h_2}\,4^{2(h_1+h_2-h_3)}\,
(2h_3-1)!
\nonu \\
&&
\times\Bigg[
\Big(U^{\,\,h_1+1,h_2+1,h_3,k}_{F}(\tfrac{1}{4})-U^{\,\,h_2+1,h_1+1,h_3,k}_{B}(\tfrac{1}{4})\Big)
[r+h_1+\tfrac{1}{2}]_{h_1+h_2-h_3-k+2}
[s+h_2+\tfrac{1}{2}]_{k}
\nonu \\
&&
+\Big(U^{\,\,h_2+1,h_1+1,h_3,k}_{F}(\tfrac{1}{4})-U^{\,\,h_1+1,h_2+1,h_3,k}_{B}(\tfrac{1}{4})\Big)
[r+h_1+\tfrac{1}{2}]_{k}
[s+h_2+\tfrac{1}{2}]_{h_1+h_2-h_3-k+2}
\,\Bigg](\Phi^{(h_3)}_{0})_{r+s}
\nonu \\
&&
+\delta^{ij}
\sum_{h_3=-1}^{h_1+h_2\,\,}\sum_{k=0}^{h_1+h_2-h_3}
q^{h_1+h_2-h_3}\,2\,(-1)^{h_1+h_2+1}\,4^{2(h_1+h_2-h_3)-4}\,
(2h_3+2)!(h_3+\tfrac{3}{2})
\nonu \\
&& 
\times\Bigg[
\Big(
U^{\,\,h_1+1,h_2+1,h_3+2,k}_{F}(\tfrac{1}{4})
+U^{\,\,h_2+1,h_1+1,h_3+2,k}_{B}(\tfrac{1}{4})
\Big)
[r+h_1+\tfrac{1}{2}]_{h_1+h_2-h_3-k}
[s+h_2+\tfrac{1}{2}]_{k} 
\nonu
\\
&&
+
\Big(
U^{\,\,h_2+1,h_1+1,h_3+2,k}_{F}(\tfrac{1}{4})
+U^{\,\,h_1+1,h_2+1,h_3+2,k}_{B}(\tfrac{1}{4})
\Big)
[r+h_1+\tfrac{1}{2}]_{k}
[s+h_2+\tfrac{1}{2}]_{h_1+h_2-h_3-k} 
\,\Bigg]\nonu \\
&& \times (\Phi^{(h_3)}_{2})_{r+s}
\nonu \\
&& 
+
\sum_{h_3=0}^{h_1+h_2+1\,\,}\sum_{k=0}^{h_1+h_2-h_3+1}
q^{h_1+h_2-h_3}\,(-1)^{h_1+h_2+1}\,4^{2(h_1+h_2-h_3)-2}\,(2h_3+1)!\,
\nonu \\
&&
\times\Bigg(
\Bigg[
\Big( U^{\,\,h_1+1,h_2+1,h_3+1,k}_{F}(\tfrac{1}{4})+U^{\,\,h_2+1,h_1+1,h_3+1,k}_{B}(\tfrac{1}{4})
\Big)
[r+h_1+\tfrac{1}{2}]_{h_1+h_2-h_3-k+1}
[s+h_2+\tfrac{1}{2}]_{k}
\nonu 
\\
&&
-\Big(U^{\,\,h_2+1,h_1+1,h_3+1,k}_{F}(\tfrac{1}{4})+U^{\,\,h_1+1,h_2+1,h_3+1,k}_{B}(\tfrac{1}{4})\Big)
[r+h_1+\tfrac{1}{2}]_{k}
[s+h_2+\tfrac{1}{2}]_{h_1+h_2-h_3-k+1}
\Bigg]  \nonu \\
&& \times (\Phi^{(h_3),ij}_{1})_{r+s}
\nonu \\
&&
+\Bigg[
\Big(U^{\,\,h_1+1,h_2+1,h_3+1,k}_{F}(\tfrac{1}{4})-U^{\,\,h_2+1,h_1+1,h_3+1,k}_{B}(\tfrac{1}{4})\Big)
[r+h_1+\tfrac{1}{2}]_{h_1+h_2-h_3-k+1}
[s+h_2+\tfrac{1}{2}]_{k}
\nonu \\
&&
-\Big(U^{h_2+1,h_1+1,h_3+1,k}_{F}(\tfrac{1}{4})-U^{h_1+1,h_2+1,h_3+1,k}_{B}(\tfrac{1}{4})\Big)
[r+h_1+\tfrac{1}{2}]_{k}
[s+h_2+\tfrac{1}{2}]_{h_1+h_2-h_3-k+1}
\Bigg]\nonu \\
&& \times (\tilde{\Phi}^{(h_3),ij}_{1})_{r+s}\Bigg),
\nonu \\
&&
\comm{(\Phi^{(h_1),i}_{\frac{3}{2}})_r}{(\Phi^{(h_2)}_{2})_m}
=
\sum_{h_3=0}^{h_1+h_2+2\,\,}\sum_{k=0}^{h_1+h_2-h_3+2}
q^{h_1+h_2-h_3}\,2\,(-1)^{h_1+h_2}\,4^{2(h_1+h_2-h_3)+1}\,(2h_3)!
\nonu \\
&&
\times\Bigg[
  \Big(T^{\,\,h_2+2,h_1+1,h_3,k}_{F}(\tfrac{1}{4})+
  \bar{T}^{\,\,h_2+2,h_1+1,h_3,k}_{B}(\tfrac{1}{4})
\Big)
[m+h_2+1]_{h_1+h_2-h_3-k+2}
[r+h_1+\tfrac{1}{2}]_{k}
\nonu \\
&& 
-\Big(\bar{T}^{\,\,h_2+2,h_1+1,h_3,k}_{F}(\tfrac{1}{4})+T^{\,\,h_2+2,h_1+1,h_3,k}_{B}(\tfrac{1}{4})
\Big)
[m+h_2+1]_{k}
[r+h_1+\tfrac{1}{2}]_{h_1+h_2-h_3-k+2}
\Bigg] \nonu \\
&& \times (\Phi^{(h_3),i}_{\frac{1}{2}})_{r+m}
\nonu \\
&& +
\sum_{h_3=0}^{h_1+h_2+1\,\,}\sum_{k=0}^{h_1+h_2-h_3+1}
q^{h_1+h_2-h_3}\,2\,(-1)^{h_1+h_2}\,4^{2(h_1+h_2-h_3)-1}\,(2h_3+2)!
\nonu \\
&&
\times\Bigg[
\Big(T^{\,\,h_2+2,h_1+1,h_3+1,k}_{F}(\tfrac{1}{4})-\bar{T}^{\,\,h_2+2,h_1+1,h_3+1,k}_{B}(\tfrac{1}{4})
\Big)
[m+h_2+1]_{h_1+h_2-h_3-k+1}
[r+h_1+\tfrac{1}{2}]_{k}
\nonu \\
&& 
+\Big(\bar{T}^{\,\,h_2+2,h_1+1,h_3+1,k}_{F}(\tfrac{1}{4})-T^{\,\,h_2+2,h_1+1,h_3+1,k}_{B}(\tfrac{1}{4})
\Big)
[m+h_2+1]_{k}
[r+h_1+\tfrac{1}{2}]_{h_1+h_2-h_3-k+1}
\Bigg]\nonu \\
&& \times (\Phi^{(h_3),i}_{\frac{3}{2}})_{r+m},
\nonu
\\
&&\comm{(\Phi^{(h_1)}_{2})_m}{(\Phi^{(h_2)}_{2})_n}
=q^{h_1+h_2}\,4^5\,C_{2,2}^{h_1,h_2}(\tfrac{1}{4})\,[m+h_1+1]_{h_1+h_2+3}\,\delta_{m+n}
\nonu \\
&& +\sum_{h_3=1}^{h_1+h_2+3\,\,}\sum_{k=0}^{h_1+h_2-h_3+3}
q^{h_1+h_2-h_3}\,2\,(-1)^{h_1+h_2+1}\,4^{2(h_1+h_2-h_3)+3}\,
(2h_3-1)! \nonu \\
&& 
\times \Bigg[
\Big(
S^{\,\,h_1+2,h_2+2,h_3,k}_{F,\,R}(\tfrac{1}{4})
-S^{\,\,h_1+2,h_2+2,h_3,k}_{B,\,R}(\tfrac{1}{4})
\Big)
[m+h_1+1]_{h_1+h_2-h_3-k+3}[n+h_2+1]_k
\nonu \\
&& 
-
\Big(
S^{\,\,h_1+2,h_2+2,h_3,k}_{F,\,L}(\tfrac{1}{4})
-S^{\,\,h_1+2,h_2+2,h_3,k}_{B,\,L}(\tfrac{1}{4})
\Big)
[m+h_1+1]_k [n+h_2+1]_{h_1+h_2-h_3-k+3}\,\,
\Bigg] \nonu \\
&& \times (\Phi^{(h_3)}_0)_{m+n}
\nonu 
\\
&& +\sum_{h_3=-1}^{h_1+h_2+1\,\,}\sum_{k=0}^{h_1+h_2-h_3+1}
q^{h_1+h_2-h_3}\,(-1)^{h_1+h_2}\,4^{2(h_1+h_2-h_3)}
\,(2h_3+2)! (h_3+\tfrac{3}{2})\nonu \\
&& \times \Bigg[
\Big(
S^{\,\,h_1+2,h_2+2,h_3+2,k}_{F,\,R}(\tfrac{1}{4})
+S^{\,\,h_1+2,h_2+2,h_3+2,k}_{B,\,R}(\tfrac{1}{4})
\Big)
[m+h_1+1]_{h_1+h_2-h_3-k+1}[n+h_2+1]_k
\nonu
+\\
&&
-
\Big(
S^{\,\,h_1+2,h_2+2,h_3+2,k}_{F,\,L}(\tfrac{1}{4})
+S^{\,\,h_1+2,h_2+2,h_3+2,k}_{B,\,L}(\tfrac{1}{4})
\Big) [m+h_1+1]_k [n+h_2+1]_{h_1+h_2-h_3-k+1}
\,\,\Bigg] \nonu \\
&& \times (\Phi^{(h_3)}_2)_{m+n}.
\label{N4algebrafor}
\eea
The $\la$ dependent coefficients appearing in (\ref{N4algebrafor}) can
be obtained
from the following expressions together with (\ref{coeff})
in \cite{AK2309}
\bea
C_{0,0}^{h_1,h_2}(\la) &= &
\frac{2\,N\,(-1)^{h_1+1}\,4^{2(h_1+h_2-4)}\,}{(2h_1-1)(2h_2-1)}\,
\sum_{i_1=0}^{h_1-1}\sum_{i_2=0}^{h_2-1}
\frac{i_1 !\,i_2 !}{(i_1+i_2+1)!} \nonu
\\
&
\times &
\Bigg[(h_1-2\lambda)(h_2-2\lambda)
  a^{i_1}(h_1,\lambda+\tfrac{1}{2})a^{i_2}
  (h_2,\lambda+\tfrac{1}{2})\nonu \\
  & - &
  (h_1-1+2\lambda)(h_2-1+2\lambda)a^{i_1}(h_1,\lambda)a^{i_2}
(h_2,\lambda)
\Bigg], \nonu \\
C_{0,2}^{h_1,h_2}(\lambda) & \equiv &
\frac{N\,(-1)^{h_1}\,4^{2(h_1+h_2)-6}\,}{(2h_1-1)}\,
\sum_{i_1=0}^{h_1-1}\sum_{i_2=0}^{h_2+1}
\frac{i_1 !\,i_2 !}{(i_1+i_2+1)!}
\nonu \\
&
\times & \Bigg[(h_1-2\lambda)
a^{i_1}(h_1,\lambda+\tfrac{1}{2})
a^{i_2}(h_2+2,\lambda+\tfrac{1}{2})\nonu \\
& + &
(h_1-1+2\lambda)
a^{i_1}\big(h_1,\lambda\big)
a^{i_2}\big(h_2+2,\lambda\big)
\Bigg],
\nonu \\
C_{\frac{1}{2},\frac{1}{2}}^{h_1,h_2}(\lambda)
& \equiv &
2\,N\,(-1)^{h_1}\,4^{2(h_1+h_2-4)}\,
\sum_{i_1=0}^{h_1}\sum_{i_2=0}^{h_2}
\frac{i_1 !\,i_2 !}{(i_1+i_2+1)!}
 \nonu \\
&
\times & \Bigg[
\beta^{i_1}(h_1+1,\lambda)
\alpha^{i_2}(h_2+1,\lambda)+
\alpha^{i_1}(h_1+1,\lambda)
\beta^{i_2}(h_2+1,\lambda)
\Bigg],
\nonu \\
C_{\frac{1}{2},\frac{3}{2}}^{h_1,h_2}(\lambda)& \equiv&
2\,N\,(-1)^{h_1+1}\,4^{2(h_1+h_2)-7}\,
\sum_{i_1=0}^{h_1}\sum_{i_2=0}^{h_2+1}
\frac{i_1 !\,i_2 !}{(i_1+i_2+1)!}
\nonu \\
&
\times & \Bigg(
\beta^{i_1}(h_1+1,\lambda)
\alpha^{i_2}(h_2+2,\lambda)-
\alpha^{i_1}(h_1+1,\lambda)
\beta^{i_2}(h_2+2,\lambda)
\Bigg),
\nonu \\
C_{1,1,\,\pm}^{h_1,h_2}(\lambda)& \equiv &
2\,N\,(-1)^{h_1+1}\,4^{2(h_1+h_2)-7}\,
\sum_{i_1=0}^{h_1}\sum_{i_2=0}^{h_2}
\frac{i_1 !\,i_2 !}{(i_1+i_2+1)!}
\nonu \\
&
\times & \Bigg[
a^{i_1}(h_1+1,\lambda+\tfrac{1}{2})
a^{i_2}(h_2+1,\lambda+\tfrac{1}{2})
\pm
a^{i_1}(h_1+1,\lambda)
a^{i_2}(h_2+1,\lambda)
\Bigg],
\nonu \\
C_{\frac{3}{2},\frac{3}{2}}^{h_1,h_2}(\lambda)& \equiv&
2\,N\,(-1)^{h_1}\,4^{2(h_1+h_2)-6}\,
\sum_{i_1=0}^{h_1+1\,\,}\sum_{i_2=0}^{h_2+1}
\frac{i_1 !\,i_2 !}{(i_1+i_2+1)!}
\nonu \\
&
\times& \Bigg[
\beta^{i_1}\big(h_1+2,\lambda\big)
\alpha^{i_2}\big(h_2+2,\lambda\big)+
\alpha^{i_1}\big(h_1+2,\lambda\big)
\beta^{i_2}\big(h_2+2,\lambda\big)
\Bigg],
\nonu \\
C_{2,2}^{h_1,h_2}(\lambda) & \equiv &
2\,N\,(-1)^{h_1+1}\,4^{2(h_1+h_2)-5}
\sum_{i_1=0}^{h_1+1\,\,}\sum_{i_2=0}^{h_2+1}
\frac{i_1 !\,i_2 !}{(i_1+i_2+1)!}
\nonu \\
&
\times & \Bigg[
a^{i_1}(h_1+2,\lambda+\tfrac{1}{2})
a^{i_2}(h_2+2,\lambda+\tfrac{1}{2})
-
a^{i_1}(h_1+2,\lambda)
a^{i_2}(h_2+2,\lambda)
\Bigg].
\label{bigC}
\eea
The structure constants appearing in (\ref{N4algebrafor})
can be determined from the following expressions in \cite{AK2309}
together with (\ref{coeff})
\bea
S^{\,\,h_1,h_2,h_3,k}_{F,\,R}(\lambda)&=&
\sum_{i_1=0}^{h_1-1}\sum_{i_2=0}^{h_2-1}\,\sum_{r=0}^{i_1+i_2}
\Bigg[\frac{(-1)^{i_1+i_2}}{(h_3+r)!}\,
a^{i_1}(h_1,\lambda+\tfrac{1}{2})
a^{i_2}(h_2,\lambda+\tfrac{1}{2})\nonu \\
&\times&
\binom{r}{h_3-1}\binom{i_2}{i_1+i_2-r}\binom{1-h_3+r}{1-
h_2 +i_2+k}
\prod_{j=0}^{r-h_3}(-r-2\lambda+j)\Bigg],
\nonu \\
S^{\,\,h_1,h_2,h_3,k}_{F,\,L}(\lambda)&=&
\sum_{i_1=0}^{h_1-1}\sum_{i_2=0}^{h_2-1}\,\sum_{r=0}^{i_1+i_2}
\Bigg[\frac{(-1)^{i_1+i_2}}{(h_3+r)!}\,
a^{i_1}(h_1,\lambda+\tfrac{1}{2})
a^{i_2}(h_2,\lambda+\tfrac{1}{2})
\nonu \\
&\times&
\binom{r}{h_3-1}\binom{i_1}{i_1+i_2-r}
\binom{1-h_3+r}{1-h_1+i_1+k}
\prod_{j=0}^{r-h_3}( -r-2\lambda+j)\Bigg],
\nonu \\
S^{\,\,h_1,h_2,h_3,k}_{B,\,R}(\lambda)&=&
\sum_{i_1=0}^{h_1-1}\sum_{i_2=0}^{h_2-1}\,\sum_{r=0}^{i_1+i_2}
\Bigg[\frac{(-1)^{i_1+i_2}}{(h_3+r)!}
\,a^{i_1}(h_1,\lambda\big)a^{i_2}(h_2,\lambda\big)
\nonu \\
&\times&
\binom{r}{h_3-1}
\binom{i_2}{i_1+i_2-r}\binom{1-h_3+r}{1-h_2+i_2+k}
\prod_{j=0}^{r-h_3}(1-r-2\lambda+j)\Bigg],
\nonu \\
S^{\,\,h_1,h_2,h_3,k}_{B,\,L}(\lambda)&=&
\sum_{i_1=0}^{h_1-1}\sum_{i_2=0}^{h_2-1}\,\sum_{r=0}^{i_1+i_2}
\Bigg[\frac{(-1)^{i_1+i_2}}{(h_3+r)!}
\,a^{i_1}(h_1,\lambda)a^{i_2}(h_2,\lambda)
\nonu \\
&\times&
\binom{r}{h_3-1}
\binom{i_1}{i_1+i_2-r}\binom{1-h_3+r}{1-h_1+i_1+k}
\prod_{j=0}^{r-h_3} (1-r-2\lambda+j)\Bigg],
\nonu \\
T_{F}^{\,\,h_1,h_2,h_3,k}(\lambda)&=&
\sum_{i_1=0}^{h_1-1}\sum_{i_2=0}^{h_2-1}\,\sum_{r=0}^{i_1+i_2}
\Bigg[\frac{(-1)^{i_1+i_2}}{(h_3+r+1)!}\,
a^{i_1}(h_1,\lambda+\tfrac{1}{2})\beta^{i_2}(h_2+1,\lambda)
\nonu \\
&\times&
\binom{r}{h_3-1}\binom{i_2}{i_1+i_2-r}\binom{1-h_3+r}{1-h_2+
i_2+k}
\prod_{j=0}^{r-h_3}(-r-2\lambda+j)
\Bigg],
\nonu \\
\bar{T}_{F}^{\,\,h_1,h_2,h_3,k}(\lambda)&=&
\sum_{i_1=0}^{h_1-1}\sum_{i_2=0}^{h_2}\,\sum_{r=0}^{i_1+i_2}
\Bigg[\frac{(-1)^{i_1+i_2}}{(h_3+r)!}\,
a^{i_1}(h_1,\lambda+\tfrac{1}{2})\alpha^{i_2}(h_2+1,\lambda)
\nonu \\
&\times&
\binom{r}{h_3}\binom{i_1}{i_1+i_2-r}\binom{-h_3+r}{1-h_1+
i_1+k}
\prod_{j=0}^{r-h_3-1}( 1-r-2\lambda+j)\Bigg],
\nonu \\
T_{B}^{\,\,h_1,h_2,h_3,k}(\lambda)&=&
\sum_{i_1=0}^{h_1-1}\sum_{i_2=0}^{h_2-1}\,\sum_{r=0}^{i_1+i_2}
\Bigg[\frac{(-1)^{i_1+i_2}}{(h_3+r+1)!}\,
a^{i_1}(h_1,\lambda)
\beta^{i_2}(h_2+1,\lambda)
\nonu \\
&\times&
\binom{r}{h_3-1}\binom{i_1}{i_1+i_2-r}\binom{1-h_3+r}{
  1-h_1+i_1+k}
\prod_{j=0}^{r-h_3}(-r-2\lambda+j)
\Bigg],
\nonu \\
\bar{T}_{B}^{\,\,h_1,h_2,h_3,k}(\lambda)&=&
\sum_{i_1=0}^{h_1-1}\sum_{i_2=0}^{h_2}\,\sum_{r=0}^{i_1+i_2}
\Bigg[\frac{(-1)^{i_1+i_2}}{(h_3+r)!}\,
a^{i_1}(h_1,\lambda)\alpha^{i_2}(h_2+1,\lambda)
\nonu \\
&\times&
\binom{r}{h_3}\binom{i_2}{i_1+i_2-r}\binom{-h_3+r}{-
h_2+i_2+k}
\prod_{j=0}^{r-h_3-1}(1-r-2\lambda+j)\Bigg],
\nonu \\
U_F^{\,\,h_1,h_2,h_3,k}(\lambda)&=&
\sum_{i_1=0}^{h_1-1}\sum_{i_2=0}^{h_2}\sum_{r=0}^{i_1+i_2}
\Bigg[\frac{(-1)^{i_1+i_2}}{(h_3+r)!}\,
\beta^{i_1}(h_1+1,\lambda)\alpha^{i_2}(h_2+1,\lambda)
\nonu \\
&\times&
\binom{r}{h_3-1}
\binom{i_2}{i_1+i_2-r}\binom{1-h_3+r}{-h_2+
i_2+k}
\prod_{j=0}^{r-h_3}(-r-2\lambda+j )\Bigg],
\nonu \\
U_B^{\,\,h_1,h_2,h_3,k}(\lambda)&=& 
\sum_{i_1=0}^{h_1-1}\sum_{i_2=0}^{h_2}\sum_{r=0}^{i_1+i_2}
\Bigg[\frac{(-1)^{i_1+i_2}}{(h_3+r)!}\,
\beta^{i_1}(h_1+1,\lambda)\alpha^{i_2}(h_2+1,\lambda)
\nonu \\
&\times&
\binom{r}{h_3-1}
\binom{i_1}{i_1+i_2-r}\binom{1-h_3+r}{1-h_1+i_1+k}
\prod_{j=0}^{r-h_3}( 1-r-2\lambda+j)
\Bigg].
\label{STRUCT}
\eea
In this paper, we only consider $\la =\frac{1}{4}$
where the structure constants behave in special forms
as in (\ref{N4algebrafor}) (the $\la$ dependent coefficients
appearing in (\ref{STRUCT}) in \cite{AK2309} become
the overall factors)
or $\la =0$ (See also Appendix $B$).

\subsection{ The subleading terms up to the $q^4$}

In order to see the extra structures in the
(anti)commutators  beyond the
lowest terms of (\ref{finalanticommcomm}), we analyze
the previous expressions in (\ref{N4algebrafor}) further.
Some of the first few terms in each (anti)commutator
appearing in (\ref{N4algebrafor})
are given by explicitly \footnote{
The weights for each component appearing in (\ref{bigPhi})
are constrained as follows 
  \bea
  \Phi_0^{(h_1)}, \qquad
  \Phi_{\frac{1}{2}}^{(h_2), i}, \qquad
  \Phi_{1}^{(h_3), i j}, \qquad
  \Phi_{\frac{3}{2}}^{(h_4), i}, \qquad
  \Phi_{2}^{(h_5)}, \qquad
  h_1, h_2 \geq 1, \qquad h_3 \geq 0,\qquad
  h_4, h_5 \geq -1 \, ,
  \nonu
  \eea
because $ \Phi_0^{(0)}$
and $\Phi_{\frac{1}{2}}^{(0), i}$
are redundant and can be written in terms of
$  \Phi_{2}^{(-1)}$
and $\Phi_{\frac{3}{2}}^{(-1), i}$
respectively.}.
\bea
&& \comm{(\Phi^{(h_1)}_{0})_m}{(\Phi^{(h_2)}_{0})_n}
=
-q^3\,\delta^{h_1,1}\delta^{h_2,2}\,\frac{\hat{c}}{96}(m-1)m\,\delta_{m+n}
+q^3\,\delta^{h_1,2}\delta^{h_2,1}\,\frac{\hat{c}}{96}(m+1)m\,\delta_{m+n}
\nonu \\
&& +
q^{4}\,\Big((h_2-1)m-(h_1-1)n \Big)\, (\Phi^{(h_1+h_2-4)}_2)_{m+n}
\nonu \\
&& +
\tt
q^3\,\delta^{h_1,1}(1-\delta^{h_2,2})\,\,4m\Big((h_2-1)m+n\Big)\, (\Phi^{(h_2-2)}_0)_{m+n}
\nonu \\
&& - 
{\tt
q^3\,\delta^{h_2,1}(1-\delta^{h_1,2})\,\,4n\Big(m+(h_1-1)n\Big)\, (\Phi^{(h_1-2)}_0)_{m+n}}
+\mathcal{O}(q^{5} )\,,
\nonu \\
&&
\comm{(\Phi^{(h_1)}_{0})_m}{(\Phi^{(h_2),i}_{\frac{1}{2}})_r}
=-q^2\,\frac{1}{8} \, (\Phi^{(h_1+h_2-2),i}_\frac{3}{2})_{m+r}
-q^4\,N^{h_1,h_2+\frac{1}{2}}_1(m,r) \, (\Phi^{(h_1+h_2-4),i}_\frac{3}{2})_{m+r}
\nonu \\
&& + \tt
q^{3}\,\delta^{h_1,1}\,4m\,\Big((h_2-\tfrac{1}{2})m+r\Big)\,
(\Phi^{(h_2-2),i}_\frac{1}{2})_{m+r}
- 
q^{3}\,\delta^{h_2,1}\,2\,(\tfrac{1}{4}-r^2)\,
(\Phi^{(h_1-2),i}_\frac{1}{2})_{m+r}
\nonu \\
&& +{ \tt
q^4\,\delta^{h_1,1}\delta^{h_2,2}\,4m(\tfrac{3}{2}\,m+r)\,(\Phi^{(-1),i}_\frac{3}{2})_{m+r}
- 
q^4\,\delta^{h_1,2}\delta^{h_2,1}\,2(\tfrac{1}{4}-r^2)\,(\Phi^{(-1),i}_\frac{3}{2})_{m+r}} 
+\mathcal{O}(q^{5} )\,,
\nonu \\
&& \comm{(\Phi^{(h_1)}_{0})_m}{(\Phi^{(h_2),ij}_{1})_n}
=-q^{2}\,\Big(h_2\,m-(h_1-1)n\Big)\,
\frac{1}{2}\, \epsilon^{i j k l}\,(\Phi^{(h_1+h_2-2),kl}_1)_{m+n}
\nonu \\
&& -
q^{4}\,\frac{4}{3}\,N^{h_1,h_2+1}_2(m,n)\,
\frac{1}{2}\, \epsilon^{i j k l}\,(\Phi^{(h_1+h_2-4),kl}_1)_{m+n}
-
{\tt
q^3\,\delta^{h_2,0}\,4n\Big(m+(h_1-1)n\Big)\,(\Phi^{(h_1-3),ij}_1)_{m+n}}
\nonu \\
&&+ {\tt
q^3\,\delta^{h_1,1}(1-\delta^{h_2,1})\,4m(h_2\,m+n)\,(\Phi^{(h_2-2),ij}_1)_{m+n}}
+\mathcal{O}(q^{5} ) \, ,
\nonu \\
&& \comm{(\Phi^{(h_1)}_{0})_m}{\Phi^{(h_2),i}_{\frac{3}{2}})_r}
=(1-{\tt \tfrac{1}{2}\delta^{h_2,-1}})\bigg(
-\frac{1}{8} \, (\Phi^{(h_1+h_2),i}_\frac{1}{2})_{m+r}
\nonu \\
&& -
q^2\,N^{h_1,h_2+\frac{3}{2}}_{1}(m,r)\,(\Phi^{(h_1+h_2-2),i}_\frac{1}{2})_{m+r}
-
q^4\,\frac{4}{3}\,N^{h_1,h_2+\frac{3}{2}}_{3}(m,r)\,(\Phi^{(h_1+h_2-4),i}_\frac{1}{2})_{m+r}
\,\bigg)
\nonu \\
&& + \tt
q\,\delta^{h_2,-1}(1-\delta^{h_1,1})\frac{1}{16}\,(\Phi^{(h_1-2),i}_\frac{3}{2})_{m+r}
+ 
q^3\,\delta^{h_2,-1}(1-\delta^{h_1,3})\frac{1}{2}\,N^{h_1,\frac{1}{2}}_{1}(m,r)\,(\Phi^{(h_1-4),i}_\frac{3}{2})_{m+r}
\nonu \\
&&- \tt
q^{3}\,\delta^{h_1,1}\delta^{h_2,1}\,2m(3m+2r)\,
(\Phi^{(-1),i}_\frac{3}{2})_{m+r}
+ 
q^{3}\,\delta^{h_1,1}\,4m\Big((h_2+\tfrac{1}{2})m+r\Big)\,
(\Phi^{(h_2-2),i}_\frac{3}{2})_{m+r}
\nonu \\
&&+ {\tt
q^{3}\,\delta^{h_2,0}(1-\delta^{h_1,2})\,2\,(r^2-\tfrac{1}{4})\,(\Phi^{(h_1-3),i}_\frac{3}{2})_{m+r}}
+\mathcal{O}(q^{5} ) \, ,
\nonu \\
&& \comm{(\Phi^{(h_1)}_{0})_m}{(\Phi^{(h_2)}_2)_n}
=
-\Bigg(\delta^{h_1, 1} \delta^{h_2, -1} \frac{2\,\hat{c}}{3}\, 4^{-4}\, m 
+ q^2\, \delta^{h_1, 3} \delta^{h_2, -1} \frac{\hat{c}}{72}\, (m + 2) (m + 1) m 
\nonu \\
&&+ q^2\, \delta^{h_1, 1} \delta^{h_2, 1} \frac{\hat{c}}{72}\, (m - 2) (m - 1) m 
+ q^2\, \delta^{h_1, 2} \delta^{h_2, 0} \frac{\hat{c}}{72}\, (m^2 - 1) m 
\nonu \\
&&+
q^4\, \delta^{h_1, 5} \delta^{h_2, -1} \frac{2\,\hat{c}}{15}\, (m + 4) (m + 3) (m + 2) (m + 1) m  
\nonu \\
&&+ q^4\, \delta^{h_1, 4} \delta^{h_2, 0} \frac{2\,\hat{c}}{15}\, (m + 3) (m + 2) (m^2 - 1) m  
\nonu \\
&& + q^4\, \delta^{h_1, 1} \delta^{h_2, 3} \frac{2\,\hat{c}}{15}\, (m - 4) (m - 3) (m - 2) (m - 1) m  
\nonu \\
&&+
q^4\, \delta^{h_1, 3} \delta^{h_2, 1} \frac{2\,\hat{c}}{15}\, (m^2 - 4) (m^2 - 1) m  
\nonu \\
&&+ q^4\, \delta^{h_1, 2} \delta^{h_2, 2} \frac{2\,\hat{c}}{15}\, (m - 3) (m - 2) (m^2 - 1) m
\Bigg)\,\delta_{m+n}
\nonu \\
&& + \Big((h_2+1)m-(h_1-1)n\Big) \, (\Phi^{(h_1+h_2)}_0)_{m+n}
\nonu \\
&& +
q^2\,(1-{\tt \delta^{h_1+h_2-2,0}})\,
\frac{4}{3}\,N^{h_1,h_2+2}_{2}(m,n)\,(\Phi^{(h_1+h_2-2)}_0)_{m+n}
\nonu \\
&& +
q^4\,(1-{\tt \delta^{h_1+h_2-4,0}})\,
\frac{16}{15}\,N^{h_1,h_2+2}_{4}(m,n)\,(\Phi^{(h_1+h_2-4)}_0)_{m+n}
\nonu \\
&& +\tt
q^3 \,\delta^{h_1,1}\,4m\Big((h_2+1)m+n\Big)\,(\Phi^{(h_2-2)}_2)_{m+n}
\nonu \\
&&- {\tt
  q^3 \,\delta^{h_2,-1}\,4n\Big(m+(h_2-1)n\Big)\,(\Phi^{(h_1-4)}_2)_{m+n}}
+\mathcal{O}(q^{5} ) \, ,
\nonu \\
&& \acomm{(\Phi^{(h_1),i}_{\frac{1}{2}})_r}{(\Phi^{(h_2),j}_{\frac{1}{2}})_s}
=
-\delta^{ij}\,\Bigg(
-q^3\,\delta^{h_1,1}\delta^{h_1,2}\,\frac{\hat{c}}{3}\,4^{-3}\,(r-\tfrac{3}{2})(r^2-\tfrac{1}{4})\,\delta_{r+s}
\nonu \\
&&+
q^3\,\delta^{h_1,2}\delta^{h_1,1}\,\frac{\hat{c}}{3}\,4^{-3}\,(r+\tfrac{3}{2})(r^2-\tfrac{1}{4})\,\delta_{r+s}
\nonu \\
&&+
q^2\,\frac{1}{4^3}\,(\Phi^{(h_1+h_2-2)}_2)_{r+s}
+q^4\,\frac{1}{8}\,N^{h_1+\frac{1}{2},h_2+\frac{1}{2}}_{1}(r,s)\,(\Phi^{(h_1+h_2-4)}_2)_{r+s}
\nonu \\
&&+
\tt
q^3\,\delta^{h_1,1}(1-\delta^{h_2,2})\,2(r^2-\tfrac{1}{4})\Big((h_2-\tfrac{1}{2})r+\tfrac{3}{2}s\Big)\,(\Phi^{(h_2-2)}_0)_{r+s}
\nonu \\
&& +
      {\tt
        q^3\,\delta^{h_2,1}(1-\delta^{h_1,2})\,2(s^2-\tfrac{1}{4})\Big(\tfrac{3}{2}\,r+(h_1-\tfrac{1}{2})s\Big)\,(\Phi^{(h_1-2)}_0)_{r+s}}\Bigg)
\nonu \\
&&-q^{2}\,\frac{1}{8}\,\Big((h_2-\tfrac{1}{2})r+(h_1-\tfrac{1}{2})s\Big)\,
(\Phi^{(h_1+h_2-2),ij}_1)_{r+s}
\nonu \\
&& -q^{4}\,\frac{1}{6}\,N^{h_1+\frac{1}{2},h_2+\frac{1}{2}}_{2}(r,s)\,
(\Phi^{(h_1+h_2-4),ij}_1)_{r+s}
\nonu \\
&&+ {\tt q^3\,\delta^{h_1,1}
  \,\frac{1}{4}\,(r^2-\tfrac{1}{4})\,\frac{1}{2}\, \epsilon^{i j k l}\,
  (\Phi^{(h_2-2),kl}_1)_{r+s}
- \tt q^3\,\delta^{h_2,1}
\,\frac{1}{4}\,(s^2-\tfrac{1}{4})\,\frac{1}{2}\, \epsilon^{i j k l}\,
(\Phi^{(h_1-2),k l}_1)_{r+s}}
+\mathcal{O}(q^{5} )\,,
\nonu \\
&&\comm{(\Phi^{(h_1),i}_{\frac{1}{2}})_r}{(\Phi^{(h_2),jk}_{1})_m}
=\Bigg( \delta^{ij}\,\Bigg(
\frac{1}{8}\,(\Phi^{(h_1+h_2),k}_{\frac{1}{2}})_{r+m}
+q^2\,N^{h_1+\frac{1}{2},h_2+1}_{1}(r,m)\,(\Phi^{(h_1+h_2-2),k}_{\frac{1}{2}})_{r+m}
\nonu \\
&&
+q^4\,\frac{4}{3}\,N^{h_1+\frac{1}{2},h_2+1}_3(r,m)\,(\Phi^{(h_1+h_2-4),k}_{\frac{1}{2}})_{r+m}
-{\tt q^3\,\delta^{h_1,1}(1-\delta^{h_2,1})\,2(r^2-\tfrac{1}{4})\,
(\Phi^{(h_2-2),k}_{\frac{3}{2}})_{r+m}}
\nonu \\
&&
-{\tt
  q^3\,\delta^{h_2,0}(1-\delta^{h_1,2})\,4m\Big(r+(h_1-\tfrac{1}{2})m\Big)\,
(\Phi^{(h_1-3),k}_{\frac{3}{2}})_{r+m}}
\Bigg)
-\delta^{ik}\,
\Bigg( j \leftrightarrow k\Bigg) \Bigg)
\nonu \\
&&+\epsilon^{ijkl}
\Bigg(
-q^2\,\Big(h_2\,r-(h_1-\tfrac{1}{2})\, m \Big)\,
(\Phi^{(h_1+h_2-2),l}_{\frac{3}{2}})_{r+m}
\nonu \\
&& -q^4\,\frac{4}{3}\,N^{h_1+\frac{1}{2},h_2+1}_2(r,m)\,
(\Phi^{(h_1+h_2-4),l}_{\frac{3}{2}})_{r+m}
-{\tt q\,\delta^{h_2,0}(1-\tfrac{1}{2}\delta^{h_1,1})\frac{m}{2}\,(\Phi^{(h_1-1),l}_\frac{1}{2})_{r+m}}
\nonu \\
&& + \tt
q^3\,\delta^{h_2,0}(1-\tfrac{1}{2}\delta^{h_1,3})\frac{4}{2h_1-3}\,N^{h_1+\frac{1}{2},1}_2(r,m)\,(\Phi^{(h_1-3),l}_\frac{1}{2})_{r+m}
\nonu \\
&&
+ \tt
q^3\,\delta^{h_1,1}(1-\tfrac{1}{2}\delta^{h_2,2})\,8(r^2-\tfrac{1}{4})(2h_2 \,r+3m)\,(\Phi^{(h_2-2),l}_\frac{1}{2})_{r+m}
\nonu \\
&&
-{\tt
  q^3\,\delta^{h_2,1}(1-\tfrac{1}{2}\delta^{h_1,2})\,8m(m^2-1)\,(\Phi^{(h_1-2),l}_\frac{1}{2})_{r+m}
\Bigg)}+\mathcal{O}(q^{5} )\,,
\nonu \\
&&\acomm{(\Phi^{(h_1),i}_{\frac{1}{2}})_r}{(\Phi^{(h_2),j}_{\frac{3}{2}})_s}
=
\delta^{ij}\,\Bigg(
\delta^{h_1, 1} \delta^{h_2, -1} \frac{\hat{c}}{6}\,4^{-5} (r + \tfrac{1}{2})
\nonu \\
&& + q^2\, \delta^{h_1, 3} \delta^{h_2, -1} \frac{\hat{c}}{18}\, 4^{-3} (r + \tfrac{5}{2}) (r + \tfrac{3}{2}) (r + \tfrac{1}{2}) 
\nonu \\
&&
+ q^2\, \delta^{h_1, 1} \delta^{h_2, 1} \,\frac{\hat{c}}{9}\, 4^{-3} (r - \tfrac{3}{2}) (r^2 - \tfrac{1}{4}) 
 +  q^2\, \delta^{h_1, 2} \delta^{h_2, 0} \,\frac{\hat{c}}{9}\, 4^{-3} (r + \tfrac{3}{2}) (r^2 - \tfrac{1}{4})
\nonu \\
&&
+q^4\, \delta^{h_1, 1} \delta^{h_2, 3} \,\frac{\hat{c}}{60}\, (r - \tfrac{7}{2}) (r - \tfrac{5}{2})
(r - \tfrac{3}{2}) (r^2 - \tfrac{1}{4}) 
+ q^4\, \delta^{h_1, 2} \delta^{h_2, 2} \,\frac{\hat{c}}{60}\, (r - \tfrac{5}{2}) (r^2 - \tfrac{9}{4}) (r^2 - \tfrac{1}{4})
\nonu \\
&& 
+q^4 \,\delta^{h_1, 3} \delta^{h_2, 1} \,\frac{\hat{c}}{60}\, (r + \tfrac{5}{2}) (r^2 - \tfrac{9}{4}) (r^2 - \tfrac{1}{4}) 
+q^4\, \delta^{h_1, 4} \delta^{h_2, 0} \,\frac{\hat{c}}{60}\, (r + \tfrac{7}{2}) (r + \tfrac{5}{2}) (r + \tfrac{3}{2}) (r^2 - \tfrac{1}{4})
\nonu \\
&& 
+q^4\, \delta^{h_1, 5} \delta^{h_2, -1} \,\frac{\hat{c}}{120}\, (r + \tfrac{9}{2}) (r + \tfrac{7}{2}) (r + \tfrac{5}{2}) (r + \tfrac{3}{2}) (r + \tfrac{1}{2})\,
\Bigg)\,\delta_{r+s}
\nonu \\
&&
+\delta^{ij}\,\Bigg(
-\Big((h_2+\tfrac{1}{2})r-(h_1 -\tfrac{1}{2})s\Big)\frac {1}{8}\, 
(1 - {\tt \tfrac {1} {2}\delta^{h_1, 1}}\,
\de^{h_2, -1})\,(\Phi_0^{(h_1 + h_2)})_{r+s}
\nonu \\
&&
-q^2 (1 - {\tt \delta^{h_1 + h_2 - 2, 0}})
\frac{1}{6}\, N^{h_1 +\frac{1}{2}, h_2 + \frac{3}{2}}_2(r,s)\, (\Phi_0^{(h_1 + h_2 - 2)})_{r+s} 
\nonu \\
&& 
- q^4 (1 - {\tt \delta^{h_1 + h_2 - 4, 0}}) \frac{2}{15}\, 
 N^{h_1 + \frac{1}{2}, h_2 + \frac{3}{2}}_4(r,s)\, (\Phi_0^{h_1 + h_2 - 4})_{r+s}
\nonu \\
&&-\tt
\delta^{h_2, -1} \frac{1}{32} \Big(r + (2 h_1 - 1) s\Big) \,(\Phi^{(h_1 - 1)}_0)_{r+s} 
\nonu \\
&& 
+\tt
q^2\,\delta^{h_2, -1} (1 - \delta^{h_1, 3}) \frac{1}{12} \, 
N^{h_1 + \frac{1}{2}, h_2 + \frac{3}{2}}_2(r, s)\, (\Phi^{(h_1 - 3)}_0)_{r+s} 
\nonu \\
&&
+\tt q^4\,\delta^{h_2, -1} (1 - \delta^{h_1, 5}) \frac{1}{15} 
N^{h_1 + \frac{1}{2}, h_2 + \frac{3}{2}}_4(r,s)\, (\Phi_0^{(h_1 - 5)})_{r+s} 
+q \,\delta^{h_2, -1} \frac{1}{128}\, (\Phi_2^{(h_1 - 2)})_{r+s}
\nonu \\
&&+ \tt q^3\, \delta^{h_2, -1} \frac{1}{16}\,N^{h_1 + \frac{1}{2}, h_2 
 + \frac{3}{2}}_1(r, s) (\Phi_2^{(h_1 - 4)})_{r+s} 
+q^3\,\delta^{h_2, 0} \,\frac{1}{4} (s^2 - \tfrac{1}{4})\, (\Phi_2^{(h_1 + h_2 - 3)})_{r+s}
\nonu \\
&&- {\tt
q^3 \,\delta^{h_1, 1}\, \frac{1}{4} (r^2 - \tfrac{1}{4}) (\Phi_2^{(h_1 + h_2 - 3)})_{r+s} }
\Bigg)
\nonu \\
&&
+\frac {1} {64} \frac{1}{2}\, \epsilon^{i j k l}\,
(\Phi_1^{(h_1 + h_2),kl})_{r + s} 
+ q^2 \,\frac{1}{8}\, N^{h_1 + \frac{1}{2}, h_2 + \frac{3}{2}}_1(r,s) 
\,
\frac{1}{2}\, \epsilon^{i j k l}\,(\Phi_1^{(h_1 + h_2 - 2), kl})_{r + s}
\nonu \\
&& 
+q^4\,\frac{1}{6}\, N^{h_1 + \frac{1}{2}, h_2 + \frac{3}{2}}_3(r, 
s) \frac{1}{2}\, \epsilon^{i j k l}\,
(\Phi_1^{(h_1 + h_2 - 4), kl})_{r + s} 
-{\tt \delta^{h_2, -1}\frac{1}{128}\,
\frac{1}{2}\, \epsilon^{i j k l}\,
  (\Phi _ 1^{(h_1 - 1), kl})_{r + s} }
\nonu \\
&&
-\tt q^2 \delta^{h_2, -1} (1 - \delta^{h_1, 2})\frac{1}{16}
N^{h_1 + \frac{1}{2}, h_2 + \frac{3}{2}}_1(r,s)\,
\frac{1}{2}\, \epsilon^{i j k l}\,
(\Phi_1^{(h_1 - 3),kl})_{r + s} 
\nonu \\
&&
-\tt
q^4 \,\delta^{h_2, -1} (1 - \delta^{h_1, 4})\frac{1}{12}
N^{h_1 + \frac{1}{2},h_2 + \frac{3}{2}}_3 (r,s)
\frac{1}{2}\, \epsilon^{i j k l}\,
(\Phi_1^{(h_1 - 5),kl})_{r + s} 
\nonu \\
&& -\tt
q\, \delta^{h_2, -1}\,\frac {1}{32}\Big(r + (2 h_1 - 1) s\Big)\, 
(\Phi^{(h_1 - 2), ij}_1)_{r + s} 
\nonu \\
&& 
+\tt q^3\, \delta^{h_2, -1}\,\frac {1} {12} \,
N^{h_1 +\frac{1}{2}, h_2 + \frac{3}{2}}_2 (r,s)\, (\Phi^{(h_1 - 4), ij}_1)_{r + s} 
\nonu \\
&& -\tt
q^3\,\delta^{h_2, 0} (r^2 - \tfrac{1}{4}) \Big(3 r + (2 h_1 - 1) s\Big)
\,(\Phi^{(h_1 - 3), ij}_1)_{r + s} 
\nonu \\
&&
-{\tt q^3\,\delta^{h_1, 1} (r^2 - \tfrac{1}{4}) \Big((2 h_2 + 1) r + 3 s\Big)\,
(\Phi^{(h2 - 2), ij} _ 1) _ {r + s}}
+\mathcal{O}(q^5)\,,
\nonu \\
&&
\comm{(\Phi^{(h_1),i}_{\frac{1}{2}})_r}{(\Phi^{(h_2)}_{2})_m}
=
\Big((h_2+1)r-(h_1-\tfrac{1}{2})m \Big)\,(\Phi^{(h_1+h_2),i}_\frac{1}{2})_{m+r}
\nonu \\
&&+
q^{2}\,\frac{4}{3}\,N^{h_1+\frac{1}{2},h_2+2}_2(r,m)\,(\Phi^{(h_1+h_2-2),i}_\frac{1}{2})_{m+r}
+
q^{4}\,\frac{16}{15}\,N^{h_1+\frac{1}{2},h_2+2}_4(r,m)\,(\Phi^{(h_1+h_2-4),i}_\frac{1}{2})_{m+r}
\nonu \\
&&+
\tt
q \,\delta^{h_2, -1} (1 - \delta^{h_1, 1}) \frac{1}{2}\, m\, (\Phi^{(h_1 - 2),i}_\frac{3}{2})_{m+r} 
\nonu \\
&& - \tt q^3 \delta^{h_2, -1} (1 - \delta^{h_1, 3}) \frac{4}{(2 h_1 - 3)} 
N^{h_1 + \frac{1}{2}, 1}_2(r, m)\,(\Phi^{(h_1 - 4),i}_\frac{3}{2})_{m+r}
\nonu \\
&&+ 
\tt q^3 \delta^{h_2, 0} (1 - \delta^{h_1, 2}) 8 m (m^2 - 1) (\Phi^{(h_1 - 3),i}_\frac{3}{2})_{m+r}
\nonu \\
&&- {\tt q^3 \delta^{h_1, 1} (1 - \delta^{h_2, 1}) \frac{4}{(h_2 - \delta^{h_2, 0}) (2 h_2 + 1)}
N^{\frac{3}{2}, h_2 + 2}_2(r, m)\,(\Phi^{(h_2 - 2),i}_\frac{3}{2})_{m+r}}
+\mathcal{O}(q^{5} )\, ,
   \nonu \\
&&\comm{(\Phi^{(h_1),ij}_{1})_m}{(\Phi^{(h_2),kl}_{1})_n}
=
\Big(\delta^{ik}\delta^{jl}-\delta^{il}\delta^{jk}\Big)
\nonu \\
&&\times
\Bigg(
q\, \delta^{h_1, 0} \delta^{h_2, 1} \frac{\hat{c}}{96}\, m (m - 1)  
- q\, \delta^{h_1, 1} \delta^{h_2, 0} \frac{\hat{c}}{96}\, m (m + 1) 
+ q^3\, \delta^{h_1, 1} \delta^{h_2, 2} \frac{\hat{c}}{6}\, m (m - 2) (m^2 - 1) 
\nonu \\
&&- q^3\, \delta^{h_1, 2} \delta^{h_2, 1} \frac{\hat{c}}{6}\, m (m + 2) (m^2 - 1) 
+ q^3\, \delta^{h_1, 0} \delta^{h_2, 3} \frac{\hat{c}}{18}\, m (m - 1) (m - 2) (m - 3) 
\nonu \\
&&- q^3\, \delta^{h_1, 3} \delta^{h_2, 0} \frac{\hat{c}}{18}\, m (m + 1) (m + 2) (m + 3) 
\Bigg)\,\delta_{m+n}
\nonu \\
&&
-\epsilon^{ijkl}\,
\,\Bigg(
\delta^{h_1, 0} \delta^{h_2, 0} \,\frac{\hat{c}}{384}\, m  
+ q^2\, \delta^{h_1, 0} \delta^{h_2, 2}\, \frac{\hat{c}}{72}\, m (m - 1) (m - 2) 
\nonu \\
&& + q^2\, \delta^{h_1, 2} \delta^{h_2, 0}\, \frac{\hat{c}}{72}\, m (m + 1) (m + 2) 
\nonu \\
&&
+ q^2\, \delta^{h_1, 1} \delta^{h_2, 1}\, \frac{\hat{c}}{72}\, m (m^2 - 1)   
+ q^4\, \delta^{h_1, 0} \delta^{h_2, 4}\, \frac{2\, \hat{c}}{15}\, m (m - 1) (m - 2) (m - 3) (m - 4) 
\nonu \\
&&
+ q^4\, \delta^{h_1, 4} \delta^{h_2, 0}\, \frac{2\, \hat{c}}{15}\, m (m + 1) (m + 2) (m + 3) (m + 4) 
\nonu \\
&& + q^4\, \delta^{h_1, 1} \delta^{h_2, 3}\, \frac{2\, \hat{c}}{15}\, m (m^2 - 1) (m - 2) (m - 3) 
\nonu \\
&&
+ q^4\, \delta^{h_1, 3} \delta^{h_2, 1}\, \frac{2\, \hat{c}}{15}\, m (m^2 - 1) (m + 2) (m + 3) 
+ q^4\, \delta^{h_1, 2} \delta^{h_2, 2}\, \frac{2\, \hat{c}}{15}\, m (m^2 - 1) (m^2 - 4) 
\Bigg)\,\delta_{m+n}
\nonu \\
&&
-\Big(\delta^{ik}\delta^{jl}-\delta^{il}\delta^{jk}\Big)
\nonu \\
&&
\times
\Bigg(
q^{2}\,\Big(h_2\,m-h_1\,n\Big)\,(\Phi^{(h_1+h_2-2)}_2)_{m+n}
+
q^{4}\,\frac{4}{3}\,N^{h_1+1,h_2+1}_{2}(m,n)\,(\Phi^{(h_1+h_2-4)}_2)_{m+n}
\nonu \\
&&
+\tt q\, \delta^{h_1, 0} (1 - \delta^{h_2, 1}) 4 m (h_2 m + n) \,(\Phi^{(h_2 - 1)}_0)_{m+n} 
\nonu \\
&& + \tt q^3\, \delta^{h_1, 0} (1 - \delta^{h_2, 3}) \frac{16}{3 (2 h_2 - 3)}
   N^{1, h_2 + 1}_3(m, n) \,(\Phi^{(h_2 - 3)}_0)_{m+n} 
\nonu \\
&&- \tt q\, \delta^{h_2, 0} (1 - \delta^{h_1, 1}) 4 n ( m + h_1 n) \,(\Phi^{(h_1 - 1)}_0)_{m+n} 
\nonu \\
&& - \tt q^3\, \delta^{h_2, 0} (1 - \delta^{h_1, 3}) \frac{16}{3 (2 h_1 - 3)}\,
   N^{h_1 + 1, 1}_3(m, n) \,(\Phi^{(h_1 - 3)}_0)_{m+n} 
\nonu \\
&&
+ \tt q^3\, \delta^{h_1, 1} (1 - \delta^{h_2, 2}) 
\frac{16}{(h_2 - 1 + \delta^{h_2, 1})
(2 h_2 - 3) (2 h_2 - 1)}\, N^{2, h_2 + 1}_3(m, n) \,(\Phi^{(h_2 - 2)}_0)_{m+n} 
\nonu \\
&&- {\tt q^3\, \delta^{h_2, 1} (1 - \delta^{h_1, 2}) 
\frac{16}{(h_1 - 1 + \delta^{h_1, 1}) (2 h_1 - 3) (2 h_1 - 1)}\, N^{h_1 + 1, 2}_3(m, n) \,(\Phi^{(h_1 - 2)}_0)_{m+n}}
\Bigg)
\nonu \\
&&
+\epsilon^{ijkl}\,\Bigg(
\Big(h_2\,m-h_1\,n\Big)\,
(\Phi^{(h_1+h_2)}_0)_{m+n}
+q^2\,(1-\de^{h_1+h_2,2} ) \,
\frac{4}{3}\,N^{h_1+1,h_2+1}_{2}(m,n)\,(\Phi^{(h_1+h_2-2)}_0)_{m+n}
\nonu \\
&&
+q^4 (1 - {\tt \delta^{h_1 + h_2, 4}}) \frac{16}{15}\, N^{h_1 + 1, h_2 + 1}_4(m, n)
\,(\Phi^{(h_1 + h_2 - 4)}_0)_{m+n} 
\nonu \\
&& + {\tt
q^3\, 2 \,\delta^{h_1, 0}  4\, m (h_2\, m + n)\,(\Phi^{(h_2 - 3)}_2)_{m+n}
-  q^3 \,\delta^{h_2, 0}  4 \,n (m + h_1\, n)\, (\Phi^{(h_1 - 3)}_2)_{m+n}}
\Bigg)
\nonu \\
&&
+ \Bigg( \delta^{ik}\,\Bigg(
-\frac{1}{8}\, (\Phi^{(h_1+h_2),jl}_1)_{m+n}
-q^{2}\,N^{h_1+1,h_2+1}_{1}(m,n)\, (\Phi^{(h_1+h_2-2),jl}_1)_{m+n}
\nonu \\
&&
-q^4\, \frac{4}{3}\,N^{h_1 + 1, h_2 + 1}_3(m, n) \,(\Phi^{(h_1 + h_2 - 4), ij}_1)_{m+n}
+ {\tt  q\, \delta^{h_1, 0} \,\frac{m}{2} \,
\frac{1}{2}\, \epsilon^{i j k l}\,  (\Phi^{(h_2 - 1), kl}_1)_{m+n}}
\nonu \\
&& + \tt q\, \delta^{h_2, 0}\, \frac{n}{2} \,
\frac{1}{2}\, \epsilon^{i j k l}\,(\Phi^{(h_1 - 1), kl}_1)_{m+n}
\nonu \\
&& + \tt  q^3\, \delta^{h_1, 0}\, \frac{2}{(h_2 - 1 + \delta^{h_2, 1})}\,
N^{1, h_2 + 1}_2(m, n) \,
\frac{1}{2}\, \epsilon^{i j k l}\,
(\Phi^{(h_2 - 3), kl}_1)_{m+n} 
  \nonu \\
  && - \tt  q^3\, \delta^{h_2, 0}\, \frac{2}{(h_1 - 1 + \delta^{h_1, 1})}\,
  N^{h_1 + 1, 1}_2(m, n) \,
\frac{1}{2}\, \epsilon^{i j k l}\,
  (\Phi^{(h_1 - 3), kl}_1)_{m+n}
   \nonu \\
   && +
         {  \tt q^3\, \delta^{h_1, 1}\, 8 m (m^2 - 1) \,
           \frac{1}{2}\, \epsilon^{i j k l}\,
           (\Phi^{(h_2 - 2), kl}_1)_{m+n}
+ 
q^3\, \delta^{h_2, 1}\, 8 n (n^2 - 1) \,
\frac{1}{2}\, \epsilon^{i j k l}\,
(\Phi^{(h_1 - 2), kl}_1)_{m+n}}
\Bigg)
\nonu \\
&&-\delta^{il}\Bigg(\,\, j\leftrightarrow k\,\,\Bigg)
-\delta^{jk}\Bigg(\,\, i\leftrightarrow j\,\,\Bigg)
+\delta^{jl}\Bigg(\,\, i\leftrightarrow j\,,\,\,k\leftrightarrow l\,\,
\Bigg) \Bigg)+\mathcal{O}(q^{5} )
\,,   
\nonu \\
&&\comm{(\Phi^{(h_1),ij}_{1})_m}{(\Phi^{(h_2),k}_{\frac{3}{2}})_r}
=
\Bigg( \delta^{ik}\,\Bigg(
-\frac{1}{8}\,(\Phi^{(h_1+h_2),j}_{\frac{3}{2}})_{m+r}
\nonu \\
&& -q^{2}\,N^{h_1+1,h_2+\frac{3}{2}}_{1}(m,r)\,(\Phi^{(h_1+h_2-2),j}_{\frac{3}{2}})_{m+r}
\nonu \\
&&
-q^4\, \frac{4}{3}\, N^{h_1 + 1, h_2 + \frac{3}{2}}_3(m,r) 
\,(\Phi^{(h_1 + h_2 - 4),j}_\frac{3}{2})_{m+r}
+ {\tt
  q\, \frac{\delta^{h_1, 0} (1 - \tfrac{1}{2} \delta^{h_2, 1})}{(h_2 + \delta^{h_2, 0})}
   N^{1, h_2 + \frac{3}{2}}_1(m, r) \,(\Phi^{(h_2 - 1),j}_\frac{1}{2})_{m+r} }
\nonu \\
&&
+ \tt
q^3\, \frac{\delta^{h_1, 0}\,8 (1 - \tfrac{1}{2} \delta^{h_2, 3})}{3 (h_2 - 1 + \delta^{h_2, 1})}
   N^{1, h_2 + \frac{3}{2}}_3(m,r) \,(\Phi^{(h_2 - 3),j}_\frac{1}{2})_{m+r}
\nonu \\
&&
+ \tt q^3\,\frac{\delta^{h_1, 1}\, 8 (1 - \tfrac{1}{2} \delta^{h_2, 2})}{(h_2 + \delta^{h_2, 0}) (h_2 - 1 + \delta^{h_2, 1}) (2 h_2 - 1)}
   N^{2, h_2 + \frac{3}{2}}_3(m,r) \,(\Phi^{(h_2 - 2),j}_\frac{1}{2})_{m+r}
\nonu \\
&&+ \tt q^3\,\frac{\delta^{h_2, 1}\,
  8 (1 - \tfrac{1}{2} \delta^{h_1, 2})}{(h_1 + \delta^{h_1, 0}) (h_1 - 1 + 
     \delta^{h_1, 1}) (2 h_1 - 3) (2 h_1 - 1)}\,
   N^{h_1 + 1, \frac{5}{2}}_3( m, r) \,(\Phi^{(h_1 - 2),j}_\frac{1}{2})_{m+r} 
\nonu \\
&&
+ \tt q^{-1}\, \delta^{h_2, -1} (1 - \tfrac{1}{2} \delta^{h_1, 0}) 4^{-2}
\,(\Phi^{(h_1),j}_\frac{1}{2})_{m+r}
+ q\, \delta^{h_2, -1} \frac{1}{2}\, N^{h_1 + 1, \frac{1}{2}}_1(m, r) 
\,(\Phi^{(h_1 - 2),j}_\frac{1}{2})_{m+r}
\nonu \\
&&+ \tt q^3\, \delta^{h_2, -1} \frac{2}{3}\, N^{h_1 + 1, \frac{1}{2}}_3(m, r) 
\,(\Phi^{(h_1 - 4),j}_\frac{1}{2})_{m+r} 
+  \delta^{h_2, -1} 4^{-2} \,(\Phi^{(h_1 - 1),j}_\frac{3}{2})_{m+r}
\nonu \\
&&+ \tt q^2\, \delta^{h_2, -1} (1 + \delta^{h_1, 2})\,\frac{1}{2}\,
N^{h_1 + 1, \frac{1}{2}}_1(m, r) \,(\Phi^{(h_1 - 3),j}_\frac{3}{2})_{m+r}
\nonu \\
&&
+ \tt
q^4\, \delta^{h_2, -1} (1 + \delta^{h_1, 4}) \frac{2}{3}\, N^{h_1 + 1, \frac{1}{2}}_3(m,r) \,(\Phi^{(h_1 - 5),j}_\frac{3}{2})_{m+r} 
\nonu \\
&&
+\tt q\, \frac{\delta^{h_2, 0}\,( 1 - \tfrac{1}{2} \delta^{h_1, 1})}{(h_1 + \delta^{h_1, 0}) (2 h_1 - 1)} N^{h_1 + 1, \frac{3}{2}}_1(m, r) 
   \,(\Phi^{(h_1 - 1),j}_\frac{1}{2})_{m+r} 
\nonu \\
&&
+ {\tt q^3\, \frac{\delta^{h_2, 0} \,
  8 (1 - \tfrac{1}{2} \delta^{h_1, 3})}{(h_1 - 1 + \delta^{h_1, 1}) (2 h_1 - 3)}
  \, N^{h_1 + 1, \frac{3}{2}}_3(m,r)
   \,(\Phi^{(h_1 - 3),j}_\frac{1}{2})_{m+r}}
\Bigg)
+\delta^{jk}\,\Bigg(
\,\, i\leftrightarrow j\,\,
\Bigg) \Bigg)
\nonu \\
&&
+\epsilon^{ijkl}\,\Bigg(
-
\Big((h_2+\tfrac{1}{2})m-h_1\,r \Big)\,(\Phi^{(h_1+h_2),l}_{\frac{1}{2}})_{m+r}
-q^{2}\,\frac{4}{3}\,\,N^{h_1+1,h_2+\frac{3}{2}}_{2}(m,r)\,(\Phi^{(h_1+h_2-2),l}_{\frac{1}{2}})_{m+r}
\nonu \\
&&
-q^4\, \frac{16}{15}\, N^{h_1 + 1, h_2 + \frac{3}{2}}_4(m, r)
\,(\Phi^{(h_1 + h_2 - 4),l}_\frac{1}{2})_{m+r} 
- {\tt
  \delta^{h_2, -1} (1 + \tfrac{1}{2} \delta^{h_1, 1})(m + 2 h_1\, r)\frac{1}{4 } 
\,(\Phi^{(h_1 - 1),l}_\frac{1}{2})_{m+r} }
\nonu \\
&&
+\tt q^2\, \delta^{h_2, -1} (1 + \tfrac{1}{2} \delta^{h_1, 3}) \frac{2}{3}\,
N^{h_1 + 1, \frac{1}{2}}_2(m, r) \,(\Phi^{(h_1 - 3),l}_\frac{1}{2})_{m+r} 
\nonu \\
&& + \tt q^4\, \delta^{h_2, -1} \frac{8}{15}\, N^{h_1 + 1, \frac{1}{2}}_4(m, r)
\,(\Phi^{(h_1 - 5),l}_\frac{1}{2})_{m+r}
\nonu \\
&&
- \tt
q\, \delta^{h_2, -1} \frac{1}{4} (m + 2 h_1\, r) \,(\Phi^{(h_1 - 2),l}_\frac{3}{2})_{m+r} 
+  q^3\, \delta^{h_2, -1} \frac{2}{3}\, N^{h_1 + 1, \frac{1}{2}}_2(m, r)
\,(\Phi^{(h_1 - 4),l}_\frac{3}{2})_{m+r} 
\nonu \\
&&
+\tt  q^3\, \delta^{h_2, 0} N^{h_1 + 1, \frac{3}{2}}_2(m, r)
\frac{4 (1 - \delta^{h_1, 2})}{(h_1 - 1 + \delta^{h_1, 1}) (2 h_1 - 1)}
   \,(\Phi^{(h_1 - 3),l}_\frac{3}{2})_{m+r} 
   \nonu \\
&& +{\tt  q\, \delta^{h_1, 0} (1 - \delta^{h_2, 0}) \frac{m}{2} \,(\Phi^{(h_2 - 1),l}_\frac{3}{2})_{m+r} 
+  q^3\, \frac{\delta^{h_1, 0} \,
  4 (1 - \delta^{h_2, 2})}{(2 h_2 - 1)}\, N^{1, h_2 + \frac{3}{2}}_2(m, r) \,(\Phi^{(
   h_2 - 3),l}_\frac{3}{2})_{m+r}} 
\nonu \\
&& +{\tt  q^3\, \delta^{h_1, 1} (1 - \delta^{h_2, 1})\, 8 (m^2 - 1) m \,(\Phi^{(h_2 - 2),l}_\frac{3}{2})_{m+r}}
\Bigg) + \mathcal{O}(q^{5} )\,,
\nonu \\
&&\comm{(\Phi^{(h_1),ij}_{1})_m}{(\Phi^{(h_2)}_2)_n}
=
\Big((h_2+1)m-h_1\,n\Big)\,(\Phi^{(h_1+h_2),ij}_1)_{m+n}
\nonu \\
&&+q^2\,\frac{4}{3}\,N^{h_1+1,h_2+2}_2(m,n),(\Phi^{(h_1+h_2-2),ij}_1)_{m+n}
+q^4\, \frac{16}{15}\, N^{h_1 + 1, h_2 + 2}_4(m, n) 
\,(\Phi^{(h_1 + h_2 - 4),ij}_1)_{m+n} 
\nonu \\
&&
-\tt q\, \delta^{h_1, 0}\, 4 m \Big((h_2 + 1) m + n\Big)\,
\frac{1}{2} \, \epsilon^{i j k l} \, (\Phi^{(h_2 - 1),kl}_1)_{m+n} 
\nonu \\
&& - q^3\, \frac{4^2\,\delta^{h_1, 0}}{3 (2 h_2 - 1)}
\, N^{1, h_2 + 2}_3(m, n)\,
\frac{1}{2}\, \epsilon^{i j k l}\,
(\Phi^{(h_2 - 3),kl}_1)_{m+n} 
\nonu \\
&&
- \tt
q^3\, \frac{ 4^2\,\delta^{h_1, 1}}{(h_2 + \delta^{h_2, 0}) (2 h_2 - 1) (2 h_2 + 1)}
\, N^{2, h_2 + 2}_3(m, n)\,
\frac{1}{2}\, \epsilon^{i j k l}\,
(\Phi^{(h_2 - 2),kl}_1)_{m+n}
\nonu \\
&&
+ \tt q\, \delta^{h_2, -1}\, 4 n (m + h_1 n)\,
\frac{1}{2}\, \epsilon^{i j k l}\,
(\Phi^{(h_1 - 2),kl}_1)_{m+n}
\nonu \\
&& + q^3\, \frac{4^2\,\delta^{h_2, -1}}{3 (2 h_1 - 3)}\, N^{h_1 + 1, 1}_3(m, n) 
\frac{1}{2}\, \epsilon^{i j k l}\,
(\Phi^{(h_1 - 4),kl}_1)_{m+n}
\nonu \\
&&
+ {\tt q^3\, \frac{16\,\delta^{h_2, 0}}{(h_1 - 1 + \delta^{h_1, 1}) (2 h_1 - 3) (2 h_1 - 1)}\,
  N^{h_1 + 1, 2}_3(m, n)\,
\frac{1}{2}\, \epsilon^{i j k l}\,
  (\Phi^{(h_1 - 3),kl}_1)_{m+n}}
+\mathcal{O}(q^{5} )\,,
\nonu \\
&&\acomm{(\Phi^{(h_1),i}_{\frac{3}{2}})_r}{(\Phi^{(h_2),j}_{\frac{3}{2}})_s}
=
\delta^{ij}\,\Bigg(
-\frac{1}{q}\, \delta^{h_1, -1} \delta^{h_2, 0}\, \frac{\hat{c}}{6}\, 4^{-5} (r - \tfrac{1}{2})  
+ \frac{1}{q}\, \delta^{h_1, 0} \delta^{h_2, -1} \,\frac{\hat{c}}{6}\, 4^{-5} (r + \tfrac{1}{2}) 
\nonu \\
&&
- q \,\delta^{h_1, -1} \delta^{h_2, 2} \,\frac{2\,\hat{c}}{9}\, 4^{-4} (r - \tfrac{1}{2}) (r - \tfrac{3}{2}) (r - \tfrac{5}{2} ) 
+ q \,\delta^{h_1, 2} \delta^{h_2, -1} \,\frac{2\,\hat{c}}{9}\, 4^{-4} (r + \tfrac{1}{2}) (r + \tfrac{3}{2}) (r + \tfrac{5}{2} )  
\nonu \\
&&- q\, \delta^{h_1, 0} \delta^{h_2, 1} \, \frac{\hat{c}}{3}\, 4^{-3} (r - \tfrac{3}{2}) (r^2 - \tfrac{1}{4}) 
+ q\, \delta^{h_1, 1} \delta^{h_2, 0}\,  \frac{\hat{c}}{3}\, 4^{-3} (r + \tfrac{3}{2}) (r^2 - \tfrac{1}{4}) 
\nonu \\
&&
- q^3\, \delta^{h_1, -1} \delta^{h_2, 4} \,\frac{2\, \hat{c}}{15}\, 4^{-2} 
(r - \tfrac{9}{2}) (r - \tfrac{7}{2}) (r - \tfrac{5}{2} ) (r - \tfrac{3}{2}) (r - \tfrac{1}{2}) 
\nonu \\
&&
+ q^3\, \delta^{h_1, 4} \delta^{h_2, -1} \,\frac{2 \,\hat{c}}{15}\,
4^{-2} (r + \tfrac{9}{2}) (r + \tfrac{7}{2}) (r + \tfrac{5}{2} ) (r + \tfrac{3}{2}) (r + \tfrac{1}{2})  
\nonu \\
&&
- q^3\, \delta^{h_1, 0} \delta^{h_2, 3} \, \frac{\hat{c}}{36} (r - \tfrac{7}{2}) (r - \tfrac{5}{2} ) (r - \tfrac{3}{2}) (r^2 - \tfrac{1}{4})  
\nonu \\
&& + q^3\, \delta^{h_1, 3} \delta^{h_2, 0} \,\frac{\hat{c}}{36} (r + \tfrac{7}{2}) (r + \tfrac{5}{2} ) (r + \tfrac{3}{2}) (r^2 - \tfrac{1}{4}) 
\nonu \\
&&
- q^3\, \delta^{h_1, 1} \delta^{h_2, 2} \,\frac{\hat{c}}{12} (r - \tfrac{5}{2} ) (r^2 - \tfrac{9}{4}) (r^2 - \tfrac{1}{4}) 
+ q^3\, \delta^{h_1, 2} \delta^{h_2, 1} \,\frac{\hat{c}}{12} (r + \tfrac{5}{2} ) (r^2 - \tfrac{9}{4}) (r^2 - \tfrac{1}{4}) 
\Bigg)\,\delta_{r+s}
\nonu \\
&&
+\delta^{ij}
\Bigg(
4^{-3}\, \,(\Phi^{(h_1 + h_2)}_2)_{r+s} 
+ q^2\,  N^{h_1 + \frac{3}{2}, h_2 + \frac{3}{2}}_1(r,s) \frac{1}{8}
\,(\Phi^{(h_1 + h_2 - 2)}_2)_{r+s}
\nonu \\
&&
+ q^4\, N^{h_1 + \frac{3}{2}, h_2 + \frac{3}{2}}_3(r,s)\frac{1}{6}
 \,(\Phi^{(h_1 + h_2 - 4)}_2)_{r+s} 
- {\tt \delta^{h_1, -1} \frac{1}{2} 4^{-3} \,(\Phi^{(h_2 - 1)}_2)_{r+s} }
\nonu \\
&&- \tt q^2\, \delta^{h_1, -1} 4^{-2} N^{\frac{1}{2}, h_2 + \frac{3}{2}}_1(r,s)
 \,(\Phi^{(h_2 - 3)}_2)_{r+s} 
- q^4\, \delta^{h_1, -1} \frac{1}{12}\, N^{\frac{1}{2}, h_2 + \frac{3}{2}}_3(r,s)
\,(\Phi^{(h_2 - 5)}_2)_{r+s} 
\nonu \\
&&
+ \tt
q^{-1}\, \delta^{h_1, -1} (1 - \delta^{h_2, 0}) \frac{1}{32} \Big((2 h_2 + 1) r + s\Big)
\,(\Phi^{(h_2)}_0)_{r+s} 
\nonu \\
&&
+ \tt q\, \delta^{h_1, -1} (1 - \delta^{h_2, 2})\frac{1}{12}\,
N^{\frac{1}{2}, h_2 + \frac{3}{2}}_2(r,s) \,(\Phi^{(h_2 - 2)}_0)_{r+s} 
\nonu \\
&&
+ \tt q^3\, \delta^{h_1, -1} (1 - \delta^{h_2, 4}) \frac{1}{15}\,
N^{\frac{1}{2}, h_2 + \frac{3}{2}}_4(r,s) \,(\Phi^{(h_2 - 4)}_0)_{r+s} 
-  \delta^{h_2, -1} \frac{1}{2} 4^{-3} \,(\Phi^{(h_1 - 1)}_2)_{r+s} 
\nonu \\
&&
- \tt q^2\, \delta^{h_2, -1} 4^{-2} N^{h_1 + \frac{3}{2}, \frac{1}{2}}_1(r,s)
\,(\Phi^{(h_1 - 3)}_2)_{r+s} -
  q^4\, \delta^{h_2, -1} \frac{1}{12}\, N^{h_1 + \frac{3}{2}, \frac{1}{2}}_3( r,s) \,(\Phi^{(h_1 - 5)}_2)_{r+s} 
\nonu \\
&&
+ \tt
  q^{-1}\, \delta^{h_2, -1} (1 - \delta^{h_1, 0}) \frac{1}{32} \Big( r + (2 h_1 + 1) s\Big)
\,(\Phi^{(h_1)}_0)_{r+s} 
\nonu \\
&&
- \tt q\, \delta^{h_2, -1} (1 - \delta^{h_1, 2}) \frac{1}{12}\,
N^{h_1 + \frac{3}{2}, \frac{1}{2}}_2(r,s) \,(\Phi^{(h_1 - 2)}_0)_{r+s} 
\nonu \\
&&
- \tt  q^3\, \delta^{h_2, -1} (1 - \delta^{h_1, 4}) \frac{1}{15}\,
N^{h_1 + \frac{3}{2}, \frac{1}{2}}_4(r,s) \,(\Phi^{(h_1 - 4)}_0)_{r+s} 
\nonu \\
&&
+\tt  q\, \delta^{h_1, 0} \frac{(1 - \delta^{h_2, 1})}{2 (h_2 + \delta^{h_2, 0}) (2 h_2 - 1)}
 \,  N^{\frac{3}{2}, h_2 + \frac{3}{2}}_2(r,s) \,(\Phi^{(h_2 - 1)}_0)_{r+s} 
\nonu \\
&&
+ \tt q^3\, \delta^{h_1, 0} \frac{4 (1 - \delta^{h_2, 3})}{3 (h_2 - 1 + \delta^{h_2, 1}) (2 h_2 - 3)}\,
   N^{\frac{3}{2}, h_2 + \frac{3}{2}}_4(r, s) \,(\Phi^{(h_2 - 3)}_0)_{r+s} 
\nonu \\
&&
- \tt q\, \delta^{h_2, 0}\frac{(1 - \delta^{h_1, 1})}{2 (h_1 + \delta^{h_1, 0}) (2 h_1 - 1)}\,
   N^{h_1 + \frac{3}{2}, \frac{3}{2}}_2(r, s) \,(\Phi^{(h_1 - 1)}_0)_{r+s}
\nonu \\
&&- \tt q^3\, \delta^{h_2, 0} \frac{4 (1 - \delta^{h_1, 3})}{3 (h_1 - 1 + \delta^{h_1, 1}) (2 h_1 - 3)}\,
   N^{h_1 + \frac{3}{2}, \frac{3}{2}}_4(r,s) \,(\Phi^{(h_1 - 3)}_0)_{r+s}
\nonu \\
&&+\tt  q^3\, \delta^{h_1, 1} \frac{4 (1 - \delta^{h_2, 2})}{(h_2 - 1 - \delta^{h_2, 1}) (h_2 - 
     \delta^{h_2, 0}) (2 h_2 - 3) (2 h_2 - 1)}
   N^{\frac{5}{2}, h_2 + \frac{3}{2}}_4(r,s) \,(\Phi^{(h_2 - 2)}_0)_{r+s} 
\nonu \\
&&
- {\tt q^3\, \delta^{h_2, 1} 
\frac{4 (1 - \delta^{h_1, 2})}{(h_1 - 1 - \delta^{h_1, 1}) (h_1 - \delta^{h_1, 0}) (2 h_1 - 3) (2 h_1 - 1)}\,
   N^{h_1 + \frac{3}{2}, \frac{5}{2}}_4(r,s) \,(\Phi^{(h_1 - 2)}_0)_{r+s}}
\Bigg)
\nonu \\
&&
+\frac{1}{8}\,\Big((h_2+\tfrac{1}{2})r-(h_1+\tfrac{1}{2})s\Big)\,
(\Phi^{(h_1+h_2),ij}_{1})_{r+s}
+q^{2}\,\frac{1}{6}\,N^{h_1+\frac{3}{2},h_2+\frac{3}{2}}_2(r,s)\,
(\Phi^{(h_1+h_2-2),ij}_{1})_{r+s}\,,
\nonu \\
&&
+ q^4\,\frac{2}{15}\,  N^{h_1 + \frac{3}{2}, h_2 + \frac{3}{2}}_4(r,s) 
\,(\Phi^{(h_1 + h_2 - 4),ij}_1)_{r+s}
\nonu \\
&&
-\tt q^3 \,\frac{1}{210} (\delta^{h_1, 1} - \delta^{h_2, 1}) 
\,N^{h_1 + \frac{3}{2}, h_2 + \frac{3}{2}}_3(r,s) \,
\frac{1}{2}\, \epsilon^{i j k l}\,
(\Phi^{(h_1 + h_2 - 3),kl}_1)_{r+s} 
\nonu \\
&&
-\tt q^{-1}\, \delta^{h_1, -1} 4^{-3} \frac{1}{2} \,
\frac{1}{2}\, \epsilon^{i j k l}\,
(\Phi^{(h_2),kl}_1)_{r+s} 
- \delta^{h_1, -1} \frac{1}{32} \Big((2 h_2 + 1) r + s\Big) \,(\Phi^{(h_2 - 1),ij}_1)_{r+s} 
\nonu \\
&&
- \tt q^2\, \delta^{h_1, -1} \frac{1}{12}\, N^{\frac{1}{2}, h_2 + \frac{3}{2}}_2(r,s) 
\,(\Phi^{(h_2 - 3),ij}_1)_{r+s} 
- q^4\, \delta^{h_1, -1} \frac{1}{15} N^{\frac{1}{2}, h_2 + \frac{3}{2}}_4(r,s) \,(\Phi^{(h_2 - 5),ij}_1)_{r+s}
\nonu \\
&&
-\tt q \, \delta^{h_1, -1} 4^{-2} N^{ \frac{1}{2}, h_2 + \frac{3}{2}}_1(r,s)
\,
\frac{1}{2}\, \epsilon^{i j k l}\,
(\Phi^{(h_2 - 2),kl}_1)_{r+s} 
y\nonu \\
&& - q^3\, \delta^{h_1, -1} \frac{1}{12}\,
N^{\frac{1}{2}, h_2 + \frac{3}{2}}_3(r,s) \,
\frac{1}{2}\, \epsilon^{i j k l}\,
(\Phi^{(h_2 - 4),kl}_1)_{r+s} 
\nonu \\
&&
-\tt q\, \delta^{h_1, 0} \frac{1}{4}
(r^2 - \tfrac{1}{4}) \,
\frac{1}{2}\, \epsilon^{i j k l}\,
(\Phi^{(h_2 - 1),kl}_1)_{r+s}
\nonu \\
&& - \tt q^3\, \delta^{h_1, 0} \frac{1}{(h_2 - 1 + \delta^{h_2, 1}) (2 h_2 - 1)}
N^{\frac{3}{2}, h_2 + \frac{3}{2}}_3(r,s) \,
\frac{1}{2}\, \epsilon^{i j k l}\,
(\Phi^{(h_2 - 3),kl}_1)_{r+s} 
\nonu \\
&&
+\tt q^3\, \delta^{h_1, 1} \frac{2}{105} (h_2 - 3) (5 + 2 h_2) (14 - h_2 + 2 h_2^2) (r^2-\tfrac{1}{4}) (r^2-\tfrac{9}{4}) \,
\frac{1}{2}\, \epsilon^{i j k l}\,
(\Phi^{(h_2 - 2),kl}_1)_{r+s} 
\nonu \\
&&- \tt q^3\, \delta^{h_2, 1} \frac{2}{105} (h_1 - 3) (5 + 2 h_1) (14 - h_1 + 2 h_1^2) (s^2 - \tfrac{1}{4}) (s^2 -\tfrac{9}{4}) \,
\frac{1}{2}\, \epsilon^{i j k l}\,
(\Phi^{(h_1 - 2),kl}_1)_{r+s}
\nonu \\
&&
+ \tt
\delta^{h_2, -1} \frac{1}{32}\, (m + (2 h_1 + 1) n) \,(\Phi^{(h_1 - 1),ij}_1)_{r+s} 
- q^2\, \delta^{h_2, -1} \frac{1}{12}\, N^{h_1 + \frac{3}{2}, \frac{1}{2}}_2(r,s)
\,(\Phi^{(h_1 - 3),ij}_1)_{r+s} 
\nonu \\
&&
-\tt q^4\, \delta^{h_2, -1} \frac{1}{15}\, N^{h_1 + \frac{3}{2}, \frac{1}{2}}_4(r,s) \,(\Phi^{(h_1 - 5),ij}_1)_{r+s} 
+ q^{-1}\, \delta^{h_2, -1} \frac{1}{8} 4^{-2} \,
\frac{1}{2}\, \epsilon^{i j k l}\,
(\Phi^{(h_1),kl}_1)_{r+s} 
\nonu \\
&&
+\tt q \, \delta^{h_2, -1} 4^{-2}\, N^{h_1 + \frac{3}{2}, \frac{1}{2}}_1(r,s) 
\,
\frac{1}{2}\, \epsilon^{i j k l}\,
(\Phi^{(h_1 - 2),kl}_1)_{r+s} 
\nonu \\
&& + q^3\, \delta^{h_2, -1} \frac{1}{12}\, N^{h_1 + \frac{3}{2}, \frac{1}{2}}_3(r,s)
\,
\frac{1}{2}\, \epsilon^{i j k l}\,
(\Phi^{(h_1 - 4),kl}_1)_{r+s} 
\nonu \\
&&+\tt q\, \delta^{h_2, 0} 4^{-1} (s^2 - \tfrac{1}{4}) \,
\frac{1}{2}\, \epsilon^{i j k l}\,
(\Phi^{(h_1 - 1),kl}_1)_{r+s} 
\nonu \\
&& + {\tt q^3\, \delta^{h_2, 0} \frac{1}{(h_1 - 1 + \delta^{h_1, 1}) (2 h_1 - 1)}
  N^{h_1 + \frac{3}{2}, \frac{3}{2}}_3(r,s) \,
\frac{1}{2}\, \epsilon^{i j k l}\,
  (\Phi^{(h_1 - 3),kl}_1)_{r+s}}
+\mathcal{O}(q^{5} )
   \,,
\nonu \\
&&\comm{(\Phi^{(h_1),i}_{\frac{3}{2}})_r}{(\Phi^{(h_2)}_2)_m}
=
\Big((h_2+1)r-(h_1+\tfrac{1}{2})m\Big)\,(1+
{\tt \de^{h_1,0}\, \de^{h_2,-1}})\,
(\Phi^{(h_1+h_2),i}_\frac{3}{2})_{r+m}
\nonu \\
&&+q^2\,\frac{4}{3}\,N^{h_1+\frac{3}{2},h_2+2}_2(r,m)\,(\Phi^{(h_1+h_2-2),i}_\frac{3}{2})_{r+m}
+q^4\,\frac{16}{15}\,N^{h_1+\frac{3}{2},h_2+2}_4(r,m)\,(\Phi^{(h_1+h_2-4),i}_\frac{3}{2})_{r+m}\,,
\nonu \\
&&
-\tt q^{-1}\, \delta^{h_1, -1}\frac{1}{2} \Big((h_2 + 1) r + \tfrac{1}{2} m\Big) 
\,(\Phi^{(h_2),i}_\frac{1}{2})_{r+m} 
+ q^{-1}\, \delta^{h_2, -1} (1 - \delta^{h_1, 0}) \frac{m}{2} \,(\Phi^{(h_1),i}_\frac{1}{2})_{r+m} 
\nonu \\
&&
-\tt \delta^{h_1, -1} (1 + \delta^{h_2, 0})\frac{1}{2} \Big((h_2 + 1) r + \tfrac{1}{2} m\Big) \,(\Phi^{(h_1 + h_2),i}_\frac{3}{2})_{r+m}
\nonu \\
&&- \tt q\, \delta^{h_1, -1} (1 -\tfrac{1}{2} \delta^{h_2, 2}) \frac{2}{3}
\,N^{\frac{1}{2}, h_2 + 2}_2(r, m) \,(\Phi^{(h_2 - 2),i}_\frac{1}{2})_{r+m} 
\nonu \\
&&
-\tt q\, \delta^{h_2, -1} (1 -\tfrac{1}{2} \delta^{h_1, 2}) \frac{4}{(2 h_1 - 1)}
\,N^{h_1 + \frac{3}{2}, 1}_2(r, m) \,(\Phi^{(h_1 - 2),i}_\frac{1}{2})_{r+m} 
\nonu \\
&&
-\tt q\, \delta^{h_1, 0} (1 - \tfrac{1}{2} \delta^{h_2, 1}) \frac{4}{(h_2 - \delta^{h_2, 0}) (1 + 2 h_2)}\,
   N^{\frac{3}{2}, h_2 + 2}_2(r, m) \,(\Phi^{(h_2 - 1),i}_\frac{1}{2})_{r+m} 
\nonu \\
&&
-\tt q\, \delta^{h_2, 0} (1 - \tfrac{1}{2} \delta^{h_1, 1}) 
\frac{4}{(h_1 - \delta^{h_1, 0}) (2 h_1 - 1) (2 h_1 + 1)}\, N^{h_1 + \frac{3}{2}, 2}_2(r, m) \,(\Phi^{(h_1 - 1),i}_\frac{1}{2})_{r+m}
\nonu \\
&&
-\tt  q^2\, \delta^{h_1, -1} \frac{2}{3} N^{\frac{1}{2}, h_2 + 2}_2(r, m)
  \,(\Phi^{(h_2 - 3,i}_\frac{3}{2})_{r+m}
\nonu \\
&&
-\tt  q^3\, \delta^{h_2, -1} (1 - \tfrac{1}{2} \delta^{h_1, 4}) \frac{4^2}{3 (2 h_1 - 3)}\,
   N^{h_1 + \frac{3}{2}, 1}_4(r, m)
   \,(\Phi^{(h_1 - 4),i}_\frac{1}{2})_{r+m}
\nonu \\
&&
-\tt  q^3\, 2
\frac{\delta^{h_1, 1}\, 4^2 (1 - \tfrac{1}{2} \delta^{h_2, 2})}{(h_2 - \delta^{h_2, 0}) (h_2 - 1 + 
       \delta^{h_2, 1})(2 h_2 - 1) (2 h_2 + 1)}\,
N^{\frac{5}{2}, h_2 + 2}_4(r, m) \,(\Phi^{(h_2 - 2),i}_\frac{1}{2})_{r+m}
\nonu \\
&&
-\tt q^3\, \delta^{h_1, -1} (1 - \tfrac{1}{2} \delta^{h_2, 4}) \frac{8}{15}\,
N^{\frac{1}{2}, h_2 + 2}_4 (r, m) \,(\Phi^{(h_2 - 4),i}_\frac{1}{2})_{r+m} 
\nonu \\
&&
-\tt  q^3\, \delta^{h_1, 0} (1 - \tfrac{1}{2} \delta^{h_2, 3}) 
\frac{32}{ 3 (h_2 - 1 - \delta^{h_2, 1}) (2 h_2 - 1)}\,
   N^{\frac{3}{2}, h_2 + 2}_4(r, m) \,(\Phi^{(h_2 - 3),i}_\frac{1}{2})_{r+m}
\nonu \\
&&
-\tt q^3\, \delta^{h_2,0} (1 - \tfrac{1}{2} \delta^{h_1, 3}) 
\frac{32}{(h_1 - 1 + \delta^{h_1, 1}) (2 h_1 - 1) (2 h_1 - 3)}
\,N^{h_1 + \frac{3}{2}, 2}_4(r, m) \,(\Phi^{(h_1 - 3),i}_\frac{1}{2})_{r+m}
\nonu \\
&&
+\tt q^3\, \delta^{h_2,1}\,2(1 -\tfrac{1}{2} \delta^{h_1, 2}) 4^3 m (m^2 - 1) (m^2 - 4) 
   \,(\Phi^{(h_1 - 2),i}_\frac{1}{2})_{r+m}
\nonu \\
&&
-{\tt q^4\, \delta^{h_1, -1} \frac{8}{15}\, N^{\frac{1}{2}, h_2 + 2}_4 (r, m) 
\,(\Phi^{(h_2 - 5),i}_\frac{3}{2})_{r+m}}
+\mathcal{O}(q^{5} )\,,
\nonu \\
&&\comm{(\Phi^{(h_1)}_2)_m}{(\Phi^{(h_2)}_2)_n}
=
\Bigg(-q^{-1}\, \delta^{h_1, -1} \delta^{h_2, 0} \frac{\hat{c}}{6}\, 4^{-2}\, (m - 1) m 
+ q^{-1}\, \delta^{h_1, 0} \delta^{h_2, -1} \frac{\hat{c}}{6}\, 4^{-2}\, (m + 1) m 
\nonu \\
&&
- q\, \delta^{h_1, -1} \delta^{h_2, 2} \frac{\hat{c}}{18}\, (m - 3) (m - 2) (m - 1) m
- q\, \delta^{h_1, 0} \delta^{h_2, 1} \frac{\hat{c}}{6}\, (m - 2) (m^2 - 1) m 
\nonu \\
&&
+ q\, \delta^{h_1, 1} \delta^{h_2, 0} \frac{\hat{c}}{6}\, (m + 2) (m^2 - 1) m 
+ q\, \delta^{h_1, 2} \delta^{h_2, -1} \frac{\hat{c}}{18}\, (m + 3) (m + 2) (m + 1) m 
\nonu \\
&&
- q^3\, \delta^{h_1, -1} \delta^{h_2, 4} \frac{8\,\hat{c}}{15}\,
(m - 5) (m - 4) (m - 3) (m - 2) (m - 1) m 
\nonu \\
&&
- q^3\, \delta^{h_1, 0} \delta^{h_2, 3} \frac{8\,\hat{c}}{9}\, (m - 4) (m - 3) (m - 2) (m^2 - 1) m 
\nonu \\
&&
- q^3\, \delta^{h_1, 1} \delta^{h_2, 2} \frac{8\,\hat{c}}{3}\, (m - 3) (m^2 - 4) (m^2 - 1) m 
+ q^3 \delta^{h_1, 2} \delta^{h_2, 1} \frac{8\,\hat{c}}{3} \,(m + 3) (m^2 - 4) (m^2 - 1) m 
\nonu \\
&&
+ q^3\, \delta^{h_1, 3} \delta^{h_2, 0} \frac{8\,\hat{c}}{9}\, (m + 4) (m + 3) (m + 2) (m^2 - 1) m 
\nonu \\
&&
+ q^3\, \delta^{h_1, 4} \delta^{h_2, -1} \frac{8\,\hat{c}}{15}\, (m + 5) (m + 4) (m + 3) (m + 2) (m + 1) m\,\Bigg)\,\delta_{m+n}
\nonu \\
&&
+\Big((h_1+1)m-(h_2+1)n\Big)\,(\Phi^{(h_1+h_2)}_2)_{m+n}
\nonu \\
&&+q^2\,\frac{4}{3}\,N^{h_1+2,h_2+2}_2(m,n)\,(\Phi^{(h_1+h_2-2)}_2)_{m+n}
+q^4\,\frac{16}{15}\,N^{h_1+2,h_2+2}_4(m,n)\,(\Phi^{(h_1+h_2-4)}_2)_{m+n}\,,
\nonu \\
&&
-\tt q^{-1}\,\Big(
\delta^{h_2, -1} (1 - \delta^{h_1, 0})
-\delta^{h_1,-1}(1 - \delta^{h_2, 0})
\Big)  \frac{2}{(2 h_1 + 2 h+2+3)}
\,N^{h_1 + 2, h_2 + 2}_1(m, n) \,(\Phi^{(h_1 + h_2 + 1)}_0)_{m+n}
\nonu \\
&&
-\tt q\,
\Big( \delta^{h_2, -1} (1 - \delta^{h_1, 2}) 
- \delta^{h_1, -1} (1 - \delta^{h_2, 2})
\Big)\frac{4^2}{3 (2 h_1 +2 h_2+ 1)} \,
   N^{h_1 + 2, h_2 + 2}_3(m, n) \,(\Phi^{(h_1 + h_2 - 1)}_0)_{m+n}
\nonu \\
&&
-\tt q^3\, \Big(\delta^{h_2, -1} (1 - \delta^{h_1, 4})
-\delta^{h_1, -1} (1 - \delta^{h_2, 4})
\Big)\frac{4^3}{15 (2 h_1+2h_2 +1)}\,
   N^{h_1 + 2, h_2 + 2}_5(m, n) \,(\Phi^{(h_1 + h_2 - 3)}_0)_{m+n}
\nonu \\
&&
-\tt q\,4^2\,\Bigg(\delta^{h_2,0}
\frac{(1 - \delta^{h_1, 1}) (2 h_1 - 3)!!}{(h_1 -\delta^{h_1, 0}) (2 h_1 + 1)!!}
-\delta^{h_1,0}
\frac{(1 - \delta^{h_2, 1}) (2 h_2 - 3)!!}{(h_2 -\delta^{h_2, 0}) (2 h_2 + 1)!!}
\Bigg)
\, N^{h_1 + 2, h_2 + 2}_3(m, n) \nonu \\
&& \times  \tt (\Phi^{(h_1 + h_2 - 1)}_0)_{m+n}
\nonu \\
&&
-\tt q^3\,\frac{2\times4^3}{3}\Bigg(\delta^{h_2, 0}  
\frac{ (1 - \delta^{h_1, 3}) (2 h_1 - 5)!!}{(h_1 + \delta^{h_1, 1} - 1) (2 h_1 - 1)!!}
-\delta^{h_1, 0} \frac{ (1 - \delta^{h_2, 3}) (2 h_2 - 5)!!}{(h_2 + \delta^{h_2, 1} - 1) (2 h_2 - 1)!!}
\Bigg)\,
\nonu \\
&& \times \tt N^{h_1 + 2, h_2 + 2}_5(m, n) 
\, (\Phi^{(h_1 + h_2 - 3)}_0)_{m+n}
\nonu \\
&&
-\tt q^3\,2\times4^3 \Bigg(
\delta^{h_2, 1}(1-\delta^{h_1,2})  \frac{ (2 h_1 - 5)!!}{(h_1 - \delta^{h_1, 0})
(h_1 - \delta^{h_1, 1} - 1) (2 h_1 + 1)!!}
\nonu \\
&&
-\tt
\delta^{h_1, 1}(1-\delta^{h_2,2})  \frac{ (2 h_2 - 5)!!}{(h_2 - \delta^{h_2, 0})
(h_2 - \delta^{h_2, 1} - 1) (2 h_2 + 1)!!}
\Bigg)
\,N^{h_1 + 2, h_2 + 2}_5(m, n) 
\,(\Phi^{(h_1 + h_2 - 3)}_0)_{m+n}\nonu \\
&& +\mathcal{O}(q^{5} )\,.
\label{finalanticommcomm1}
\eea
The parameter $q$ is small quantity but not equal to zero.
The higher order terms
which are not written in (\ref{finalanticommcomm1})
can be read off from (\ref{N4algebrafor}).
Here $\hat{c}$ is given by $\hat{c}=6 N$
\footnote{We use the typewriter fonts in (\ref{finalanticommcomm1})
  for $h_1, h_2 =-1,0,1,2$. The mode dependent function
  $N^{h_1,h_2}_h(m,n)$ is defined in (\ref{Nphi}). }.

\subsection{The Jacobi identity
\label{jjacobi}}

We have checked that the Jacobi identity
for (\ref{finalanticommcomm}) is satisfied for $q=0$.
We would like to check the Jacobi identity for nonzero $q$.
We can explicitly
check the Jacobi identity of (\ref{finalanticommcomm1}).
That is, in (\ref{finalanticommcomm}), by substituting
the condition $q=0$ into this relation,
we can check explicitly that the Jacobi identity is satisfied.
Then how do we check the Jacobi identity
for small but nonzero $q$?
The power of $q$ is given by four
in the first commutator of (\ref{finalanticommcomm}).
How do we check the Jacobi identity having this
commutator up to the $q^4$?

Let us calculate
the Jacobi identity with
the condition $h_1, h_2, h_3 \geq 2$ for simplicity
\footnote{We can also check the Jacobi identity if we relax
this condition.}
\bea
&&
\comm{(\Phi^{(h_1 )}_0)_{m}}{\comm{(\Phi^{(h_2)}_0)_{n}}{(\Phi^{(h_3 )}_\frac{1}{2})_{r}}}
+\comm{(\Phi^{(h_3 )}_\frac{1}{2})_{r}}{\comm{(\Phi^{(h_1 )}_0)_{m}}{(\Phi^{(h_2)}_0)_{n}}}
\nonu \\
&& -\comm{(\Phi^{(h_2 )}_0)_{n}}{\comm{(\Phi^{(h_1)}_0)_{m}}{(\Phi^{(h_3 )}_\frac{1}{2})_{r}}}
\,,
\label{jacobi}
\eea
which should be equal to zero for associativity of the algebra.
The minus sign in the last term of (\ref{jacobi})
comes from the change of the two operators inside the commutator. 

The first term of (\ref{jacobi}), by using
the (anti)commutators in (\ref{finalanticommcomm1})
explicitly, 
can be obtained 
\bea
&&
\comm{(\Phi^{(h_1 )}_0)_{m}}{\comm{(\Phi^{(h_2)}_0)_{n}}{(\Phi^{(h_3 )}_\frac{1}{2})_{r}}}=
\nonu \\
&&
q^8\, \frac{2}{3} 
\,A(h_2,h_3)_{n,r}\,(\delta^{h_1,2} \delta^{ h_2 + h_3-4,2}-1)
\,
(\delta^{ h_1 + h_2 + h_3-4,4}-1)\,N^{h_1, h_2 + h_3-\frac{5}{2}}_3(m,n + r)
\,\nonu \\
&& \times (\Phi^{(h_1 + h_2 + h_3-8),i}_\frac{1}{2})_{m+n+r}
\nonu \\
&&
+q^7\,A(h_2,h_3)_{n,r}\,\Big((n + r)^2-\tfrac{1}{4}\Big) 
    \delta^{h_2 + h_3-4, 0 } (\delta^{h_1,2}-1) 
\,(\Phi^{(h_1+h_2+h_3-7),i}_\frac{3}{2})_{m+n+r} 
\nonu \\
&&
+q^6\,\frac{1}{6}  
\bigg(
3\, A(h_2,h_3)_{n,r}\, \,N^{h_1, h_2 + h_3-\frac{5}{2}}_1(m,n + r) 
+ (1 - \delta^{ h_1,2} \delta^{h_2 + h_3-2,2}) \,
\nonu \\
&& \times N^{h_1, h_2 + h_3-\frac{1}{2}}_3(m, n + r)\bigg)
\, (\Phi^{(h_1 + h_2 + h_3-6),i}_\frac{1}{2})_{m+n+r}
\nonu \\
&&
+q^4\, \frac{1}{16} \bigg(
A(h_2,h_3)_{n,r}\, + 
    2 N^{h_1, h_2 + h_3-\frac{1}{2}}_1(m, n + r)\bigg)
    \,(\Phi^{( h_1 + h_2 + h_3-4),i}_\frac{1}{2})_{m+n+r} 
\nonu \\
&&+ q^2\,\frac{1}{64}  \,(\Phi^{(h_1 + h_2 + h_3-2),i}_\frac{1}{2})_{m+n+r}
\,,
\label{firstone}
\eea 
where we have the mode dependent function
\bea
&&
A(h_2,h_3)_{n,r} \equiv
\nonu \\
&& (5 - 7 h_2 + 2 h_2^2 - 12 h_3 + 16 h_2 h_3 - 4 h_2^2 h_3 + 4 h_3^2 - 
 4 h_2 h_3^2 + 4 n^2 - 12 h_3 n^2 + 8 h_3^2 n^2 - 24 n r
\nonu \\
&&+ 16 h_2 n r + 24 h_3 n r - 16 h_2 h_3 n r + 12 r^2 - 20 h_2 r^2 + 
 8 h_2^2 r^2)\,.
\label{inter}
\eea

The second term of (\ref{jacobi}) with the help of
(\ref{finalanticommcomm1}) becomes
\bea
&&\comm{(\Phi^{(h_3 )}_\frac{1}{2})_{r}}{\comm{(\Phi^{(h_1 )}_0)_{m}}{(\Phi^{(h_2)}_0)_{n}}}=
\Big((h_2-1) m -(h_1-1) n\Big)
\nonu \\
&&
\times \Bigg(
-q^8\,\frac{16}{15} 
( \delta^{h_1 + h_2 + h_3-4,4}-1) 
\,N^{ h_3+\frac{1}{2} , h_1 + h_2-2}_4(r,m + n) 
\,(\Phi^{(h_1 + h_2 + h_3-8),i}_\frac{1}{2})_{m+n+r} 
\nonu \\
&&
+ q^7\,\frac{4  (\delta^{h_1 + h_2-4,1}-1) \delta^{h_3,1}}{(2 h_1 + 2 h_2-7) (h_1 + h_2-4 - \delta^{h_1 + h_2-4,0})}
\,   N^{\frac{3}{2}, h_1 + h_2-2}_2(r, m + n)
      \,(\Phi^{( h_1 + h_2.+h_3-7),i}_\frac{3}{2})_{m+n+r}) 
\nonu \\
&&- 
 q^7\,8 ( m + n-1) (m + n) (1 + m + n)
 \delta^{h_1 + h_2-4,0} ( \delta^{ h_3,2}-1) \,
 (\Phi^{(h_1+h_2+h_3-7),i}_\frac{3}{2})_{m+n+r}
\nonu \\
&&+ 
 q^6\,\frac{4}{3}   \,
 N^{h_3+\frac{1}{2}, h_1 + h_2-2}_2(r, m + n) \,(\Phi^{(h_1 + h_2 + h_3-6),i}_\frac{1}{2})_{m+n+r} 
\nonu \\
&&
+q^4\, \frac{1}{2}  
(m - 2 h_3 m + n - 2 h_3 n - 6 r + 2 h_1 r + 2 h_2 r) \,(\Phi^{(h_1 + h_2 + h_3-4),i}_\frac{1}{2})_{m+n+r}\,\Bigg)
\,.
\label{secondone}
\eea

The third term of (\ref{jacobi})
can be obtained by taking
$h_1 \leftrightarrow h_2$ and $m \leftrightarrow n$
from (\ref{firstone}).

Therefore, we determine
the Jacobi identity for (\ref{jacobi})
by summing over (\ref{firstone}), (\ref{secondone}) and
(\ref{firstone}), where
$h_1$ replaced by $h_2$ and $m$ replaced by $n$
and vice versa, together with (\ref{inter}).
We can check that
there are no 
$q^2$ terms and $q^4$ terms.
Therefore, we have checked the Jacobi identity
up to $q^4$ terms on this particular three operators.
However, there exist $q^6$ terms, $q^7$ and $q^8$ terms.
By ignoring these terms (higher orders in $q$),
the Jacobi identity satisfies. Note that
the parameter is a small quantity and nonzero \footnote{
By considering the more terms in (\ref{finalanticommcomm1}),
the above nonzero higher order terms in $q$ can be calculated explicitly
and the coefficients appearing in  these terms should vanish
although we did not do it. In principle, we can check the Jacobi
identity order by order in the parameter $q$ in this way.}.


\subsection{The OPEs}

For convenience, we present the
OPEs in the antiholomorphic sector (by using the unusual unbarred
notations for the complex coordinates)
corresponding to (\ref{finalanticommcomm}) as follows:
\bea
\Phi^{(h_1)}_{0}(z)\,\Phi^{(h_2)}_{0}(w)
&=&
\frac{1}{(z-w)^2}\,q^4\,(h_1+h_2-2)\,\Phi^{(h_1+h_2-4)}_{2}(w)
\nonu \\
& + & \frac{1}{(z-w)}\,q^4\,(h_1-1)\,\partial\Phi^{(h_1+h_2-4)}_{2}(w)
+ \cdots \, ,
\nonu \\
\Phi^{(h_1)}_{0}(z)\,\Phi^{(h_2),i}_{\frac{1}{2}}(w)
&=&
-\frac{1}{(z-w)}\,q^{2}\,
\frac{1}{8}\,\Phi^{(h_1+h_2-2),i}_\frac{3}{2}(w)
+ \cdots \,,
\nonu \\
\Phi^{(h_1)}_{0}(z)\,\Phi^{(h_2),ij}_{1}(w)
&=&
-\frac{1}{(z-w)^2}\,q^2\,(h_1+h_2-1)\,
\frac{1}{2} \epsilon^{i j k l} \, \Phi^{(h_1+h_2-2),k l}_1(w)
\nonu \\
&- & \frac{1}{(z-w)}\,q^2\,(h_1-1)\,
\frac{1}{2} \epsilon^{i j k l} \,
\pa \, \Phi^{(h_1+h_2-2),k l}_1(w)
+ \cdots \,,
\nonu \\
\Phi^{(h_1)}_{0}(z)\,\Phi^{(h_2),i}_{\frac{3}{2}}(w)
&=&
-\frac{1}{(z-w)}\,\frac{1}{8}\,\Phi^{(h_1+h_2),i}_\frac{1}{2}(w)
+ \cdots \,,
\nonu \\
\Phi^{(h_1)}_{0}(z)\,\Phi^{(h_2)}_{2}(w)
&=&
\frac{1}{(z-w)^2}\,(h_1+h_2)\,\Phi^{(h_1+h_2)}_{0}(w)
+\frac{1}{(z-w)}\,(h_1-1)\,\partial\Phi^{(h_1+h_2)}_{0}(w)
\nonu \\
&+& \cdots \,,
\nonu \\
\Phi^{(h_1)i}_{\frac{1}{2}}(z)\,\Phi^{(h_2),j}_{\frac{1}{2}}(w)
&=&
-\frac{1}{(z-w)^2}\,q^2\,\frac{1}{8}\,(h_1+h_2-1)\,\Phi^{(h_1+h_2-2),ij}_{1}(w)
\nonu \\
&
-&
\frac{1}{(z-w)}\,\Bigg[ q^{2}\,\delta^{ij}\,\frac{1}{64}\,\Phi^{(h_1+h_2-2)}_{2}
+q^2\,\frac{1}{8}\,(h_1-\tfrac{1}{2})\,\partial\Phi^{(h_1+h_2-2),ij}_{1}
\Bigg](w)
\nonu \\
&+& \cdots \,,
\nonu \\
\Phi^{(h_1)i}_{\frac{1}{2}}(z)\,\Phi^{(h_2),jk}_{1}(w)
&=&
-\frac{1}{(z-w)^2}\,q^2\,(h_1+h_2-\tfrac{1}{2})\,\epsilon^{ijkl}\,\Phi^{(h_1+h_2-2),l}_{\frac{3}{2}}(w)
\nonu \\
&
+&
\frac{1}{(z-w)}\,\Bigg[\frac{1}{8}\,\Big(\delta^{ij}\,\Phi^{(h_1+h_2),k}_{\frac{1}{2}}
-\delta^{ik}\,\Phi^{(h_1+h_2),j}_{\frac{1}{2}}\Big)
\nonu \\
& - &
q^2\,(h_1-\tfrac{1}{2})\,\epsilon^{ijkl}\,\partial\Phi^{(h_1+h_2-2),l}_{\frac{3}{2}}
\Bigg](w) + \cdots \,,
\nonu \\
\Phi^{(h_1)i}_{\frac{1}{2}}(z)\,\Phi^{(h_2),j}_{\frac{3}{2}}(w)
&=&
-\frac{1}{(z-w)^2}\,\delta^{ij}\,\frac{1}{8}\,(h_1+h_2)\,\Phi^{(h_1+h_2)}_{0}(w)
\nonu \\
&
-&
\frac{1}{(z-w)}\,\Bigg[ \delta^{ij}\,\frac{1}{8}\,(h_1-\tfrac{1}{2})\,\partial\Phi^{(h_1+h_2)}_{0}
  -\frac{1}{64}\,
\frac{1}{2} \epsilon^{i j k l}\,
\Phi^{(h_1+h_2), k l}_{1}
\Bigg](w) + \cdots \,,
\nonu \\
\Phi^{(h_1)i}_{\frac{1}{2}}(z)\,\Phi^{(h_2)}_{2}(w)
&=&
\frac{1}{(z-w)^2}\,(h_1+h_2+\tfrac{1}{2})\,\Phi^{(h_1+h_2),i}_{\frac{1}{2}}(w)
\nonu \\
& + &
\frac{1}{(z-w)}\,(h_1-\tfrac{1}{2})\,
\partial\Phi^{(h_1+h_2),i}_{\frac{1}{2}}(w) + \cdots \,,
\nonu \\
\Phi^{(h_1),ij}_{1}(z)\,\Phi^{(h_2),kl}_{1}(w)
&=&
\frac{1}{(z-w)^2}(h_1+h_2)\Bigg[
-q^2(\delta^{ik}\delta^{jl}-\delta^{il}\delta^{jk})\Phi^{(h_1+h_2-2)}_2
+\epsilon^{ijkl}\Phi^{(h_1+h_2)}_0
\Bigg](w)
\nonu \\
&
+& \frac{1}{(z-w)}\,\Bigg[
-q^2\,(\delta^{ik}\delta^{jl}-\delta^{il}\delta^{jk})\,h_1\,\partial\,\Phi^{(h_1+h_2-2)}_2
+h_1\,\epsilon^{ijkl}\,\partial\,\Phi^{(h_1+h_2)}_0
\nonu \\
&
-& \frac{1}{8}\,
\bigg(
\delta^{ik} \Phi^{(h_1+h_2),jl}_1
-\delta^{il} \Phi^{(h_1+h_2),jk}_1
-\delta^{jk} \Phi^{(h_1+h_2),il}_1
+\delta^{jl} \Phi^{(h_1+h_2),ik}_1\bigg)
\Bigg](w)\,,
\nonu \\
\Phi^{(h_1),ij}_{1}(z)\,\Phi^{(h_2),k}_{\frac{3}{2}}(w)
&=&
-\frac{1}{(z-w)^2}\,(h_1+h_2+\tfrac{1}{2})\,\epsilon^{ijkl}\,\Phi^{(h_1+h_2),l}_\frac{1}{2}(w)
\nonu \\
&
-& \frac{1}{(z-w)}\,\Bigg[
h_1\,\epsilon^{ijkl}\,\partial\Phi^{(h_1+h_2),l}_\frac{1}{2}
+\frac{1}{8}\,\Big(
\delta^{ik}\,\Phi^{(h_1+h_2),j}_\frac{3}{2}-\delta^{jk}\,\Phi^{(h_1+h_2),i}_\frac{3}{2}
\Big)
\Bigg](w) \nonu \\
& + & \cdots \,,
\nonu \\
\Phi^{(h_1),ij}_{1}(z)\,\Phi^{(h_2)}_{2}(w)
&=&
\frac{1}{(z-w)^2}\,(h_1+h_2+1)\,\Phi^{(h_1+h_2),ij}_1(w)
+\frac{1}{(z-w)}\,h_1\,\partial\Phi^{(h_1+h_2),ij}_1(w)
\nonu \\
& + & \cdots \,,
\nonu \\
\Phi^{(h_1),i}_{\frac{3}{2}}(z)\,\Phi^{(h_2),j}_{\frac{3}{2}}(w)
&=&
\frac{1}{(z-w)^2}\,\frac{1}{8}\,(h_1+h_2+1)\,\Phi^{(h_1+h_2),ij}_1(w)
\nonu \\
&
+& \frac{1}{(z-w)}\,\Bigg[
\delta^{ij}\,\frac{1}{64}\,\Phi^{(h_1+h_2)}_2+\frac{1}{8}\,(h_1+\tfrac{1}{2})\,\partial\Phi^{(h_1+h_2),ij}_1
\Bigg](w) + \cdots \,,
\nonu \\
\Phi^{(h_1),i}_{\frac{3}{2}}(z)\,\Phi^{(h_2)}_{2}(w)
&=&
\frac{1}{(z-w)^2}\,(h_1+h_2+\tfrac{3}{2})\,\Phi^{(h_1+h_2),i}_{\frac{3}{2}}(w)
+\frac{1}{(z-w)}\,(h_1+\tfrac{1}{2})\,\partial\Phi^{(h_1+h_2),i}_{\frac{3}{2}}(w)
\nonu \\
& +& \cdots \, ,
\nonu \\
\Phi^{(h_1)}_{2}(z)\,\Phi^{(h_2)}_{2}(w)
&=&
\frac{1}{(z-w)^2}\,(h_1+h_2+2)\,\Phi^{(h_1+h_2)}_{2}(w)
+\frac{1}{(z-w)}\,(h_1+1)\,\partial\Phi^{(h_1+h_2)}_{2}(w)
\nonu \\
& +& \cdots \,.
\label{opeversion}
\eea
For the higher order terms appearing in (\ref{finalanticommcomm1}),
the similar higher singular terms can be added into (\ref{opeversion}).
Of course, by using the standard conformal field  theory,
we can determine the (anti)commutators in (\ref{finalanticommcomm})
from (\ref{opeversion}).

\subsection{ The $w_{1+\infty}^{2,2}[\la]$ algebra}

We present  the $w_{1+\infty}^{2,2}[\la]$ algebra
for lower $q$ terms in the generic $\la$
\bea
&& \comm{(\Phi^{(h_1)}_{0})_m}{(\Phi^{(h_2)}_{0})_n}
= 
q^{4}\,\Big((h_2-1)m-(h_1-1)n \Big)\, f^{h_1,h_2}_{0,0,2}(\la)\,(\tilde{\Phi}^{(h_1+h_2-4)}_2)_{m+n}
\nonu \\
&&+
q^{2}\,\Big((h_2-1)m-(h_1-1)n \Big)\, \frac{2(h_1+h_2-1)(4\la-1)}{(2h_1-1)(2h_2-1)}\,(\Phi^{(h_1+h_2-2)}_0)_{m+n}
\nonu \\
&&+
q^{4}\,N^{h_1,h_2}_2(m,n)\,  (4\la-1)\,f^{h_1,h_2}_{0,0,0}(\la)\,(\Phi^{(h_1+h_2-4)}_0)_{m+n}
\, ,
\nonu \\
&& \comm{(\Phi^{(h_1)}_{0})_m}{(\Phi^{(h_2),i}_{\frac{1}{2}})_r}
= 
-\frac{1}{8}\,q^{2}\,
(\tilde{\Phi}^{(h_1+h_2-2),i}_\frac{3}{2})_{m+r} 
\nonu \\
&&+
q^2 \,\Big((h_2-\tfrac{1}{2})m-(h_1-1)r\Big)\, \frac{4(h_2-1)(h_1+h_2-1)(4\la-1)}{(2h_1-1)(2h_2-1)(2h_1+2h_2-3)}\,(\Phi^{(h_1+h_2-2),i}_\frac{1}{2})_{m+r}\, ,
\nonu \\
&&
\comm{(\Phi^{(h_1)}_{0})_m}{(\Phi^{(h_2),ij}_{1})_n} 
= 
-q^{2}\,\Big(h_2\,m-(h_1-1)n \Big)
\,
\frac{1}{2} \, \epsilon^{ijkl} (\Phi^{(h_1+h_2-2),kl}_1)_{m+n}
\nonu \\
&&+
q^{2}\,\Big(h_2\,m-(h_1-1)n \Big)
\,
\frac{(4\lambda-1)}{(2h_1-1)}\, (\Phi^{(h_1+h_2-2),ij}_1)_{m+n}\, ,
\nonu \\
&& \comm{(\Phi^{(h_1)}_{0})_m}{(\tilde{\Phi}^{(h_2),i}_{\frac{3}{2}})_r}
= 
-\frac{1}{8}\,(\Phi^{(h_1+h_2),i}_\frac{1}{2})_{m+r} 
 \, ,
\nonu \\
&& \comm{(\Phi^{(h_1)}_{0})_m}{(\tilde{\Phi}^{(h_2)}_{2})_n}
=   \Big((h_2+1)m-(h_1-1)r \Big)\,(\Phi^{(h_1+h_2)}_0)_{m+n}\, ,
\nonu \\
&& \acomm{(\Phi^{(h_1),i}_{\frac{1}{2}})_r}{(\Phi^{(h_2),j}_{\frac{1}{2}})_s}
=  -\frac{1}{64}\,q^2 \, \delta^{ij}\,
(\tilde{\Phi}^{(h_1+h_2-2)}_{2})_{r+s} 
\nonu \\
&&-
q^2 \,\delta^{ij}\,N^{h_1+\frac{1}{2},h_2+\frac{1}{2}}_1(r,s)\,
\frac{(h_1+h_2-1)(4\lambda-1)}{2(2h_1-1)(2h_2-1)(2h_1+2h_2-3)}\,(\Phi^{(h_1+h_2-2)}_0)_{r+s}\, 
\nonu \\
&&- \frac{1}{8}
\,
q^{2}\,\Big((h_2-\tfrac{1}{2})r-(h_1-\tfrac{1}{2})s \Big)\,(\Phi^{(h_1+h_2-2),ij}_{1})_{r+s}
\nonu \\
&&+ 
q^{2}\,\Big((h_2-\tfrac{1}{2})r-(h_1-\tfrac{1}{2})s \Big)\,\frac{(4\lambda-1)}{8(2h_1-1)(2h_2-1)}\,\frac{1}{2}\epsilon^{ijkl}(\Phi^{(h_1+h_2-2),kl}_{1})_{r+s}\, ,
\nonu \\
&& \comm{(\Phi^{(h_1),i}_{\frac{1}{2}})_r}{(\Phi^{(h_2),jk}_{1})_m}
=  \delta^{ij}\,
\frac{1}{8}\,(\Phi^{(h_1+h_2),k}_{\frac{1}{2}})_{r+m}
-\delta^{ik}\,\frac{1}{8}\,
(\Phi^{(h_1+h_2),j}_{\frac{1}{2}})_{r+m}
\nonu \\
&&-  q^2 \, \epsilon^{ijkl}\,
\Big(h_2\,r-(h_1-\tfrac{1}{2})m \Big)\, (\tilde{\Phi}^{(h_1+h_2-2),l}_{\frac{3}{2}})_{r+m}
\nonu \\
&&- q^2 \, \epsilon^{ijkl}\,
N^{h_1+\frac{1}{2},h_2+1}_1(r,m)\,\frac{4(h_2-1)\,(4\lambda-1)}{(2h_1-1)(2h_2-1)(2h_1+2h_2-3)}\, (\Phi^{(h_1+h_2-2),l}_{\frac{1}{2}})_{r+m}
\, ,
\nonu \\
&&
\acomm{(\Phi^{(h_1),i}_{\frac{1}{2}})_r}{(\tilde{\Phi}^{(h_2),j}_{\frac{3}{2}})_s}
=  \frac{1}{64}\,
\frac{1}{2} \, \epsilon^{ijkl} \,
(\Phi_{1}^{(h_1+h_2),kl})_{r+s} \, 
\nonu \\
&& - \frac{1}{8}\, \delta^{ij}\,
\Big((h_2+\tfrac{1}{2})r-(h_1-\tfrac{1}{2})s\Big)\, (\Phi^{(h_1+h_2)}_{0})_{r+s} \, ,
\nonu \\
&& \comm{(\tilde{\Phi}^{(h_1)}_{2})_m}{(\Phi^{(h_2),i}_{\frac{1}{2}})_r}
=  
\Big((h_2-\tfrac{1}{2})m-(h_1+1)r \Big)\,(\Phi^{(h_1+h_2),i}_\frac{1}{2})_{m+r} \, ,
\nonu \\
&& \comm{(\Phi^{(h_1),ij}_{1})_m}{(\Phi^{(h_2),kl}_{1})_n}
=  -q^{2}\,(\delta^{ik}\delta^{jl}-\delta^{il}\delta^{jk})
\,\bigg[
(h_2\,m-h_1\,n)\,(\tilde{\Phi}^{(h_1+h_2-2)}_2)_{m+n} 
\nonu \\
&&+
N^{h_1+1,h_2+1}_2(m,n)\,\frac{8\,(4\la-1)}{(2h_1-1)(2h_2-1)(2h_1+2h_2-3)}\,(\Phi^{(h_1+h_2-2)}_0)_{m+n} 
\bigg]
\nonu \\
&& +  \epsilon^{ijkl}\,(h_2\,m-h_1\,n)\,
(\Phi^{(h_1+h_2)}_0)_{m+n}
\nonu \\
&& +  \frac{1}{8} \, \bigg[\, -\delta^{ik}\,(\Phi^{(h_1+h_2),jl}_{1})_{m+n}
+\delta^{il}\,(\Phi^{(h_1+h_2),jk}_{1})_{m+n}
\nonu \\
&& +  \delta^{jk}\,(\Phi^{(h_1+h_2),il}_{1})_{m+n}
-\delta^{jl}\,(\Phi^{(h_1+h_2),ik}_{1})_{m+n}\,\bigg]\,,
\nonu \\
&& \comm{(\Phi^{(h_1),ij}_{1})_m}{(\tilde{\Phi}^{(h_2),k}_{\frac{3}{2}})_r}
= 
-\frac{1}{8}\,\delta^{ik}\,(\tilde{\Phi}^{(h_1+h_2),j}_{\frac{3}{2}})_{m+r}
+\frac{1}{8}\,\delta^{jk}\,(\tilde{\Phi}^{(h_1+h_2),i}_{\frac{3}{2}})_{m+r}
\nonu \\
&& -
\Big((h_2+\tfrac{1}{2})m-h_1\,r \Big)\frac{(4\la-1)}{(2h_2+1)(2h_1+2h_2+1)}
\nonu \\
&& \times
\Big(
\delta^{ik}\,(\Phi^{(h_1+h_2),j}_{\frac{1}{2}})_{m+r}
-\delta^{jk}\,(\Phi^{(h_1+h_2),i}_{\frac{1}{2}})_{m+r}\Big)
\nonu \\
&&-  \epsilon^{ijkl}\,
\Big((h_2+\tfrac{1}{2})m-h_1\,r \Big)\,(\Phi^{(h_1+h_2),l}_{\frac{1}{2}})_{m+r} \, ,
\nonu \\
&& \comm{(\Phi^{(h_1),ij}_{1})_m}{(\tilde{\Phi}^{(h_2)}_{2})_n}
=  \Big((h_2+1)m-h_1\,n\Big)\,(\Phi^{(h_1+h_2),ij}_1)_{m+n}\, ,
\nonu \\
&&
\acomm{(\tilde{\Phi}^{(h_2),i}_{\frac{3}{2}})_r}{(\tilde{\Phi}^{(h_2),j}_{\frac{3}{2}})_s}
=  \frac{1}{64}\,\delta^{ij}\,
(\tilde{\Phi}^{(h_1+h_2)}_{2})_{r+s}
\nonu \\
&&+\delta^{ij}\,N^{h_1+\frac{3}{2},h_2+\frac{3}{2}}_{1}(r,s)\frac{(4\la-1)(h_1+h_2+1)}{2(2h_1+1)(2h_2+1)(2h_1+2h_2+1)}\,(\Phi^{(h_1+h_2)}_0)_{r+s}
\nonu \\
&&+  \frac{1}{8}\,\Big((h_2+\tfrac{1}{2})r-(h_1+\tfrac{1}{2})s\Big)\,
(\Phi^{(h_1+h_2),ij}_{1})_{r+s}
\nonu \\
&&-  \Big((h_2+\tfrac{1}{2})r-(h_1+\tfrac{1}{2})s\Big)\,\frac{(4\la-1)}{8(2h_1+1)(2h_2+1)}\,
\frac{1}{2}\epsilon^{ijkl}(\Phi^{(h_1+h_2),kl}_{1})_{r+s}\, ,
\nonu \\
&&\comm{(\tilde{\Phi}^{(h_1)}_{2})_m}{(\tilde{\Phi}^{(h_2),i}_{\frac{3}{2}})_r}
=  \Big((h_2+\tfrac{1}{2})m-(h_1+1)r \Big)\,
(\tilde{\Phi}^{(h_1+h_2),i}_\frac{3}{2})_{m+r}
\nonu \\
&&+N^{h_1+2,h_2+\frac{3}{2}}_1(m,r)\,
\frac{4\,h_1\,(4\la-1)}{(2h_1+1)(2h_2+1)(2h_1+2h_2+1)}\,(\Phi^{(h_1+h_2),i}_\frac{1}{2})_{m+r}
\, ,
\nonu \\
&& \comm{(\tilde{\Phi}^{(h_1)}_{2})_m}{(\tilde{\Phi}^{(h_2)}_{2})_n}
= 
\Big((h_2+1)m-(h_1+1)n\Big)\,(\tilde{\Phi}^{(h_1+h_2)}_2)_{m+n}
\nonu \\
&& +
N^{h_1+2,h_2+2}_2(m,n)\,\frac{8\,(4\la-1)}{(2h_1+1)(2h_2+1)(2h_1+2h_2+1)}\,(\Phi^{(h_1+h_2)}_0)_{m+n}\, ,
\label{wgenericla}
\eea
where $\la$ dependent coefficients are given by
%
\bea
&& \Big((h_2-1)m-(h_1-1)n \Big)\, f^{h_1,h_2}_{0,0,2}(\la) \equiv
\nonu \\
&&    \sum_{k=0}^1(-1)^{h_1+h_2}\frac{4(2h_1+2h_2-6)!}{(2h_1-1)(2h_2-1)}
\Bigg[
\bigg(
(h_1 + h_2 - 3 + 2 \lambda) (h_1 - 2 \lambda) (h_2 - 2 \lambda) 
S_{F,R}^{h_1, h_2, h_1 + h_2 - 2,k}(\lambda) 
\nonu \\
&& + (h_1 + h_2 - 2 - 2 \lambda) (h_1 - 1 + 2 \lambda) 
(h_2 - 1 + 2 \lambda) 
S_{B,R}^{h_1, h_2, h_1 + h_2 - 2, k}(\lambda)\bigg) \nonu \\
&& \times [m + h_1 - 1]_{1 -k} [n + h_2 - 1]_{k} 
\nonu \\
&& -\bigg ((h_1 + h_2 - 3 + 2 \lambda) (h_1 - 2 \lambda) (h_2 - 2 \lambda)
 S_{F,L}^{h_1, h_2, h_1 + h_2 - 2,  k}( \lambda) 
 \nonu \\
 && + (h_1 + h_2 - 2 - 2 \lambda) (h_1 - 1 + 2 \lambda)
\nonu \\
&& \times (h_2 - 1 + 2 \lambda) 
S_{B,L}^{h_1, h_2, h_1 + h_2 - 2, k}(\lambda)\bigg) [m + h_1 - 1]_k [n + h_2 - 1]_{1 - k}
\Bigg],
\nonu \\
&& N^{h_1,h_2}_2(m,n)\,  (4\la-1)\,f^{h_1,h_2}_{0,0,0}(\la) \equiv
\nonu \\
&&
\sum_{k=0}^3  (-1)^{h_1 + h_2 + 1} 4^5 
\frac{(2 h_1 + 2h_2 - 9)!}{2 (2 h_1 - 1) (2 h_2 - 1)} \Bigg[
\bigg((h_1 - 2 \lambda) (h_2 - 2 \lambda) 
S_{F,R}^{h_1, h_2,  h_1 + h_2 - 4, k}(\lambda) 
\nonu \\
&&
- (h_1 - 1 + 2 \lambda) (h_2 - 1 + 2 \lambda) 
S_{B,R}^{h_1, h_2, h_1 + h_2 - 4,k}( \lambda)\bigg)
[m + h_1 - 1]_{3 - k} [n + h_2 - 1]_k 
\nonu \\
&& - \bigg(
(h_1 - 2 \lambda) (h_2 - 2 \lambda) S_{F,L}^{h_1, h_2,  h_1 + h_2 - 4,k}(\lambda)
- (h_1 - 1 + 2 \lambda) (h_2 - 1 +  2 \lambda) 
S_{B,L}^{h_1, h_2, h_1 + h_2 - 4,  k}(\lambda)\bigg) 
\nonu \\
&&
\times [m + h_1 - 1]_k [n + h_2 - 1]_{3 - k}\Bigg].
\label{Coeffcoeff}
\eea
Obviously, at $\la =\frac{1}{4}$,
the above (anti)commutators (\ref{wgenericla})
reduce to (\ref{finalanticommcomm}) together with
$f^{h_1,h_2}_{0,0,2}(\la=\frac{1}{4})=1$ in (\ref{Coeffcoeff}).
The Jacobi identity for generic $\la$ is also
satisfied as in Appendix
\ref{jjacobi}. The higher order terms in $q$ appearing in
(\ref{wgenericla}) depend on
the $\la$.


\section{
The extension of ${\cal N}=2$ $SO(2)$
superconformal algebra:
the 
${\cal N}=2$ supersymmetric
$W_{1+\infty}^{K,K}[\la =0]$ algebra}

Finally, in this Appendix,
some details appearing in section $5$ are
presented.

\subsection{The 
${\cal N}=2$ supersymmetric
$W_{1+\infty}^{K,K}[\la =0]$ algebra}

We present the 
${\cal N}=2$ supersymmetric
$W_{1+\infty}^{K,K}[\la =0]$ algebra
from Appendix $D$ in \cite{AK2309}
by using the relations (\ref{simple}) and (\ref{Phih}) 
%
%
\bea
&&
\comm{(\Phi^{(h_1)}_2)_{m}}{(\Phi^{(h_2)}_2)_{n}}
= 
\nonu \\
&& q^{h_1+h_2}\,(-1)^{h_1+h_2}\,K\,\Big(c_{W_F}^{h_1+2,h_2+2}(0)+c_{W_B}^{h_1+2,h_2+2}(0) \Big)[m+h_1+1]_{h_1+h_2+3}\,\delta_{m+n}
\nonu \\
&&
+\frac{1}{2}\sum_{h_3=-1}^{h_1+h_2+1\,} \sum_{k=0}^{\,h_1+h_2-h_3+1}
(4q)^{h_1+h_2-h_3}\,(-1)^{h_1+h_2} (2h_3+3)! \, 
\nonu \\
&& \times \bigg[
\Big((1-\de^{h_3,-1})\, S_{F,R}^{\,\,h_1+2,h_2+2,h_3+2,k}(0)
+ (1+\de^{h_3,-1})\,S_{B,R}^{\,\,h_1+2,h_2+2,h_3+2,k}(0)\Big)
\,\nonu \\
&& \times [m+h_1+1]_{h_1+h_2-h_3-k+1}[n+h_2+1]_k
\nonu \\
&&
-
\Big( (1-\de^{h_3,-1})\,S_{F,L}^{\,\,h_1+2,h_2+2,h_3+2,k}(0)
+ (1+\de^{h_3,-1})\,S_{B,L}^{\,\,h_1+2,h_2+2,h_3+2,k}(0)\Big)
\, \nonu \\
   && \times [m+h_1+1]_{k}[n+h_2+1]_{h_1+h_2-h_3-k+1}
\bigg]\, (\Phi^{(h_3)}_2)_{m+n}
\nonu \\
&&
+\sum_{h_3=0}^{h_1+h_2+2\,} \sum_{k=0}^{\,h_1+h_2-h_3+2}
(4q)^{h_1+h_2-h_3}\,(-1)^{h_1+h_2} (2h_3+1)! \, 
\nonu \\
&& \times \bigg[
\Big(S_{F,R}^{\,\,h_1+2,h_2+2,h_3+1,k}(0)
- S_{B,R}^{\,\,h_1+2,h_2+2,h_3+1,k}(0)\Big)
\,[m+h_1+1]_{h_1+h_2-h_3-k+2}[n+h_2+1]_k
\nonu \\
&&
-
\Big(S_{F,L}^{\,\,h_1+2,h_2+2,h_3+1,k}(0)
- S_{B,L}^{\,\,h_1+2,h_2+2,h_3+1,k}(0)\Big)
\, [m+h_1+1]_{k}[n+h_2+1]_{h_1+h_2-h_3-k+2}
\bigg]\nonu \\
&& \times (\Phi^{(h_3)}_1)_{m+n},
\nonu \\
&& \comm{(\Phi^{(h_1)}_2)_{m}}{(\Phi^{(h_2)}_1)_{n}}
=
\nonu \\
&& q^{h_1+h_2}\,(-1)^{h_1+h_2+1}\,K\,2\,\Big(
(1+\de^{h_2,0})\,c_{W_F}^{h_1+2,h_2+1}(0)-
(1-\de^{h_2,0})\,c_{W_B}^{h_1+2,h_2+1}(0) \Big)
\nonu \\
&& \times      [m+h_1+1]_{h_1+h_2+2}\,\delta_{m+n}
\nonu \\
&&
+\frac{1}{4}\sum_{h_3=-1}^{h_1+h_2\,} \sum_{k=0}^{\,h_1+h_2-h_3}
(4q)^{h_1+h_2-h_3}\,(-1)^{h_1+h_2+1} (2h_3+3)! \, 
\nonu \\
&& \times \bigg[
\Big((1+\de^{h_2,0})\, (1-\de^{h_3,-1})\,S_{F,R}^{\,\,h_1+2,h_2+1,h_3+2,k}(0)
-(1-\de^{h_2,0})\,(1+\de^{h_3,-1})\, S_{B,R}^{\,\,h_1+2,h_2+1,h_3+2,k}(0)\Big)
\nonu \\
      && \times [m+h_1+1]_{h_1+h_2-h_3-k}[n+h_2]_k
\nonu \\
&& 
-
\Big((1+\de^{h_2,0})\,(1-\de^{h_3,-1})\,S_{F,L}^{\,\,h_1+2,h_2+1,h_3+2,k}(0)
- (1-\de^{h_2,0})\,S_{B,L}^{\,\,h_1+2,h_2+1,h_3+2,k}(0)\Big)
\nonu \\
      && \times [m+h_1+1]_{k}[n+h_2]_{h_1+h_2-h_3-k}
\bigg]\, (\Phi^{(h_3)}_2)_{m+n}
\nonu \\
&&
+\frac{1}{2}\sum_{h_3=0}^{h_1+h_2+1\,} \sum_{k=0}^{\,h_1+h_2-h_3+1}
(4q)^{h_1+h_2-h_3}\,(-1)^{h_1+h_2+1} (2h_3+1)! \, 
\nonu \\
&& \times \bigg[
\Big((1+\de^{h_2,0})\,S_{F,R}^{\,\,h_1+2,h_2+1,h_3+1,k}(0)
+ (1-\de^{h_2,0})\,S_{B,R}^{\,\,h_1+2,h_2+1,h_3+1,k}(0)\Big)
\nonu \\
      && \times [m+h_1+1]_{h_1+h_2-h_3-k+1}[n+h_2]_k
\nonu \\
&&
-
\Big( (1+\de^{h_2,0})\,S_{F,L}^{\,\,h_1+2,h_2+1,h_3+1,k}(0)
+ (1-\de^{h_2,0})\,S_{B,L}^{\,\,h_1+2,h_2+1,h_3+1,k}(0)\Big)
\nonu \\
      && \times [m+h_1+1]_{k}[n+h_2]_{h_1+h_2-h_3-k+1}
\bigg]\, (\Phi^{(h_3)}_1)_{m+n},
\nonu \\
&&
\acomm{(\Phi^{(h_1),-}_\frac{3}{2})_{r}}{(\Phi^{(h_2),+}_\frac{3}{2})_{s}}=
q^{h_1+h_2}\,K\,c_{Q \bar{Q}}^{h_2+1,h_1+1}(0)\,[r+h_1+\tfrac{1}{2}]_{h_1+h_2+2}\,\delta_{r+s}
\nonu \\
&&
+\sum_{h_3=-1}^{h_1+h_2\,} \sum_{k=0}^{\,h_1+h_2-h_3}
(4q)^{h_1+h_2-h_3}\,(-1)^{h_1+h_2+1} (2h_3+3)! \, 
\nonu \\
&&
\times \bigg[
(1-\de^{h_3,-1})\,U_{F}^{\,\,h_2+1,h_1+1,h_3+2,k}(0)\,[r+h_1+\tfrac{1}{2}]_{k}[s+h_2+\tfrac{1}{2}]_{h_1+h_2-h_3-k}
\nonu \\
&& 
+(1+\de^{h_3,-1})\,U_{B}^{\,\,h_2+1,h_1+1,h_3+2,k}(0)\,[r+h_1+\tfrac{1}{2}]_{h_1+h_2-h_3-k}[s+h_2+\tfrac{1}{2}]_{k}
\bigg]\,(\Phi^{(h_3)}_2)_{r+s}
\nonu \\
&& +
\sum_{h_3=0}^{h_1+h_2+1\,} \sum_{k=0}^{\,h_1+h_2-h_3+1}
(4q)^{h_1+h_2-h_3}\,2(-1)^{h_1+h_2+1} (2h_3+1)! \, 
\nonu \\
&&
\times \bigg[
U_{F}^{\,\,h_2+1,h_1+1,h_3+1,k}(0)\,[r+h_1+\tfrac{1}{2}]_{k}[s+h_2+\tfrac{1}{2}]_{h_1+h_2-h_3-k+1}
\nonu \\
&& -U_{B}^{\,\,h_2+1,h_1+1,h_3+1,k}(0)\,[r+h_1+\tfrac{1}{2}]_{h_1+h_2-h_3-k+1}[s+h_2+\tfrac{1}{2}]_{k}
\bigg]\,(\Phi^{(h_3)}_1)_{r+s},
\nonu \\
&&\comm{(\Phi^{(h_1)}_2)_{m}}{(\Phi^{(h_2),+}_\frac{3}{2})_{r}}=
\sum_{h_3=0}^{h_1+h_2+1\,} \sum_{k=0}^{\,h_1+h_2-h_3+1}
(4q)^{h_1+h_2-h_3}\,(-1)^{h_1+h_2+1} (2h_3+2)! \, 
\nonu \\
&& \times \bigg[
T_{F}^{\,\,h_1+2,h_2+1,h_3+1,k}(0)\,[m+h_1+1]_{h_1+h_2-h_3-k+1}[r+h_2+\tfrac{1}{2}]_{k}
\nonu \\
&& -T_{B}^{\,\,h_1+2,h_2+1,h_3+1,k}(0)\,[m+h_1+1]_{k}[r+h_2+\tfrac{1}{2}]_{h_1+h_2-h_3-k+1}
\bigg]\,(\Phi^{(h_3),+}_\frac{3}{2})_{m+r} \, ,
\nonu \\
&& \comm{(\Phi^{(h_1)}_2)_{m}}{(\Phi^{(h_2),-}_\frac{3}{2})_{r}}=
\sum_{h_3=-1}^{h_1+h_2+1\,} \sum_{k=0}^{\,h_1+h_2-h_3+1}
(4q)^{h_1+h_2-h_3}\,(-1)^{h_1+h_2+1} (2h_3+2)! \, 
\nonu \\
&& \times \bigg[
\bar{T}_{F}^{\,\,h_1+2,h_2+1,h_3+1,k}(0)\,[m+h_1+1]_{k}[r+h_2+\tfrac{1}{2}]_{h_1+h_2-h_3-k+1}
\nonu \\
&&
-\bar{T}_{B}^{\,\,h_1+2,h_2+1,h_3+1,k}(0)\,[m+h_1+1]_{h_1+h_2-h_3-k+1}[r+h_2+\tfrac{1}{2}]_{k}
\bigg]\,(\Phi^{(h_3),-}_\frac{3}{2})_{m+r} \, ,
\nonu \\
&& \comm{(\Phi^{(h_1)}_1)_{m}}{(\Phi^{(h_2),+}_\frac{3}{2})_{r}}=
\frac{1}{2}\sum_{h_3=0}^{h_1+h_2\,} \sum_{k=0}^{\,h_1+h_2-h_3}
(4q)^{h_1+h_2-h_3}\,(-1)^{h_1+h_2} (2h_3+2)! \, 
\nonu \\
&& \times \bigg[
(1+\de^{h_1,0})\,
  T_{F}^{\,\,h_1+1,h_2+1,h_3+1,k}(0)\,[m+h_1]_{h_1+h_2-h_3-k}[r+h_2+\tfrac{1}{2}]_{k}
\nonu \\
&&
+(1-\de^{h_1,0})\,
T_{B}^{\,\,h_1+1,h_2+1,h_3+1,k}(0)\,[m+h_1]_{k}[r+h_2+\tfrac{1}{2}]_{h_1+h_2-h_3-k}
\bigg]\,(\Phi^{(h_3),+}_\frac{3}{2})_{m+r} \, ,
\nonu \\
&& \comm{(\Phi^{(h_1)}_1)_{m}}{(\Phi^{(h_2),-}_\frac{3}{2})_{r}}=
\frac{1}{2}
\sum_{h_3=-1}^{h_1+h_2\,} \sum_{k=0}^{\,h_1+h_2-h_3}
(4q)^{h_1+h_2-h_3}\,(-1)^{h_1+h_2} (2h_3+2)! \, 
\nonu \\
&& \times \bigg[
(1+\de^{h_1,0})\,\bar{T}_{F}^{\,\,h_1+1,h_2+1,h_3+1,k}(0)\,[m+h_1]_{k}[r+h_2+\tfrac{1}{2}]_{h_1+h_2-h_3-k}
\nonu \\
&&
+(1-\de^{h_1,0})\,\bar{T}_{B}^{\,\,h_1+1,h_2+1,h_3+1,k}(0)\,[m+h_1]_{h_1+h_2-h_3-k}[r+h_2+\tfrac{1}{2}]_{k}
\bigg]\,(\Phi^{(h_3),-}_\frac{3}{2})_{m+r} \, ,
\nonu \\
&&
\comm{(\Phi^{(h_1+1)}_1)_{m}}{(\Phi^{(h_2+1)}_1)_{n}}=q^2\,4\,\comm{(\Phi^{(h_1)}_2)_{m}}{(\Phi^{(h_2)}_2)_{n}}\, , \qquad h_1, h_2 \geq 1 \,,
\nonu \\
&&\comm{(\Phi^{(0)}_1)_{m}}{(\Phi^{(h)}_1)_{n}}=
q^{h}\,(-1)^{h}\,4^2\,K\,c_{W_F}^{1,h+1}(0)\,[m]_{h+1}\,\delta_{m+n}
\nonu \\
&&
+\frac{1}{4}\sum_{h_3=-1}^{h-1\,} \sum_{k=0}^{\,h-h_3-1}
(4q)^{h-h_3}\,(-1)^{h} (2h_3+3)! \, 
\nonu \\
&& \times \bigg[
(1-\delta^{h_3,-1})S_{F,R}^{\,\,1,h+1,h_3+2,k}(0)
\,[m]_{h-h_3-k-1}[n+h]_k
\nonu \\
&&
-
\Big((1-\delta^{h_3,-1})S_{F,L}^{\,\,1,h+1,h_3+2,k}(0)
\,[m]_{k}[n+h]_{h-h_3-k-1}
\bigg]\,(\Phi^{(h_3)}_2)_{m+n}
\nonu \\
&&
+\frac{1}{2}\sum_{h_3=0}^{h\,} \sum_{k=0}^{\,h-h_3}
(4q)^{h-h_3}\,(-1)^{h} (2h_3+1)! \, 
\nonu \\
&& \times \bigg[
S_{F,R}^{\,\,1,h+1,h_3+1,k}(0)
\,[m]_{h-h_3-k}[n+h]_k
-
S_{F,L}^{\,\,1,h+1,h_3+1,k}(0)
\,[m]_{k}[n+h]_{h-h_3-k}
\bigg]\,(\Phi^{(h_3)}_1)_{m+n}\,,
\label{n2appendix}
\eea
where the central terms are
\bea
c_{W_F}(h_1,h_2)& \equiv &  N \sum_{i_1=0}^{h_1-1}\sum_{i_2=0}^{h_2-1}
(-1)^{h_2-1}\,4^{h_1+h_2-4}
a^{i_1}(h_1,\lambda+\tfrac{1}{2})a^{i_2}(h_2,\lambda+\tfrac{1}{2})
\frac{i_1 ! i_2 !}{(i_1+i_2+1)!}\,,\nonumber
\\
c_{W_B}(h_1,h_2)& \equiv & N\,\sum_{i_1=0}^{h_1-1}\sum_{i_2=0}^{h_2-1}
(-1)^{h_2}\,4^{h_1+h_2-4}\,
a^{i_1}(h_1,\lambda)a^{i_2}(h_2,\lambda)
\frac{i_1 !\,i_2 !}{(i_1+i_2+1)!},
\label{threec}
\\
c_{Q\bar{Q}}(h_1,h_2)& \equiv & N\,\sum_{i_1=0}^{h_1-1}\sum_{i_2=0}^{h_2}
2\,(-1)^{h_2+1}\,4^{h_1+h_2-2}
\,\beta^{i_1}(h_1+1,\lambda)
\alpha^{i_2}(h_2+1,\lambda)\frac{i_1 !\,i_2 !}{(i_1+i_2+1)!}.
\nonumber
\eea
The previous relation (\ref{STRUCT}) is used in (\ref{n2appendix}) and
the relations in (\ref{coeff}) are used in (\ref{threec}).


\subsection{The subleading terms up to the $q^2$}

We present the leading and subleading terms from (\ref{n2appendix})
\bea
\comm{(\Phi^{(h_1)}_2)_{m}}{(\Phi^{(h_2)}_2)_{n}}&= &
\tt
-\frac{1}{q}\,\frac{c}{48}\,K\,(m^2-m)\,\delta^{h_1,-1}\delta^{h_2,0}\delta_{m+n}
+\frac{1}{q}\,\frac{c}{48}\,K\,(m^2+m)\,\delta^{h_1,0}\delta^{h_2,-1}\delta_{m+n}
\nonu \\
&-& \tt
\frac{c}{72}\,K\,(m-2)(m-1)m\,\delta^{h_1,-1}\delta^{h_2,1}\delta_{m+n}
\nonu \\
&- & \tt
\frac{c}{72}\,K\,(m+2)(m+1)m\,\delta^{h_1,1}\delta^{h_2,-1}\delta_{m+n}
\nonu \\
&
+& \frac{c}{24}\,K\,(m^3-m)\,\delta^{h_1,0}\delta^{h_2,0}\delta_{m+n}
\nonu \\
& - &
{\tt q\,\frac{c}{90}\,K\, (m-3)(m-2)(m-1)m\,\delta^{h_1,-1}\delta^{h_2,2} \delta_{m+n}  } 
\nonu \\
&
+&
{\tt q\,\frac{c}{90}\,K\, (m+3)(m+2)(m+1)m\,\delta^{h_1,2}\delta^{h_2,-1} \delta_{m+n}  } 
\nonu \\
&
-&
{\tt q^2\,\frac{c}{105}\,K\, (m-4)(m-3)(m-2)(m-1)m\,\delta^{h_1,-1}\delta^{h_2,3} \delta_{m+n}  } 
\nonu \\
&
-&
{\tt q^2\,\frac{c}{105}\,K\, (m+4)(m+3)(m+2)(m+1)m\,\delta^{h_1,3}\delta^{h_2,-1} \delta_{m+n}  } 
\nonu
\\
&
+& q^2\,\frac{c}{27}\,K\, (m-2)(m-1)m(m+1)(m+2)\,\delta^{h_1,1}\delta^{h_2,1} \delta_{m+n} 
\nonu \\
&
+& \Big((h_2+1)m-(h_1+1)n\Big)\,(\Phi^{(h_1+h_2)}_2)_{m+n}
\nonu \\
&
+& q\,\frac{1}{4(2h_1+1)(2h_2+1)(2h_1+2h_2+1)}\,
N^{h_1+2,h_2+2}_{2}(m,n)
\nonu \\
& \times &
\Big( (\Phi_1^{(h_1+h_2-1)})_{m+n}-2q\,\delta^{h_1+h_2-2,-1}\,(\Phi^{(-1)}_2)_{m+n}\Big)
\nonu \\
&
+&
q^2\,\frac{(2h_1+2h_1^2+2h_2+6h_1 h_2+4h_1^2 h_2+2h_2^2+4h_1 h_2^2-1)}{6(2h_1+1)(2h_2+1)(2h_1+2h_2+1)}\,
\nonu \\
& \times & N^{h_1+2,h_2+2}_{2}(m,n)
\,
(\Phi_2^{(h_1+h_2-2)})_{m+n}
\nonu \\
& + & \tt \delta^{h_1,-1}\,m\Big((h_2+1)m+n\Big)\,
\nonu \\
& \times & \tt
\Big((1+\delta^{h_2,1})\,q\, (\Phi^{(h_2-2)}_2)_{m+n}-\frac{1}{2}\,(\Phi^{(h_2-1)}_1)_{m+n}\Big)
\nonu \\
& - & {\tt \delta^{h_2,-1}\,n\Big(m+(h_1+1)n\Big)\,
\Big((1+\delta^{h_1,1})\, q\, (\Phi^{(h_1-2)}_2)_{m+n}-\frac{1}{2}\,(\Phi^{(h_1-1)}_1)_{m+n}\Big)}
\nonu \\
&
-& \tt
\delta^{h_1,-1}\,q^2\,\frac{h_2(h_2-1)}{6(2h_2-3)(2h_2-1)(2h_2+1)}\,N^{1,h_2+2}_{3}(m,n)\,(\Phi^{(h_2-3)}_1)_{m+n}
\nonu \\
&
+& \tt
\delta^{h_2,-1}\,q^2\,\frac{h_1(h_1-1)}{6(2h_1-3)(2h_1-1)(2h_1+1)}\,N^{h_1+2,1}_{3}(m,n)\,(\Phi^{(h_1-3)}_1)_{m+n}
\nonu \\
& + & \mathcal{O}(q^3)\,,
\nonu \\
\comm{(\Phi^{(h_1)}_2)_{m}}{(\Phi^{(h_2)}_1)_{n}}&=&
{\tt
\frac{1}{q}\,\frac{c}{24}\,K\,m\,\delta^{h_1,-1}\delta^{h_2,0}\delta_{m+n}
+\frac{c}{24}\,K\,(m^2-m)\,\delta^{h_1,-1}\delta^{h_2,1}\delta_{m+n}}
\nonu \\
& + & {\tt 
q\,\frac{c}{36}\,K\,(m-2)(m-1)m\,\delta^{h_1,-1}\delta^{h_2,2}\delta_{m+n} }
\nonu \\
& - &
q\,\frac{c}{36}\,K\,(m^2-1)m\,\delta^{h_1,0}\delta^{h_2,1}\delta_{m+n}
\nonu \\
& + &
\tt
q^2\,\frac{c}{45}\,K\,(m-3)(m-2)(m-1)m\,\delta^{h_1,-1}\delta^{h_2,3}\delta_{m+n}
\nonu \\
&
+& \Big(h_2\,m-(h_1+1)n\Big)\,\bigg(
(\Phi^{(h_1+h_2)}_1)_{m+n}
\nonu \\
& - & {\tt 2q\,\delta^{h_1,-1}\delta^{h_2,1}\,(\Phi^{(h_1+h_2-1)}_2)_{m+n}}
+2q\,\delta^{h_2,0}(1-\delta^{h_1,0})\,(\Phi^{(h_1+h_2-1)}_2)_{m+n}
\bigg)
\nonu \\
&
+&
q^2\,\frac{(-2h_1^2-2h_2-2h_1 h_2+4h_1^2 h_2+2h_2^2+4h_1 h_2^2-1)}{6(2h_1+1)(2h_2+1)(2h_1+2h_2-1)}\,
\nonu \\
& \times & N^{h_1+2,h_2+1}_{2}(m,n)
\,
\Big(1-\frac{3\,\delta^{h_2,0}}{2h_1^2+1}\Big)(\Phi_1^{(h_1+h_2-2)})_{m+n}
\nonu \\
& - & \tt  \delta^{h_1,-1}\,2m\Big(h_2\,m+n\Big)\,
\Big((1+\delta^{h_2,2})\, q^2\, (\Phi^{(h_2-3)}_2)_{m+n}-q\,\frac{1}{2}\,(\Phi^{(h_2-2)}_1)_{m+n}\Big)
\nonu \\
& + & \mathcal{O}(q^3)\,,
\nonu \\
\acomm{(\Phi^{(h_1),-}_\frac{3}{2})_{r}}{(\Phi^{(h_2),+}_\frac{3}{2})_{s}}&=&
\tt
-\frac{1}{q}\,\frac{c}{12}\,K\,(r-\tfrac{1}{2})\,\delta^{h_1,-1}\delta^{h_2,0}\delta_{r+s}
-\frac{c}{18}\,K\,(r-\tfrac{3}{2})(r-\tfrac{1}{2})\,\delta^{h_1,-1}\delta^{h_2,1}\delta_{r+s}
\nonu \\
&
+&
\frac{c}{6}\,K\,
(m^2-\tfrac{1}{4})\,\delta^{h_1,0}\delta^{h_2,0}\delta_{r+s}
\nonu \\
& - & \tt
q\frac{2\,c}{45}\,K\,(r-\tfrac{5}{2})(r-\tfrac{3}{2})(r-\tfrac{1}{2})\,\delta^{h_1,-1}\delta^{h_2,2}\delta_{r+s}
\nonu \\
& - & \tt
q^2\frac{4\,c}{105}\,K\,(r-\tfrac{7}{2})(r-\tfrac{5}{2})(r-\tfrac{3}{2})(r-\tfrac{1}{2})\,\delta^{h_1,-1}\delta^{h_2,3}\delta_{r+s}
\nonu \\
&
+&
q^2\frac{4\,c}{27}\,K\,(r^2-\tfrac{9}{4})(r^2-\tfrac{1}{4})\,\delta^{h_1,1}\delta^{h_2,1}\delta_{r+s}
\nonu \\
&
+& 2\,(\Phi^{(h_1+h_2)}_2)_{r+s}
\nonu \\
&-& 2\,\Big(  (h_2+\tfrac{1}{2})r-(h_1+\tfrac{1}{2})s\Big)\,
 \bigg((\Phi^{(h_1+h_2)}_1)_{r+s}
\nonu \\
&
+& q\,(1-
\delta^{h_1+h_2,0}+{\tt 4\delta^{h_1,-1}\delta^{h_2,1}})
\frac{2}{(2h_1+1)(2h_2+1)}\,(\Phi^{(h_1+h_2-1)}_2)_{r+s}
\bigg)
\nonu \\
&
+&
q\,\frac{2(h_1+h_2+1)}{(2h_1+1)(2h_2+1)(2h_1+2h_2+1)}\,N^{h_1+\frac{3}{2},h_2+\frac{3}{2}}_{1}(s,r)\,
\nonu \\
& \times & (\Phi^{(h_1+h_2-1)}_1)_{r+s}
\nonu \\
&
+&
q^2\,\frac{4(h_1+h_1^2+h_2+3h_1 h_2+2h_1^2 h_2+h_2^2+2h_1 h_2^2)}{(2h_1+1)(2h_2+1)(2h_1+2h_2+1)}\,N^{h_1+\frac{3}{2},h_2+\frac{3}{2}}_{1}(s,r)
\nonu \\
&
\times &
(1-\delta^{h_1+h_2,1}+{\tt \tfrac{3}{2}\delta^{h_1,-1}\delta^{h_2,2}})(\Phi^{(h_1+h_2-2)}_2)_{r+s}
\nonu \\
&
+&
q^2\,\frac{2(h_1^2+h_1 h_2+2h_1^2 h_2+h_2^2+2h_1 h_2^2-1)}{3(2h_1+1)(2h_2+1)(2h_1+2h_2-1)}\,N^{h_1+\frac{3}{2},h_2+\frac{3}{2}}_{2}(s,r)\,
\nonu \\
& \times & (\Phi^{(h_1+h_2-2)}_1)_{r+s}
+ \mathcal{O}(q^3)\,,
\nonu \\
\comm{(\Phi^{(h_1)}_2)_{m}}{(\Phi^{(h_2),+}_\frac{3}{2})_{r}}&=&
\Big((h_2+\tfrac{1 {\tt-\delta^{h_1,-1}}}{2})\,m-(h_1+1)r\Big)\,
(\Phi^{(h_1+h_2),+}_\frac{3}{2})_{\,\,\,m+r}
\nonu \\
&
-&
q\,\frac{h_1(1-\tt{\delta^{h_1,-1}h_2} )}{(2h_1+1)(2h_2+1)(2h_1+2h_2+1)}
\,N^{h_1+2,h_2+\frac{3}{2}}_{1}(m,r)\,
\nonu \\
& \times & (\Phi^{(h_1+h_2-1),+}_\frac{3}{2})_{\,\,\,m+r}
\nonu \\
&
+& q^2\,\frac{(h_1^2+h_1 h_2+2h_1^2 h_2+h_2^2+2h_1 h_2^2-1)
\big(1- {\tt \frac{ 3\delta^{h_1,-1}}{2h_2-1}}\big)}{3(2h_1+1)(2h_2+1)(2h_1+2h_2-1)}\,
\nonu \\
& \times & N^{h_1+2,h_2+\frac{3}{2}}_{2}(m,r)
\,
(\Phi^{(h_1+h_2-2),+}_\frac{3}{2})_{m+r}
+ \mathcal{O}(q^ 3)
\,,
\nonu \\
\comm{(\Phi^{(h_1)}_2)_{m}}{(\Phi^{(h_2),-}_\frac{3}{2})_{r}}&=&
\Big((h_2+\tfrac{1\tt{+\delta^{h_1,-1}}}{2})\,m-(h_1+1)r\Big)\,
(\Phi^{(h_1+h_2),-}_\frac{3}{2})_{m+r}
\nonu \\
&
+& q  \,\frac{h_1(1+\tt{\delta^{h_1,-1}h_2} )}{(2h_1+1)(2h_2+1)(2h_1+2h_2+1)}
\,N^{h_1+2,h_2+\frac{3}{2}}_{1}(m,r)\,
\nonu \\
& \times & (\Phi^{(h_1+h_2-1),-}_\frac{3}{2})_{m+r}
\nonu \\
&+ &
q^2\,\frac{(h_1^2+h_1 h_2+2h_1^2 h_2+h_2^2+2h_1 h_2^2-1)
\big(1+ {\tt \frac{ 3\delta^{h_1,-1}}{2h_2-1}}\big)}{3(2h_1+1)(2h_2+1)(2h_1+2h_2-1)}\,
\nonu \\
& \times & N^{h_1+2,h_2+\frac{3}{2}}_{2}(m,r)
\,
(\Phi^{(h_1+h_2-2),-}_\frac{3}{2})_{m+r}
+ \mathcal{O}(q^3)\,,
\nonu \\
\comm{(\Phi^{(h_1)}_1)_{m}}{(\Phi^{(h_2),+}_\frac{3}{2})_{r}}&=&
(\Phi^{(h_1+h_2),+}_\frac{3}{2})_{m+r}
-q\,\Big(\frac{2}{(2h_2+1)(2h_1+2h_2+1)}-2\,\delta^{h_1,0}\Big)
\nonu \\
& \times &
\Big((h_2+\tfrac{1}{2})m-h_1\,r\Big)\,(\Phi^{(h_1+h_2-1),+}_\frac{3}{2})_{m+r}
\nonu \\
&
+& q^2\,
\frac{2(-h_1+h_1^2-h_1 h_2+2h_1^2 h_2-h_2^2+2h_1 h_2^2)
\big(1-  \frac{ \delta^{h_1,0}}{h_2^2+\delta^{h_2,0}}\big)}{(2h_1-1)(2h_2+1)(2h_1+2h_2-1)}
\nonu \\
&\times& 
N^{h_1+1,h_2+\frac{3}{2}}_{1}(m,r)\,(\Phi^{(h_1+h_2-2),+}_\frac{3}{2})_{m+r}
+\mathcal{O}(q^3)\,,
\nonu \\
\comm{(\Phi^{(h_1)}_1)_{m}}{(\Phi^{(h_2),-}_\frac{3}{2})_{r}}&=&
-(\Phi^{(h_1+h_2),-}_\frac{3}{2})_{m+r}
-q\,\Big(\frac{2}{(2h_2+1)(2h_1+2h_2+1)}-2\,\delta^{h_1,0}\Big)
\nonu \\
& 
\times &
\Big((h_2+\tfrac{1}{2})m-h_1\,r\Big)\,(\Phi^{(h_1+h_2-1),-}_\frac{3}{2})_{m+r}
\nonu \\
&
-& q^2\,
\frac{2(-h_1+h_1^2-h_1 h_2+2h_1^2 h_2-h_2^2+2h_1 h_2^2)
\big(1-  \frac{ \delta^{h_1,0}}{h_2^2+\delta^{h_2,0}}\big)}{(2h_1-1)(2h_2+1)(2h_1+2h_2-1)}
\nonu \\
&\times&
N^{h_1+1,h_2+\frac{3}{2}}_{1}(m,r)\,(\Phi^{(h_1+h_2-2),-}_\frac{3}{2})_{m+r}+\mathcal{O}(q^3)\,,
\nonu \\
\comm{(\Phi^{(h_1)}_1)_{m}}{(\Phi^{(h_2)}_1)_{n}}&
 = & \frac{c}{6}\,K\,m\,\delta^{h_1,0}\delta^{h_2,0}\, \delta_{m+n}
+q^2\,\frac{c}{6}\,K\,(m^3-m)\,\delta^{h_1,1}\delta^{h_2,1}\delta_{m+n}
\nonu \\
&
+& q \,\delta^{h_1,0}\,2h_2\,m\,(\Phi^{(h_2-1)}_1)_{m+n}
-q\,\delta^{h_2,0}\,2h_1\,n\,(\Phi^{(h_1-1)}_1)_{m+n}
\nonu \\
&
+& q^2\,4\Big(h_2\,m-h_1\,n\Big)\,(1-\delta^{h_1+h_2,1})
\,(\Phi^{(h_1+h_2-2)}_2)_{m+n}
+\mathcal{O}(q^3)\,.
\label{subleading}
\eea
The newly obtained results
can be denoted by typewriter fonts
\footnote{When either $\Phi_2^{(h)}$ or
$\Phi_{\frac{3}{2}}^{(h),-}$ with $h=-1$ occurs on the left hand sides
of the (anti)commutators, the expressions having ${\tt \de^{h_1,-1}}$ or
${\tt \de^{h_2,-1}}$ on the right hand sides
are denoted by using the typewriter fonts.}.
Note that
the structure constants depend on
the $h_1$ and $h_2$ as well as the mode dependent
function $N$ defined in (\ref{Nphi}).

\subsection{The Jacobi identity}

We present the Jacobi identity by looking at
the previous results in (\ref{subleading}).
In (\ref{genn2case}), the $q$ dependent term appears in the last
commutator relation. We would like to check whether
the Jacobi identity for (\ref{genn2case}) is satisfied or not.
If we substitute the condition $q \neq 0$ into these
relations, the Jacobi identity is not satisfied.
This implies that we need to consider other 
$q$ terms on the right hand sides in (\ref{genn2case}).
Let us consider the following Jacobi identity
having the last commutator of (\ref{genn2case})
\bea
&&\comm{(\Phi^{(h_1)}_1)_{m}}{\comm{(\Phi^{(h_2)}_1)_{n}}{(\Phi^{(h_3),+}_\frac{3}{2})_{r}}}
-\comm{(\Phi^{(h_2)}_1)_{n}}{\comm{(\Phi^{(h_1)}_1)_{m}}{(\Phi^{(h_3),+}_\frac{3}{2})_{r}}}
\nonu \\
&& +
\comm{(\Phi^{(h_3),+}_\frac{3}{2})_{r}}{\comm{(\Phi^{(h_1)}_1)_{m}}{(\Phi^{(h_2)}_1)_{n}}}\, ,
\label{Jacobiexp}
\eea
which should vanish.
The first term of 
(\ref{Jacobiexp}), by using the (anti)commutators in
(\ref{subleading}),  can be written as
\bea
&&
\comm{(\Phi^{(h_1)}_1)_{m}}{\comm{(\Phi^{(h_2)}_1)_{n}}{(\Phi^{(h_3),+}_\frac{3}{2})_{r}}}
=  (\Phi_\frac{3}{2}^{(h_1 + h_2 + h_3), +})_{m+n+r} 
\nonu \\
&&
- q\,\Bigg(\frac{(1 + 2 h_3) (m + n)  - 
       2 (h_1 + h_2) r}{(1 + 2 h_3) (1 + 2 h_1 + 2 h_2 + 2 h_3)} - 
      \Big((1 + 2 h_2 + 2 h_3) m - 2 h_1 (n + r)\Big) \delta^{h_1,0}
\nonu \\
&&- (n + 2 h_3 n - 2 h_2 r) \delta^{h_2,0}\Bigg) 
      \,(\Phi_\frac{3}{2}^{(h_1 + h_2 + h_3 - 1),+})_{m+n+r}
\nonu \\
&&   + q^2 \Bigg(
(\Big(\tfrac{1}{2} + h_3) n - h_2 r\Big)
\Big((h_2 + h_3-\tfrac{1}{2}) m - h_1 (n + r)\Big)
\nonu \\
&&
\times
\bigg(
\frac{2}{(-1 + 2 h_2 + 2 h_3) (-1 + 2 h_1 + 2 h_2 + 2 h_3)}
-2 \delta^{h_1,0}\bigg)
\bigg(\frac{2}{(1 + 2 h_3) (1 + 2 h_2 + 2 h_3)} -2 \delta^{h_2,0}\bigg)
\nonu \\
&&+ 2 (-h_1 + h_1^2 - h_1 h_2 + 2 h_1^2 h_2 - 
h_2^2 + 2 h_1 h_2^2 - h_1 h_3 \nonu \\
&& + 2 h_1^2 h_3 - 2 h_2 h_3 + 
             4 h_1 h_2 h_3 - h_3^2 + 2 h_1 h_3^2)
\nonu \\
&&\bigg(1 - \frac{\delta^{h_1,0}}{(h_2 + h_3)^2 +\delta^{h_2 + h_3,0}} \bigg)
\frac{N^{h_1+1, h_2 + h_3+\frac{3}{2}}_1(m,n + r)}{
(-1 + 2 h_1) (1 + 2 h_2 + 2 h_3) (-1 + 2 h_1 + 2 h_2 + 2 h_3)}
\nonu \\
&&+\frac{2(-h_2 + h_2^2 - h_2 h_3 + 2 h_2^2 h_3 - h_3^2 + 2 h_2 h_3^2)}{(-1 + 2 h_2) (1 + 2 h_3) (-1 + 2 h_2 + 2 h_3)}
            \Big(1 - \frac{\delta^{h_2,0}}{h_3^2 + \delta^{h_3,0}}\Big)
            \, N^{h_2+1, h_3+\frac{3}{2}}_1(n,r) 
            \Bigg)\,
            \nonu \\
&& \times (\Phi_\frac{3}{2}^{h_1 + h_2 + h_3 - 2, +})_{m+n+r} 
\nonu \\
&&
-q^3\,\frac{2}{(-1 + 2 h_2 + 2 h_3)}
    \Bigg(
  (-1 + 2 h_1 - h_1^2 + 2 h_2 - 5 h_1 h_2 + 2 h_1^2 h_2 - h_2^2 + 
           2 h_1 h_2^2 + 2 h_3 - 5 h_1 h_3 
\nonu \\
&& + 2 h_1^2 h_3 - 2 h_2 h_3 + 4 h_1 h_2 h_3 - h_3^2 + 2 h_1 h_3^2) 
\Big((h_3+\tfrac{1}{2}) n - h_2 r \Big) \nonu \\
&& \times
\Big(\frac{2}{(1 + 2 h_3) (1 + 2 h_2 + 2 h_3)}-2 \delta^{h_2,0}\Big) 
\nonu \\
&& 
\times
\Big(1 - \frac{\delta^{h_1,0}}{(-1 + h_2 + h_3)^2+\delta^{-1 + h_2 + h_3,0}} \Big)
\frac{1}{(-1 + 2 h_1) (-3 + 2 h_1 + 2 h_2 + 2 h_3)}
\nonu \\
&& \times N^{h_1+1, h_2 + h_3+\frac{1}{2}}_1(m,n + r)
\nonu \\
&&
+ (-h_2 + h_2^2 - h_2 h_3 + 2 h_2^2 h_3 - h_3^2 + 2 h_2 h_3^2)
   \Big(( h_2 + h_3-\tfrac{3}{2}) m - h_1 (n + r)\Big) 
\nonu \\
&&
\times
\Big(\frac{2}{(-3 + 2 h_2 + 2 h_3) (-3 + 2 h_1 + 2 h_2 + 2 h_3)} - 2 \delta^{h_1,0}\Big) 
 \Big(1 - \frac{\delta^{h_2,0}}{h_3^2 + \delta^{h_3,0}}\Big)
 \frac{1}{(-1 + 2 h_2) (1 + 2 h_3)}
\nonu \\
&&
\times
N^{h_2+1,h_3+\frac{3}{2}}_1(n,r) \Bigg)
\,(\Phi_\frac{3}{2}^{(h_1 + h_2 + h_3 -3), +})_{m+n+r}
+
q^4\,\Bigg(
\nonu \\
&& 
\frac{4 (-h_2 + h_2^2 - h_2 h_3 + 2 h_2^2 h_3 - h_3^2 +  2 h_2 h_3^2)}{(-1 + 2 h_1) (-1 + 2 h_2) (1 + 2 h_3) (-3 + 2 h_2 + 2 h_3) (-1 + 2 h_2 + 2 h_3) (-5 + 2 h_1 + 2 h_2 + 2 h_3)}
\nonu \\
&&
\times
(-4 + 9 h_1 - 3 h_1^2 + 4 h_2 - 9 h_1 h_2 + 2 h_1^2 h_2 - 
        h_2^2 + 2 h_1 h_2^2 + 4 h_3 - 9 h_1 h_3 + 2 h_1^2 h_3 - 2 h_2 h_3
\nonu \\
&&+ 4 h_1 h_2 h_3 - h_3^2 + 2 h_1 h_3^2) 
\Big(1 - \frac{\delta^{h_2,0}}{h_3^2 + \delta^{h_3,0}}\Big)
\Big(1 -\frac{\delta^{h_1,0}}{(-2 + h_2 + h_3)^2+ \delta^{h_2 + h_3-2,0}} \Big)
\nonu \\
&&
N^{h_1+, h_2 + h_3-\frac{1}{2}}_1(m, n + r) N^{h_2+1, h_3+\frac{3}{2}}_1(n, r)
\Bigg) \,(\Phi_\frac{3}{2}^{(h_1 + h_2 + h_3 - 4), +})_{m+n+r} \, .
\label{firstterm}
\eea
The second term of (\ref{Jacobiexp})
can be obtained by taking
$h_1 \leftrightarrow h_2$ and $m \leftrightarrow n$
in (\ref{firstterm}).
The third term of 
(\ref{Jacobiexp}) can be written as
\bea
&&
\comm{(\Phi^{(h_3),+}_\frac{3}{2})_{r}}{\comm{(\Phi^{(h_1)}_1)_{m}}{(\Phi^{(h_2)}_1)_{n}}}
= 
\nonu \\
&&
q^2 \,2 \,
\Bigg(
\delta^{h_1,0}\,
h_2 m \Big((h_3+\tfrac{1}{2}) (m + n) + r - h_2 r\Big) 
\Big(\frac{2}{(1 + 2 h_3) (-1 + 2 h_2 + 2 h_3)}-2\, \delta^{h_2-1,0}\Big)
\nonu \\
&&
- \delta^{h_2,0}\,  h_1 n \Big(( h_3+\tfrac{1}{2}) (m + n) + r - h_1 r \Big)
\Big(\frac{2}{(1 + 2 h_3) (-1 + 2 h_1 + 2 h_3)} - 2 \delta^{h_1-1,0}\Big)
\nonu \\
&&
+ 2 (h_2 m - h_1 n)
\Big(( h_1 + h_2-1) r -\tfrac{1}{2}(m + n) (1 + 2 h_3 - \delta^{h_1 + h_2-2,-1})\Big)
(1 -\delta^{h_1+h_2,1}) 
\Bigg)
\nonu \\
&&\times (\Phi_\frac{3}{2}^{(h_1 + h_2 + h_3 -2), +})_{m+n+r}
\nonu \\
&&
+ q^3
\Bigg(
\frac{(1 - \delta^{h_1+h_2,1})(1-\delta^{h_1 + h_2 - 2, -1} h_3)\, 4(h_2 m - h_1 n) (h_1 + h_2 - 2)}{(-3 + 2 h_1 + 2 h_2) (1 + 2 h_3) (-3 + 2 h_1 + 2 h_2 + 2 h_3)}\,
       N^{h_1 + h_2, h_3 + \frac{3}{2}}_1(m + n, r)
\Bigg)
\nonu \\
&&\times(\Phi_\frac{3}{2}^{(h_1 + h_2 + h_3 - 3), +})_{m+n+r} 
\nonu \\
&&-q^4\,
\Bigg(\frac{(1-\delta^{h_1 + h_2, 1}) 4 (h_2 m -  h_1 n)}{3 (-3 + 2 h_1 + 2 h_2)(1 + 2 h_3)(-5 + 2 h_1 + 2 h_2 + 2 h_3)}
(3 - 4 h_1 + h_1^2 - 4 h_2 + 2 h_1 h_2 + h_2^2 
\nonu \\
&&
+ 6 h_3 - 7 h_1 h_3 + 2 h_1^2 h_3 - 7 h_2 h_3 + 4 h_1 h_2 h_3 + 2 h_2^2 h_3 - 
3 h_3^2 + 2 h_1 h_3^2 + 2 h_2 h_3^2)
\nonu \\
&& \times 
\Big(1 - \frac{3 \delta^{h_1 + h_2 - 2,-1}}{(2 h_3 - 1)}\Big)
\,
N^{h_1 + h_2,h_3 + \frac{3}{2}}_2(m + n, r) 
\Bigg)
\,(\Phi_\frac{3}{2}^{(h_1 + h_2 + h_3 - 4), +})_{m+n+r}
\nonu \\
&&
- q \,\delta^{h_1,0} \,2 h_2\, m \, (\Phi_\frac{3}{2}^{(h_2 + h_3 - 1),+})_{m+n+r} 
+ q \,\delta^{h_2,0}\, 2 h_1\, n \,(\Phi_\frac{3}{2}^{(h_1 + h_3 - 1), +})_{m+n+r}
\nonu \\
&&
- q^3\,\Bigg(
\frac{\delta^{h_1,0}\, 4 h_2\, m \,
(2 - 3 h_2 + h_2^2 + 3 h_3 - 5 h_2 h_3 + 2 h_2^2 h_3 - 3 h_3^2 + 2 h_2 h_3^2) }{(-3 + 2 h_2) (1 + 2 h_3) (-3 + 2 h_2 + 2 h_3)}
\Big(1 - \frac{\delta^{h_2 - 1,0}}{h_3^2 + \delta^{h_3, 0}}\Big)
\nonu \\
&&
\times N^{h_2, h_3 + \frac{3}{2}}_1(m + n, r)
\Bigg)
\,(\Phi_\frac{3}{2}^{(h_2 + h_3 - 3), +})_{m+n+r} 
\nonu \\
&&+q^3
\Bigg(
\frac{\delta^{h_2,0} \,4 h_1\, n \, (2 - 3 h_1 + h_1^2 + 3 h_3 - 5 h_1 h_3 +  2 h_1^2 h_3 - 3 h_3^2 + 2 h_1 h_3^2) }{(-3 + 2 h_1) (1 + 2 h_3) (-3 + 2 h_1 + 2 h_3)}
\Big(1 - \frac{\delta^{h_1 - 1,0}}{h_3^2 + \delta^{h_3, 0}}\Big)
\nonu \\
&&
\times N^{h_1, h_3 + \frac{3}{2}}_1(m + n, r)
\Bigg)
\, (\Phi_\frac{3}{2}^{(h1 + h3 - 3), +})_{m+n+r} \, .
\label{thirdterm}
\eea
Then we obtain
the final expression for the Jacobi identity in (\ref{Jacobiexp})
by adding (\ref{firstterm}), (\ref{firstterm}) where
$h_1$ replaced by $h_2$ and $m$ replaced by $n$
and vice versa and (\ref{thirdterm}).
It turns out that
there are no $q$ independent terms,
$q$ terms and $q^2$ terms.
Therefore, the coefficient of $q^2$ terms
is identically zero and up to this order
the Jacobi identity is satisfied.
However, there are $q^3$ terms and $q^4$ terms.
By ignoring these terms (and recalling that
$q$ is a small quantity), the Jacobi identity is satisfied.
We expect that by considering more terms in (\ref{subleading}),
the above nonzero higher order terms in $q$ can be checked explicitly
and the coefficients of these terms vanish.


\subsection{The 
${\cal N}=2$ supersymmetric
$W_{1+\infty}^{K,K}[\la =0]$ algebra in terms of the known structure constants}

For
\bea
h \geq 0 \qquad \mbox{in} \,\, \Phi^{(h)}_{1,2},
\qquad
h \geq 0 \qquad \mbox{in} \,\, \Phi^{(h)}_{\frac{3}{2}},
\label{COND1}
\eea
appearing on the left hand sides of
(\ref{n2appendix}) we can write down (\ref{n2appendix}) as 
\bea
&&
\comm{(\Phi^{(h_1)}_2)_{m}}{(\Phi^{(h_2)}_2)_{n}}
=
q^{h_1+h_2}\,\delta^{h_1,h_2}\frac{K\,c}{6}\,\big(c_{W_F}^{h_1+2}+c_{W_B}^{h_1+2}\big)[m+h_1+1]_{2h_1+3}\,\delta_{m+n}
\nonu \\
&&
+\sum_{h_3=0}^{h_1+h_2+1} 
q^{h_1+h_2-h_3}\,\frac{\big(1+(-1)^{h_1+h_2+h_3} \big)}{4}
\, \Big(p_{F,\,h_1+h_2-h_3}^{h_1+2,h_2+2}(m,n)+p_{B,\,h_1+h_2-h_3}^{h_1+2,h_2+2}(m,n)
\Big)\,(\Phi^{(h_3)}_2)_{m+n}
\nonu \\
&&
+\sum_{h_3=0}^{h_1+h_2+2} 
q^{h_1+h_2-h_3}\,\frac{\big(1-(-1)^{h_1+h_2+h_3} \big)}{8}
\nonu \\
&& \times
\Big(p_{F,\,h_1+h_2-h_3+1}^{h_1+2,h_2+2}(m,n)-p_{B,\,h_1+h_2-h_3+1}^{h_1+2,h_2+2}(m,n)
\Big)\,(\Phi^{(h_3)}_1)_{m+n}\,,
\nonu \\
&& \comm{(\Phi^{(h_1)}_2)_{m}}{(\Phi^{(h_2)}_1)_{n}}
=
q^{h_1+h_2}\,\delta^{h_1,h_2-1}\frac{K\,c}{3}\,\big(c_{W_F}^{h_1+2}-c_{W_B}^{h_1+2}\big)[m+h_1+1]_{2h_1+3}\,\delta_{m+n}
\nonu \\
&&
+\sum_{h_3=0}^{h_1+h_2} 
q^{h_1+h_2-h_3}\,\frac{\big(1-(-1)^{h_1+h_2+h_3} \big)}{2}
\nonu \\
&& \times
\Big( (1+ \de^{h_2,0})\,
p_{F,\,h_1+h_2-h_3-1}^{h_1+2,h_2+1}(m,n)-
(1- \de^{h_2,0})\,p_{B,\,h_1+h_2-h_3-1}^{h_1+2,h_2+1}(m,n)
\Big)\,(\Phi^{(h_3)}_2)_{m+n}
\nonu \\
&&
+\sum_{h_3=0}^{h_1+h_2+1} 
q^{h_1+h_2-h_3}\,\frac{\big(1+(-1)^{h_1+h_2+h_3} \big)}{4}
\nonu \\
&& \times \Big((1+ \de^{h_2,0})\,p_{F,\,h_1+h_2-h_3}^{h_1+2,h_2+1}(m,n)+
(1- \de^{h_2,0})\,p_{B,\,h_1+h_2-h_3}^{h_1+2,h_2+1}(m,n)
\Big)\,(\Phi^{(h_3)}_1)_{m+n}\,,
\nonu \\
&& \acomm{(\Phi^{(h_1),-}_\frac{3}{2})_{r}}{(\Phi^{(h_2),+}_\frac{3}{2})_{s}}=
q^{h_1+h_2}\,\delta^{h_1 h_2}\frac{K\,c}{6}\,c_{Q}^{
h_1+1}\,[r+h_1+\tfrac{1}{2}]_{2(h_1+1)}\,\delta_{r+s}
\nonu \\
&&
+\sum_{h_3=0}^{h_1+h_2}
q^{h_1+h_2-h_3}\,\frac{(-1)^{h_1+h_2+h_3}}{2}
 \Big(
o_{F,\,h_1+h_2-h_3}^{h_1+\frac{3}{2},h_2+\frac{3}{2}}(s,r)
+o_{B,\,h_1+h_2-h_3}^{h_1+\frac{3}{2},h_2+\frac{3}{2}}(s,r)
\Big)\,(\Phi^{(h_3)}_2)_{r+s}
\nonu \\
&& +
\sum_{h_3=0}^{h_1+h_2+1} 
q^{h_1+h_2-h_3}\,\frac{(-1)^{h_1+h_2+h_3+1}}{4}  \, 
\Big(
o_{F,\,h_1+h_2-h_3+1}^{h_1+\frac{3}{2},h_2+\frac{3}{2}}(s,r)
-o_{B,\,h_1+h_2-h_3+1}^{h_1+\frac{3}{2},h_2+\frac{3}{2}}(s,r)
\Big)\,(\Phi^{(h_3)}_1)_{r+s} \, ,
\nonu \\
&& \comm{(\Phi^{(h_1)}_2)_{m}}{(\Phi^{(h_2),+}_\frac{3}{2})_{r}}=
\sum_{h_3=0}^{h_1+h_2+1} 
q^{h_1+h_2-h_3}\,(-1)^{h_1+h_2+h_3} 
\nonu \\
&& \times
\Big(q_{F,\,h_1+h_2-h_3}^{h_1+2,h_2+\frac{3}{2}}(m,r)+q_{B,\,h_1+h_2-h_3}^{h_1+2,h_2+\frac{3}{2}}(m,r)\Big) \times (\Phi^{(h_3),+}_\frac{3}{2})_{m+r} \, ,
\nonu \\
&& \comm{(\Phi^{(h_1)}_2)_{m}}{(\Phi^{(h_2),-}_\frac{3}{2})_{r}}=
\sum_{h_3=0}^{h_1+h_2+1} 
q^{h_1+h_2-h_3}\nonu \\
&& \times
\Big(q_{F,\,h_1+h_2-h_3}^{h_1+2,h_2+\frac{3}{2}}(m,r)+q_{B,\,h_1+h_2-h_3}^{h_1+2,h_2+\frac{3}{2}}(m,r)\Big)\,(\Phi^{(h_3),-}_\frac{3}{2})_{m+r} \, ,
\nonu \\
&& \comm{(\Phi^{(h_1)}_1)_{m}}{(\Phi^{(h_2),+}_\frac{3}{2})_{r}}=
\sum_{h_3=0}^{h_1+h_2} 
q^{h_1+h_2-h_3}\,2(-1)^{h_1+h_2+h_3+1}\nonu \\
&& \times
\Big((1+ \de^{h_2,0})\,q_{F,\,h_1+h_2-h_3-1}^{h_1+1,h_2+\frac{3}{2}}(m,r)-
(1+ \de^{h_2,0})\,q_{B,\,h_1+h_2-h_3-1}^{h_1+1,h_2+\frac{3}{2}}(m,r)\Big)\,(\Phi^{(h_3),+}_\frac{3}{2})_{m+r} \, ,
\nonu \\
&& \comm{(\Phi^{(h_1)}_1)_{m}}{(\Phi^{(h_2),-}_\frac{3}{2})_{r}}=
\sum_{h_3=0}^{h_1+h_2} 
q^{h_1+h_2-h_3}\nonu \\
&& \times
2\,
\Big((1+ \de^{h_1,0})\,q_{F,\,h_1+h_2-h_3-1}^{h_1+1,h_2+\frac{3}{2}}(m,r)-
(1+ \de^{h_1,0})\,q_{B,\,h_1+h_2-h_3-1}^{h_1+1,h_2+\frac{3}{2}}(m,r)\Big)\,(\Phi^{(h_3),-}_\frac{3}{2})_{m+r} \, ,
\nonu \\
&& \comm{(\Phi^{(h_1)}_1)_{m}}{(\Phi^{(h_2)}_1)_{n}}
=
q^{h_1+h_2}\,\delta^{h_1,h_2}\frac{2\,K\,c}{3}\,\big(c_{W_F}^{h_1+1}+c_{W_B}^{h_1+1}\big)[m+h_1]_{2h_1+1}\,\delta_{m+n}
\nonu \\
&&
+\sum_{h_3=0}^{h_1+h_2-1} 
q^{h_1+h_2-h_3}\,\big(1+(-1)^{h_1+h_2+h_3} \big)
\nonu \\
&&
\times \Big(
(1+\delta^{h_1,0})(1+\delta^{h_2,0})p_{F,\,h_1+h_2-h_3-2}^{h_1+1,h_2+1}(m,n)
+
(1-\delta^{h_1,0})(1-\delta^{h_2,0})p_{B,\,h_1+h_2-h_3-2}^{h_1+1,h_2+1}(m,n)
\Big)\nonu \\
&& \times (\Phi^{(h_3)}_2)_{m+n}
\nonu \\
&&
+\sum_{h_3=0}^{h_1+h_2} 
q^{h_1+h_2-h_3}\,\frac{\big(1-(-1)^{h_1+h_2+h_3} \big)}{2}
\nonu \\
&&
\times
\Big(
(1+\delta^{h_1,0})(1+\delta^{h_2,0})p_{F,\,h_1+h_2-h_3-1}^{h_1+1,h_2+1}(m,n)
-
(1-\delta^{h_1,0})(1-\delta^{h_2,0})p_{B,\,h_1+h_2-h_3-1}^{h_1+1,h_2+1}(m,n)
\Big)\nonu \\
&& \times (\Phi^{(h_3)}_1)_{m+n}\,,
\label{simpleexp}
\eea
where the central terms are given by
\footnote{We also have $c_{W_{F}}^{h=1}=c_{W_{B}}^{h=1} =\frac{1}{8}$. }.
\bea
c_{W_{F}}^h & = & \frac{2^{2(h-3)}\big((h-1)!\big)^2}{(2h-3)!!(2h-1)!!}\,,\quad
c_{W_{B}}^h=\frac{2^{2(h-3)}(h-2)!h!}{(2h-3)!!(2h-1)!!}
\,,\quad
c_{Q}^h=\frac{2^{2(h-1)}(h-1)!h!}{\big((2h-1)!!\big)^2}\,,
\label{cthree}
\eea
and
\bea
c=6\,N\,.
\label{central6N}
\eea
Note that the number of free fields having $SU(N)$
fundamental (or antifundamental) indices
for $(\beta,\ga)$ and $(b,c)$ system in
(\ref{fundOPE}) is given by $N$.
This $N$ dependence appears in (\ref{central6N}).
In (\ref{cthree}), the previous relations in (\ref{threec})
are used.
The mode dependent functions appearing on the
right hand sides in (\ref{simpleexp})
are given by as follows:
\bea
p_{\mathrm{F}}^{h_1h_2 h}(m,n)
&
=&\frac{1}{2(h+1)!}\,\phi^{h_1,h_2}_{h}(0,\textstyle{-\frac{1}{2}})
\,N^{h_1,h_2}_{h}(m,n),
\nonu\\
p_{\mathrm{B}}^{h_1h_2 h}(m,n)
&
=&\frac{1}{2(h+1)!}\,\phi^{h_1,h_2}_{h}(0,0)
\,N^{h_1,h_2}_{h}(m,n),
\nonu\\
q_{\mathrm{F}}^{h_1h_2 h}(m,r) 
&
=&\frac{(-1)^h}{4(h+2)!}\Big(
(h_1-1)\,\phi^{h_1,h_2+\frac{1}{2}}_{h+1}(0,0)
\nonu \\
& - & (h_1-h-3)\,\phi^{h_1,h_2+\frac{1}{2}}_{h+1}(0,\textstyle{-\frac{1}{2}})
\Big)
\,N^{h_1,h_2}_{h}(m,n),
\nonu\\
q_{\mathrm{B}}^{h_1h_2 h}(m,r) 
&
=&\frac{-1}{4(h+2)!}\Big(
(h_1-h-2)\,\phi^{h_1,h_2+\frac{1}{2}}_{h+1}(0,0)-(h_1)\,\phi^{h_1,h_2+\frac{1}{2}}_{h+1}(0,\textstyle{-\frac{1}{2}})
\Big)
\,N^{h_1,h_2}_{h}(m,n),
\nonu\\
o_{\mathrm{F}}^{h_1h_2h}(r,s) 
&
=&\frac{4(-1)^h}{h!}\Big(
(h_1+h_2-1-h)\,\phi^{h_1+\frac{1}{2},h_2+\frac{1}{2}}_{h}(
\textstyle{\frac{1}{2}},\textstyle{-\frac{1}{4}})
\nonu \\
& - & (h_1+h_2-\frac{3}{2}-h)\,\phi^{h_1+\frac{1}{2},h_2+\frac{1}{2}}_{h+1}(
\textstyle{\frac{1}{2}},\textstyle{-\frac{1}{4}})
\Big)\,  N^{h_1,h_2}_{h-1}(m,n),
\nonu\\
o_{\mathrm{B}}^{h_1h_2h}(r,s) 
&
=&-\frac{4}{h!}\Big(
(h_1+h_2-2-h)\,\phi^{h_1+\frac{1}{2},h_2+\frac{1}{2}}_{h}(
\textstyle{\frac{1}{2}},\textstyle{-\frac{1}{4}})
\nonu \\
& - &
(h_1+h_2-\frac{3}{2}-h)\,\phi^{h_1+\frac{1}{2},h_2+\frac{1}{2}}_{h+1}(\textstyle{\frac{1}{2}},\textstyle{-\frac{1}{4}})
\Big)
 \,  N^{h_1,h_2}_{h-1}(m,n).
\label{modedependence}
\eea
Furthermore, let us introduce
the following quantities appearing in (\ref{modedependence})
\bea
N^{h_1,h_2}_{h}(m,n)
&
=&
\sum_{l=0 }^{h+1}(-1)^l
\left(\begin{array}{c}
h+1 \\  l \\
\end{array}\right)
[h_1-1+m]_{h+1-l}[h_1-1-m]_l
\nonu \\
      & \times & [h_2-1+n]_l [h_2-1-n]_{h+1-l},
\nonu\\
\phi^{h_1,h_2}_{h}(x,y)
&
= &
{}_4 F_3
 \Bigg[
\begin{array}{c}
  -\frac{1}{2}-x-2y, \frac{3}{2}-x+2y, -\frac{h+1}{2}+x,
  -\frac{h}{2} +x \\
-h_1+\frac{3}{2},-h_2+\frac{3}{2},h_1+h_2-h-\frac{3}{2}
\end{array} ; 1
  \Bigg].
\label{Nphi}
\eea
The generalized hypergeometric function, 
with $4$ upper arguments $a_i$,  $3$ lower arguments $b_i$
and variable $z$, is 
defined as the series
\bea
 {}_4 F_3
 \Bigg[
\begin{array}{c}
a_1, a_2, a_3, a_4 \\
b_1,b_2,b_3
\end{array} ; z
  \Bigg] 
=
\sum_{n=0 }^{\infty}
\frac{(a_1)_n (a_2)_n (a_3)_n (a_4)_n}
{(b_1)_n (b_2)_n (b_3)_n}
\frac{z^n}{n!}\,.
\label{defF43}
\eea
Because there is a relation between
the arguments and the variable, $b_1+b_2+b_3 = a_1+a_2+a_3+a_4
+z$ for the $\phi^{h_1,h_2}_{h}(x,y)$ in (\ref{Nphi}),
the infinite series (\ref{defF43}) for this
particular case can terminate \cite{PRSnpb}. 

Then
when the left hand sides of (\ref{n2appendix})
contain
\bea
\Phi_2^{(-1)},
\qquad
\Phi^{(-1),-}_{\frac{3}{2}},
\label{exception}
\eea
the following
(anti)commutators do not have any simple forms with the known
structure constants
\bea
&& \comm{(\Phi^{(-1)}_2)_{m}}{(\Phi^{(h)}_2)_{n}}\,,
\qquad
\comm{(\Phi^{(-1)}_2)_{m}}{(\Phi^{(h)}_1)_{n}}\,,
\qquad
\qquad
\acomm{(\Phi^{(-1),-}_\frac{3}{2})_{r}}{(\Phi^{(h),+}_\frac{3}{2})_{s}}\,,
\nonu \\
&& \comm{(\Phi^{(-1)}_2)_{m}}{(\Phi^{(h),\pm}_\frac{3}{2})_{r}}\,,
\qquad
\comm{(\Phi^{(h)}_2)_{m}}{(\Phi^{(-1),\pm}_\frac{3}{2})_{r}}\,,
\qquad
\comm{(\Phi^{(h)}_1)_{m}}{(\Phi^{(-1),-}_\frac{3}{2})_{r}}\,,
\label{finalequation}
\eea
where the modes for above generators
(\ref{exception}) appear.
These (\ref{finalequation}) can be obtained from the previous results in
(\ref{n2appendix}).


\end{document}